\documentclass[doublespace,nopageskip]{UVAthesis} 

\usepackage{lipsum}
\usepackage{amsmath}
\usepackage{wrapfig}
\usepackage{graphicx}
\DeclareUnicodeCharacter{0302}{}

\title{A Novel Computational Thermodynamics Framework with Intrinsic Chemical Short-Range Order}

\author{Chuliang Fu}

\homeaddress{Eureka Springs, Kansas}

\firstpriordegree{Bachelor of Science}

\firstpriorinstitution{Yunnan University}

\firstpriordate{2017}

\program{Materials Science} 

\degree{Doctor of Philosophy} 

\thesis{Dissertation}

\graddate{August 2023} 

\dedication{Dedicated to all ordinary people with Caritas and Mathēmatikós to help the others and solve the challenge.}

\acknowledge{
First of all, I would like to thank my dissertation supervisor, Professor Bi-Cheng Zhou, for his support throughout my research project. His knowledge and advice have been invaluable in shaping the direction of my work and ensuring academic rigor in materials science.
\\I am greatly indebted to all the dissertation committee members in our university: Professor Sean Agnew, Professor Leonid Zhigilei, Professor Eugene Kolomeisky, and Professor Tao Sun. I could still remember Professor Agnew's claim about “Materials Science IS Defects” in his class. I gradually understood more about this after these years. I strengthened my understanding of computational materials science in Professor Zhigilei's class and try to combine the theoretical physicists' understanding of statistical physics with the help of Professor Kolomeisky's course. Professor Tao Sun kindly helped and guided me a lot from the experimental perspective, though I have not yet done any experiments. 
\\I want to thank Professors Soffa, Zangari, and Floro who patiently provided me with confidence in this area when I entered the MSE department. Before that, I had never learned materials science seriously. I want to express my highest respect to Professor Soffa. I could still remember what Professor Soffa told me such as, \textit{``Did you learn something from the exam?"} and \textit{``You are my mathematician."}. That really impressed me. Professor Floro has always been very kind addressing my any request, when I met difficulty, and I really want to express my gratitude for his kindness. Professor Zangari taught me thermodynamics, which is the center of this dissertation. I would also thank Professor Opila for her constant interest in my current work and for always proposing interesting questions during my qualifying exam and my proposal defense.
\\Many professors from other departments taught me a lot: They are Professor Leonid Petrov from the mathematics department, Professors Jundong Li, Tom Fletcher, and Mona Zebarjadi from the electrical engineering department, and Professor Mohammad Mahmoody from the computer science department. It's their efforts in class which gave me wider scope and help me see the wider connections in the world from a different perspective.
\\It’s my labmates  that makes me stronger to face the challenge in daily life, they are: Kang Wang, Du Cheng, Yuanchen Gao, Rahim Zaman in our group; and many friends sharing the same 303 ``fish bowl" office: Hao Huang, Shaoning Geng, Lingyu Guo, Miao He, Chaobo Chen, Yuan Xu, Yunkai Sun, Maxim Shugaev, Mikhail Arefev, and Mukil Ayyasamy. I also want to thank many people including but not limited to Qiyuan Lin,
Naiming Liu, Shelby Fields, Rebecca Skelton Marshall, Mark Wischhusen, 
Adrian Savovici, and
Christian Walters as well as many others, including the staff in our department.
\\My friends have accompanied me from a very young age, and we now have more than ten years of friendship, I am thanking of Mengzhuo You and Zhuoli Jin in California, Hong Shen in Shanghai, and Libin Pan in Hangzhou. Luckily, even after more than ten years or even twenty years, we are still taking care of each other. Their support is very precious to me as I walk along this step onto a very special journey.
\\I want to thank my roommates during these years including Peng Wang, Song Wang, Zheng Huang, Junwu Bai, Ruoyu Zhang, Beichen Wang, Xiaobao Zhang, Junchao Zhu, Xinjian Liu, and Jingze Diao. When I experienced a panic attack in the emergency room, the doctor mentioned that, \textit{``you really have many supportive relationships"}. I think I do have this with their efforts. I also want to thank Shuai Li, Qianhong Zhao, Shuman Sun, and Tingyang Meng for their support during daily life.
\\I would like to share my respect for my students, including Hongzhou Jiang, who is the most brilliant student I have ever met during my teaching internship before I came here for my PhD degree. Their warm welcome really inspired me again when I did a teaching internship in their middle school and planned to be a math teacher for middle school students. But finally, I applied to graduate school and constantly heard the good news in their life during these years. I wish them all the best.
\\I want to thank my friend Yushun Dong. Since Yushun is the actual vocalist in my daily life, it motivated and encouraged me a lot to help me discover my talents in singing and accept my own voice. He really treats the singing very seriously, and this attitude activated me. I want to thank Dr. Jing Ma, the TA who helped me the most. You really encouraged me and made me stronger to reach inner peace.
\\I shared an excellent summer with many other friends when I visited the quantum measurement group last summer: Zhantao Chen, Ryotaro Okabe, Tongtong Liu,  Abhijatmedhi Chotrattanapituk, Thanh Nguyen, Manasi Mandal, Hatice Zeynep Gokturk, Xiang Fu, etc. It was really an exciting journey with their help and Boston is really a beautiful city.
\\I also want to mention some other names who have supported me including Qi Zhang, Hanqi Wen, Rebecca Little, Yirui Gui, etc.
\\Many doctors have supported me during these years. I want to thank Dr. Baozhen Xie from CAPS for taking care of my mental health and all my primary care doctors from UVA student health, including Tracy Buni, Margaret Zayas, Shawn Pennetti, Drs. Charles Friel (UVA hospital digestive department),  James Shorten(UVA Foot and Ankle Center), Chukwueloka Obionwu(MIT urgent care), and Kamau Karanja(MIT medical), and all my dentists from Charlottesville dental care. 
\\I am here to express my highest gratitude to Zhou Shen, Charlie Zhou, who is an exceptional and brilliant singer, invariably with positive emotion and optimism. His \textit{Big Fish} would always be my relaxing choice.
\\I would like to acknowledge many big names who inspired me, such as Professor Chen-Ning Yang. His master thesis motivates this work and I have learned a lot from his literature about the taste of physics. I am also influenced a lot by two great applied mathematicians, Professor Chia-Chiao Lin, and Professor Weinan E. Their great work help me decide the direction of my life: to understand and explore the real world with applied mathematics in hand.
\\I would also like to express my special gratitude to Professor Mingda Li. I would have become a secondary school teacher without you, that really would also be a good life journey but not as amazing as the one I am now pursuing. I guess I really got the luck the Doraemon gave me and enjoy all I have.
\\Finally, I want to share my gratitude to My family, particularly my mom and dad. Thank you for all your support in helping me really grow up.}

\abstract{Exploiting chemical short-range order (SRO) is a promising new avenue for manipulating the properties of alloys. However, existing computational thermodynamic modeling frameworks do not account for the possibility of chemical SRO in multicomponent ($\geq 3$) alloys. CALPHAD is a leading method for modeling and calculating phase equilibria in materials. Still, the prevailing solution model used in CALPHAD, the sublattice model, is an empirical mean-field model based on Bragg-Williams (ideal entropy of mixing) approximation. This makes CALPHAD inadequate for properly describing order-disorder transformations or chemical SRO in alloys. First-principles calculations of phase diagrams, using the cluster variation method(CVM), or cluster expansion method(CEM), can describe SRO but are generally limited to binary or ternary systems due to the large number of configuration variables. 
Here, we propose to develop a hybrid framework by marrying the unique advantages of CVM and CALPHAD by incorporating chemical SRO into CALPHAD using a cluster-based solution model. The most crucial technique is the Fowler-Yang-Li transform which can decompose the cumbersome cluster probabilities in CVM into fewer site/point probabilities of the basis cluster, thereby considerably reducing the number of variables that must be minimized for multicomponent ($\geq 3$) systems. Modern, efficient algorithms are employed to minimize the non-linear cluster-based free energy functions. Prototype phase diagrams of the fcc AB binary system have been calculated as benchmark tests for different models. The phase diagram calculated from FYL-CVM possesses the same topology as that obtained from CVM and has a good balance of accuracy and efficiency. We also included free energy contributions from vibrational, elastic, and electronic contributions using reduced-order models. We observed that all the physical contributions clearly influence the order-disorder phase boundaries.
We have also implemented our current method in the Cu-Au, a typical fcc ordering system, to test its effectiveness in real materials. We created a workflow to determine the cluster energies for the Cu-Au system. The result is a new set of parameters that are regularized for the CVM-CALPHAD modeling. The calculated phase diagram agrees well with experimental phase boundary data. The importance of including non-configurational contributions to free energy  (vibrational, elastic, and electronic contributions) is revealed and satisfies the physics of the real system. The SRO parameters in the Cu-Au system based on the tetrahedron cluster probability are visualized in the composition-temperature space. Similarly, we have generalized the algorithms of SRO parameters to any multicomponent($\geq 3$) alloys in the disordered phase and present the demonstrated results for the Cu-Au-Ag system. In summary, our proposed CVM-CALPHAD modeling framework enables the chemical SRO to be exploited for systems exhibiting order-disorder transformations of the solid solution using a novel cluster-based model, which balances both accuracy and efficiency and incorporates more physics into CALPHAD.}

\begin{document}

  \frontmatter
  \maketitle
  \tableofcontents

	\listoffigures
	\listoftables

 
\nomenclature{NLP}{Natural Language Processing}

\nomenclature{$M_\odot$}{The mass of the Sun ($1.989 \times 10^{30}$ kg)}

	\mainmatter

\chapter{Introduction} \label{ch:introduction}
 
    \section{Motivation} \label{se:background}
  In recent years, the metallurgy community has seen a blossoming in the research field of high entropy alloys(HEAs) or complex concentrated alloys (CCAs) \autocite{miracle2017critical,george2019high}. In recent studies, CCAs have demonstrated excellent performance in mechanical properties \cite{george2020high}, radiation resistance \cite{el2019outstanding}, thermoelectricity \cite{jiang2021high}, corrosion resistance \cite{scully2020controlling}, etc. These excellent materials' performance would make CCAs one of the most essential materials for the future sustainable society. On the other hand, CCAs bring immense opportunities for property tuning due to their vast composition space. This huge composition space of CCAs brings the materials design community opportunities and challenges.

As some researchers limit their CCAs restrict to a single-phase random solid solution, but recent studies have found that order-disorder transformations \autocite{liang2018high,miracle2020refractory} and chemical short-range order (SRO) \autocite{ma2018chemical,singh2015atomic} are common in CCAs. The terms order and disorder here are not in a structural sense (amorphous vs. crystalline) but are used to describe local chemical interactions/arrangements. 
  
  It is non-trivial to experimentally characterize SRO in CCAs as SRO is an extremely local phenomenon (up to a few nearest-neighbor shells), but recently Zhang et al. \autocite{zhang2020short} verified the existence of chemical SRO in a medium-entropy NiCoCr alloy directly through energy-filtered transmission electron microscopy. CCAs also possess unusual dislocation behavior \cite{ma2020unusual}, as the lattice friction resulting from composition inhomogeneity causes dislocation pile-ups. As it has also been reported that SRO can affect the stacking-fault energy and dislocation mobility \autocite{ding2019tuning} in CCAs, it would be a major factor in controlling mechanical properties. Therefore, exciting opportunities exist in exploiting SRO for both phase stability and property enhancement in the context of CCAs. Based on all those observations mentioned above, we can conclude that: with the CCAs coming into the research spotlight, atomic-scale order is being realized by the materials community as a new dimension for property manipulation \cite{nohring2020design}. However, sufficient tools for reliably predicting SRO still do not exist. As the SRO will become a new degree of freedom for alloy processing and design, the science of CCAs and SRO presents such new challenges for the alloy theory community.
     \begin{figure}
    \centering
\includegraphics[width=0.7\textwidth]{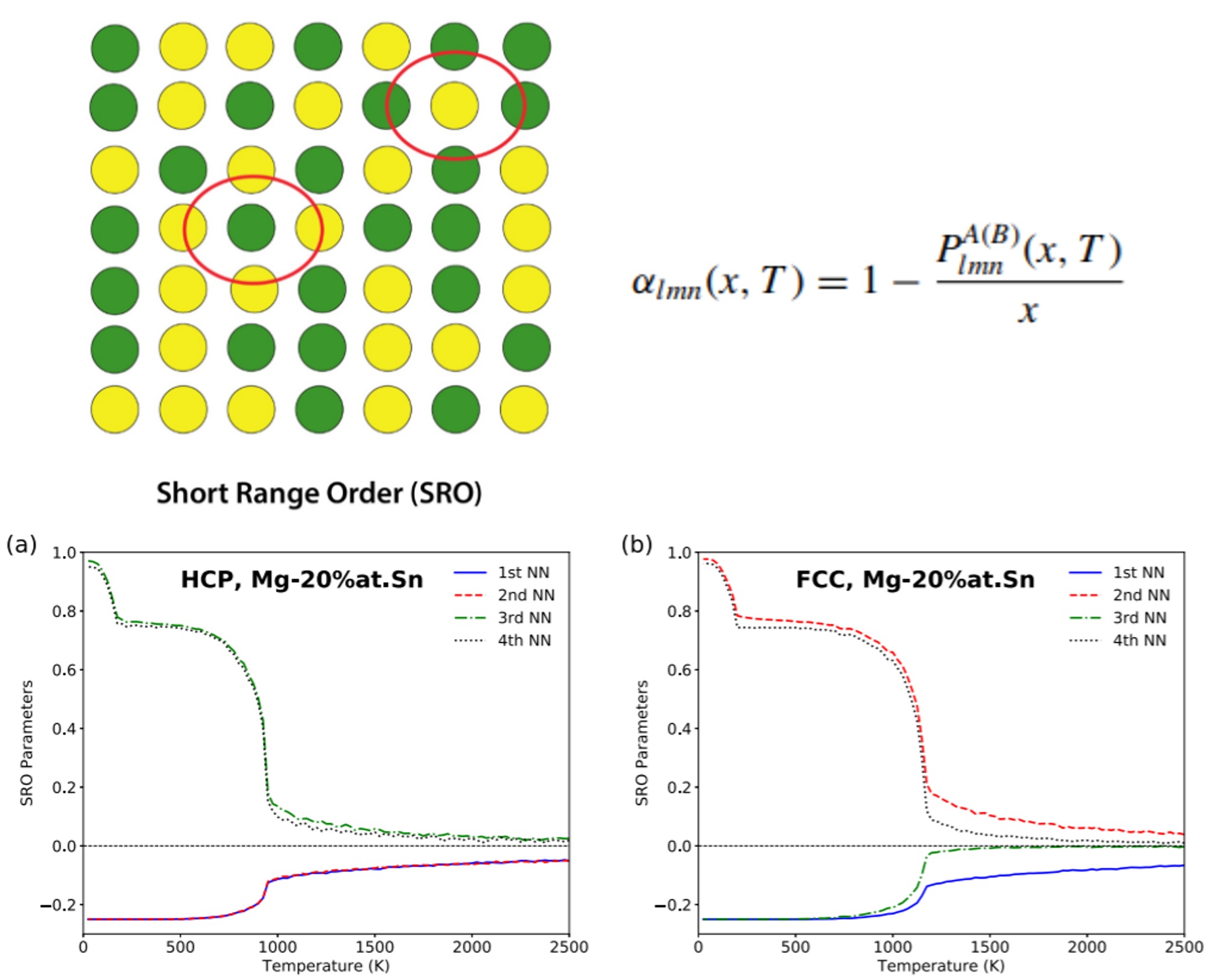}
    \caption{The illustration figure for chemical SRO. The SRO v.s. temperature curves are from \cite{wang2020generalizing}.}
    \label{fig:SRO}
\end{figure}
  
Alloy thermodynamics plays a central role in understanding phase stability and phase transformations in materials which would be necessary to understand the SRO.  Over the years, the field of computational thermodynamics has seen tremendous progress \cite{van2013methods}. Many different statistical-mechanics-based methods, such as the CVM \cite{de1979configurational,kikuchi1951theory,mohri2017cluster} and CEM \cite{de1994cluster} were developed which allows the in silico prediction of configurational thermodynamics of alloys. However, these methods are generally limited to binary or ternary systems due to many configuration variables. In addition, phase diagrams calculated by these methods are not considered quantitatively accurate enough for technological applications because the chemical accuracy of first-principles inputs limits them \cite{aldegunde2016quantifying}. Known for its simplicity and efficiency, the CALculation of PHAse Diagram (CALPHAD) method is a leading phenomenological method for thermodynamic modeling and calculations of complex phase equilibria in materials \cite{sundman2007computational,liu2016computational}. Instead of relying only on first-principles inputs, experimental thermochemical and phase equilibrium data are also often combined with theoretical data to parameterize a self-consistent thermodynamic model in CALPHAD modeling. Thermodynamic databases of realistic multicomponent alloys can thus be established. These databases are extremely valuable because they can be used to predict phase equilibria and plot complex phase diagrams and serve as inputs for kinetic models such as the phase-field model. 
 
However, the current solid solution model in CALPHAD, the sublattice model, is inadequate for properly describing order-disorder transformations or chemical SRO in alloys. The sublattice model only involves the long-range order within the sublattice but cannot describe the SRO without the unphysical parameters. It is thus highly desirable to directly use a solution model with intrinsic SRO for CALPHAD. In the history of alloy theory, many such models have been developed. The quasi-chemical model \cite{blander1987thermodynamic} and the subsequent modified quasi-chemical model \cite{pelton2000modified}, developed by Blander, Pelton, and co-workers, consider the pair interactions but ignore other types of short-range interactions. Yang \cite{yang1945generalization,yang1947general} and Li \cite{li1949quasi,li1949quasi2} derived a generalized quasi-chemical model of order in superlattices (the Yang-Li approximation). CVM, a direct approximation method to the free energy pioneered by Kikuchi \cite{kikuchi1951theory}, captures the ordering structures by the cluster formalism. However, the equilibrium minimization in CVM causes a substantial computational burden, especially for multicomponent systems. In recent years, cluster expansion coupled with the Monte Carlo method \cite{van2013methods,de1994cluster} has been widely adopted to calculate phase diagrams for ordering systems. However, this method relies on first-principle data and cannot be combined with experimental data to produce quantitatively accurate phase diagrams. Meanwhile, the wide use of correlation functions in cluster expansions has overshadowed CVM approximations based on the explicit formulation of the linear equations relating different cluster configurations because the intricacy of these constraints renders such approximations cumbersome to formulate for large clusters \cite{balabin2000thermodynamics}. The Yang-Li approximation \cite{yang1945generalization,yang1947general,li1949quasi,li1949quasi2} was later revived by Oates to produce the cluster/site approximation (CSA) \cite{oates1996cluster,oates1999improved}. The key advantage of CSA compared to CVM is that the number of minimizing variables to approach equilibrium is dramatically reduced. This feature makes CSA a candidate for the solution model suitable for CALPHAD. However, the Yang-Li approximation has a major assumption that the selected clusters must be of non-interference, which leads to significant inaccuracy \cite{oates1996cluster}. To improve CSA’s accuracy, Oates inserted an adjustable parameter into the CSA entropy formula without clear physical meaning to mimic the contribution of the cluster interactions \cite{oates1999improved}. Recently, van de Walle et al. \cite{van2017software} proposed an SRO correction in the context of high-throughput CALPHAD using truncated polynomial expansions of the CVM pairwise entropy, to make full use of the high-throughput results of special quasirandom structures \cite{zunger1990special}.

It must be emphasized that SRO plays a significant role in relative phase equilibrium, although its contribution to the alloy’s total free energy difference is usually small compared with differences caused by the crystal structure change \cite{sigli1985theoretical}. Equilibria between two different crystal structures usually involve phases that are far from the ordering transition on their respective sublattices. Thus, SRO effects are commonly incorporated into the sublattice models using a phenomenological expansion of the excess free energy with reciprocal interaction parameters \cite{sundman1998thermodynamic}. This empirical approach to the description of configurational free energy has serious limitations. The most significant one is that the configurational entropy cannot be properly approximated by a polynomial expansion over extended temperature and composition ranges \cite{sigli1985theoretical}. Whereas using a cluster-based solution model connects CALPHAD with the mainstream statistical mechanics and naturally accounts for SRO in CALPHAD. Moreover, suppose a statistical-mechanics-based model is to be used to describe the configurational contribution to free energy. In that case, the non-configurational contributions to the free energy (vibrational and elastic) also needs to be considered \cite{de1979configurational}. There is a pressing need to move from the traditional mean-field model to more statistical-mechanics-based models for CALPHAD, especially in the context of the emerging CCAs and the observation of SRO in them. The current two main theoretical challenges for achieving this goal are:

1)	Computational efficiency of some advanced models such as CVM is relatively low to be directly used in CALPHAD modeling, while another first-principles-based method such as CEM would even be limited due to both chemical accuracy of the first-principles input and the computing power. 

2)	The non-configurational (vibrational, elastic, electronic) contributions to the free energy must be included computationally inexpensively. Due to its high cost, even vibrational entropy is rarely considered in standard CVM or cluster expansion procedures.

    \section{Contributions}
To deal with the challenge we mentioned, we push our efforts on the theoretical and computational directions to all the principles, data, and algorithm sides.

From the principles side, we reorganize the theoretical foundation for the cluster-based thermodynamics models: CVM \cite{kikuchi1951theory}, CSA \cite{oates1996cluster} with a uniformed framework, and construct the model combined the positive sides of both models. The model is constructed based on the grand canonical ensembles for the basic cluster, then combine with the CVM-introduced hierarchical free energy expansion structure to determine the thermodynamics quantities for the sub-cluster. After constructing the unified theory framework, we impose the Fowler-Yang-Li transform \cite{yang1945generalization,yang1947general,li1949quasi,li1949quasi2}, the key technique to reduce the number of variables of the CVM free energy functional to construct a novel cluster-based thermodynamic framework with intrinsic chemical SRO. This framework would have the big advantage to balance both computational accuracy and efficiency. With the help of a reconstructed unified model, further physics is introduced naturally, such as the vibrational contribution, elastic contribution, and electronic contribution. 

From the data side, this motivates us to reconsider CALPHAD in the AI era. CALPHAD could be considered as a machine learning(ML) task specialized for thermodynamics, and it should be renamed as DataMining4Thermodynamics or even AI4Thermodynamics under the AI era. Then what we do is exactly to find out a proper learning model to better fit the multi-modal \cite{baltruvsaitis2018multimodal} thermodynamic data for the solid solution with the intrinsic chemical SRO. Besides, as we introduced a cluster-based model, this also introduced a new type of thermodynamic data, cluster energy. this cluster energy is defined through our developed thermodynamic model and a workflow to estimate the cluster energy is constructed by combining the first-principles calculation to benefit CALPHAD modeling.

From the algorithm side, we deal with several challenges, such as equilibrium calculation, phase boundary tracking, and parameter optimization. The contribution is mainly for the equilibrium calculation. Due to its hierarchical structure, the thermodynamic function here for (FYL-)CVM is non-convex. This non-convexity is a challenge, as the searching algorithm is easy to track into local minima/saddle points and cannot reach the global minima from the theoretical computer science perspective. As a result, the equilibrium calculation here in this dissertation is reconsidered under the optimization perspective but not just a problem about solving non-linear equations. The previous natural iteration method \cite{kikuchi1974superposition} is essentially a fixed-point scheme, while the constantly used Newton–Raphson method with the correlation function as the variables is a second-order derivative method. We designed the accelerated/stochastic gradient descent method to deal with this minimization for FYL-CVM, which is acceptable with the random initial value and would provide insightful observation of the free energy minimization during the heat process. This gradient descent method only requires the first-order gradient calculation and could be naturally matched with GPU-accelerated differentiable computing, such as JAX \cite{jax2018github}. The phase boundary tracking could be considered as the enhanced sampling for the ``rare event". We have tested the binary search and would try to combine the thermodynamic relation along the phase boundary in the chemical potential to enhance the efficiency of the sampling. Parameter optimization is a challenge for the cluster-based model. Here we would point out that there exists a bilevel optimization \cite{colson2007overview} for the parameter optimization: the inside optimization is the equilibrium minimization. In contrast, the outside optimization is the optimization for the phase boundaries data and thermochemical data to reach the optimized cluster energy. The cluster energy would cause the ``entanglement" between the inside and outer optimization. This is related to the challenge of meta-learning for the hyper-parameter within the ML community \cite{franceschi2018bilevel}. 

Combined with all the progress from the fundamental principle, the data, and the algorithms, we apply the developed computational thermodynamic framework onto a typical FCC ordering system, Cu-Au system to make a thermodynamic assessment. To achieve the realistic calculation, we also develop the workflow to combine the first-principles calculation data into the framework, which could be automatically performed in the future with the workflow automation tool.

\section{Dissertation Outline} \label{Outline}
In Chapter 2, the background is introduced for the essential techniques. Chapter 3 derives the whole theoretical framework for the configurational contribution as the foundation to introduce the chemical SRO. In Chapter 4, the non-configurational contribution will be introduced into the framework. The vibrational contribution and electronic contribution would be introduced through the coarse graining formalism. The elastic contribution would be considered separately.

    We would deal with the algorithms and workflow in Chapter 5 to achieve all these fundamental sets up. Three different types of algorithms are to be discussed: equilibrium minimization calculation, phase boundary detection, and parameter optimization. We would also discuss the constructed workflow for determining cluster energy based on first-principles calculations. In Chapter 6, we applied all the computational frameworks onto one typical FCC ordering system Cu-Au system and list the future directions, and summarize all the work in Chapter 7 at the end.
    
    \chapter{Background} \label{ch:lit_review}
    \section{CALPHAD}
    \subsection{Current CALPHAD: First-Principles Data and Models}

    CALPHAD stands for CALculation of PHAse Diagrams, a methodology introduced in 1970 by Larry Kaufman is perhaps one of the earliest methods integrating physical modeling and learning from data to digitize the thermodynamics of materials with phase diagrams, which is the graphical representations of the distributions of the phases under the different external conditions \autocite{kaufman1970computer}. There are several excellent review articles for CALPHAD already \cite{sundman2007computational,liu2016computational}, but here is a brief review of CALPHAD in history to make the whole dissertation self-contained and coherent. The CALPHAD workflow can be roughly divided into the following sections: data capture, construction of the thermodynamic model, optimization of the model through updating the undetermined parameters, database generation, and application to many cases such as phase stability prediction \autocite{hubbard1996thermodynamic,liu2001thermodynamics}, phase-field modeling \autocite{zhu2002linking,wu2004simulating}, precipitation simulation \autocite{xia2016precipitation,PERRON201416}, etc. 

    From the data perspective, CALPHAD makes use of the experimental data while recently, more and more first-principles data have come to help to provide accurate thermochemical prediction \cite{liu2009first}. First-principles calculations based on the density functional theory(DFT) \cite{kohn1965self} were taken to predict the finite-temperature thermodynamic properties of metallic compound phases. It usually involves the static energy, the vibrational contribution coming from phonon, and the electronic free energy:
\begin{equation}
\begin{split}
    F(V,T) = E_0(V) + F_{vib}(V,T) + F_{ele}(V,T)
    \end{split}
    \label{eq:finite}
\end{equation}
where $E_0$ represents the energy at 0K, which could be determined through DFT calculation. $F_{vib}$ is the vibrational free energy mostly could be determined based on quasi-harmonic approximation \cite{wang2004thermodynamic,shang2010first}. $F_{ele}$ is the electronic free energy.  It takes the temperature dependence of volume into consideration by
performing the harmonic approximation at volumes varying from the equilibrium volume of the
system. The non-harmonic nature of the potential energy is taken into account by extrapolating
these harmonic contributions at different volumes to become volume dependence. With the help of force
constant approach under this quasi-harmonic approximation, the Helmholtz energy of the metallic compound system is now
described with an additional volume dependence as
\begin{equation}
\begin{split}
    F_{vib}(V,T) = k_B T \int_{0}^{\infty} ln\left[2sinh\left(\frac{\hbar\omega}{2k_B T}\right)\right]g(\omega,V)d\omega
    \end{split}
    \label{eq:phonon}
\end{equation}
where
$\omega$
represents the phonon frequency and $g(\omega,V)$ is the phonon DOS at the $\omega$ and volume $V$.
$F_{ele}$ is the electronic free energy for a metallic compound, which could be determined with the DFT calculated density of states:
\begin{equation}
\begin{split}
    F_{ele}(V,T) = E_{ele} -TS_{ele}
    \end{split}
    \label{eq:ele}
\end{equation}
while the energy $E_{ele}$ and $S_{ele}$ can both be expressed as an integral format with a density of states:
\begin{equation}
\begin{split}
    E_{ele} = \int^{+\infty}_{-\infty} n(\epsilon,V)f(\epsilon,T)    \epsilon d\epsilon + \int^{\mu}_{-\infty} n(\epsilon,V)   \epsilon d\epsilon
    \end{split}
    \label{eq:elee}
\end{equation}
where $\mu$ is the chemical potential, $n(\epsilon,V)$ is the electronic density at energy $\epsilon$ and volume $V$, $f$ is the Fermi-Dirac distribution.
\begin{equation}
\begin{split}
    S_{ele} = -k_B \int n(\epsilon,V) \left[f(\epsilon,T)lnf(\epsilon,T)+(1-f(\epsilon,T))ln(1-f(\epsilon,T))\right] d\epsilon
    \end{split}
    \label{eq:eles}
\end{equation}

This formalism could help determine the thermochemical properties of most single elements and metallic compound compounds. For a solid solution, it could be possible to gain some calculated DFT energy as input based on special quasirandom structures \cite{zunger1990special}. Besides, the first-principles calculation is not limited to DFT. For example, \textit{ab initio} molecular dynamics \cite{car1985unified} could come to determine the thermodynamics for solid, glass, and liquid phase \cite{jiang2020ab}.

After collecting the data from the first-principles calculation and experiments, the next step is determining the thermodynamic model. The main value of the thermodynamic model is that once the Gibbs energy description
has been optimized for each phase in the system, the functions can be extrapolated and generalized into other thermodynamic conditions or other systems where experimental data does not exist. Since the thermodynamics quantities are correlated with each other through some theoretical transformation, we could select Gibbs free energy as the main function to be optimized with data. In the CALPHAD community, the Gibbs energy is often refined to be expressed in the following
temperature-dependent polynomial:
\begin{equation}
\begin{split}
G-H^{SER} =  a+bT+cTlnT+dT^2+eT^{-1} 
\end{split}
\label{eq:ser}
\end{equation}
where $a$, $b$, $c$, $d$, and $e$ are model parameters evaluated in the parameter optimization with the temperature dependent free energy data. The left hand side of the equation refers to that the Gibbs energy is defined with respect to a standard element reference
state (SER) which is defined as the stable structure at 298.15 K and 1 atm. This is determined in the SGTE pure elements database \cite{dinsdale1991sgte}. This model with only several empirical temperature-dependent terms is widely used for the metallic compound.
For the solid solution phase,  people use the sublattice model or say, compound energy formalism \autocite{hillert2001compound}. The essence of this sublattice model is the Bragg-Williams approximation \cite{bragg1934effect}, which makes use of the ideal mixing for the configurational entropy. If we take the binary system as an example, the molar Gibbs energy of a solution phase of atoms A and B under the sublattice model is
given by:
\begin{equation}
\begin{split}
G_m^{\phi} = x_A G^{\phi}_A + x_B G^{\phi}_B +RT(x_Alnx_A + x_Blnx_B) + G^{\phi}_{xs}
\end{split}
\label{eq:sbl}
\end{equation}
where $x_A$ and $x_B$ are the mole fractions of $A$ and $B$, respectively,
$G^{\phi}_A$ and
$G^{\phi}_B$ are the Gibbs
energies of pure A and pure B in the structure $\phi$, respectively, and $G^{\phi}_{xs}$ is the excess Gibbs energy. The $RT(x_Alnx_A + x_Blnx_B)$ reflects the ideal mixing of the configurational entropy. The excess Gibbs energy is modeled with a Redlich-Kister polynomial to make the correction \cite{redlich1948algebraic}:
\begin{equation}
\begin{split}
G^{\phi}_{xs} = x_Ax_B \sum_{k=0}^{K} L^{\phi,k}_{A,B}(x_A-x_B)^{k}
\end{split}
\label{eq:rk}
\end{equation}
where $K$ is the order of this polynomial determined case by case, $L^{\phi,k}_{A,B}$ represents the non-ideal interactions between A and B and is usually defined with a
linear temperature dependence:
\begin{equation}
\begin{split}
L^{\phi,k}_{A,B} = A^{\phi,k} + B^{\phi,k} T
\end{split}
\label{eq:rk2}
\end{equation}
where $A^{\phi,k}$ and $B^{\phi,k}$  are model parameters to be evaluated. Since it takes Bragg-Williams approximation as the key component, one of the key issues is that the sublattice model can only describe the ideal mixing properly without SRO. People are trying to incorporate SRO effects into the sublattice models by means of a phenomenological expansion
of the excess free energy with reciprocal interaction parameters \cite{sundman1998thermodynamic,sundman2018review}. However, the reciprocal interaction parameters are implemented artificially and would lead to the inconsistency between the data and the thermodynamic model.  

The thermodynamic model for the liquid phase meets a similar challenge in describing the SRO. To account for the SRO in the liquid phase, the associated solution model is developed by assuming some hypothetical ``molecular" existed in the liquid phase to represent the SRO. A better model that captures the SRO in the liquid phase is the modified quasi-chemical model \cite{pelton2000modified,pelton2001modified,chartrand2001modified}. But both models are relatively simple to reflect the more complex structure of the liquid phase.

With the help of the first-principles calculation, available experimental data, developed thermodynamic model, and the corresponding algorithms, CALPHAD got much progress in the development of the materials database during the past 50 years. However, we consider current CALPHAD still has several challenges: First, lack of high-quality data. As we have mentioned, the current state-of-the-art CALPHAD uses experimental data for phase equilibrium information and first-principles data for thermochemical information. Performing first-principles calculations for thermochemical data is convenient but still requires lots of phase equilibrium data from the experiments, which requires lots of effort. Second, the commonly used thermodynamic model is simple but not reflects the physics. The widely used thermodynamic model has a simple mathematical form, as the CALPHAD requires a speedy computational response. However, the parameters of the thermodynamic models don't reflect much physical meaning but a mathematical construction. Third, due to the possible inconsistency caused by the multiple data sources, it’s hard to determine the optimized modeling automatically but with lots of fine-tuning work.

    \subsection{CALPHAD in AI Era}
    
   While huge progress in many scientific disciplinaries is made by AI, especially the neural-network (NN) based ML recently, it’s especially effective for the multiscale phenomena with the curse of dimensionality \autocite{weinan2020integrating} such as protein folding \autocite{jumper2021highly}, fluid mechanics \autocite{brunton2020machine} and quantum many-body simulations \autocite{carleo2017solving}. Though thermodynamics is typically not considered as a multiscale problem, its complexity still comes from the massive amount of the degree of freedom. 
   In this section, we want to mention several thoughts about the connection between CALPHAD and AI and provide insight into how AI would shape CALPHAD. In particular, we point out CALPHAD is mathematically equivalent to a specialized supervised ML task. 

  \begin{figure}
    \centering
\includegraphics[width=1\textwidth]{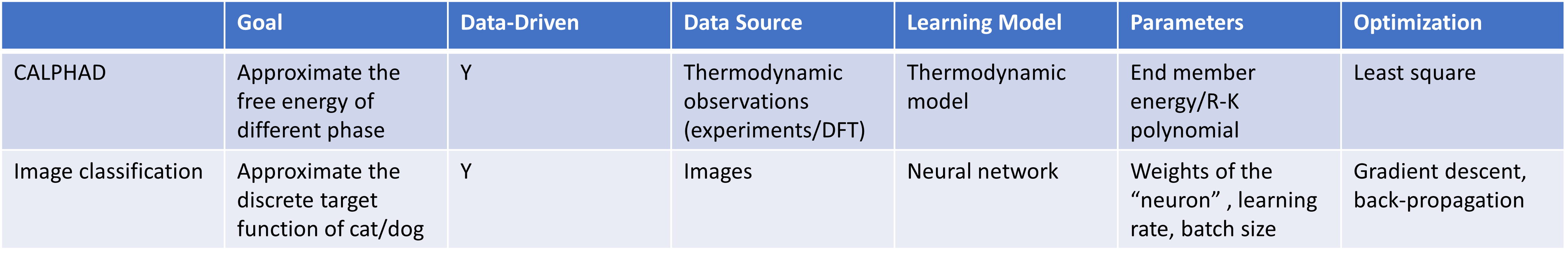}
    \caption{Comparison between CALPHAD and image classification, a typical supervised learning task in computer vision.}
    \label{fig:compare}
\end{figure}
   The target of CALPHAD could be roughly summarized as to approximate several thermodynamic functions: $F_p(x,T)$ for $p$ different phases, where $x$ is the composition and $T$ is the temperature. Of course, you could generalize these functions under different thermodynamic conditions and different thermodynamic potentials. To better approximate the thermodynamic functions, we could make use of some specific parameterized thermodynamic model to formulate the corresponding function $F_p(x,T)$, then make use of experimental data or first-principles calculated data based on the observation of phase equilibrium state and thermochemical properties to determine these parameters. These data actually indicates the observed value of thermodynamic function $D\left[F_p(x_i,T_i)\right] = c_i$ to help to determine the parameters, where $D$ is the defined cost function to transfer the free energy into the phase boundary data or thermochemical data. The final step is to determine the parameters based on the data, which could be considered as an optimization problem.

    Mathematically, supervised learning could be formulated like this: given the dataset $S=\{x_j, y_j = f^{*}(x_j)\}$, and the target is to approximate $f^*$ for the dataset's represented features/phenomena as accurately as we can with the help of dataset \cite{ma2020towards,weinan2020integrating}. If $f^*$ takes continuous values, this is called a regression problem. If $f^*$
takes discrete values. This is called a classification problem.

Based on this comparison, we notice that CALPHAD is exactly a regression problem of supervised learning. The thermodynamic model with the undetermined parameters is precisely the $f^*$ in supervised learning. The dataset of CALPHAD is these first principles calculated data and experimental measured data for better observing the thermodynamic properties of the specific materials system. The goal is to approximate an accurate model $f^*$ to describe the thermodynamics of this materials system completely. As a result, CALPHAD could be smoothly mapping to a regressive supervised learning problem, making CALPHAD closer to DataMining4Thermodynamics. One comparison between CALPHAD and image classification is presented in figure \ref{fig:compare}, here image classification is one of the most typical supervised machine learning tasks.

   On the other hand, there are several unique points for CALPHAD compared to many other common supervised learning in the computer science community. 

1. It's a few-shot learning. Few-shot learning means the ML with a small size dataset \cite{wang2020generalizing}. The size of the thermodynamic dataset is always limited, especially for multi-component systems. While CALPHAD could actually be considered as a data mining task, highly accurate data is always required for CALPHAD. However, the data is difficult to reach without the assistance from the experiments and high accurate first-principles calculations, especially for the multicomponent cases which construct the thermodynamic function at higher-dimensional space and require much more data \cite{chen2018database}. 
    
2. Multi-modal data input. 
In the ML community, the multi-modal usually refers to natural
language and visual signals \autocite{baltruvsaitis2018multimodal}. From CALPHAD side, it could be considered multi-modal learning, as many parts of different thermodynamic quantities could be considered as one ``modal" such as free energy, phase boundaries, entropy, activity, etc. We could organize the dataset for all of them from different experimental or computational techniques. They could be roughly separated into two parts: thermochemical data and phase boundary data. Since the source of thermodynamic data is different, one challenge for CALPHAD is coordinating the modeling to compromise different ``modals". But on the other side, it is not a challenge for CALPHAD to deal with such multi-modal learning since the different thermodynamic quantities could be closely connected to each other with the thermodynamic relation.

3. Multi-objective optimization. Since we have a multi-modal dataset, the parameter optimization is actually a multi-objective optimization with multiple objective functions. The critical issue is that the experimental data from different sources and first-principles data may be inconsistent to have some contradicting error bars, and minimizing all objective functions simultaneously becomes a challenge. Then this may require some trade-offs to balance these uncertainties. One possible way to address this issue is to perform the uncertainty quantification \cite{otis2017high,bocklund2019espei,ury2022generalized}.

4. The fundamental thermodynamic laws and phase rule constrains the modeling. This is a standard feature for all AI4Science problems. The related ML model has to satisfy the physics-related constraint. One such common constraint is the symmetry \cite{weinan2020integrating,batzner20223}, while here, the constraint is the thermodynamics.

As a result, two aspects could be improved a lot in the AI era, the data source and the thermodynamic model.

One of the critical challenges of the current CALPHAD is the lack of the necessary data, including thermochemical and phase boundary data. However, experimental data acquisition is always a complex process. As a result, the first-principles calculations based on DFT have been widely equipped to supplement the experimental data and reach the acceptable accuracy for thermochemical data \autocite{wang2004ab,liu2009first}. Nevertheless, required phase boundary data is still hard to be obtained directly from the first-principles calculations. The main challenge to calculate the phase boundaries is it requires high computational accuracy: an error of 1 meV/atom in Gibbs energy of a condensed matter phase may cause an error around 10 K in transition temperature \autocite{zhu2017efficient,dorner2018melting}.
The error of the current state-of-the-art phase boundary estimation fully based on DFT without ML is around 100K \autocite{zhu2017efficient,zhu2020performance} and is far away from replacing the experimental measurement to provide the predictions for technological purpose. One way to deal with this accuracy issue is to go beyond DFT to attempt more accurate first principles methods \autocite{sola2009melting,dorner2018melting,rang2019first}. However, it is usually accompanied by a substantial computational cost.
With the help of current NN-based ML, it is possible to tackle the \textit{ab initio} thermodynamic calculation by more efficiently constructing DFT functionals approaching the chemical accuracy, and several recent attempts have been done along this path \autocite{bogojeski2020quantum, chen2020deepks}. However, we need to consider solid solution \autocite{van2002automating} and liquid phase with the SRO \autocite{becker2014thermodynamic}, while in realistic experiments, anharmonicity \autocite{fultz2010vibrational} and defects \autocite{rogal2014perspectives} always exist in materials, only DFT-level electronic structure calculation cannot easily handle all these defective or complex structures and need the more extensive scale simulation. That’s exactly why we need the multiscale modeling to reach the required accuracy with ML as a tool to across the scale gap \autocite{weinan2020integrating}. One recent progress is the satisfied phase diagrams produced by DeePMD, a molecular dynamics simulation with the ML trained \textit{ab initio} potentials \autocite{zhang2018deep,niu2020ab}. It can be hypothesized that if we connect all the state-of-the-art computational methods at the different scales with the help of ML, it would have the chance to produce accurate enough high-throughput phase boundaries data based on first principles calculations and resolve the lack of experimental data for CALPHAD. 

CALPHAD needs easy-to-use to depict the stability of the different phases. Here the thermodynamic model means to develop a description for the free energy of a phase as a function of some degrees of freedom related to our interest. We consider the thermodynamic model within CALPHAD itself could be seen as a kind of specialized ML model. It is trained with a limited amount of the collected data to identify the correct phase stability, while CALPHAD modeling here could be seen as a supervised learning task. Recently CALPHAD community expects to impose more physics into CALPHAD and replace the empirical model to improve its robust \autocite{oates1996putting}. However, the current popular NN-based ML models 
usually require massive amounts of data, and that’s precisely one of the challenges of CALPHAD, lack of high-quality data. Even with a larger dataset, an NN-based model usually cannot consider physics constraints such as thermodynamic law and phase rule. Some recent work uses non-NN-based methods within the CALPHAD community but on specific cases such as the uncertainty quantifications \autocite{otis2017high,paulson2019quantified}. The rapidly developing few-shot learning  \autocite{10.1145/3386252} may be proper for this scenario, but the related attempt is still lacking. 

Some other topics are out of data and models in CALPHAD under the ML perspective. Some currently overlooked directions within CALPHAD, such as better optimization algorithms and better database management, may also become important, especially for multi-component systems in the future. On the other hand, we would consider implementing more physical learning models such as CVM to improve the compatibility between the data and the thermodynamic model for CALPHAD under AI era.

    \section{Cluster-Based Thermodynamic Model}
    \subsection{Kikuchi's CVM}
CVM is created to deal with cooperative phenomena at early 1950s by Ryoichi Kikuchi \cite{kikuchi1951theory,kikuchi1951theory2,kurata1953theory3}. The original motivation for Kikuchi to develop CVM perhaps might be that people had already noticed that the high order interaction or say correlation would be significant when quantifying the configurational entropy during the order-disorder transition. After Bragg-Williams' original work \cite{bragg1934effect} concerning the problem of order-disorder but before Kikuchi, Bethe \cite{bethe1935statistical} had already developed his pair approximation which could be considered as a pairwise version of the general CVM. Takagi \cite{takagi1941statistical} clarifies the relation between Bethe's work and the order-disorder transition in the binary alloy. As a result, Kikuchi's contribution is to continue along this path and take use of larger ``figure" or say basic cluster in our current understanding and clarify the combinatorial relation between basic clusters and sub-cluster's contributions and coupling this idea systematically into the free energy variational functional. 

Although Kikuchi had explored the hierarchical relation between the cluster and sub-cluster, it experienced a long time to understand fully. Later Morita \cite{morita1957cluster,morita1972general,morita1984consistent} presented the reformulation of the Kikuchi's original formula, and then Schlijper \cite{schlijper1983convergence} noticed that the formulation due to Morita is related to the Mobius inversion formula. An \cite{an1988note} revealed how the Mobius function plays an essential role in the formulation, while Morita \cite{morita1990cluster} rewrote his formulation after them. The key idea is to introduce the inclusion-exclusion principle with Mobius inversion formula \cite{rota1964foundations}. As a result, this mathematical hierarchy structure guarantees every part of the crystal would only be considered once and only once. These long-time clarification efforts better simplify the original CVM.

Since CVM is a variational model and the stable and metastable state correspond to the global and local minima of the functional, it also requires one convenient algorithm to reach these minima. It actually took people a long time to find it out since it could be clearly noticed that the CVM functional is non-convex \cite{pelizzola2005cluster}. One of the essential algorithms in developing CVM is the natural iteration method(NIM) \cite{kikuchi1974superposition}. During the development of CVM, people gradually noticed that the variables in CVM actually have some redundancy. An alternative way to deal with the variables is to transform it into correlation function formalism which could be loosely considered a linear transform between an arbitrary coordinate system into an orthogonal coordinate system \cite{sanchez1984generalized,de1994cluster}. After this transformation, the CVM community noticed that it's also OK to use the Newton-Raphson method to calculate the solution of these non-linear equations based on the first derivative of the CVM's free energy functional. However, it is reported that the Newton-Raphson method is fast but would have some convergence issues with the different assigned initial values \cite{mohri2013cluster}. Besides, introducing the correlation function formalism might be considered the origin of the widely used CEM \cite{de1994cluster}.

 CVM has a lot of applications. It is not only applied in the order-disorder transition of alloy system \cite{kikuchi1980theoretical,wei1987first,asta1993theoretical,lim1994cvm,de1994cluster} and oxides \cite{kikuchi1988calculation,tetot1994evaluation} but also in magnetic system \cite{morita1966application,finel1986phase,lawrence1986chemical}. It is also applied to study the interface and boundary energy as well \cite{kikuchi1979theory,kikuchi1980grain,kikuchi1987grain}.
We also want to mention that Kikuchi generalized the essence of his CVM into kinetics. Following the same cluster-subcluster relation to determine one free energy functional for optimization along the kinetic path, he developed the path probability method\cite{kikuchi1966path}. This kinetic modeling method would help study the SRO effect in the kinetic process. Besides, in recent years, people also develop the continuous displacement cluster variation method(CDCVM) to take lattice vibration and local atomic distortion into consideration \cite{kikuchi1997continuous}.
Readers could also check some other significant reviews \cite{de1994cluster,finel1994cluster,mohri2013cluster,mohri2017cluster,KIKUCHI200233} to understand more about the CVM.

Actually, not only physics or materials science people care about CVM but also some computer scientists.
 Statistical mechanics is strongly connected with information science since it steps into history with Shannon's significant contribution to the information theory with entropy \cite{shannon1948mathematical}. As essentially CVM is mainly used to deal with the interrelationship, it is not unexpected to see CVM could be used to determine the inference between the different information sources.
 
 The inference problem we mentioned is in the probabilistic graph model \cite{koller2009probabilistic} with the belief propagation method as the algorithm to determine the marginal/conditional probabilistic distribution. The equivalence between the Bethe
approximation and the belief propagation method was recognized in 1988 \cite{pearl1988probabilistic}. As we have mentioned, Bethe approximation could be considered a simplified version of CVM, it is natural to generalize its corresponding version in a probabilistic graph model with the help of CVM.
As a result, CVM is introduced to gain the insight to develop the generalized belief propagation to study the probabilistic inference \autocite{yedidia2000generalized,yedidia2005constructing}. After that, CVM is taken into computer science related areas in some applications such as computer vision \cite{freeman2000learning}, decoding of error–correcting codes \cite{kabashima2004statistical}, bioinformatics \cite{krogh2001predicting}, etc.
  
 Readers with further interests could check this review work clarify the relation between CVM and the graphical probabilistic graph model in 2005
 \cite{pelizzola2005cluster}. Furthermore, There is one exciting intersection between CVM from the physical side and the probabilistic graph model from the information side, spin glass. Interested readers may want to consult an engaging textbook \cite{zhou2015spin} which exactly covered this topic.

    \subsection{Yang-Li Approximation and Cluster/Site Approximation}
Before the creation of the cluster-variation method, there were several attempts to deal with the order-disorder transition as well. After Bragg-Williams original work \cite{bragg1934effect} with their Bragg-Williams approximation which only considers the ideal mixing without any atomic correlation, Bethe \cite{bethe1935statistical} developed the Bethe approximation, which took pairwise interaction into consideration and could be considered as the simplest version of CVM. Later, Fowler and Guggenheim developed the quasi-chemical method \cite{fowler1940statistical}, originally devised
for the theory of regular solutions, which applies equally
well to the theory of superlattices with long-range order. This could be considered as one stage further of Bragg-Williams' work \cite{bragg1934effect}, and compared to Bethe's work \cite{bethe1935statistical}, it also has its mathematical simplicity.

Later Chen-Ning Yang stepped further with his master thesis work by generalizing Fowler and Guggenheim's quasi-chemical method into the cluster and Yin-Yuan Li continued his work to develop it further to calculate the phase diagram of Cu-Au \cite{yang1945generalization,yang1947general,li1949quasi,li1949quasi2}. 
Kikuchi and his coauthor \cite{kurata1953theory3} considered Yang-Li approximation could be considered as a similar generalization compared to CVM. However, it has some tiny subtle differences, which we will clarify in the following chapters.
After several decades, Oates revived this method as he sought a simplified method to replace CVM \cite{oates1996cluster}. The reason is the computational efficiency and complexity of CVM is relatively large for phase diagram calculation. He noticed Yang and Li's work and found out their method could actually reduce the dimensionality of the variational space of CVM. It means we could solve less number of equations to reach the equilibrium state. Oates also renamed this method as cluster/site approximation(CSA) and took one simplification by manually adjusting the contribution ratio between the cluster and site \cite{oates1999improved}. Later, it was used in many system coupling with CALPHAD methods such as Ni-Al \cite{zhang2003application}, Cd-Mg \cite{zhang2001cluster}, Cu-Au-Ag \cite{cao2007thermodynamic}, etc. However, this method hasn't been widely used yet, perhaps due to the model's accuracy and explainability issues.
More details could be checked in an excellent thesis \cite{cao2006application}, which is related to this CSA method.

\chapter{Configurational Contribution: Fowler-Yang-Li-CVM\protect \footnote{The content of this chapter is mainly adapted from my work in arxiv: https://arxiv.org/abs/2306.15384, which would be submitted to Acta Materialia.}} 

\label{ch:conf}

\section{Introduction}

 Based on what we mentioned for the motivation of this dissertation and all we reviewed,  we start to answer a natural question to resolve the challenge we identified in the first Chapter: can the chemical SRO and other physical contributions to the free energy be incorporated into a novel free energy solution model which satisfies all the requirements of CALPHAD?

In this Chapter, we addressed this question by proposing a novel cluster-based and versatile thermodynamic model that enables the prediction of chemical SRO and fulfills all the requirements of the thermodynamic model within CALPHAD. The key to this novel cluster-based solution model is a mathematical transform called the Fowler-Yang-Li (FYL) transform, which is coined by Oates \cite{oates2007configurational} to honor Fowler \cite{fowler1940statistical}, Yang \cite{yang1945generalization,yang1947general}, and Li \cite{li1949quasi,li1949quasi2} who pioneered this idea. According to Fowler’s theory of gaseous atom/molecule equilibrium \cite{fowler1940statistical}, if the equilibrium constant and the atom/molecule mass balance are known, the Helmholtz energy can be expressed in terms of the atom concentrations rather than the molecular concentrations. This idea was later applied to model order-disorder in alloys by Yang \cite{yang1945generalization,yang1947general} and Li \cite{li1949quasi,li1949quasi2}. The application of the FYL transform to CVM was hinted in Oates’ later paper \cite{oates2007configurational} but never entirely derived and realized. The number of minimizing variables in CVM is significantly reduced after the FYL transform. The reduction of the minimizing variables is from the order of $n^s$  to $n \times s$, which makes it easy to extend to multicomponent alloys. The mathematical formalism of CVM has inspired us to insert the cluster/sub-cluster related contributions into CSA to correct the Yang-Li approximation \cite{yang1945generalization,yang1947general} used in it. Computational efficiency and physical soundness are thus well balanced in the proposed FYL-CVM model, of which a detailed derivation is presented in the following section. After taking the prototype AB binary system as the benchmark system and comparing it with other solid solution thermodynamic models, we observed the proposed FYL-CVM can correctly capture the exact topology of the phase diagram and provide the correct trend for the thermodynamic quantities with a clear physics picture and much less computational burden.

\section{Theory of Configurational Contribution}
\subsection{Model Derivation of FYL-CVM}
The derivation in this section demonstrates the proposed steps to obtain the novel cluster-based model. The overall free energy of phase $\phi$ in an alloy containing n components is expressed as:
\begin{equation}
\begin{split}
    G^{\phi} = \sum_{i=1}^n x_iG_i^{\phi} + \Delta_{mix}G^{\phi}_{conf} + \Delta_{mix}G^{\phi}_{non-conf}
    \end{split}
    \label{eq:ch3eq1}
\end{equation}
where $G_i^\phi$ is the lattice stability of element i and will be taken from the pure element database \cite{dinsdale1991sgte}. Here $ \Delta_{mix}G^{\phi}_{conf}$ represents the configurational free energy of mixing while $\Delta_{mix}G^{\phi}_{non-conf}$ represents the non-configurational free energy of mixing.

In this section, we take the FCC binary alloy as the paradigm to demonstrate how the mechanisms work for this configuration contribution briefly. Then we discuss in detail for some particular parts in the following sections to extend it into the most generalized case with rigorous mathematical derivation and proper physics.
To proceed, let us consider our definite example: that of an FCC solution for a binary AB alloy, and the tetrahedron is adopted as the basis cluster. Neglecting the ambient pressure contribution to the free energy, based on the statistical mechanics definition of the grand canonical ensemble, we have:
\begin{equation}
\begin{split}
 \Delta_{mix}G^{\phi}_{conf} = \Delta F_{mix} = \Delta \Omega(x,T) + \Delta \mu N
    \end{split}
    \label{eq:ch3eq2}
\end{equation}
$x$ represents the composition, $T$ the temperature, $\Delta \Omega$ the grand potential (or the Landau potential) of mixing, $\Delta \mu$ the chemical potential change on mixing, and N the total number of atoms. Grand canonical ensemble is generally used to deal with a case with an unconstraint number of particles. Here we select the grand canonical ensemble aiming to illustrate its deeper physical meaning for such cluster variation idea. In fact, the original derivation given by CN Yang is based on canonical ensemble \cite{yang1945generalization,yang1947general}, which can also be generalized into our proposed model equivalently, but we think the physics is more clear in our derivation. Besides, the proof of Yang’s original idea \cite{yang1945generalization,yang1947general} needs to apply complicated mathematical techniques, which is unnecessary in the current grand canonical ensemble. We will briefly discuss this later, and the readers will see that key ideas of them match with each other very well.

The first step is introducing a CVM-like hierarchical formalism \cite{finel1994cluster} to expand the total free energy as a weighted summation of the free energy contributions based on the clusters and subclusters. Such a CVM-like hierarchical formalism aims to deal with the overcounting problem during the statistical calculation of the clusters. This is because when we consider every possible cluster, it leads to some sharing with other clusters. The basic procedure is that this kind of expansion applies to the basis cluster or maximum cluster first, assuming the basis cluster is the most extensive range of the SRO considered. To subtract the overlapping between the basis clusters, we perform the statistical calculation on its subclusters formulated by the sharing between the basis clusters. This kind of subtraction usually involves even smaller overlapping to formulate smaller subclusters of the subclusters, then we do this kind of operation back and forth iteratively. The final goal is to ensure each part of the crystal’s configurational contribution is calculated and only once. Kikuchi developed his conventional CVM in around 50s last century \cite{kikuchi1951theory} but finally this kind of formalism is optimized with the mathematical inclusion and exclusion principle later to perform such calculation automatically and systematically and morphed into the current version \cite{finel1994cluster,an1988note,morita1990cluster}. The detailed derivation should be referred to previous literature. We would point out the related formalism.

Based on the reason above, we can expand the $\Delta \Omega$ and $\Delta \mu$ terms in equation\eqref{eq:ch3eq2} in a CVM-like cluster-subcluster hierarchical formalism \cite{finel1994cluster}, we have:
\begin{equation}
\begin{split}
 \Delta F_{mix} = \Delta \Omega(x,T) + \Delta \mu N = \sum_{\alpha \in \{all cluster types\}} a_{\alpha} \left[\Omega_{\alpha}(x,T) + \gamma_{\alpha}\Delta \mu_{\alpha}N\right] + \mu_{ideal}N
    \end{split}
    \label{eq:ch3eq3}
\end{equation}
where $\Omega_{\alpha}$ is the grand potential for cluster type $\alpha$, $a_{\alpha}$ and $\gamma_{\alpha}$ are coefficients ensuring that there is no overcounting in various clusters’ contributions. The mathematical foundation of $a_{\alpha}$ and $\gamma_{\alpha}$ in CVM is the exclusion-inclusion principle \cite{rota1964foundations} with the Möbius inversion \cite{an1988note,morita1984consistent}. $\gamma_{\alpha}$ is equal to the number of clusters $\alpha$ shared by one site divided by the multiplicity $m$ for a specific subcluster $\alpha$ in the basis cluster (for the basis cluster, $m=1$). $a_{\alpha}$ can be 1 or -1 and is determined by the Möbius inversion formula \cite{an1988note,morita1984consistent}. $\mu_{ideal}=\frac{k_B T}{S}\sum_{n,s} x_n^s ln(x_n^s)$ is the chemical potential of ideal solution, where $k_B$ is the Boltzmann constant, $x_n^s$ is the site fraction of the $s$ site in the basis cluster for component $n$, and $S$ is the total number of sites in the basis cluster, divide $s$ is to make sure of the number of particles is conserved for this reference as the high-temperature limit. $\Delta \mu_{\alpha}=\mu_{\alpha}-\mu_{\alpha,ideal}$ is the excess chemical potential of the $\alpha$ cluster, and $\mu_{\alpha,ideal}=\sum_{n,s} x_n^s ln(x_n^s)$ for cluster $\alpha$ with cluster size $s$, site $x^s$ should involve in this cluster $\alpha$. This term characterizes the contribution of such cluster configuration compared to the reference setup at the high-temperature limit. At the high-temperature limit, we should have $\mu_{\alpha} = \mu_{\alpha,ideal}$, and this leads to the ideal free energy of mixing $\Delta F_{ideal-mix} (x,T)=Nk_B T \sum_{n} x_n ln(x_n)$, as all the site should share the same site distribution under the high-temperature limit and cancel with pre-factor $\frac{1}{S}$.

  \begin{figure}
    \centering
\includegraphics[width=1\textwidth]{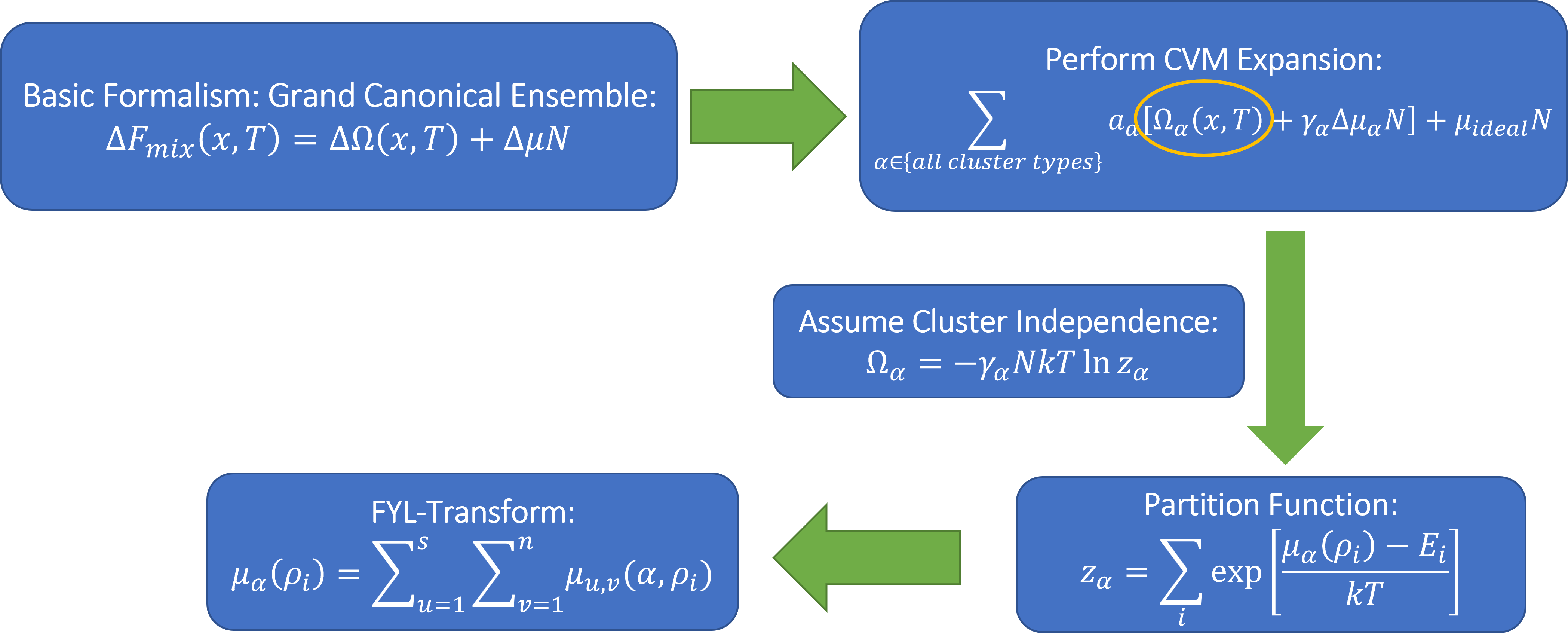}
    \caption{Illustration of the derivation process: we start from the grand canonical ensemble with the CVM expansion to deal with the hierarchical cluster-subcluster contribution systematically to completely describe the configurational contribution. This kind of CVM expansion defines a variational functional to reach the equilibrium of the state by performing the minimization. Assuming the clusters are independent of each other, we can write the partition function. Finally, we perform the key FYL-Transform on the chemical potential to rewrite the partition function to simplify the variables of this whole system.}
    \label{fig:illustrationFYLCVMderivation}
\end{figure}

For the grand potential $\Omega_{\alpha}$ of the basis tetrahedron $\alpha$, we have $\Omega_{\alpha} = -k_B T ln Z_{\alpha}$, where $Z_{\alpha}$ is the grand canonical partition function of $\alpha$. With the tetrahedron basic cluster treatment with $\alpha$, we simplify all the many-body interactions among the crystal structure into an independent averaged basic cluster. We have $Z_{\alpha}=z_{\alpha}^{\gamma N}$, where $z_{\alpha}$ is the grand canonical partition function for a single cluster $\alpha$, and the total number of cluster $\alpha$ within the crystal is $\gamma N$. Therefore, $\Omega_{\alpha} = -\gamma N k_B T ln z_{\alpha}$. We will discuss more about this in the following subsections. All mentioned above are generally applicable formulas. Then we take the FCC solid solution of binary AB alloy as the example and adopt the 1st nearest-neighbor tetrahedron as the basis cluster. Based on the properties of the grand canonical ensemble, we have:
\begin{equation}
\begin{split}
 z_{\alpha} = \sum_{ijkl} exp\left[\frac{{\color{red}\mu_{\alpha}(\rho_{ijkl})}-{\color{blue}\epsilon_{ijkl}}}{k_B T}\right]
    \end{split}
    \label{eq:ch3eq4}
\end{equation}
where $\rho_{ijkl}$ denotes the probability of occurrence of a cluster consisting of atomic species A or B occupying the indicated sites $i$, $j$, $k$, $l$ in the cluster. $ijkl$ is the notation to represent the collection set of cluster configurations. Regarding the tetrahedron cluster in an AB binary alloy, ijkl can possibly be: AAAA, AAAB, AABA, ABAA, BAAA, BBAA, ABBA, AABB, BABA, ABAB, BAAB, BBBA, ABBB, BBAB, BABB, and BBBB. With respect to these cluster probabilities, the free energy functional must be minimized at equilibrium in conventional CVM. The calculated cluster probability $\rho_{ijkl}$ can reflect the chemical SRO. $\color{blue}\epsilon_{ijkl}$ is the tetrahedron-type cluster energy which can be seen as the parameter under the CALPHAD framework. In contrast, $\color{red}\mu_{\alpha}(\rho_{ijkl})$ is the variable of the minimization to reach the equilibrium state.

From the point of view of the grand canonical ensemble, the number of specific clusters is undetermined, but the chemical potentials of these clusters should be fixed. However, this is precisely how CVM is designed to approach equilibrium. Recall the key to reach chemical equilibrium is the chemical potentials reach their corresponding equilibrium values. So here, we enforced the number of the basis clusters to be fixed. The chemical potentials should be considered variables during the variational calculation when the system reaches equilibrium by minimizing the free energy. That’s the reason why such a grand canonical ensemble can reveal the physical essence of the CVM. Though conventional CVM uses the cluster probabilities as the variables during the variational calculation to reach equilibrium, it’s equivalent to using the clusters' chemical potentials (activities). The idea is to perform a mathematical transformation based on this kind of partition function expression. The following subsection demonstrated a direct connection between the proposed FYL-CVM and conventional CVM.

Next, the FYL transform is performed on the cluster chemical potential $\mu_{\alpha}(\rho_{ijkl})$. The essence of the Fowler-Yang-Li (FYL) transform is to decompose the chemical potential with specific cluster dependence in cluster partition function into the summation of individual chemical potentials of each component on every site of the basis cluster as the site and component dependence. The former depends on the cluster probabilities, while the latter on the site variables.
\begin{equation}
\begin{split}
 \mu_{\alpha}(\rho_{ijkl}) \stackrel{FYL-transform} \longrightarrow \sum_{s} \sum_{n} \mu_{n}^{(s)} = \mu_{A}^{(1)} + \mu_{B}^{(1)}+\mu_{A}^{(2)}+\mu_{B}^{(2)}+\mu_{A}^{(3)}+\mu_{B}^{(3)}+\mu_{A}^{(4)}+\mu_{B}^{(4)}
    \end{split}
    \label{eq:ch3eq5}
\end{equation}
where $\mu_{n, n \in\{i,j,k,l\}}=k_BTln\lambda^{(s)}_n$ if the site $s$ is occupied by component $n$; $\mu_{n, n \in\{i,j,k,l\}}=0$ if $s$ is occupied by a component other than $n$, or say A or B depends which one is selected as the reference. This operation selects the reference of the chemical potentials to reduce the number of variables. This is because only the relative value is meaningful in adjusting the statistical distribution to reach the equilibrium, and we can reach this system's real degree of freedom in this way. Note here we are only treating binary alloy, so half of the site chemical potentials can be eliminated. The site variable $\lambda_n^{(s)}$ is regarded as the absolute activity of component n on-site $S$ \cite{baxter2016exactly}. Set B as a reference, which means $\mu_B^{(s)}=0(\lambda_B^{(s)}=1, \lambda_A^{(s)}=\lambda^{(s)})$ for all $s$ from $ijkl$, we have:
\begin{equation}
\begin{split}
 \mu_{\alpha}(\rho_{ijkl}) =\mu_{A}^{(1)} + \mu_{A}^{(2)}+\mu_{A}^{(3)}+\mu_{A}^{(4)} = k_B T ln(\lambda^{(1)}\lambda^{(2)}\lambda^{(3)}\lambda^{(4)})
    \end{split}
    \label{eq:ch3eq6}
\end{equation}

The cluster partition function $z_\alpha$ and the corresponding cluster probability $\rho_ijkl$ can thus be expressed as:
\begin{equation}
\begin{split}
 z_{\alpha}  =\sum_{ijkl} \lambda^{(1)}\lambda^{(2)}\lambda^{(3)}\lambda^{(4)}exp\left[\frac{-\epsilon_{ijkl}}{k_B T}\right]
    \end{split}
    \label{eq:ch3eq7}
\end{equation}

\begin{equation}
\begin{split}
\rho_{ijkl}=\frac{1}{z_{\alpha}}\lambda^{(1)}\lambda^{(2)}\lambda^{(3)}\lambda^{(4)}exp\left[\frac{-\epsilon_{ijkl}}{k_B T}\right]
    \end{split}
    \label{eq:ch3eq8}
\end{equation}

With the FYL transform, the configuration variables in cluster calculations of free energy are changed from cluster probabilities$\rho_{ijkl}$ to considerably fewer site probabilities $\rho^{(s)}$. The minimizing variables are reduced from the order of $n^s$ in CVM to the order of $n\times s$ in FYL-CVM, where n is the number of components and s the number of sites in the basis cluster, as shown in figure \ref{fig:illustrationFYLCVM}.

Using the mass balance relations: $x_A^{(1)} = \sum_{jkl} \rho_{Ajkl}$; $x_A^{(2)} = \sum_{ikl} \rho_{iAkl}$;$x_A^{(3)} = \sum_{ijl} \rho_{ijAl}$; $x_A^{(4)} = \sum_{ijk} \rho_{ijkA}$, the equilibrium site fractions $x_A^{s}$ for A on site $s$ can be related to the site variables $\lambda^{(s)}$ by:
\begin{equation}
\begin{split}
x_A^{s} = \frac{\lambda^{(s)}}{z_{\alpha}} \frac{\partial z_{\alpha}}{\partial \lambda^{(s)}} = \frac{\partial ln z_{\alpha}}{\partial ln \lambda^{(s)}} = \frac{k_B T \partial ln z_{\alpha}}{k_B T \partial ln \lambda^{(s)}} = - \frac{1}{\gamma_{\alpha}N}\frac{\partial \Omega_{\alpha}}{\partial \mu^{(s)}_{A}}
\end{split}
\label{eq:ch3eq9}
\end{equation}
combine equation \eqref{eq:ch3eq6} with the expression of $\mu_{\alpha,ideal}$, we have
\begin{equation}
\begin{split}
\gamma_{\alpha} \Delta \mu_{\alpha} N = \gamma_{\alpha}\left(\mu_{\alpha} -\mu_{\alpha,ideal}\right) N = \gamma_{\alpha} N \left( \sum_{s} k_B T x_{A}^{(s)} - k_BT\sum_{n,s} x_n^s ln x_n^s \right)
    \end{split}
    \label{eq:ch3eq10}
\end{equation}
substitute equation \eqref{eq:ch3eq10} and $\Omega_{\alpha}$ into \eqref{eq:ch3eq3}, and perform the final free energy function for an FCC AB alloy is obtained:

Equilibrium is obtained by minimizing the site variable $\lambda^{(s)}$ with respect to the total free energy of mixing under the constraint of probability conservation. $\lambda^{(p)}$ is the site variable of pair clusters and is related to $\lambda^{(s)}$.
\begin{equation}
\begin{split}
\Delta F_{mix} &= 2Nk_B T \left(\sum_s x_A^{s} ln\lambda^{(s)} -ln z_{\alpha} \right) -Nk_B T\sum_{p_1,p_2 \in {ijkl},p_1 \neq p_2} \left(\sum_{pair} x_A^{p} ln \lambda^{(p)} - ln z_{pair}\right) \\&+ \frac{5Nk_BT}{4} \sum _{s} \left(x_A^sln x_A^s + (1-x_A^s)ln(1-x_A^s)\right)
    \end{split}
    \label{eq:ch3eq11}
\end{equation}
Note the subcluster’s probabilities are entirely determined by the basis cluster probability, which is reasonable as the correct physics picture is about the statistical mixing of the basis cluster to emerge the SRO, and the subclusters’ distribution should be pre-determined by the basis clusters. So we actually don’t need to adjust any variables related to the subclusters to reach the equilibrium but correlate each subclusters’ term with the basis clusters’ variables. All the sub-cluster (pair) words should be reformulated by the basis cluster distributions in the calculation, which can be inferred from the above-mentioned mass balance relations. Then the free energy contributed by such distribution, we can rewrite the formula into a CVM-like formula as it’s equivalent to the case with cluster distribution as the variables. Another point we want to mention up to this point is that this kind of expression unifies the expression for both the ordered and disordered phases. At the same time, conventional CVM provides two separate formulas to describe the ordered phase and disordered phase by imposing the sublattice into the disordered formula. We can control the symmetry of the set of variables $\lambda^{(s)}$ to determine which kind of phase it belongs to. This point will be discussed more in the following sections.

\subsection{FYL transform}
In the original papers of Yang and Li \cite{yang1945generalization,yang1947general}, they have already implemented this transform implicitly but haven't formulated and clarified the FYL transform. When Oates revived this method as the Cluster Site Approximation (CSA) \cite{oates1996cluster}, he concluded their ideas as transforming the cluster probabilities into the site probabilities to save the variables. Here we can see the essence of FYL transform is to compress the cluster chemical potentials with cluster dependence into the averaged site chemical potentials among the basis cluster with site dependence:
\begin{equation}
\begin{split}
 \mu_{\alpha} (\rho_i) = \sum_s \sum_n \mu_n^{(s)}
    \end{split}
    \label{eq:ch3eq12}
\end{equation}
Here $s$ and $n$ are the indices for the different sites and chemical species.

One of the results based on the imposed mass balance relation is we can write the site fractions $x_n^{(s)}$ as the derivative of the grand potential free energy with respect to the corresponding chemical potentials:

\begin{equation}
\begin{split}
x_n^{(s)} = -\frac{1}{\gamma_{\alpha}N}\frac{\partial \Omega_{\alpha}}{\partial \mu_n^{(s)}}
    \end{split}
    \label{eq:ch3eq13}
\end{equation}
\begin{figure}
    \centering
\includegraphics[width=1\textwidth]{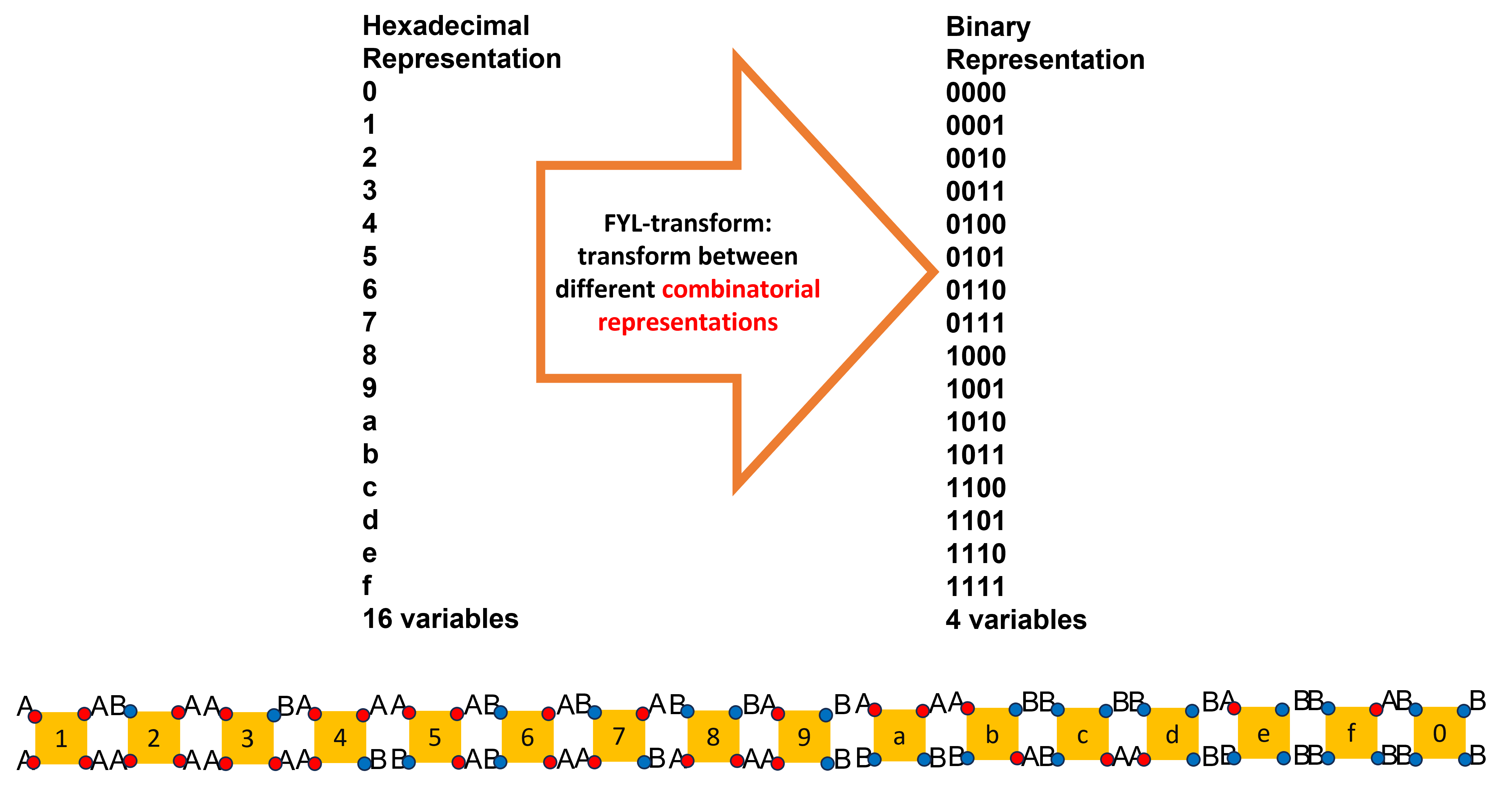}
    \caption{The illustration of FYL-transform in a hexadecimal-to-binary conversion understanding.}
    \label{fig:BinaryT}
\end{figure}
This reveals that the site fraction and site chemical potential are thermodynamic conjugate variables. After that, we can insert it back into the partition function and correlate the chemical potential with the activity $\mu_n^{(s)} =k_   BT ln\lambda^{(s)}_n$, and we impose the probability conservation by setting the activity of one selected component as 1 no matter which site it is. Because we are only interested in the relative value to control the total chemical potential as the probability conservation imposed, setting up a reference for our activities at each site. This transformation could be roughly considered as a hexadecimal-to-binary conversion with the help of 4 sites for the binary case. See figure \ref{fig:BinaryT}. The essence of this illustration is that we could understand FYL-transform as a way to compress the original combinatorial complex variational space into some site-occupation-dependent representation to reduce the complexity. Then it's just like the hexadecimal-to-binary conversion for representing the cluster chemical potential in the binary case. 

It is worthwhile to mention that while original CVM is designed to deal with the disordered phase and gets the corresponding formula for the ordered phase by imposing the sublattices \cite{colinet2001applications,mohri1985binary}, FYL-CVM provides a unified framework to analyze which phase the calculated equilibrium state belongs to automatically based on the symmetry of the site chemical activities. Here the symmetry means whether these variables equal or not after the minimization to approach the equilibrium state. As the symmetry of the chemical potential for the sites determines whether their corresponding site probabilities within the basis cluster are the same, it also determines whether the distributions among the sites are equal or not. The site variables naturally serve as the symmetry-adapted order parameters \cite{natarajan2017symmetry}. The physics picture is that: after we select the basis cluster, what we are doing is to choose an averaged local structure to represent the whole crystal. We can call such statistically averaged local structure of the basis clusters among the whole crystal as the representative local structure to represent the global structure. Then the symmetry of the site probabilities would determine the symmetry of such local structure to determine the symmetry of the whole crystal. We thus can analyze the symmetry among the site activities as the variables during the minimization to expose the symmetry of the crystal so that determine which phase it is. That’s because the solid phase classification between order-disorder transition can be all determined by the symmetry of the crystal structure, which has been well-known for many years based on Landau’s second order phase transition theory \cite{landau2013statistical,sethna2021statistical}. Of course, it implies that we must carefully select an appropriate basis cluster that can represent all possible ordering structures of the crystal. This usually is equivalent to requesting the basis cluster should be large enough.

The (generalized) quasi-chemical method \cite{yang1945generalization} is not just composed of FYL transform. FYL transform is just a mathematical technique that decomposes the chemical potentials of the cluster to site chemical potentials. While in the quasi-chemical method, besides FYL transform, one must also assume non-interference or total independence of the clusters. This is not the case for sites in the clusters as they still have correlations as measured by different cluster energies.

\subsection{Relationship between CVM and FYL-CVM}

The key difference between conventional CVM and our FYL-CVM model based on Yang-Li’s approximation \cite{yang1945generalization,yang1947general} is that the traditional CVM originated from Kikuchi \cite{kikuchi1951theory} deals with the energy and entropy parts separately to produce the free energy but Yang-Li approximation or say CSA utilize the energy to get the free energy based on statistical mechanics relation directly. However, we can observe that these two formalisms can be related by inserting FYL-CVM’s cluster probability with its transformation. Starting from the FYL-CVM’s general form, we have:
\begin{equation}
\begin{split}
F(x,T) &= \sum_{\alpha} \left(a_{\alpha} \gamma_{\alpha} N k_B T\left(\sum_{n,s}(x_n^{(s)}ln \lambda_n^{(s)}-lnz_{\alpha})\right)\right) + constant_{site}Nk_BT\sum_{n,s} x_n^{(s)}lnx_n^{(s)}
    \end{split}
    \label{eq:ch3eq14}
\end{equation}
For CVM, it can be expressed as \cite{finel1994cluster,pelizzola2005cluster}:
\begin{equation}
\begin{split}
F(x,T) &= \sum_{\alpha} \left(a_{\alpha} \gamma_{\alpha} N \left(\sum_{t}\rho_t \epsilon_{t\alpha} + k_B T \sum_{t} \rho_{t\alpha} ln \rho_{t\alpha}\right)\right) + a_{site} \gamma_{site} Nk_BT\sum_{n,s} x_n^{(s)}lnx_n^{(s)}
    \end{split}
    \label{eq:ch3eq15}
\end{equation}

Note that for the disordered phase, the site contribution is equivalent among each site so that the high-temperature limit terms of the basis clusters and many subclusters collapse together to get the single point approximation term. The coefficients due to the overlapping counting should be the same between conventional CVM and FYL-CVM, as they make use of the same combinatorial method and could both approach the correct high-temperature limit, so we should conclude $constant_{s_{site} }=a_{site} {\gamma}_{site}$ if we can prove the following relation:
\begin{equation}
\begin{split}
k_B T \left(\sum_{j=1}^s \sum_{i=1}^n x_{i,j} ln \lambda_{i,j}-lnz_{\alpha}\right) = \left(\sum_{t} \rho_t\epsilon_t + k_B T \sum_t \rho_t ln \rho_t\right)
    \end{split}
    \label{eq:ch3eq16}
\end{equation}

For each cluster type $\alpha$. We want to conclude this is correct under a transformation performed on the cluster probabilities. We just insert: $\rho_{t\alpha} = \frac{(\prod_{j=1}^s \xi_j(t))exp(-\epsilon_{t\alpha}/k_B T)}{z_{\alpha}}$ in the RHS.

To see it we must operate all the calculations between CVM and FYL-CVM on the same type of cluster $\alpha$, and we can get rid of the $\alpha$ notation and only check the general form:
\begin{equation}
\begin{split}
RHS &= \sum_{t} \rho_t (\epsilon_{t} + k_B T ln\rho_{t}) = \sum_t \rho_t \left(\epsilon_{t} + k_B T ln \frac{(\prod_{j=1}^s \xi_j(t))exp(-\epsilon_{t\alpha}/k_B T)}{z_{\alpha}}\right) \\
&= \sum_t\rho_t\left(\epsilon_t + k_B T \left(\sum_j ln(\xi_j(t))\right)-\frac{\epsilon_t}{k_B T}-lnz\right)\\
&=k_BT \left( \sum_t \rho_t \sum_j ln (\xi_j(t)) -\sum_t \rho_t lnz\right)\\
&= k_B T\left(\sum_{j=1}^s \sum_{i=1} ^ n x_{i,j}ln \lambda_{i,j} - lnz\right) \\
&=LHS
    \end{split}
    \label{eq:ch3eq17}
\end{equation}

The final equality is based on these two equalities:
\begin{equation}
\begin{split}
\sum_{t} \rho_t \left(\sum_{j} ln (\xi_j(t))\right) = \sum_{j=1}^{s} \sum_{i=1}^n x_{i,j} ln \lambda_{i,j}
    \end{split}
    \label{eq:ch3eq18}
\end{equation}
That’s because we should have the mass balance relation:
\begin{equation}
\begin{split}
x_{i,j} = \sum_t \rho_t \delta_{i,j}
    \end{split}
    \label{eq:ch3eq19}
\end{equation}
Here $\delta_{i,j}$ means component $i$ occupy site $j$. Notice that we have $\xi_j(t) = \lambda_{i,j}$ if $t_j = i, i = 1,..., n-1$. $\xi_j(t) =\lambda_{n,j} = 1$, if $t_j = n$. Where $s$ is the number of sites within this cluster and $n$ is the number of components in this system. As a result, it’s just a rearrangement of the summation. Another equality is expressed as:
\begin{equation}
\begin{split}
 \sum_t \rho_t ln z = ln z
    \end{split}
    \label{eq:ch3eq20}
\end{equation}
which is the consequence of probability conservation. 
So we can conclude that FYL-CVM is nothing but the result of performing a transformation on the cluster probabilities within CVM (or say the chemical potentials mentioned in our main content and they are actually equivalent under the equilibrium, to see this insert the general form of cluster probability $\rho_{t\alpha}=\frac{exp((\mu_{t\alpha}-\epsilon_{t\alpha})/k_BT)}{z_{\alpha}}$, which we named as FYL transform, to reformulate it into the formalism with the averaged site probabilities among the basis clusters as the variables.

\section{Application to the FCC Prototype system}

\subsection{Phase Diagram}
In this section, we apply our newly proposed method to the prototype hypothetical AB FCC system and present our calculated results compared with existing thermodynamic modeling.  Using the tetrahedron cluster approximation mentioned above, we assume this prototype system is symmetrical for the hypothetical elements (A and B). It is because the standard tetrahedron cluster can be seen as the building block of such an FCC system and is enough to capture the symmetry of the ordering phase we want to study here: $L1_0$, $L1_2$, and disordered FCC phase.

We set up the cluster energy of this prototype AB FCC system for FYL-CVM by taking the symmetrical pair energy $\epsilon_{AA}=\epsilon_{BB}=k_B$ and $\epsilon_{AB}=\epsilon_{BA}=-k_B$, $k_B$ is the Boltzmann constant. Based on this pair energy setting, we construct the tetrahedron cluster energy by counting the energy of the bonding pairs within the corresponding cluster: $\epsilon_{AAAA}=\epsilon_{BBBB}=6k_B$, $\epsilon_{AAAB}=\epsilon_{AABA}=\epsilon_{ABAA}=\epsilon_{BAAA}=\epsilon_{BBBA}=\epsilon_{BBAB}=\epsilon_{BABB}=\epsilon_{ABBB}=0k_B$, $\epsilon_{AABB}=\epsilon_{ABAB}=\epsilon_{BABA}=\epsilon_{BBAA}=\epsilon_{ABBA}=\epsilon_{BAAB}=-2k_B$. Here the cluster energies are all chemical energy whose physical picture is the bonding potential energy but not vibrational related energy. As a result, there is no temperature-dependent term. We must mention that this kind of construction is a simplified physical picture that may not precisely reflect the reality that FYL-CVM could depict. The reason is all this kind of construction is only based on the bonding formalism without the real specific tetrahedron term. This means the multi-body interactions beyond the pair interaction have not been included. However, FYL-CVM or any CVM-like formalism are able to take this kind of multi-body interactions into account by the specific cluster energy inputs. However, as we must compare with previously calculated results in the literature, we must be consistent with their settings. This point will be discussed later in this chapter. 

\begin{figure}
    \centering
\includegraphics[width=0.7\textwidth]{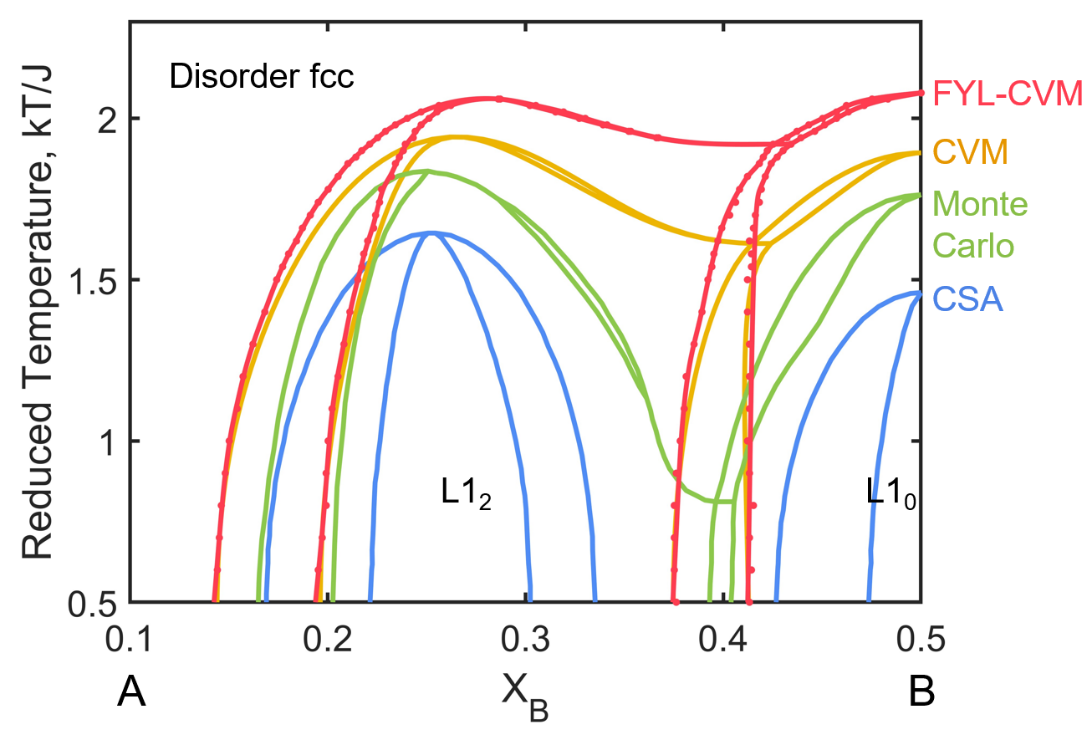}
    \caption{Calculated prototype ordering phase diagrams of AB binary alloy on FCC lattice by 4 different models (FYL-CVM, CVM, Monte Carlo, and CSA), The CVM data is based on \cite{kikuchi1974superposition,sanchez1982comparison,polgreen1984monte}; The CSA phase diagram data is adapted from \cite{li1949quasi2,oates1996cluster}; The Monte Carlo data is based on \cite{ackermann1986ordering,inden2001atomic,polgreen1984monte,ferreira1998evaluating}.}
    \label{fig:ABPD}
\end{figure}

The calculated phase diagram is presented in figure \ref{fig:ABPD}. The Monte Carlo calculated results are based on the Ising model from previous works \cite{ackermann1986ordering,inden2001atomic}. This presented MC result is preferred as it correctly captures the invariant temperature to produce a more reasonable phase diagram. The phase diagram from CVM is from \cite{kikuchi1974superposition,sanchez1978fee}. The phase diagram of CSA is adapted based on Li’s original work \cite{li1949quasi2}, as the CSA is fundamentally the same as Yang and Li’s generalized quasi-chemical approximation \cite{oates1996cluster,oates1999improved}.  All the energy settings between different models, including the Monte Carlo method, are carefully matched to each other to keep consistency. The smoothened phase diagram boundaries from FYL-CVM are the fitting results based on the original data points after the numerical minimization.

  \begin{figure}
    \centering
\includegraphics[width=1\textwidth]{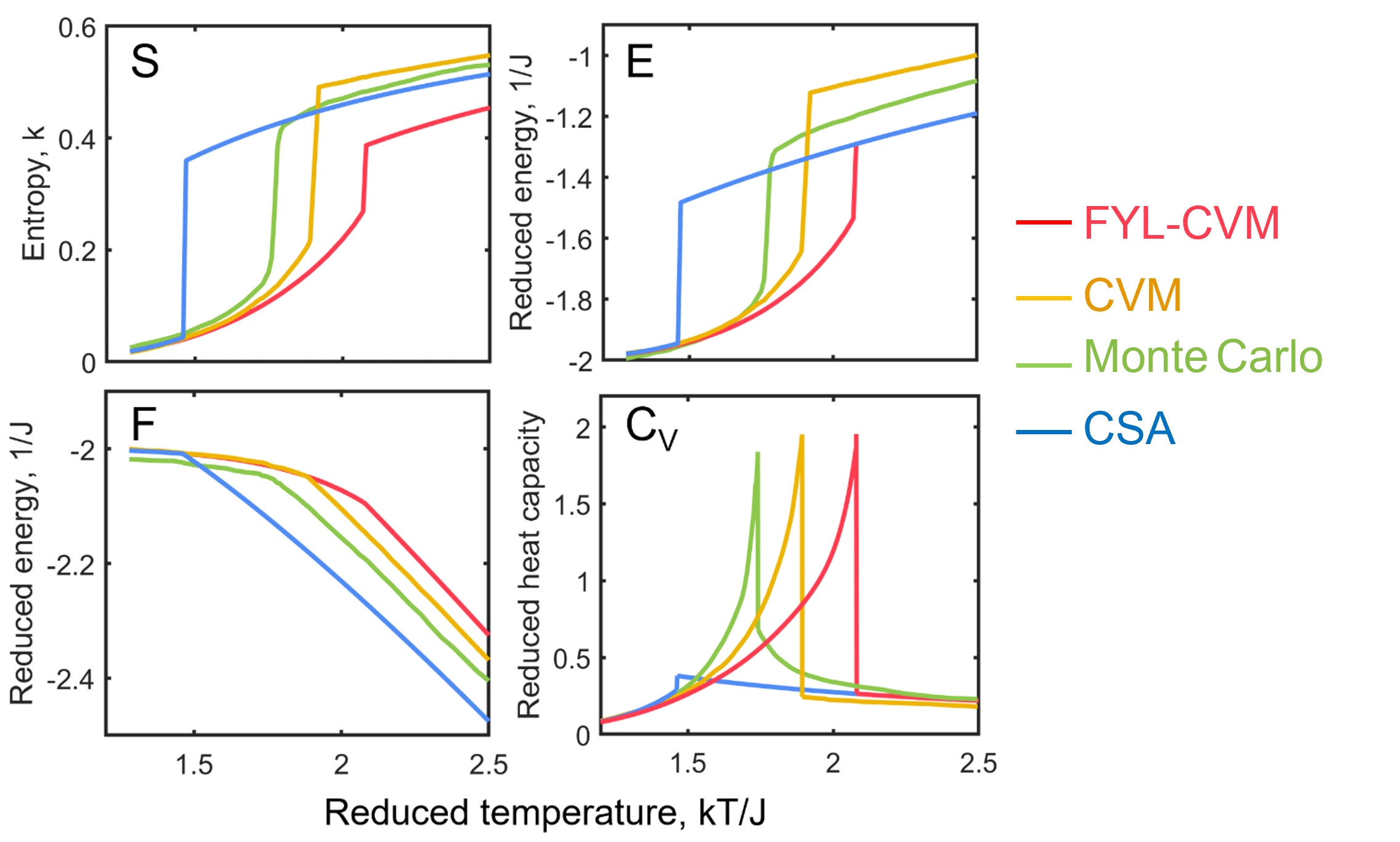}
    \caption{Calculated E, S, F, and $C_V$ for an $A_{0.5}B_{0.5}$ alloy by 4 different models.The CVM data is based on \cite{kikuchi1974superposition,sanchez1982comparison,polgreen1984monte}; The CSA thermodynamic data is reproduced from \cite{oates1996cluster,oates1999improved}; The Monte Carlo data is based on \cite{ackermann1986ordering,inden2001atomic,polgreen1984monte,ferreira1998evaluating}. }
    \label{fig:ABTP}
\end{figure}

\begin{figure}
    \centering
\includegraphics[width=0.7\textwidth]{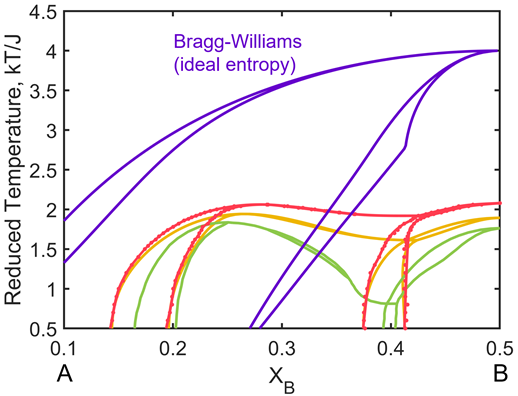}
    \caption{Prototype ordering phase diagrams of AB binary alloy on FCC lattice by compound energy formalism. This is originally adapted from \cite{shockley1938theory} and the data is reproduced with Thermo-Calc2020a \cite{andersson2002helander}.}
    \label{fig:ABPD2}
\end{figure}

Generally, we would like to take the Monte Carlo-produced results as the benchmark. FYL-CVM correctly captures the essential topology of the phase diagram and the CVM and Monte Carlo methods. The results of CSA lost the critical invariant temperature, and the disordered phase reached low temperatures. This indicates a significant error of the CSA that it can’t even capture the correct topology of the phase diagram. Interestingly, the results between CVM and FYL-CVM are very consistent. The phase diagram of FYL-CVM is just consistently shifted a bit compared with CVM. This indicates that FYL-CVM keeps the critical feature of the original CVM. However, it essentially reduces its number of variables and simplifies minimization. The results of the thermodynamic model used in conventional CALPHAD, the compound energy formalism (CEF) \cite{shockley1938theory}, is also presented in figure \ref{fig:ABPD2}. Though CEF has been developed with further added parameters to capture the SRO, we are trying to compare all the models under the parameter-free condition to demonstrate the physics picture clearly and consistently. When comparing with the ordering phase diagram of the AB prototype system produced by the CEF, we can see all cluster-based phase diagrams not only produce better transition temperature estimations with mainly better configurational entropy calculations but also capture the topology of phase diagram closer to the numerical MC results. This indicates the necessity to develop the cluster-based analytical model for CALPHAD, as the intrinsic SRO affects the topology of the phase diagram and the transition temperatures.

  \begin{figure}
    \centering
\includegraphics[width=1\textwidth]{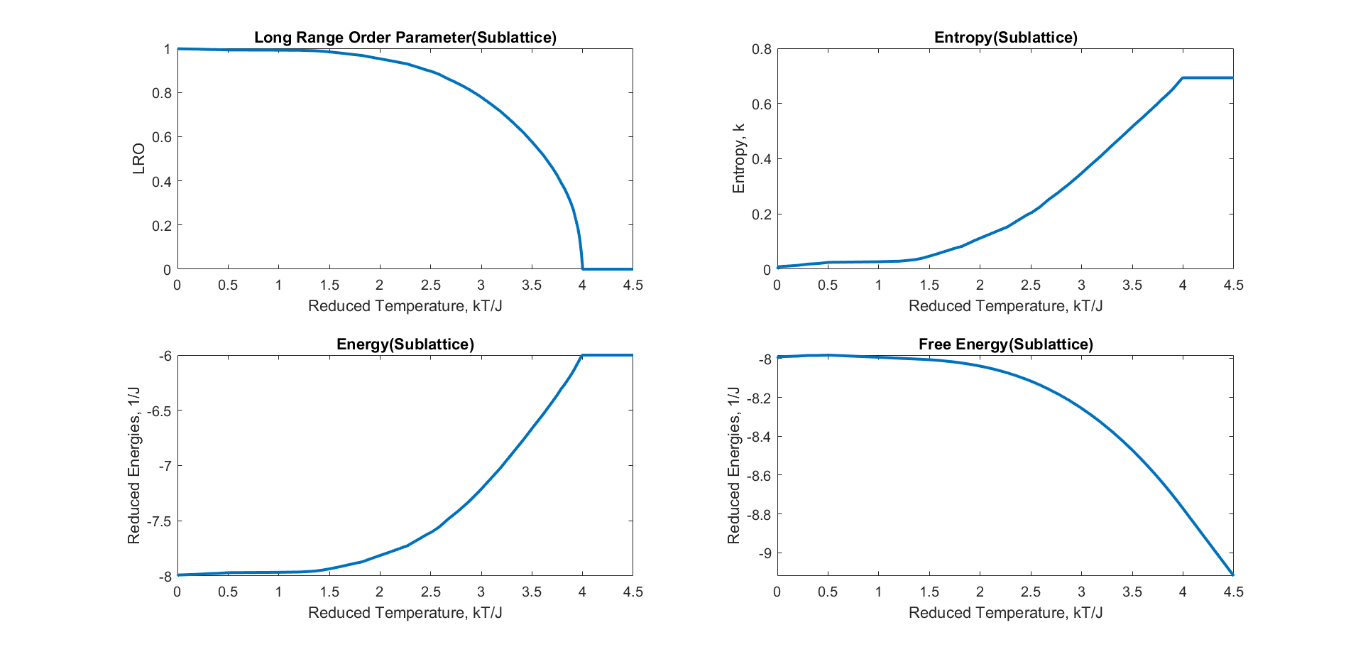}
    \caption{Calculated thermodynamic properties based on CEF for the prototype AB binary alloy on FCC lattice with 0.5A and 0.5B composition, and the data is reproduced from \cite{shockley1938theory}.}
    \label{fig:BWTP}
\end{figure}

\subsection{Thermodynamic Properties}

We also compare the detailed thermodynamic data produced from different computational models. The numerical results from MC were taken as the benchmark. All the calculated results are presented in figure \ref{fig:ABTP}. These results are calculated for the $A_{0.5} B_{0.5}$ composition at the FCC lattice with the previous prototype AB binary system setting. The free energy, energy, and entropy data are adapted from one more recent paper \cite{ferreira1998evaluating} The heat capacity data from MC is adapted from \cite{polgreen1984monte}, as the recent MC work mentioned doesn’t provide much heat capacity data but their results are consistent with each other. All the CVM data is from the same paper \cite{polgreen1984monte}. All CSA and FYL-CVM thermodynamic data are reproduced consistently by us, as their analytical model is much simpler to perform than the other two. The CSA data reproduced here matches the previous results from Oates \cite{oates1996cluster,oates1999improved}. The heat capacity data is produced by taking the derivative of the internal energy with respect to the temperature. The reduced transition temperatures between $L1_0$ phase and disordered phase from different methods are MC transition temperature T=1.736; FYL-CVM transition temperature T=2.08; CVM transition temperature T=1.89, CSA transition temperature T=1.46. More order-disorder transition temperatures in different thermodynamic models for prototype AB binary system can be found in Table \ref{table:TT4AB} below.

\begin{table}[h!]
\centering
\begin{tabular}{||c||c||c||} 
\hline
Thermodynamic model & $L_{12}$  & $L_{10}$\\ 
\hline\hline
Bragg-Williams [63] &3.28 &	4\\
Bethe-Peierls [64] & 1.78 & $\backslash$\\
Quasi-Chemical [65] & $\backslash$ & 3.57\\
CSA [36] & 1.64 & 1.46\\
FYL-CVM (this work)	& 2.06 & 2.08\\
CVM [57] & 1.94 & 1.89\\
Monte Carlo [66] & 1.87 & 1.76\\
Monte Carlo [67] & 1.84 & 1.76\\
\hline
\end{tabular}
\caption{Order-disorder transition temperatures in different thermodynamic models for prototype AB binary system.}
\label{table:TT4AB}
\end{table}

\begin{figure}
    \centering
\includegraphics[width=1\textwidth]{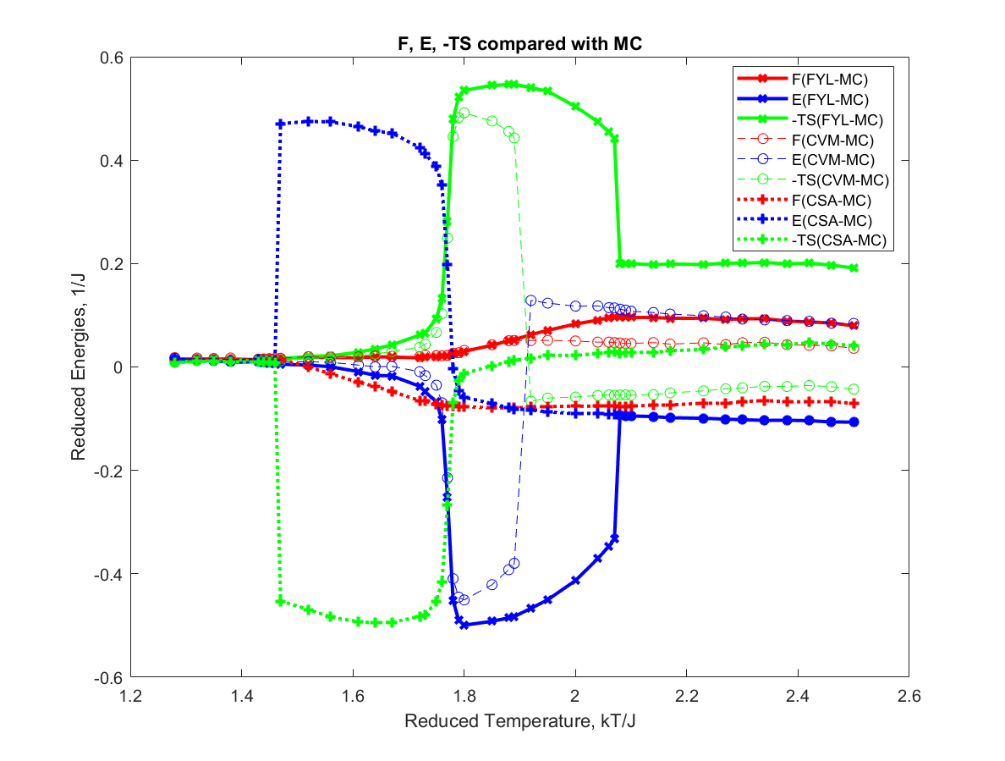}
    \caption{Error cancellation in CVM and FYL-CVM. The red line is for free energy, the blue line is for the internal energy and the green line is for the negative production between temperature and entropy.}
    \label{fig:abec}
\end{figure}

Compared to the thermodynamic data with temperature dependence produced based on the sublattice model \cite{shockley1938theory} presented in figure \ref{fig:BWTP}, all the cluster-based models can reflect the effects of SRO. The long-range order parameter data is adapted from the original work \cite{shockley1938theory}. The energy and entropy data is calculated based on the original formulas that depict the relation between the long-range order parameters and the entropy/energy. The free energy is calculated based on the entropy and energy dependence on the temperature. For the CEF model, the configurational entropy of the disordered phase directly approaches the Bragg-Williams approximation, which is the high-temperature limit with random mixing. CSA, CVM, Ising model-based MC, and FYL-CVM can all gradually capture the SRO in the configurational entropy approaching the high-temperature limit. The long-range order parameters control the phase transition within the CEF model.
Within this comparison, conventional CVM keeps its highest accuracy among the analytical models if we select the MC results as the benchmark. We can observe the deviation between the CVM and FYL-CVM as a consistent shift. FYL-CVM overestimates the ordered phase's stability, leading to such a shifting deviation. However, the minimization of FYL-CVM is much more practical than the conventional CVM, especially for CCAs with many components, due to reducing the number of variables during the minimization. The heat capacity plot demonstrates that CSA fails to get the heat capacity correctly due to underestimating the transition temperature from CSA. Another intriguing point is at the high-temperature disordered phase, the energy curves of FYL-CVM and CSA almost overlap. This is because of the input setting of the cluster energy. As we set up the cluster energy as the bonding-dependent quantity, this makes the non-interference assumption would not cause any error in such a case. Under the non-interference assumption, the bonding pairs are still considered by only taking the tetrahedron cluster with no edge sharing. As a result, when we perform the minimization, FYL-CVM and CSA would reach identical cluster distributions due to the current internal energy settings and the similar FYL-transform and produce the same total internal energy. We will discuss more about this cluster energy setting later. Another interesting observation about CSA is it produces the best estimation of the entropy within the disordered phase compared with the MC entropy. It is also observed in Oates’ original work about the modification of CSA \cite{oates1999improved}, the deviations in the CSA results are no worse than those found from the CVM in this particular case for configurational entropy. This might be due to the same reason mentioned above about the initial cluster energy setting. Notice that the accuracy of thermodynamic properties alone is not the whole story. This is because of the error cancellation effect of CVM first reported in Ferreira, Wolverton, and Zunger’s previous work \cite{ferreira1998evaluating}. We present the errors between each analytical cluster-based model and the MC results in figure \ref{fig:abec} to observe it clearly. CVM overestimates the internal energy and entropy at the high temperature. This behavior leads to the error cancellation of free energy, expressed as $F=E-TS$, and finally reduces the free energy error for CVM. In fact, all the cluster-based models possess such error cancellation effect between energy and entropy. This phenomenon reveals that it is not practical to attempt to optimize the model’s energy and entropy terms separately to produce better results as the error cancellation effect exists \cite{ferreira1998evaluating}. But we should holistically revise the original CVM model, like the FYL-CVM model proposed here.

The thermodynamic features of order-disorder transition deserve to be discussed more. Generally, the order-disorder transition should be considered as the second-order transition \cite{landau2013statistical,soffa2014diffusional}. It shouldn’t have the discontinuous latent heat jump, and any first-order derivative of the chemical potential shouldn’t have any discontinuity but just the discontinuity in the second-order derivative, such as the heat capacity. However, here every cluster-based analytical model treats the phase transition just like the first-order transition, and the latent heats all exist. The main reason for this energy discontinuity comes from the initial assumption within the cluster-based model. Notice that we set up the basis cluster as the cutoff of the range for the chemical effect and perform the mean-field approximation among the basis cluster, which limits and defines the SRO. This limited correlation length makes the CVM, CSA and FYL-CVM perform poorly near the critical point. That’s because we know that the second-order transition point is a singularity and the correlation length of the fluctuation very close to the crucial point would diverge to infinity and definitely go beyond the size of the basis cluster \cite{landau2013statistical}. This is the critical reason all the cluster-based analytical models cannot precisely capture the second-order transition feature near the transition point. However, as we mainly focus on the global behaviors of the phase diagram and thermodynamic properties, this deficiency near the local critical point is unavoidable and acceptable.

\section{Discussion}

\subsection{Comparison with CEF, QC, CVM, CSA, and CE}

In this section, let’s clarify the relationship between different mainstream cluster-based models. Before that, we should mention a little bit about the differences between the current thermodynamic solid solution model, the compound energy formalism, CEF \cite{sundman2018review}, and the cluster-based model. The critical difference is whether there is any intrinsic SRO. The lattice is artificially separated in CEF to describe the different long-range order. The atoms can only be significantly influenced within the range of the several nearest neighbor atoms, as the chemical effect is not a long-range effect like the elastic field but a short-range effect. To include the SRO, people must introduce the extra reciprocal parameters into CEF \cite{sundman1998thermodynamic}. However, the cluster-based models differ and include the critical crystal structure information nearly absent in CEF. Within the short range, the combination of the basis cluster structure is finite and pre-determined, and we can mix the cluster instead of the atoms to describe the behavior of the whole crystal. Then the solid solution is not the mixture of the atoms but the mixture of the basis cluster structure, which can be seen as the building block. Then this kind of cluster-based formalism can capture the intrinsic SRO automatically, as it inserts such basis cluster information. As a result, the cluster-based model can provide a much better prediction of configurational thermodynamics without the extra parameters.

FYL transform is a fundamental hypothesis in the original quasi-chemical (QC) model (mass action relation + non-interference) by Fowler and Guggenheim and the generalized quasi-chemical model by Yang and Li. The term “quasi-chemical” is coined because of the similarity to the mass action relation in chemical reaction equilibrium. However, in the original quasi-chemical model, the expansion is limited to pairs. Yang and Li extended to clusters but kept the non-interference assumption. FYL transform takes the essence of the quasi-chemical model by adopting the mass action law-like treatment but combines it with Kikuchi’s CVM, which completely treats the interference effect. The accuracy lost due to the FYL transform (which resulted in the difference in the FCC prototype phase diagram compared to CVM) is simply due to the compression of the independent variable space, not the CVM correlation entropy (inclusion-exclusion formula).

The CVM is originated from Kikuchi \cite{kikuchi1951theory}. CVM is a systematic recursive correction algorithm to calculate the configurational entropy of a lattice by the basis cluster and its subclusters \cite{kikuchi1951theory,de1979configurational,mohri2017cluster}. In our model, we rederive and reformulate the CVM free energy formalism from the grand canonical ensemble and the corresponding mean-field approximation of the cluster in the previous sections. The barrier of the computational burden of the CVM-like formalism during the minimization can then be resolved quickly with such transform, especially for CCAs that have at least four or five components. The difference in the calculated phase diagrams in CVM can be attributed primarily to the truncation of the entropy functional to different degrees of accuracy. Another point is CALPHAD requires a handy thermodynamic model to build up the corresponding database, and this simplification would be meaningful to make the CVM-CALPHAD feasible for database building. 

The cluster site approximation (CSA) \cite{oates1996cluster} revived by Oates should be attributed to CN Yang’s original work \cite{yang1945generalization,yang1947general}. CSA already has FYL transform. In fact, Oates coined the term “FYL transform” \cite{oates2007configurational}. However, the main difference between FYL-CVM and Oates’ CSA \cite{oates1996cluster} is the non-interference assumption \cite{yang1945generalization}. It means the original CSA only deals with non-interferent clusters and captures SRO within the basis cluster. However, it is hard to quantify the contributions due to the cluster overlapping as they are treated separately. The modified CSA \cite{oates1999improved} adopts one fitting factor $\gamma$ to mimic the overlapping interactions, albeit it still uses the incorrect non-interference assumption. Oates modified the original CSA with this adjusted parameter to control the ratio between basis cluster contribution and point contribution or, say, high-temperature limit term. This does get better accuracy. However, the introduced parameter is unphysical, and developing it into a more general or flexible formalism for multicomponent systems is hard. 

\begin{figure}
    \centering
\includegraphics[width=1\textwidth]{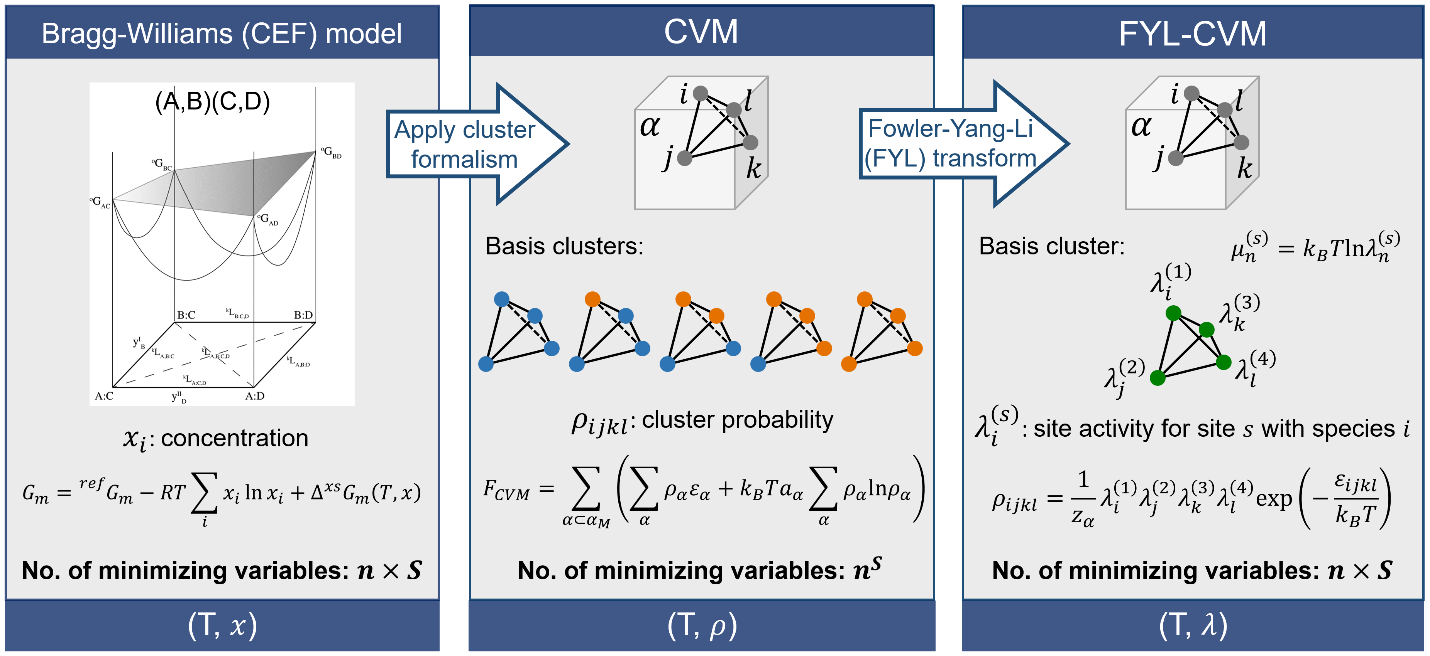}
    \caption{Illustration of the relationship between the models: sublattice model, CVM, and the proposed FYL-CVM.}
    \label{fig:illustrationFYLCVM}
\end{figure}

This non-interference assumption ignores the contribution from the cluster, which may lead to double-counting or overlapping. That’s why the original Yang’s work \cite{yang1945generalization,yang1947general} or CSA-related work \cite{oates1996cluster,oates1999improved} always only has the basis cluster contribution and the point contribution. This assumption implies that subclusters are not allowed. This is because the FYL transform is adopted before Kikuchi’s CVM \cite{kikuchi1951theory} was invented to include the cluster-subcluster contribution systematically. By taking the CVM-like formalism within FYL-CVM, our FYL-CVM model allows the overlapping between the clusters and quantifies the overlapping using the inclusion-exclusion principle \cite{rota1964foundations} to improve the accuracy. Then we can discard such non-interference assumption, which only allows point-like sharing of the basis clusters by including only part of the basis clusters but can currently deal with the overlapping between the basis clusters systematically. In addition, we also conclude that FYL-CVM takes advantage of both the conventional CVM and the original CSA by combining their key features, as is illustrated in figure \ref{fig:illustrationFYLCVM}.

Finally, the Cluster Expansion (CE) coupled with Monte Carlo (MC) is discussed as the end of this Section \cite{van2013methods,de1994cluster,wu2016cluster}. One of the connections between CE and CVM is dealing with the internal energy expansion with effective cluster interactions (ECIs) \cite{sanchez1984generalized,asta1991effective,wolverton1991effective}. As this energy expansion already expands the Hamiltonian, it becomes possible to depict the complete thermodynamic properties of the system by CE itself with the Monte Carlo sampling based on the first-principles data. This CE+MC formalism provides a relatively accurate phase diagram based on the DFT and the intrinsic chemical SRO introduced by the cluster expansion.

It’s perhaps unnecessary to compare CE+MC and FYL-CVM, as CE+MC is a fully numerical and first-principles method, while FYL-CVM is an analytical method. But they still have the exact origin: expand the energy with a cluster-based formalism. Unlike FYL-CVM, CE expresses the energy with cluster formalism by the ECI and the correlation functions. In contrast, FYL-CVM expresses similar expansion with the probability and the combinatorial formalism to a variation functional to approach the equilibrium by performing minimization analytically. Another difference is FYL-CVM only involves the basis cluster and its subcluster’s contributions to the free energy based on its combinatorial hierarchical expression. In contrast, CE can involve more clusters with first-principles calculation data to fit the ECIs.

When we observe many calculated ECIs with cluster dependence, it is expected that only several more compact clusters would usually have the significant ECIs \cite{ozolicnvs1998first,puchala2013thermodynamics,natarajan2016early,wu2016lithium,geng2017first,wang2020first}. Compared to FYL-CVM, if we take the 1st nearest neighbor tetrahedron as the basis cluster, it will only involve 1st nearest neighbor pair and the basic tetrahedron. The triplet is missing because its combinatorial formalism does not involve triplet overlapping. As a result, the calculated ECI from CE may indicate the cutoff of the basis cluster selection of the FYL-CVM can be relatively small. Though CE + MC method provides reasonably accurate results based on the first-principles calculation and its sound physical origin, these methods are generally limited to binary or ternary systems due to their large number of configuration variables. In addition, phase diagrams calculated by these methods are not considered quantitatively accurate enough for technological applications because the chemical accuracy of first-principles inputs limits them \cite{aldegunde2016quantifying}. This indicates it is necessary to develop an analytical simplified cluster model such as the proposed FYL-CVM model.

  \begin{figure}
    \centering
\includegraphics[width=0.6\textwidth]{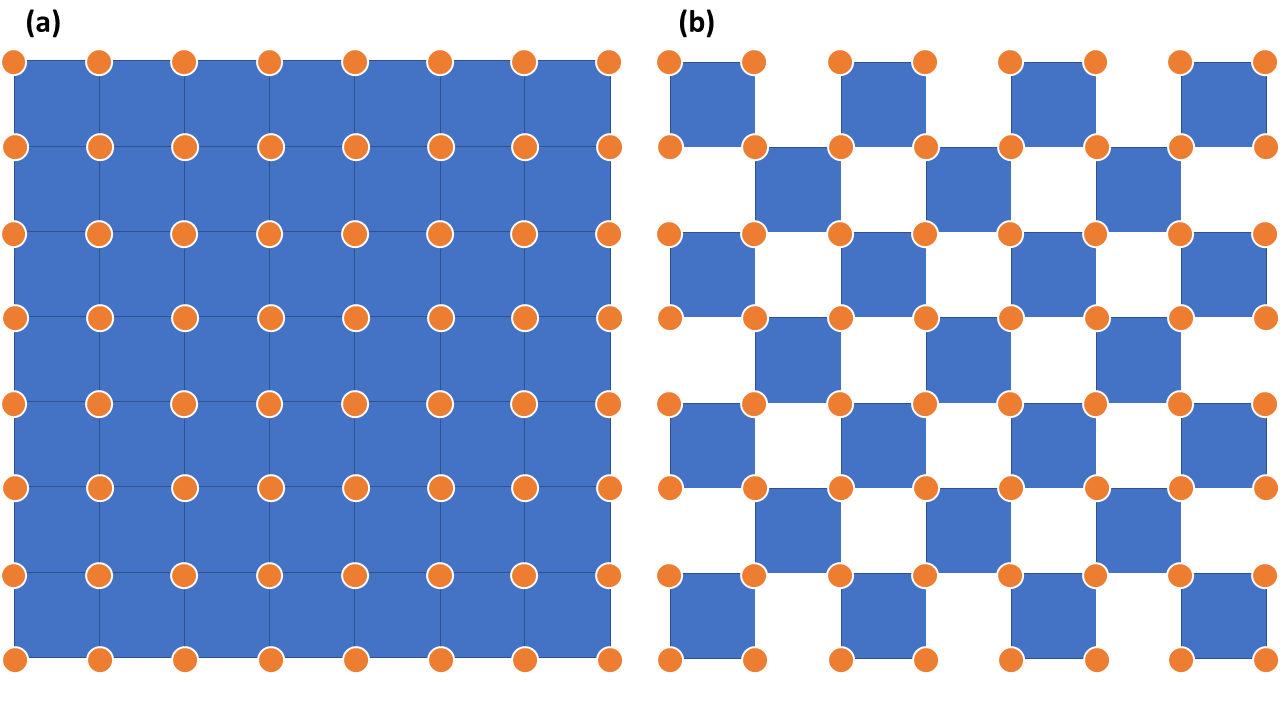}
    \caption{The illustration diagram to distinguish the non-interference case with 2-dimensional square lattice. Here the orange circles represent the atoms, and the blue squares represent the square cluster interactions taken into counting. (a) is the counting strategy for CVM and FYL-CVM, one has to take all the square clusters and consider their overlapping pairs. (b) is the counting strategy for CSA based on non-interference assumption, which only considers the square clusters with only corner sharing. }
    \label{fig:illustrationnoninterference}
\end{figure}

\subsection{Advantages of FYL-CVM}
Through the FYL-CVM model, we have put more physics, primarily intrinsic SRO, into CALPHAD while maintaining its practicality and efficiency. It advances the field by leveraging statistical mechanics to yield a more physical description of configurational entropy and by opening the door to cluster-based CALPHAD database development. The advantages of introducing more physics into solution models are apparent: correct phase diagram topology, correct lattice structures, correct order-disorder behavior, etc. The recognition of these properties in the models should give more confidence in data extrapolation and involve use of fewer fitting parameters. Advantages of the proposed FYL-CVM also include superior computational efficiency and scaling behavior compared to CVM. The minimizing variables are reduced from the order of $n^s$ in CVM to the order of $n \times s$ in FYL-CVM, making extending FYL-CVM to multicomponent systems easy. It brings many benefits to the CALPHAD modeling, such as avoiding spurious ordering caused by the unphysical ordering behavior of sublattice model, a more convenient model for vacancy thermodynamics, and removal of hypothetical endmembers \cite{oates2007compound}. The configurational and non-configurational (vibrational, elastic) contributions to free energy will be modeled separately, gaining insights into their respective effects on phase stability. The versatility of the new solution model means it applies well to metallic, ionic, and semiconductor alloys while offering the possibility of considering larger clusters and/or more components within the CVM hierarchy. This kind of cluster-based free energy model is naturally more suitable for describing the thermodynamics of coherent interfaces and inhomogeneous systems, like the anti-phase boundary energy \cite{finel1994cluster} or the Cahn-Hilliard gradient energy term \cite{lass2006correlation}.

The FYL-CVM model includes the critical crystal structure information nearly absent in sublattice models. FYL-CVM also provides a unified description of order-disorder free energy instead of separating it into ordered and disordered parts as in the conventional CALPHAD. In CVM, sublattices are still needed to describe the long-range order (LRO) \cite{mohri1985binary}. While in FYL-CVM, LRO can be inferred by observing the symmetry of site variables $\lambda_s$ after minimization without the need of predefined sublattices. For example, the tetrahedron in disordered FCC phase would have $\lambda_1=\lambda_2=\lambda_3=\lambda_4$, but $\lambda_1=\lambda_2 \neq \lambda_3=\lambda_4$ in ordered L12 phase. The site variables naturally serve as the symmetry-adapted order parameters \cite{natarajan2017symmetry}.
A problem of the conventional CALPHAD method is that being based on classical thermodynamics, it may miss all the metastable, microstructural constituents, and transient phases that are at the core of the most successful engineering alloys \cite{tsao2017high}. CVM-CALPHAD will unleash the potential of CALPHAD to model these metastable and transient structures/phases. The novel FYL-CVM model developed here should be able to describe simultaneous ordering and clustering (phase separation) reactions by capturing the nuances of the curvature in free energy vs. composition (G-x) diagrams, such as those in the technologically important Ni-Al, Ni-Ti, and Al-Li alloys identified by Soffa and Laughlin \cite{soffa1989decomposition}.
A so-called CVM-CALPHAD framework was proposed previously \cite{colinet2001applications}, but it was still considered too complicated for the multicomponent alloys using CALPHAD. After the detailed benchmark tests on the prototype AB system, we conclude the following table below to elaborate the proposed FYL-CVM’s key features compared to the previous analytical solid solution thermodynamic model.

A remark is made here on the choice of the optimum basis cluster. The choice of the basis cluster should ideally include the longest interaction distance. But in practice, it may be enough to make the basis cluster have all the major ordering types shown on the phase diagram, such as $L1_2$ and $L1_0$ for the FCC phase. The basis cluster size does not equal the domain size observed in experiments. The basis cluster should capture the ordering type exhibited in the ordered domain.

\begin{table}[h!]
\centering
\begin{tabular}{||c || c|| c|| c|| c||} 
 \hline
$\backslash$  & Sublattice & CSA & FYL-CVM & CVM \\ 
 \hline\hline
Include intrinsic SRO? &	No &	Yes &	Yes &	Yes\\
Extendable to multicomponent? &	Yes &	Yes &	Yes &	No\\
Number of variables &	High &	Low	& Low &	Very High\\
Accuracy &	Low for SRO &	Low &	High &	Very High\\
Computational cost &	Very Low	& Low	& Low &	Very High\\
Usability/Versatility &	Very High &	Low	& High	& Low\\
 \hline
\end{tabular}
\caption{Comparisons of the expected performance of different analytical solution models.}
\label{table:MODELCOMPARE}
\end{table}

\subsection{Energy from Cluster Expansion}
The pairwise energy may be problematic because it limits the interaction range, though people test this pair energy to benchmark the model. In Kikuchi’s original idea, he mainly tried to deal with the configurational entropy part but not the energy. Even until his later work \cite{kikuchi1974superposition} and some recent review also point it out that \cite{colinet2001applications}, people still write the corresponding energy with this expression:
\begin{equation}
\begin{split}
 E = 2N \sum_{ijkl} \rho_{ijkl} \epsilon_{ijkl}
    \end{split}
    \label{eq:5flow1}
\end{equation}
Where $\rho_{ijkl}$ is the probability of the cluster $ijkl$ and $\epsilon_{ijkl}$ is the corresponding cluster energy, usually expressed as the summation of the bonding energy.

This can be correct only when this energy is adequate or reduced. The corresponding basis cluster would also have the overlapping issue, which can be resolved by the CVM-like expansion, which Kikuchi initially only wants to apply on the entropy term. As a result, the bonding energy setting here is the correct way to capture all the pair interactions. Still, it can’t use the tetrahedron to capture the higher-order interactions. However, as we deal with all this systematically within the grand canonical ensemble coupling cluster expansion formalism outside of the free energy directly, we have to account for the energy and entropy altogether with such expansion. Actually, such issue within the original CVM is systematically resolved in the computer science community and statistical physics community when they borrow Kikuchi’s idea to develop their region base model, probabilistic graphical model, and the generalized belief propagation method \cite{yedidia2000generalized,pelizzola2005cluster}. Besides, recent CVM calculation deal with energy based on the ECI based on CE calculation, and this problem then can be resolved \cite{mohri2017cluster,colinet2001applications}.

Back to the cluster energy concept, all the tetrahedron energy in the literature were calculated by the summation of pairwise bonding energy, which is also the reason why here all calculation setting can be consistent with the Ising model but not the generalized Ising model with expression:

\begin{equation}
\begin{split}
E = \sum_{ij}J_{ij} S_i S_j + \sum_{ijk} J_{ijk}S_i S_j S_k + \sum_{ijkl}J_{ijkl} S_i S_j S_k S_l + ...
    \end{split}
    \label{eq:5flow2}
\end{equation}

which would capture the interaction beyond the pairwise nearest neighbor and is the foundation of the CE. The physical picture depicted by the bonding energy-dominated case does not precisely reflect reality. That’s why CSA and FYL-CVM would have the same internal energy at the high temperature, and CSA even has a better estimation for the entropy compared to all the models here. The reason is if only pairwise bonding energy is involved, the total energy is only about counting the total number of different types of bonding. In such cases, the non-interference assumption would not affect the results if the minimized probability cluster distribution were the same. The reason is the bonding contribution can be fully included only once, even under the non-interference assumption. But when the tetrahedron cluster energy, which includes multi-body interaction, is used, the non-interference assumption would ignore some energy caused by such 4-atom interaction from the ignored tetrahedron cluster as it doesn’t allow sharing except point sharing. As a result, we think the phase diagram produced from the current MC still does not fully reflect the whole physical picture. However, adding the multi-body interactions would be helpful to demonstrate the better-matched results with FYL-CVM. This is expected as it can capture such multi-body interaction with consistent cluster energy as the input. However, in the real alloy case, determining the cluster energy based on first-principles calculations is non-trivial. We will discuss this in the following chapters.

\section{Conclusions}
We have now constructed the framework for the configurational contribution of the solid solution. We expect this new FYL-CVM model to open a new routine to design the multicomponent materials such as CCAs with the intrinsic SRO to capture the essence of the physics without the arbitrary parameters while saving the computational burden with the FYL transform.

The key features of the configurational contribution of this novel cluster-based model, FYL-CVM, are:

	It captures the intrinsic SRO with clearly physical meaning, the mixture of clusters within the solid solution phase.
 
	It keeps the advantages of CSA to use about $(n-1)\times s$ parameters to achieve the minimization to reach the equilibrium states. CVM will use about $n^{s-1}$ independent variables (without symmetry considerations yet) to do the same thing.
 
	It introduces the reasonable cluster-subcluster hierarchical formalism based on CVM to improve the accuracy of comparing the original CSA without adding other unphysical parameters. Based on the demo calculation on the prototype AB binary alloy, we illustrate the superiority of FYL-CVM to capture the correct topology of the phase diagram and the correct trend of the temperature dependence of the thermodynamic quantities.
 
	Our work generalized the particular situations in CSA or the original Yang-Li’s approximation about the different species of the clusters and the multicomponent alloys into the most generalized grand canonical ensemble formalism rigorously in mathematics with clear physical meaning and elegant physics picture to cancel the non-interference assumption.

	We illustrate the mechanism based on symmetry to unify the expression systematically for ordered and disordered phases which is generalized very naturally.
 
This FYL-CVM provides free energy efficiently and accurately with clear physics picture and rigorous mathematics. We look forward to taking it as the solid solution model within the CALPHAD. However, there are still several problems left or more problems emerge after this foundation work to construct such a novel model and formalism. Unlike the configurational contribution, how to deal with the non-configurational contribution, such as the elastic contribution? The cluster energy is a little vague, and the degrees of freedom can go beyond the configurational degree of freedom, such as the vibrational contribution and electronic contribution. How to derive and deal with this under the current framework? How to parameterize and determine the cluster energy accurately and efficiently for an actual alloy case study? As the first-principles method becomes part of the computational thermodynamics and the experimental data is always scarce within CALPHAD \cite{liu2009first}, how can we make use of the first principle calculation to determine these parameters or at least to provide a reasonable initial value? As the variational calculation or say, the minimization calculation actually can be seen as a mathematically optimization problem, we can conclude with a simple calculation that both CVM and FYL-CVM actually are the non-convex problems \cite{pelizzola2005cluster} so that may not precisely converge to the global minimum in some cases, how to design better and more efficient minimization algorithm for it? We want to discuss several exciting problems mentioned above in the following chapters.


\chapter{Non-Configurational Contributions: Vibrational, Elastic, and Electronic Effects}\label{ch:nonconf}
\section{Introduction}

In this Chapter, we would introduce non-configurational contributions to the solid solution thermodynamics, mainly involving vibrational, elastic, and electronic effects. While people have already studied many alloy systems focusing on their chemical configurational contribution, the non-chemical configurational contributions are usually overlooked. These contributions could include but are not limited to atomic vibrations, elastic deformation, electronic structure, magnetism, etc. Compared to the chemical configurational contribution, the non-configurational contribution is usually considered relatively small for order-disorder transition. After all, the order-disorder transition is dominated by the chemical configurational transition itself. On the other hand, it is rare to introduce the non-configurational contribution into the order-disorder transition consideration, as introducing a non-configurational contribution usually costs a lot of resources to deal with.

However, several recent studies indicate that non-configurational contribution would significantly affect the phase stability and order-disorder transformation, especially for the CCAs \cite{wallace2021modeling,ma2015ab}. Overlooking these non-configurational contributions may lead to a significant deviation in phase stability. Here we would like to concentrate mainly on the vibrational and elastic contributions due to their typical appearance in all kinds of materials.

The vibrational contribution to phase stability has been studied a lot since several years ago \cite{van2002effect,fultz2010vibrational}. It is indicated that the atomic vibrations would significantly affect the phase stability \cite{ozolicnvs1998first,ozolicnvs2001large,burton2006first,hua2018first}, the solid solubility at high temperature \cite{shulumba2016lattice} and the precipitation sequences \cite{wolverton2001entropically}. The effects caused by the elastic contribution on the phase stability and order-disorder transformation seem not obvious. However, it would still play an essential role in the phase diagram, which would enlarge the mixing regime \cite{ferreira1987effect,ferreira1988chemical}.

Based on all previous observations mentioned above, it is implied that while we proposed the cluster-based solution model to connect CALPHAD with the mainstream statistical mechanics and naturally account for the chemical SRO in CALPHAD, the non-configurational contributions to the (vibrational and elastic) free energy also needs to be considered \cite{de1979configurational}. As we want to put more physics into CALPHAD to make the versatile thermodynamic model capable of capturing different originated effects \cite{oates1996putting}, it is theoretically natural to generalize the current model to include all these non-configurational contributions mentioned above. Besides, it is also required that the non-configurational (vibrational, elastic) contributions to the free energy must be included in a computationally inexpensive and practical way to couple with the CALPHAD method cheaply.

In this Chapter, we will continue discussing the thermodynamic framework derived in the previous Chapter and further develop it to include more contributions mainly the vibrational and the elastic contributions, beyond the only chemical configurational contribution. The vibrational free energy will be modeled by the coarse graining process into the configurational cluster energy \cite{garbulsky1994effect}. In contrast, the elastic energy will be modeled in the compositional-dependent mean-field term, which is motivated from the $\epsilon-G$ method proposed by Zunger \cite{ferreira1988chemical}. Of course, other different sources of contributions would affect the ordering of solid solutions, such as electronic and magnetic contributions, etc. However, electronic contribution is generally much smaller and could be obtained directly through the DFT calculation easily while magnetic contribution is much more sophisticated and perhaps would lead to the interplay between chemical SRO and magnetic SRO. We would leave the magnetic part to further discussion.

Combining all these strategies to include the vibrational, electronic, and elastic contributions into the proposed cluster-based solid solution thermodynamics model, we take the prototype AB binary system to test how the vibrational and the elastic models currently taken here would affect the thermodynamic calculation and the phase diagrams. The observed results satisfy our expectations and indicate the huge potential of the proposed cluster-based model to provide physics meaningful adjustment for the thermodynamic assessment of the CCAs in the future coupling with CALPHAD method.

\section{Theoretical Model of Non-Configurational Contribution}

\subsection{Model for Vibrational Contribution}

The vibrational contribution can be seen as the thermal kinetic excitation of the atoms contributed by the phonon, which is the quantization of the cooperative atomic motion. We consider vibrational contribution a weak chemical dependence contribution because there is evidence that the vibrational difference between the ordered phase and disordered phase affects the configurational order-disorder transformation case by case. 

Here we perform the “coarse graining” processing of the partition function invented by Ceder \cite{ceder1991alloy,ceder1993derivation}, then the weak chemical configurational dependence contribution can be smoothly inserted into the configurational mechanism by adding one more temperature-dependent energy term. From the statistical mechanics perspective, the vibrational contribution under the configurational formalism would generate more microstates due to the more degrees of freedom from the displacements of the atoms:
\begin{equation}
\begin{split}
z_{\alpha} = \sum_{\sigma} exp\left(\frac{\mu_{\sigma}-\epsilon_{\sigma}}{k_B T}\right) \to \sum_{\sigma} \sum_{v} exp\left(\frac{\mu_{\sigma}-\epsilon_{\sigma,v}}{k_B T}\right) 
    \end{split}
    \label{eq:4-1}
\end{equation}

Then we assume the vibrational energy can be separated from the chemical configurational energy, and we sum over all the vibrational microstates first which are associated with the same chemical configurational microstates:
\begin{equation}
\begin{split}
z_{\alpha} = \sum_{\sigma} \sum_{v} exp\left(\frac{\mu_{\sigma}-\epsilon_{\sigma,v}}{k_B T}\right) = \sum_{\sigma} \left(exp\left(\frac{\mu_{\sigma}-\epsilon_{\sigma}}{k_B T}\right)\left(\sum_{v(\sigma)}exp\left(-\frac{\epsilon_{v(\sigma)}}{k_B T}\right)\right)\right)
    \end{split}
    \label{eq:4-2}
\end{equation}
 Depending on the specific cluster configuration, the inside represents a canonical sub-ensemble inside the specific microstate. Finally, we have
\begin{equation}
\begin{split}
F_{vib,\sigma} &= -k_B T ln \left(\sum_{v(\sigma)}exp\left(-\frac{\epsilon_{v(\sigma)}}{k_B T}\right)\right) = -k_B T ln Z_{vib(\sigma)}\\
z_{\alpha} &= \sum_{\sigma} exp\left(\frac{\mu_{\sigma}- (\epsilon_{\sigma}+F_{vib,\sigma})}{k_B T}\right)
    \end{split}
    \label{eq:4-3}
\end{equation}
Here the physics behind this mathematical transform is the “coarse graining” of the partition function which integrates out the “fast” degrees of freedom (e.g., vibrations) before considering ”slower” ones (e.g., configurational changes) \cite{ceder1993derivation}. The double sum of the partition function reflects the coupling between the vibrational degree of freedoms and the specific “slower” configurational cluster decorations. However, there would be many local energy minima in the neighborhood of the specific ideal cluster microstates caused by the vibrations. Then the summation with a canonical sub-ensemble provides an optimized constrained (with the specific “slower” configurational degree of freedom) averaged free energy term to determine the statistics caused by the vibrational contribution to estimate the most probable microstate in the neighborhood of the phase space associated with the ideal configuration without the vibrations. This configuration associated with vibrational free energy for the cluster defines the constrained state in phase space to remain in the neighborhood of the ideal configuration.

 We must remind the readers that we have to assume the vibration is harmonic to lead to the local energy minima. If the vibration has an extreme case of anharmonicity, it may lead to no existence of local minimum. However, this “coarse graining” is an entirely mathematical transformation that is just a summation of any micro-states. As a result, there is no limitation to applying this “coarse graining” mathematically for this kind of instability case. We can still assign the constrained free energy, though it is difficult to be defined. It might lead to remarkably high free energy and even negative entropy. Further discussion can refer to \cite{van2002effect}. An intriguing discussion is in its Appendix F.

The next step is to select a concrete form for $F_{vib,\sigma}$  with clear physical meaning to fit the CALPHAD. The selection is flexible and should be simple to couple with the CALPHAD. Recall three main origins of the vibrational effects on phase stability: the “bond proportion” effect, the volume effect, and the size mismatch effect \cite{van2002effect}. The “bond proportion” effect comes from the change in the number of types of bonding during the order-disorder transition, resulting in a change in its vibrational entropy. This effect can be identified from the changes of the phonon density of states (DOS) during the order-disorder transition. This effect has been both experimentally \cite{anthony1994magnitude,fultz1995phonon} and computationally \cite{althoff1997vibrational} confirmed. The volume effect comes from the thermal expansion and the anharmonicity would lead to the vibrational entropy for this effect. This volume effect would lead to a whole shift of the phonon DOS. However, it is believed that the vibrational entropy difference originating from the anharmonicity should have a smaller effect compared to the harmonic contribution \cite{van2002effect}. It’s due to the canceling effect between the anharmonic contribution and the vibrational enthalpy. Experimental measurements also support this observation as they have not identified the volume effect as the major origin of the vibrational entropy difference. The size mismatch effect is due to the different size atoms' ability to experience compressive or tensile stress due to the constraint of the surrounding chemical environment \cite{van2002effect}. This would lead to the locally stiffer or softer regions leading to the vibrational entropy difference.

As a result, we use the “bond proportion model” to take the vibrational contribution into account due to its simplicity. The main assumption comes from the simplification of the origin of the vibrational entropy. First, it assumes only the “bonding” or say the nearest neighbor force constant contributed to the vibrational entropy. This can be justified by the calculation based on the accuracy of the nearest neighbor spring model is believed sufficient for the practical phase diagram calculation \cite{van2002effect,garbulsky1996contribution,van1998first}. Another assumption may lead to more deviations. It’s about ignoring the volume effect and especially the size mismatch effect. That is because, in the original bond proportion model, every type of bond is assumed to have an intrinsic independent stiffness. This would somehow lead to severe deviations and is precisely correlated to the other source of the vibrational entropy difference, the volume effect, and in particular, the size mismatch effect. Actually, the theoretical calculations indicate the stiffness would change substantially as a function of its environment \cite{van2000first,morgan2000vibrational}, which indicates the simple “bond proportion model” would lead to a relatively large deviation due to the size mismatch or, say the local environment. However, the vibrational contribution under our current framework is naturally local environment-dependent; the vibrational free energy inserted is based on the specific outer cluster configurations. As a result, it’s possible to consider the size mismatch at least partially under our framework and couple the bond proportion model into our framework conveniently and more physics-meaningfully. 

To illustrate how the bond proportion model implement into our framework, we first assume each atom vibrates independently and the motions in three dimensions are decoupled just like in the Einstein model. It can be intuitively understood as each atom's mass has three independent springs. Then it is natural to assume all the vibrational modes with the specific interatomic springs only depend on the 1NN pair of atoms. Note that for the different crystal lattices, the number of nearest neighbors is not necessarily the number of vibrational modes. Hence, we need a ratio to ensure the total number is consistent. For an FCC crystal, each atom has 12 nearest neighbors, and the total number of nearest neighbors is 6N, as N is the number of atoms. The total number of vibrational modes is $3N$ for three dimensions. Then the ratio is $\eta=3N/6N=1/2$ to associate the bonding frequency. Then we have the explicit form for binary alloy AB.
\begin{equation}
\begin{split}
Z_N = \left(\frac{k_B T}{\hbar \omega_{AB}}\right)^{\eta N_{AB}}\left(\frac{k_B T}{\hbar \omega_{AA}}\right)^{\eta N_{AA}}\left(\frac{k_B T}{\hbar \omega_{BB}}\right)^{\eta N_{BB}}
    \end{split}
    \label{eq:4-4}
\end{equation}
Where the $N_AB$, $N_AA$, $N_BB$ are the numbers of the corresponding bonding among the crystal. $\omega_AB$, $\omega_AA$ and $\omega_BB$ are the corresponding frequencies for different bonds. It can be easily generalized into the multi-component case by adding more possible frequencies with different bonding. Notice the current partition function is for the whole crystal, we can limit it to the specific cluster configuration.
\begin{equation}
\begin{split}
Z_N = \left(\frac{k_B T}{\hbar \omega_{AB}}\right)^{\eta N_{AB,\sigma}}\left(\frac{k_B T}{\hbar \omega_{AA}}\right)^{\eta N_ {AA,\sigma}}\left(\frac{k_B T}{\hbar \omega_{BB}}\right)^{\eta N_{BB,\sigma}}
    \end{split}
    \label{eq:4-5}
\end{equation}

Where the $N_{AB,\sigma}$, $N_{AA,\sigma}$, $N_{BB,\sigma}$ are the numbers of the corresponding bonding among the cluster configuration $\sigma$, which can be AAAA, AAAB, AABA, ABAA, BAAA, BBAA, ABBA, AABB, BABA, ABAB, BAAB, BBBA, ABBB, BBAB, BABB, and BBBB for the binary case. Then we get the free energy for this canonical sub-ensemble:

\begin{equation}
\begin{split}
F_{vib, \sigma} = -k_B T \left(\eta N_{AA,\sigma} ln \frac{k_B T}{\hbar \omega_{AA}} + \eta N_{AB,\sigma} ln \frac{k_B T}{\hbar \omega_{AB}} + \eta N_{BB,\sigma} ln \frac{k_B T}{\hbar \omega_{BB}}\right)
    \end{split}
    \label{eq:4-52}
\end{equation}
As the FYL-CVM’s CVM-like expansion would make sure each part of the crystal should be taken into account once and only once, we should generalize this vibrational contribution free energy term into all types of cluster and subcluster to make sure all the vibration contribution has been fully considered. Then insert back into the partition function for the specific type of cluster, then we can observe how vibrational effects participate in this thermodynamic model: 
\begin{equation}
\begin{split}
z_{\alpha} = \sum_{\sigma} exp\left(\frac{\mu_{\sigma}- (\epsilon_{\sigma}+ F_{vib,\sigma})}{k_B T}\right)
    \end{split}
    \label{eq:4-6}
\end{equation}

All the previous discussion is about how to implement the simple bond proportion model into our cluster-based configurational formalism. Then we are considering whether this formalism can be coupled with the local chemical and volume effects under the current framework.

If we include the local chemical environment into consideration or say the size mismatch effect, we can directly set up all the parameters $\omega_{AA}$, $\omega_{AB}$, $\omega_{BB}$ become local cluster configuration dependent: $\omega_{AB,\sigma}$, $\omega_{AA,\sigma}$, $\omega_{BB,\sigma}$. However, all these would increase the number of the needed extra parameters to be determined in the CALPHAD formalism to increase the burden of the thermodynamic modeling. People should consider how to balance simplicity and functionality under this flexible formalism in the future.

    \subsection{Illustration of Electronic Contribution}
Following the same coarse graining approach, we can even expand the partition function to include the electronic contribution by inserting more electronic dependent microstates:$z_{\alpha} = \sum_{\sigma} \sum_{v} \sum_{e}exp(\frac{\mu_{\sigma}-(\epsilon_{\sigma}+\epsilon_{v}+\epsilon_{e})}{k_B T})$. Then performing the same coarse graining, we can formulate the electronic microstate contribution into a cluster configuration dependent on free energy to cover the summation over the electronic microstates and insert this back into the cluster energy term. However, while the electron contribution is negligible compared to other contributions for the chemical structure transition, we will focus on discussing the vibrational part here in this Chapter, as promised. But we must mention that the electronic contribution could be reached freely through DFT calculation to get the density of state to calculate the electronic free energy with the fixed density of state approximation and then incorporate it into the CEM and finally into the cluster energy. The workflow details to get the electronic contribution is discussed in the next Chapter.

    \subsection{Model for Elastic Contribution}

Unlike vibrational contribution, elastic contribution is considered fully compositional dependent but not configurational dependent. We are assuming the equilibrium molar volume depends on composition but not on the state of order \cite{ferreira1988chemical}. People have already tried to separate this contribution from the configurational mechanism and can observe that this assumption works very well. Before we derive the elastic contribution for our mechanism, let’s see how to include the non-configurational contribution out of the previous configurational formalism derived with the straightforward mean-field approximation. The idea is to directly perform the mean-filed approximation among the atoms, which means we treat them as compositional dependence with a further degree of freedom.

Our main goal is to see how elastic contribution fits into this mechanism. The initial idea originated from a derived simplified equation based on Zunger $\epsilon-G$ method to separate the chemical configurational contribution and elastic contribution to consider the effect coming from elastic contribution for order-disorder transition. This elastic contribution mainly comes from the volume mismatch which would increase the system's energy. 

However, here we require a multi-component version to generalize it. Here we could provide an intuitive and simpler generalization of the original Zunger’s derivation based on the previously derived compositional dependent formalism to reproduce the exact same simplified expression we want. That’s because the microstate at one atom only depends on the species of the atoms, so we have the partition function here $z_{elastic,s}=exp(-\frac{E_s}{k_B T})$, where $s$ labels type of the specie of the atom. Then we could have $F_{elastic}=E_{elastic}=\sum_{s}n_s k_B T ln z_{elastic,s}=N\sum_{s} x_s E_{s}$. The reason why here $F=E$ is this formalism would not generate any entropy, as all possible configurational microstates are equivalent to each other due to their compositional dependence.
Then the problem remains how to define this $E_s$. We can expand this energy as 
\begin{equation}
\begin{split}
E_s = \sum_{r\neq s} w_r \Delta V_r
    \end{split}
    \label{eq:4-8}
\end{equation}
Where $r$ represents different species of atoms, $\Delta V_r$ is the corresponding volume change of type $j$ atoms per mol when adding one s atom, $w_r$ is the corresponding strain energy density. Then transform the $\Delta V_r$ into the atomic size level, $\Delta V_r = N_A \delta v x_r $, where $\Delta v_r$ is the volume change per atom and $x_r$ is the fraction of this element. Finally, we have:
\begin{equation}
\begin{split}
E_s = N \sum_{s} x_s E_s = N \sum_{rs,r\neq s} x_r x_s (N_A w_r \Delta v_r + N_A w_s \Delta v_s) = N \sum_{i\neq j} \Omega_{ij} x_i x_j  
    \end{split}
    \label{eq:4-9}
\end{equation}
Here we assume $\Omega_{ij}$ an averaged constant to represent $(N_A w_i \Delta v_i + N_A w_j \Delta v_j)$.for any possible pair of species. We want to insert this into our thermodynamic model for the elastic contribution. The parameterization of $\Omega_{ij}$ in CALPHAD based on the first-principles method is non-trivial. Still, it could reach the initial value whether through Vegard's law or the derived equation based on \cite{ferreira1988chemical}. However, as we simplified it with only some parameters, we might use the R-K polynomial to parametrize these factors.

based on our derivation, it’s adding one parabolic-like positive energy into the original free energy with compositional dependence. This can be understood as the difference between different phases is become smaller which means we would look forward to seeing the mixing region would become larger. We actually would see this in the later AB prototype alloys’ benchmark test.

It is worth mentioning that this derived formula of the elastic energy matches the similar one derived based on the mixed space cluster expansion for the constituent strain energy with the same $x(1-x)$ linear dependence. Notice that constituent strain energy is the direction-dependent strain energy for depicting the morphology of the microstructure, while the elastic energy for the phase stability can be seen as their average \cite{wang2023generalization}. This observation again confirmed what we derived about the composition-dependent formula for elastic energy.

Finally, we want to mention that after constructing the elastic energy term, we must subtract the corresponding value in the original configurational cluster energy term to keep the energy conservation. This means we separate the volume-dependent energy outside the configurational formalism since it has a fixed volume requirement. Then the cluster energy left is the actual chemical energy which dominates the cluster probability partition to determine the crystal configuration distribution.

\section{Result and Analysis}

\subsection{Vibration Contribution}
To observe how the added vibrational contribution affects the order-disorder transition, we select AB prototype FCC binary system as the platform to perform the benchmark test. We set up all the chemical configurational cluster energy, the same as the setting in the previous chapter to ensure both are comparable. 

To consider the vibrational contribution, we can select the bond proportional model we mentioned before to set up the frequency for the three types of bonding: AA, BB, AB. Here we can take the frequency term as the linear temperature-dependent energy term representing the linear temperature-dependent relation of the vibrational energy indicated by the equipartition theorem at the high temperature solid \cite{landau2013statistical}. It’s a rough assumption but here we only need one vibrational energy format to test this model. Later we would also notice the setting of the form of the frequency is not the essence in the current bond proportional model coupling with our cluster-based thermodynamic framework. Notice here that we force all the expressions mentioned later would have the form $\hbar \omega_{ij}=Constant(k_B T)$ format and we leave $\hbar$ and $k_B T$ out but only mention $\omega_{ij}=Constant$ to make the illustration clear.

In order to observe how the vibration affects the chemical configurational contribution, we can first target the order-disorder transformation along the fixed composition $A_{0.5} B_{0.5}$. Then we test different combinations of the values for the $\hbar \omega_{AA}$, $\hbar \omega_{BB}$, $\hbar \omega_{AB}$ to observe how the order-disorder transformation happens and calculate the related thermodynamic quantities.

\begin{figure}
    \centering
\includegraphics[width=1\textwidth]{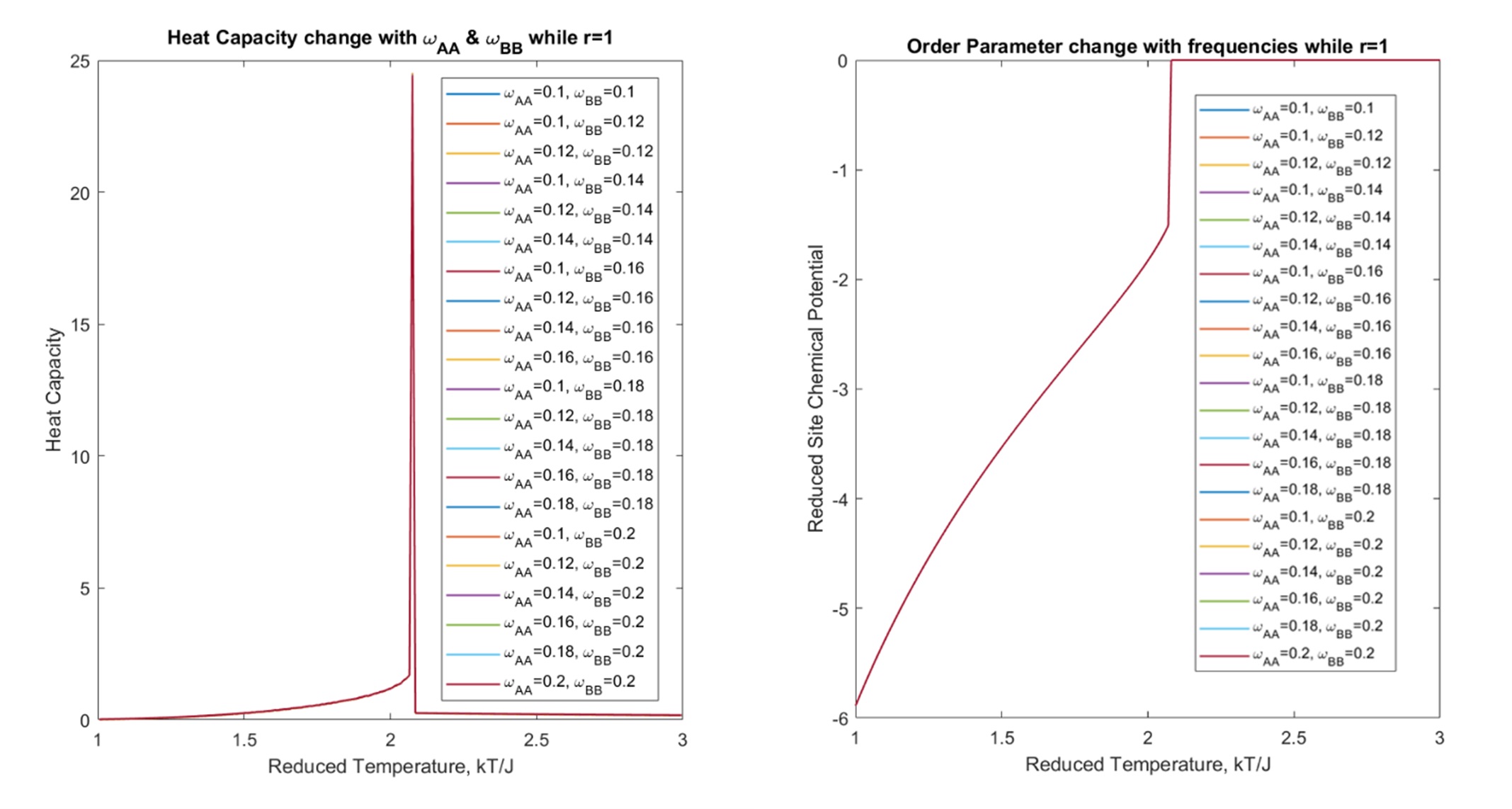}
    \caption{The heat capacity and the reduced site chemical potential at $A_{0.5}B_{0.5} $ with $r=1$.}
    \label{fig:vibTP1}
\end{figure}

We define the relation constant between these three parameters as $r=\frac{\omega_{AB}}{\sqrt{\omega_{AA}\omega_{BB}}}$ to help us illustrate the benchmark test. First of all, we can test the different relations between the frequencies. We mainly want to observe two cases: 1. Keep  $r=1$ and change the values of $\omega_{AA}$ and $\omega_{BB}$, 2. Change $r$ to observe how different $r$ would affect the thermodynamic quantities. We fix the composition at the $A_{0.5}$ $B_{0.5}$ to see how these thermodynamics change with the temperature to observe the order-disorder transformation behavior along this $L1_0$ ordering composition.

Firstly, we present the results when we keep $r=1$. When $r=1$, the configurational ordering transformation would not be disturbed by any vibrational contribution in figure \ref{fig:vibTP1}. We present the reduced site chemical potential of the first site as the order parameter and see how it changes with the different frequencies. The reason it can be seen as the order parameter is that the symmetry of the site chemical potential must be constrained, allowing it to reach 0 at this particular composition. Here the reduced means it would reduce to $\mu/k_B T$ to be a dimensionless quantity. This means the system comes to the disordered phase when the reduced site chemical potentials all reach 0. The observation from the order parameters and the heat capacity could conclude that when $r=1$, the atomic vibration described by the bond proportion model would not affect the order-disorder transformation.
\begin{figure}
    \centering
\includegraphics[width=1\textwidth]{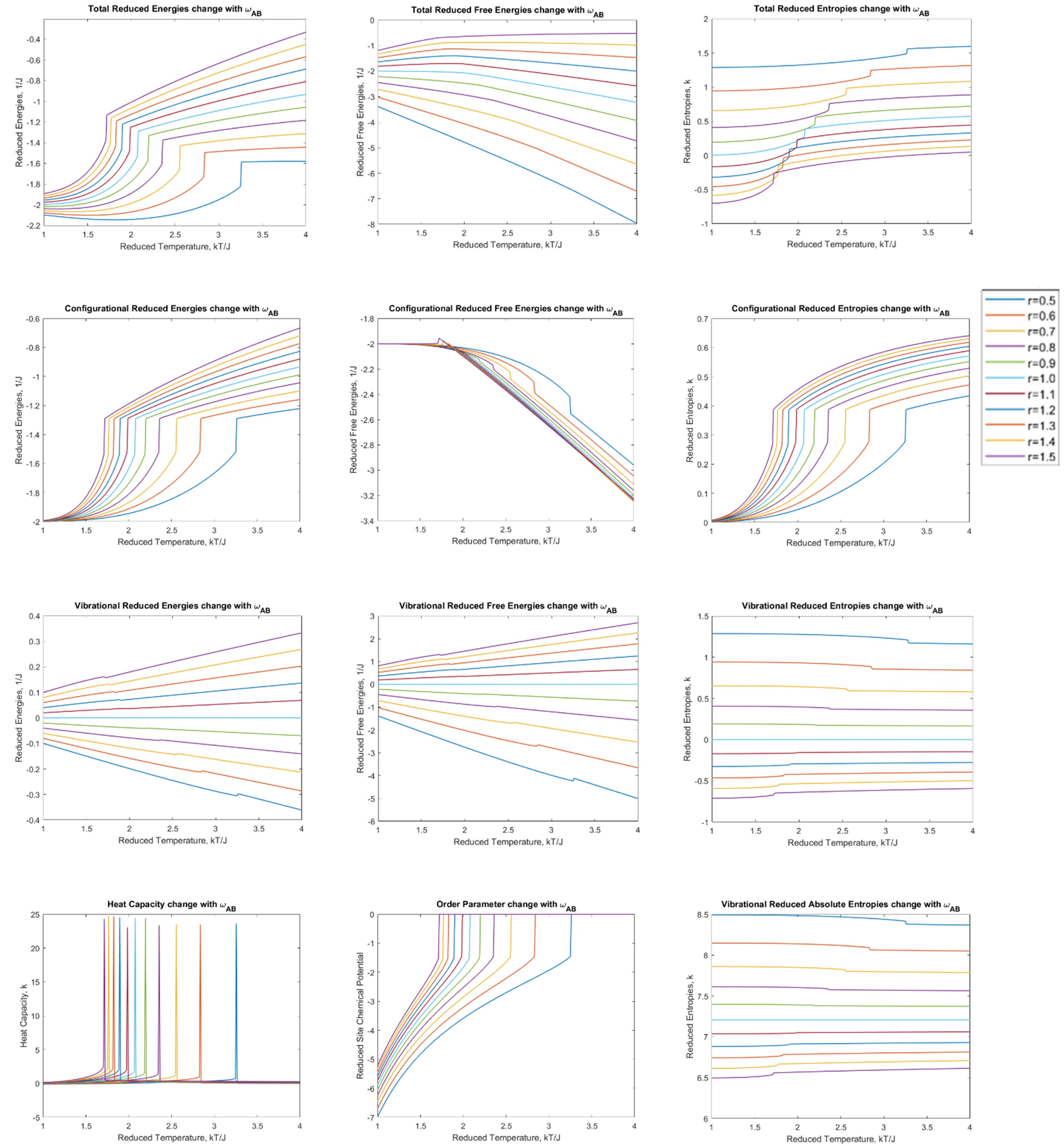}
    \caption{The thermodynamic properties when $r \neq 1$ for $A_{0.5} B_{0.5}$.}
    \label{fig:vibTP2}
\end{figure}
Then we fix $\omega_{AA}=\omega_{BB}=1$ and change the ratio $r$ to observe how the vibrational contribution represented by this ratio affects the configurational transformation along the composition $A_{0.5} B_{0.5}$. The results are presented in figure \ref{fig:vibTP2}. Many exciting features are revealed in this series of results about the related free energy, energy, entropy, heat capacity, and order parameter. First, we can observe that the vibrational contribution would shift the order-disorder transition temperature from the different directions: if $r<1$, the transition temperature increases; when $r>1$, the transition temperature decreases. Another significant observation is about the negative vibrational mixing entropy. In fact, it has already been reported in some experimental work before, and is concluded that this phenomenon should attribute to the stiffening of bonds \cite{fultz2010vibrational,delaire2004negative}. 

However, this doesn’t mean it would violate the thermodynamic third law. They should be allowed to have the negative part because they are all the mixing entropy. We provide the results for the absolute vibrational entropy presented in the figure, it reveals that the absolute vibrational entropy dominates the whole entropy and has a large enough absolute value during the whole temperature range and guarantees that the whole system would follow the fundamental thermodynamic third law.

Besides, we can also observe that the vibrational contribution would distort the pure configurational contribution in different directions due to the different values of $r$. However, we can notice that the pure configurational part contributes more “difference” for the internal energy and the entropy during the transformation, indicating that it’s still the configurational part that dominates this chemical configurational transformation rather than the vibrational contribution. Notice the “difference” mentioned is due to our approximation on this FYL-CVM, while the order-disorder transition should be a second-order transition. But this “difference” is still useful here because this actually measures how “fast” or say “rapid” for the change of the thermodynamic quantities during the phase transformations. The results presented show that the dominant “difference” still comes from the configurational part. The total free energy would continue here while the other two kinds of free energy discontinue. This is because the vibrational free energy is calculated inside the FYL-CVM mechanism, so we have experienced this approximation to have such as discontinuity. The total free energy curve is calculated due to the variational minimization calculation so that it keeps continuous. Finally, the configurational part has to have discontinuities due to energy conservation.

These previous computational observations reveal some exciting features of this bond proportion model-defined vibrational contribution. We can even make it further through simple analytical observations to see these properties more clearly. Starting from the vibrational mixing energy, we can analytically calculate all these based on the different cluster configurations. Then we should have three non-zero symmetry equivalent cluster configurations:
\begin{equation}
\begin{split}
F_{vib,AABB,mix} &= -2k_B T \eta ln\left(\frac{\hbar \omega_{AA} \hbar \omega_{BB}}{\hbar^2 \omega_{AB}^2}\right)\\
F_{vib,ABBB,mix} &= -\frac{3}{2}k_B T \eta ln\left(\frac{\hbar \omega_{AA} \hbar \omega_{BB}}{\hbar^2 \omega_{AB}^2}\right)\\
F_{vib,AABB,mix} &= -\frac{3}{2}k_B T  \eta ln\left(\frac{\hbar \omega_{AA} \hbar \omega_{BB}}{\hbar^2 \omega_{AB}^2}\right)
    \end{split}
    \label{eq:4-10}
\end{equation}

We can observe that all three different types of the tetrahedron configuration actually only depend on the ratio $\rho = \frac{\hbar \omega_{AA} \hbar \omega_{BB}}{\hbar^2 \omega_{AB}^2} = \frac{1}{r^2}$ due to its intrinsic set up of this bond proportion model. This explains why we observed only the ratio determining which direction shifts the phase boundary between the ordered and disordered phases. Intuitively we consider three types of cluster energy during the mixing: AAAB, AABB, ABBB. When the vibrational contribution is negligible. After considering the atomic vibration, it would increase or reduce some specific energy altogether due to the vibration. If the cluster energy is increased with the $\rho<1$ and $r>1$, the disordered phase would be easier to form, and the transition temperature would be reduced. On the other hand, if the cluster energy is decreased uniformly with the $\rho>1$ and $r<1$, the disordered phase would be difficult to form with the higher cluster energy stability, and the transition temperature would increase.

However, we should mention that this theoretical observation is only based on the binary system and the tetrahedron basic cluster approximation for the bond proportion model. This may be varied due to the multi-component case and the different selection of the basic cluster. We would observe this in future work when dealing with the realistic multi-component system.

  \begin{figure}
    \centering
\includegraphics[width=0.7\textwidth]{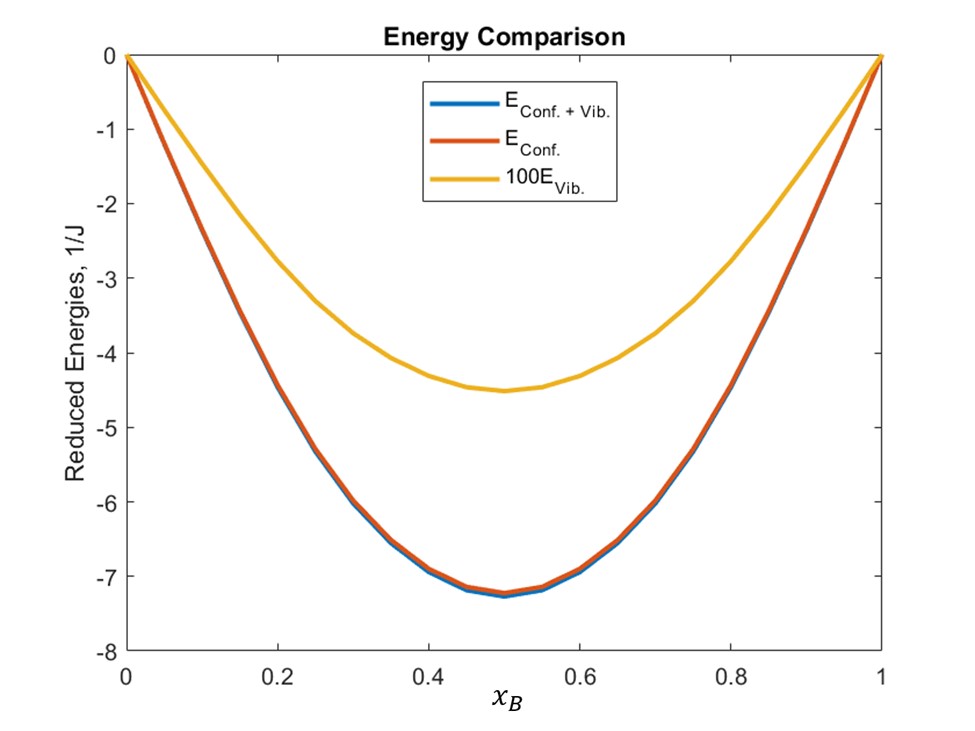}
    \caption{The different energy contributions for the whole mechanism at $T=2.5$ and set up the vibrational settings as: $\hbar\omega_{AA}=\hbar\omega_{BB}=0.1k_B T$, we take $\omega_{AB}=0.9\sqrt{\omega_{AA}\omega_{BB}} )=0.09k_B T$.}
    \label{fig:vibE}
\end{figure}
  \begin{figure}
    \centering
\includegraphics[width=0.7\textwidth]{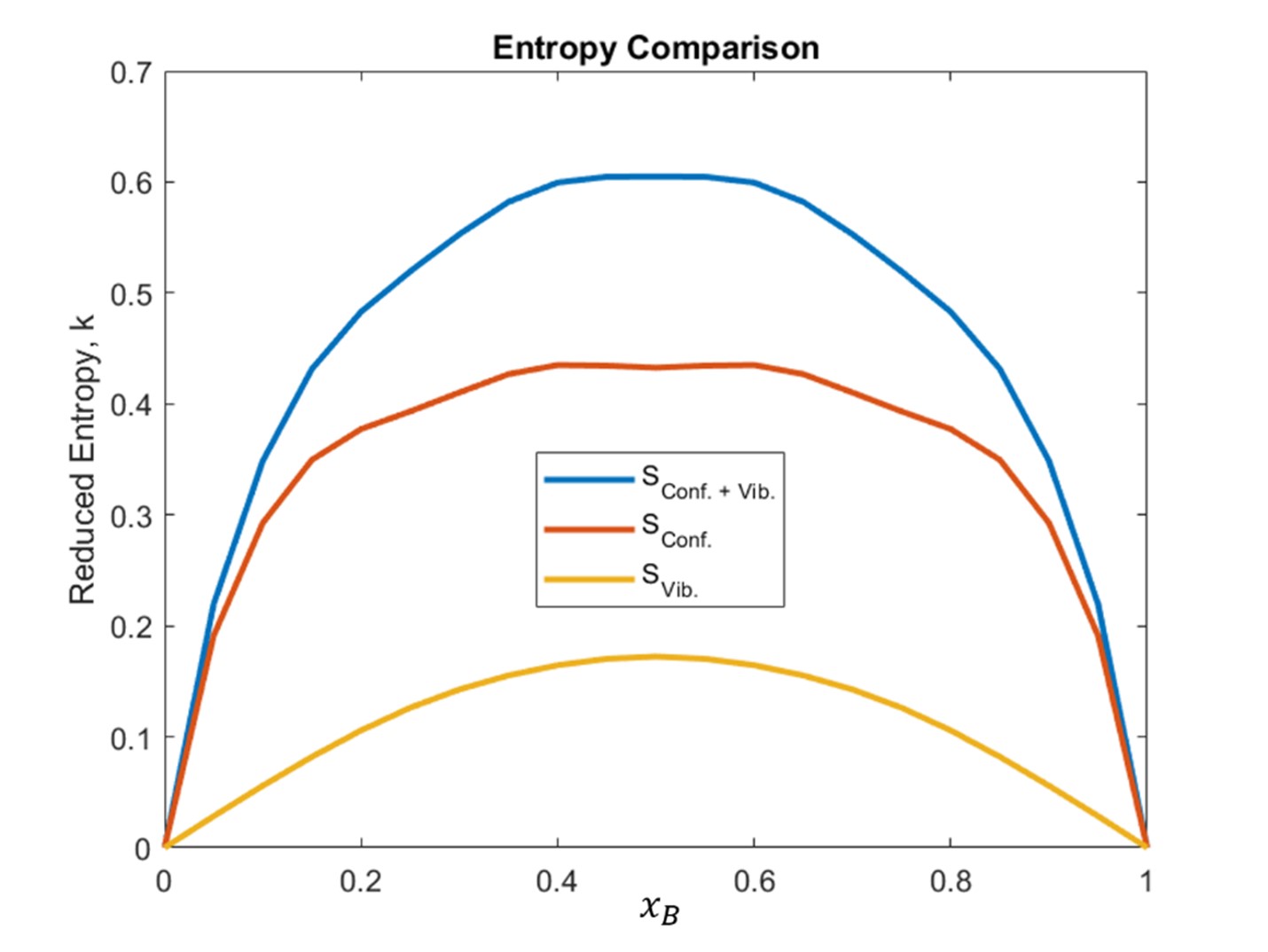}
    \caption{The different entropy contributions for the whole mechanism at $T=2.5$ and set up the vibrational settings as: $\hbar\omega_{AA}=\hbar\omega_{BB}=0.1k_B T$, we take $\omega_{AB}=0.9\sqrt{\omega_{AA}\omega_{BB}} )=0.09k_B T$.}
    \label{fig:vibS}
\end{figure}

We can also compare how the atomic vibration and chemical configuration contribute to the disordered phase's phase stability by directly comparing the thermodynamic quantities among the whole composition range. We select the vibrational related quantity: $\hbar\omega_{AA}=\hbar\omega_{BB}=0.1k_B T$, we take $\omega_{AB}=0.9\sqrt{\omega_{AA}\omega_{BB}} )=0.09k_B T$, then we keep the same chemical configurational settings but take the mixing formalism to make sure all of these are comparable. We fix the reduced temperature as $T=2.5$, and then observe how the curves represented by different sources vary along the composition. The results are presented in figure \ref{fig:vibE} and figure \ref{fig:vibS}.

The vibrational contribution of the mixing entropy is obvious and would change the total entropy a lot among the whole composition range. Compared to the mixing entropy, the mixing energy from the atomic vibration is relatively small, and we must enlarge it by ×100. This indicates that mixing entropy is the main routine of the vibrational contribution to affect the thermodynamics of the system. At the same time, the vibration's mixing energy would remain almost the same during the mixing process. This indicates that vibrational entropy is the primary source to impact the phase stability at the high temperature compared to the vibrational energy. Overall the trend of the curve presented here qualitatively matches some early results about the comparison between the different sources of the thermodynamics based on the Debye model \cite{asta1993theoretical}.

After figuring out the thermodynamic quantity calculation, we can deal with the phase diagram calculation with the vibrational contribution. We keep the chemical cluster energy the same as the one in the previous chapter as before to make sure all the results are comparable and set up the vibration-related parameters as: r=0.9, 0.95, 1, 1.05, 1.1 to observe how the ratio between different frequencies affects the phase diagram. 

All the phase diagrams and the related thermodynamic energies are calculated through the enumerated sampling necessarily in dense chemical potentials $\lambda_i$ for different phases with different symmetry to reach the minimum to get the equilibrium free energy with the composition as the constraint. After sampling the interested composition and temperature range to get enough free energy data, we use the convex hull calculation to estimate the phase boundaries over the whole temperature range between the different phases to formulate the phase diagram. A more efficient minimization algorithm is in progress to be developed, but the produced thermodynamic calculation based on the more efficient algorithm should be close to the current one.
  \begin{figure}
    \centering
\includegraphics[width=0.7\textwidth]{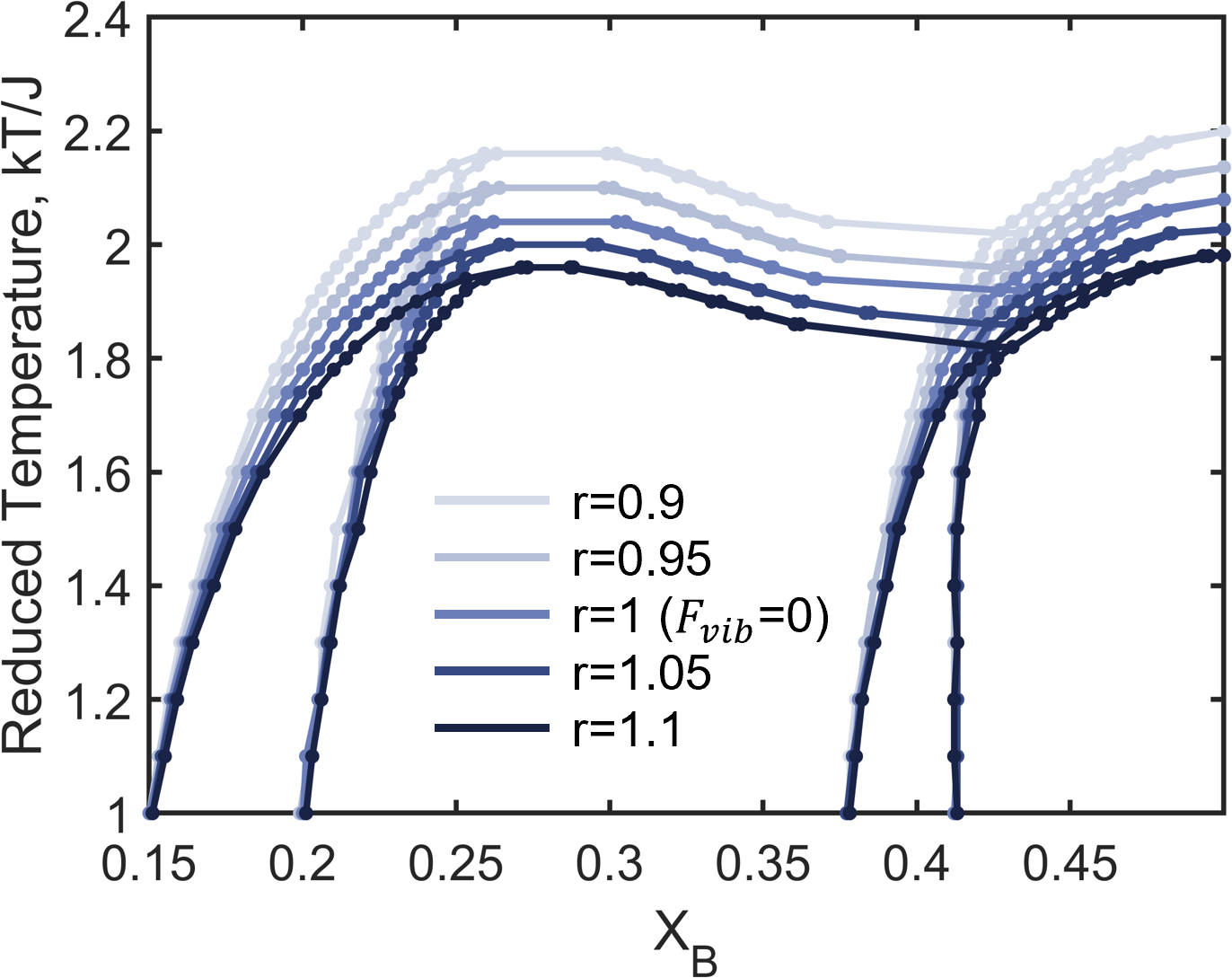}
    \caption{The phase diagram with different vibrational contribution at different ratio $r$ in the AB prototype system.}
    \label{fig:vib1}
\end{figure}
  \begin{figure}
    \centering
\includegraphics[width=0.7\textwidth]{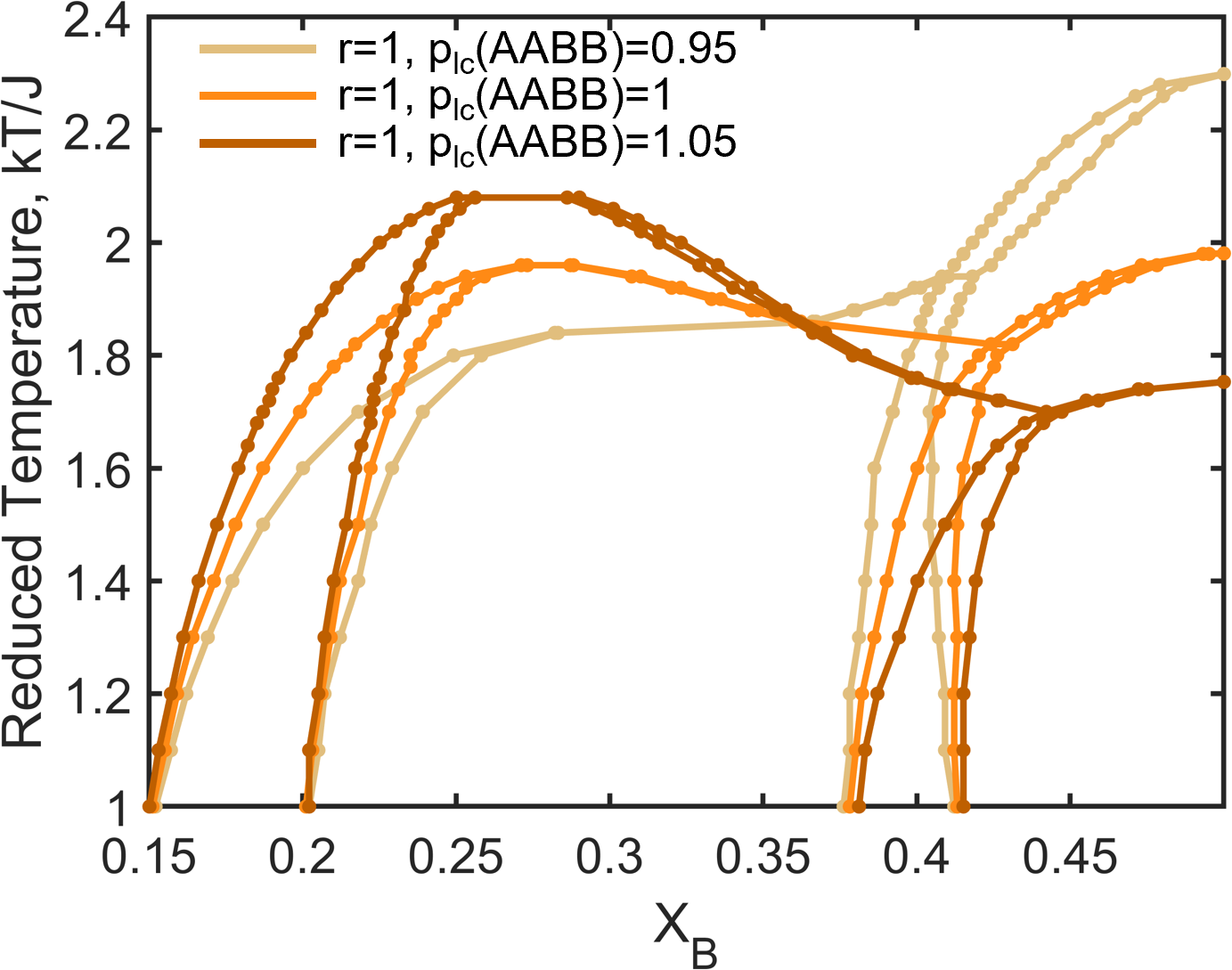}
    \caption{The phase diagram with a different vibrational contribution at the same ratio $r$ but different local chemical parameter$p_{lc}$ in the AB prototype system.}
    \label{fig:vib2}
\end{figure}
The phase diagrams including the vibrational contribution are presented in the figure \ref{fig:vib1}. We can observe that the vibrational contributed phase diagram would shift after the original pure chemical configurational contributed phase diagram. This means all this vibrational contribution affects the phase stability similarly so that the phase boundaries shift uniformly among the composition range with the current settings.

The previous calculation is about the simple bond proportion model that fixes the bond frequencies in every cluster environment. As mentioned above, the possibility of letting the vibrational frequencies have the local cluster configuration dependence to represent the effects is caused by the size mismatch or the local chemical environment. To test it, we can multiply one local cluster-dependent parameter $p_{lc}$ onto the frequency term: $\omega_{AB}\to p_{lc} \omega_{AB}$. This parameter effectively measures how the local cluster environment affects the frequency $\omega_{AB}$ when the AB bond lies in different environments. As mentioned above, the cluster vibrational free energy within this bond proportional model only depends on the ratio $\rho$. Equivalently this generalization extends the original one to let the $\rho$ have the local cluster environment dependence. 

We take a straightforward test to present this generalization on the benchmark prototype AB FCC binary system at the end of this section to show how flexible and versatile this current proposed model is. We set up all the chemical configurational parameters the same as the previous one and then set up the frequencies’ ratio as $r=1.1$ and set the previously defined parameter $p_{lc} (AAAB type)=p_{lc} (ABBB type)=1$ and make the $p_{lc} (AABB)=0.95$ or $1.05$. The results are presented in figure \ref{fig:vib2}.

We can observe that only a tiny change of the local chemical-related parameters would lead to a big nonuniform shift of the phase boundaries over the whole composition on the phase diagram. This nonuniform shift changes the phase boundaries of the different ordering with the different directions which implies the local chemical effects would somehow obviously affect the phase stability. Compared to the original one, one key feature is the shifted directions around $L1_0$ and $L1_2$ are very different. Here we are trying to incorporate the local chemical or say the size mismatch effect into the bond proportion model. All this observation matches the previous computational results which mentioned the importance and necessity of including the local chemical effect and the possibility of revising the bond proportion model to make it fit for the local environment \cite{morgan1998local,morgan2000vibrational}.

\subsection{Elastic Contribution}
In this section, we calculate how the inserted elastic term affects the thermodynamics under the proposed framework. This composition-dependent elastic term is a parabolic positive energy term out of the configurational contribution so that it would not affect the specific order-disorder transition temperature under the fixed composition but shift the boundary of the mixing region between different phases definitely. 

  \begin{figure}
    \centering
\includegraphics[width=0.7\textwidth]{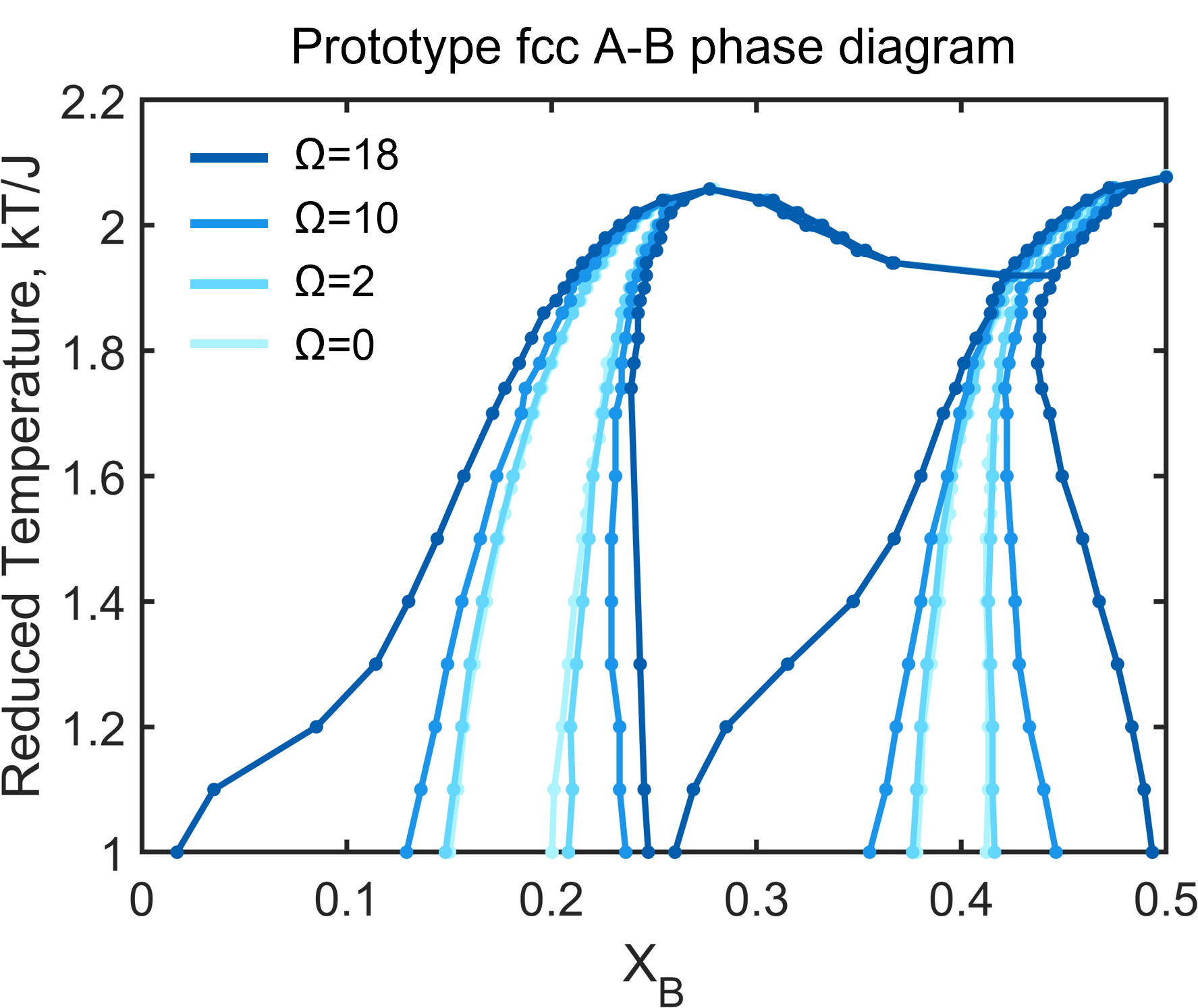}
    \caption{The phase diagram with different elastic constants in the AB prototype system.}
    \label{fig:elasticPD}
\end{figure}

Using the AB prototype FCC binary alloy as the benchmark test platform, we can clearly observe how this elastic energy term affects the phase boundaries. We keep all the configurational cluster energy settings as before, but add one more compositional-dependent positive energy term out of the configurational formalism. In the derived formula, the constant $\Omega$ should have the unit, so we impose this elastic related quantity as $\Omega=\Omega_0 k_B$, we can assume $\Omega_0$ is a pure dimensionless constant, and $k_B$ would implement the physical meaning as the energy into it. The corresponding phase diagram for the prototype AB system is presented in figure \ref{fig:elasticPD}.

Based on the results presented, we can observe that the elastic energy would enlarge the mixing region of the ordered and disordered phase more and more obviously when enlarging the $\Omega_0$ which characterizes the strength of the elastic energy of the system. This results from the parabolic shape of the elastic energy due to the mismatch strain effect. The phase at the center of the phase diagram would increase more energy, while the phase at the edge of the phase diagram would increase less energy. This behavior caused by the elastic energy would reduce the energy difference between the phase at the center and the phase at the edge, which would broaden the mixing region.

  \begin{figure}
    \centering
\includegraphics[width=0.7\textwidth]{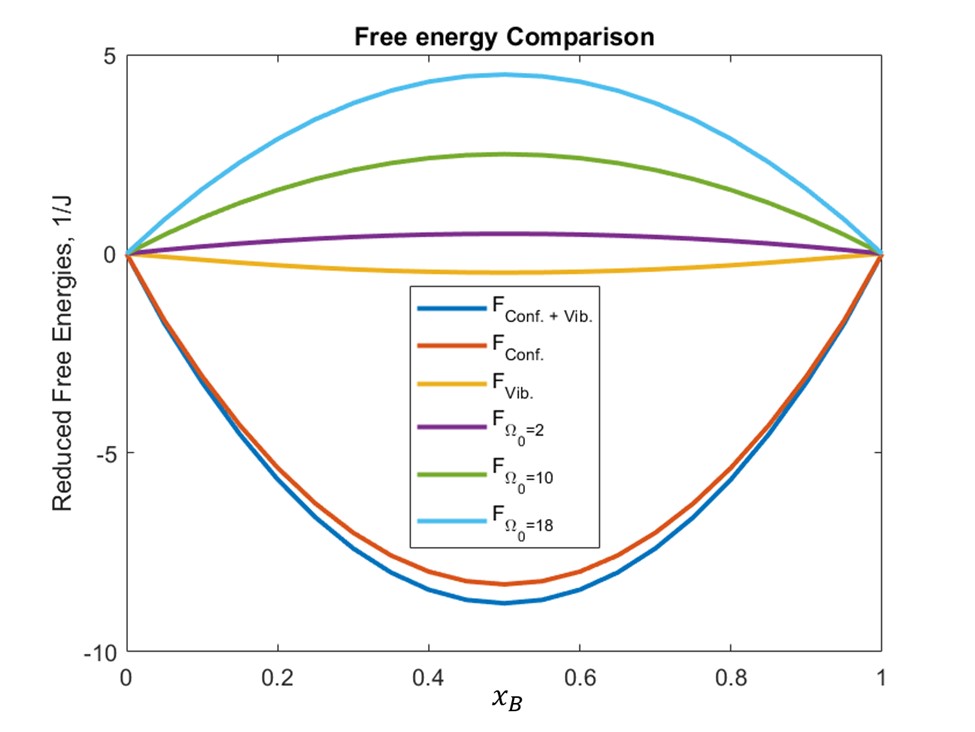}
    \caption{The free energy among the whole composition with both elastic and vibration contribution in the AB prototype system. The vibrational setting is $\hbar \omega_{AA}=\hbar \omega_{BB}=0.1k_B T$, and take $\omega_{AB}=0.9\sqrt{(\omega_{AA} \omega_{BB} )}=0.09k_B T$, the elastic setting is different elastic constant $\Omega_0=2,8,18$, temperature is set up at $T=2.5$.}
    \label{fig:vibelasticF}
\end{figure}

Finally, we can compare the different sources of the free energy of the disordered phase to see the importance of the elastic contribution during the phase stability calculation. We keep selecting the vibrational related quantity: $\hbar \omega_{AA}=\hbar \omega_{BB}=0.1k_B T$, and take $\omega_{AB}=0.9\sqrt{(\omega_{AA} \omega_{BB} )}=0.09k_B T$. We keep the same chemical configurational settings but take the mixing formalism to make sure all of these are comparable. We present the elastic-related thermodynamic quantities when $\Omega_0=2,8,18$ to make the comparison. We fix the reduced temperature as $T=2.5$. Then we can directly compare the mixing of free energy from the different sources over the full composition range. 

The calculated results are presented in figure \ref{fig:vibelasticF}. We can notice that firstly the mixing of vibrational free energy is obvious to be observed, which indicates the importance of the vibrational contribution. For the elastic contribution, we notice that this positive parabolic energy contributed is comparable to the chemical configurational and vibrational contributions. This matches the previous observation in the realistic alloy system, which indicates the importance of the significant elastic contribution and shows a similar behavior \cite{ferreira1988chemical,asta1993theoretical}.

\section{Discussion}
In this Chapter, we have generalized our previous thermodynamic model with intrinsic SRO to include the non-configurational degrees of freedom to see how contributions out of the chemical configurational contribution affect the ordering behavior for the multi-component system. The configurational and non-configurational (here concentrated vibrational, electronic, elastic) contributions to free energy are modeled separately, gaining insights into their respective effects on phase stability. 

For the vibrational contribution, we use the “coarse graining” to take the vibrational contribution under the FYL-CVM formalism to see if this weak configurational-dependent contribution would affect the phase stability. After reviewing all three possible sources of the vibrational entropies, the “bond proportion” effect, the volume effect, and the size mismatch effect, we select the “bond proportion model” as the compromise of the accuracy and convenience for the further coupling with the CALPHAD method. Under the prototype AB FCC binary system as the benchmark test, we can observe the atomic vibration does affect the configurational order-disorder transformation very much just as we expected. The trend of how atomic vibration affects the order-disorder transformation is also detailed and studied in the selected bond proportion model computationally and analytically. The results presented here satisfied the expectations and matched the conclusions within the previous literature: the thermodynamic quantities and the phase diagrams would be shifted almost uniformly based on the ratio between the frequencies $\rho = \frac{\omega_{AA}\omega_{BB}}{\omega_{AB}}$.  Interestingly, if we modify the bond proportion model slightly to let the related frequency become the cluster environment dependent but not just fix it, the corresponding phase diagram would be largely changed. This somehow may indicate the importance of the size mismatch effect and the local environment for the atomic vibrations. This adjustment of the implemented model also indicates the potential of the current framework to include not only the “bond proportion” effect of the atomic vibration but also at least partially simulate the volume effect or the size mismatch effect.

We separate this assumed composition-dependent term for the elastic contribution and provide a multicomponent parabolic simple form to express this elastic energy. The key feature of this elastic contribution is it would enlarge the mixing region while keeping the topology of the phase diagram determined by the configurational contribution. This makes the phase diagram closer to the realistic experimental one and the calculated elastic energy among the whole composition on this benchmark test prototype system is qualitatively comparable with another source of free energy and matched the previous observation in the literature as we mentioned above. 

Though we have detailed discussed the current framework for the configurational and non-configurational contribution from the different sources of the contribution to the phase stability, there are still a lot of open questions or say challenges remain. We have to say it even generates more questions and leave more necessary exploration for the future work after the establishment of this framework.

The versatility of the new solution model means it applies well to metallic, ionic, and semiconductor alloys while offering the possibility of considering larger clusters and/or more components within the CVM hierarchy. Here we only consider the case for the FCC due to its simplicity, however, generalization into the different lattice systems and the different basic cluster approximation should be natural and easy to reach. Similar case for the materials out of the metallic alloys. As this model is designed universally, it would also be suitable for other multi-component systems to capture the intrinsic SRO as well. As the strength of the SRO may vary due to the different materials, for example, the ionic ceramic system may have much stronger bonding and SRO, the features of the results might be different from those presented here, and the approximation may also be replaced, or imposed limit the current framework to make it properly designed for some specific materials.

The minimization algorithm is non-trivial as the free energy functional of the CVM-like expansion is non-convex, so the minimization to approach the equilibrium is a typical constraint non-convex optimization and has some specific method to deal with that \cite{pelizzola2005cluster}. This indicates the possibility of being trapped into some local minima instead of the global minima over the landscape with the gradient descent or Newton method. However, as the meaningful optimal point should correlate to the symmetry of the parameters to represent the realistic phase, we may also implement some symmetric constraints to avoid the meaningless local or global minima. This would be an exciting problem to be pursued as a next step.

The coupling between this framework and the first principles calculation should be natural and necessary. Determining the energy parameters based on the first-principles method is the next problem for future work to deal with the realistic alloy system, especially for the most straightforward chemical configurational cluster energy. DFT can estimate all the cluster energy but may need some tricky to provide more meaningful chemical cluster energy. This is precisely what we want to figure out when we deal with the realistic alloy system. We can use the first-principles method to estimate the initial value to help the CALPHAD modeling for the vibrational and elastic energy parameters. However, due to the simplification during considering these two sources of contributions, the detailed consideration for approaching a reasonable initial value for these vibrational and elastic parameters would be left to the future.

Besides, the connection between the current framework and other potential contributions coming from the external fields and other intrinsic physical degrees of freedom such as the magnetic SRO is still a challenge to be considered. However, we have already shown the possibility of combining them in a composition-dependent term like the elastic contribution or the coarse graining by considering the vibration as the sub-degree of freedom of the chemical configuration. This reveals the huge potential to combine other exciting factors which affect the phase stability by this statistical cluster framework.

Here we didn’t discuss too much about the electronic contribution but would directly introduce the workflow to add this contribution in the next Chapter, as it’s generally negligible for such macroscopic thermodynamic properties. However, it is worth mentioning that we noticed the developed FYL-CVM can, on the other hand, provide the predicted local cluster probability for the electron transport calculation under the non-local coherent potential approximation (NLCPA) framework \cite{rowlands2009short}. This would be very helpful to study how configurational disorder and the chemical SRO affects the electron transport and the electronic structure of the multicomponent materials such as the CCAs in future work \cite{mu2018electronic}.
While the challenges would be targeted on in future work, the presented results have already shown the current cluster-based framework would open a new possible routine to depict the detailed thermodynamics of the solid solutions phase for the multicomponent materials. We believe the FYL-CVM and the associated proposed computational thermodynamic framework coupling CALPHAD would benefit a lot for the materials science community. 

The key feature of the current framework is that it can conveniently be generalized into the multicomponent and keep the clear physics, especially the intrinsic chemical SRO inside it. This is the result of applying the FYL-transform mentioned in the previous Chapter. After equipping more physical contributions, including the atomic vibration and the elastic deformation, under the framework within this work, this feature has been adapted. It should be followed when including even other types of intrinsic physics, such as magnetic contribution, in the future. As a result, it is possible for us to achieve the balance between accuracy, multiple physics, and efficiency under this framework for the CALPHAD method. This is the key advantage from the fundamental perspective while we are trying to include more physics into the current framework.

From the application from the industry perspective, one more advantage is that the physics meaning of all the parameters and how it shifts the phase boundaries are clearly observed and satisfy the expectation in this work. This could benefit the thermodynamic assessment of the solid solution phase for the realistic system in the future to save the state-of-art optimization procedure to fit the experimental data. It is because the traditional CALPHAD method requires back and forth many times adjustments of several parameters, while some of them may even have no physical meaning. These back-and-forth parameter adjustments are necessary to produce the proper phase diagrams integrated with all first principles and experimental phase equilibria or thermal chemical data. However, the current framework perhaps would be helpful to save the efforts during this procedure. Firstly, as all the parameters used in this framework have a clear physics meaning, the first-principles calculations should determine at least the initial value, and the value range should be comparable with the first-principles data. Secondly, the change of the phase boundaries based on the change parameters represents the vibrational or the elastic contribution expected by the prototype systems' previous observations due to their clear physical intention. Adjusting the parameters during the assessment should imply the physical meaning and is helpful for people to get the correct intuition to produce the required phase diagram. The topology of the phase diagram will be primarily determined by the cluster chemical energy terms, which can have a relatively close initial value based on the first-principles calculations. The vibrational and elastic terms are used to remove discrepancies with experimental phase equilibria. All these observations indicate that this cluster-based framework is exactly suitable for coupling with the CALPHAD method from both the fundamental and the application sides.

\section{Conclusions}
We summarize this Chapter as further developing the thermodynamic model proposed in the previous Chapter to include other sources of physics that affect chemical ordering. For the weak configurational dependence contribution, we use the “coarse graining” process and implement the bond proportion model for vibrational contribution. For the compositional dependence contribution, we separate it from the configurational term and use mean-field approximation among the atoms directly to take the $\epsilon-G$ method. The benchmark test based on the prototype AB system reveals the power of the currently developed formalism, as the calculated results matched the trend of how the vibrational and elastic contribution affects the ordering phase stability.  

Based on all the current models, calculations, and discussions, we can conclude that:

1) Multiple different physics, including atomic vibrations, electronic excitation, and elastic deformation, can be conveniently and effectively implemented into the cluster-based thermodynamic model proposed in the previous Chapter successfully to study how different physics-originated contributions affect the phase stability and make the proposed thermodynamic framework completely capture the whole physics.

2) The atomic vibration implemented through the coarse graining would affect the phase stability case by case, mainly determined by the different bond frequencies under the local cluster environment. At the same time, we integrated the generalized bond proportion model into this cluster-based thermodynamic model.

3) The elastic contribution implemented through the composition-dependent parabolic formula would mainly affect the phase stability by enlarging the mixing region of the phases.

4) Integrating all these contributions with the intrinsic chemical SRO in the proposed cluster-based thermodynamic framework would be helpful for future CALPHAD modeling due to its balance between accuracy, multi-physics, and efficiency.

Overall, we are looking forward to seeing how this multi-physics-originated computational thermodynamic framework with intrinsic SRO developed in this work would contribute to the computational thermodynamics for the multi-component ordering solid solution by coupling with the CALPHAD method in the future to benefit the community.

\chapter{Workflow with Algorithms for Phase Diagram Calculation} 
 \label{ch:algo}
 \section{Introduction}

 \begin{figure}
    \centering
\includegraphics[width=1\textwidth]{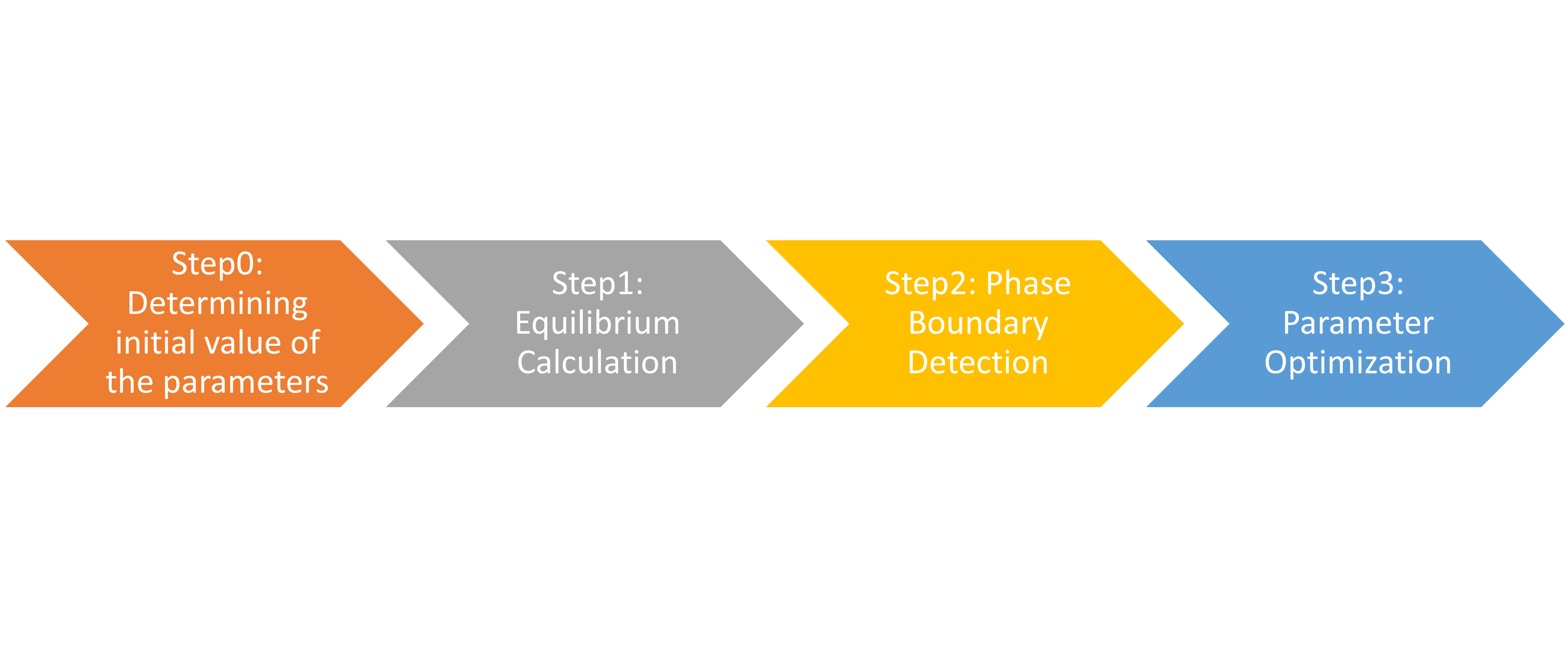}
    \caption{The four steps workflow to get the phase diagram discussed in this Chapter.}
    \label{fig:workflow4}
\end{figure}
After constructing the fundamental theoretical framework, we must develop the corresponding workflow with algorithms to make the framework computable to get the phase diagram. Mainly three steps are involved in our variational framework in CALPHAD: equilibrium calculation, phase boundary detection, and parameter optimization. Before that, we need to determine the initial value of the parameters. One illustration diagram is presented in figure \ref{fig:workflow4}. Our focus would be the equilibrium calculation since developing the algorithm for the phase boundary detection and the parameter optimization need to coordinate with the current existing algorithms.

First, we will construct the workflow to obtain the cluster energy. Determining the cluster energy based on the first-principles data is a challenge after constructing this thermodynamic framework. It involves the chemical contribution, which would play a role in the configuration-dependent free energy. With all these considerations, we finally clarify the connection between the DFT calculated data and the cluster energy as well to build up a workflow to cooperate with the DFT calculations to get the cluster energy efficiently. This would benefit automatically performing high-throughput calculations with the current FYL-CVM thermodynamic model in the future.

The equilibrium calculation is unique for these variational thermodynamics models of the solid solution. It requires minimization of the free energy function to determine the equilibrium state \cite{kikuchi1974superposition}. The computational efficiency of such equilibrium calculation usually is the bottleneck for 
the practical applications \cite{oates1996cluster}. We developed the corresponding minimization algorithms for FYL-CVM to deal with this bottleneck. 

After the single equilibrium calculations, a sampling algorithm is required to infer the phase boundaries based on the single-point equilibrium calculations. The main challenge is that the existence of the phase boundary could be considered a kind of ``rare event". We need an efficient way to sample the composition-temperature space $x-T$ or chemical potential-temperature space $\mu-T$ to get the phase diagram. On the other hand, these phase boundary detection algorithms should be stable enough to resist the potential inaccurate results caused by the non-convexity of the free energy functional \cite{pelizzola2005cluster}.

With the help of equilibrium calculation and the phase boundary detection algorithm, the phase diagram could be constructed conveniently. The next step for CALPHAD modeling is to combine the experimental and first-principles data and the thermodynamic model to perform the parameter optimization to reach the exact thermodynamic descriptions of the target system. For the solid solution phase, parameter optimization is usually challenging and must be performed manually. The difficulty may come from the unphysical thermodynamic model and the primitive parameter optimization algorithm. However, to automatically make the parameter optimization of CALPHAD, we still require an algorithm to optimize the parameters for the variational model based on the specific system's experimental and first-principles data. In our current framework, these energetic parameters could mainly attribute to the cluster energy-related parameters to determine the specific thermodynamic for the required system. Many challenges impede the development of automatical optimization algorithms as we concluded it is multi-object and multi-modal. However, it's still necessary to have an algorithm to drive the direction to adjust the parameters. Through this direction, we finally noticed that all the parameter optimization with the variational thermodynamic model should be considered as the bilevel optimization \cite{franceschi2018bilevel} and propose a potential algorithm to deal with this issue.

 \section{Step0: Workflow for determining Initial Value of the CALPHAD Parameters based on First-Principles Calculations}
The goal of CALPHAD is to construct a learning model to better and adequately make use of the input data. A better model that satisfies the inside physics would provide better-fitting results. As we have proposed a fundamental physical model based on FYL-CVM, the data is another significant aspect of CALPHAD. The acceleration of the corresponding database is necessary for future database construction. For the experimental data, a high-throughput experimental technique is the way to explore. We must consider the automation tool for the ab initial calculation data to accelerate the computation. As a result, constructing the workflow is significant for our future database construction of this model. In the current model, two kinds of contributions could be considered necessary based on the first-principle calculation data. They are the configurational contribution and electronic contribution. They could be directly based on several DFT calculations. The elastic contribution might have the chance to be determined from the first-principles calculation based on the original $\epsilon-G$ method \cite{ferreira1988chemical}, but we would leave it to future exploration.

  \begin{figure}
    \centering
\includegraphics[width=0.8\textwidth]{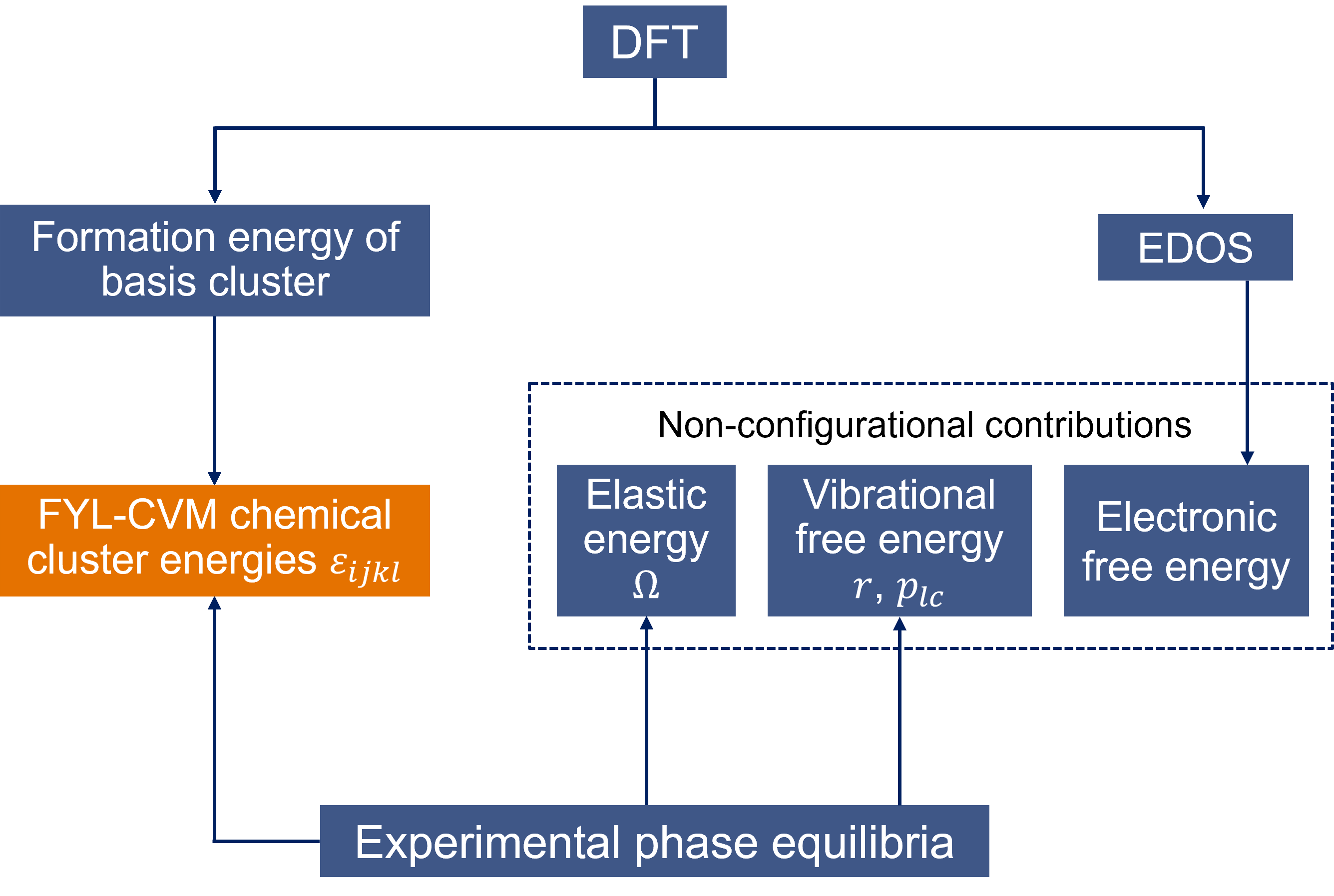}
    \caption{Illustration of the workflow }
    \label{fig:workflow}
\end{figure}

 \subsection{Configuration Contributions}

The origin of cluster energy has been discussed in Chapter \ref{ch:conf}. Here, we dig deeper into the workflow details to reach the DFT cluster energy.

To get the cluster energy from the first-principles calculation, we could use the CEM to get the ECI and then transform this into the cluster energy. Both (FYL-)CVM and CEM expanded the total energy into the cluster-based description. The difference is (FYL-)CVM decomposes the total energy into the dot product between the basic cluster energy and the cluster probability, which is a probabilistic perspective. CEM also decomposes the total energy into several weighted summations over all kinds of clusters but in a correlation function perspective. If we assume both of them are expanded within the tetrahedron approximation level, we have the energy expansion:
\begin{equation}
\begin{split}
 &E_{FYL-CVM} = 2N \sum_{ijkl} \rho_{ijkl} \epsilon_{ijkl}\\
  &E_{CEM} = E_0+\sum_i J_i S_i +\sum_{ij}J_{ij} S_i S_j +\sum_{ijk}J_{ijk} S_i S_j S_k +\sum_{ijkl}J_{ijkl} S_i S_j S_k S_l=\sum_I J_I m_I \xi_I 
    \end{split}
    \label{eq:confwf1}
\end{equation}
Here $ijkl$ represent different cluster configuration, $p_{ijkl}$ is the cluster probability, $E_{ijkl}$ is the corresponding tetrahedron cluster energy, $2N$ is the coefficient to keep the particle conservation, $I$ represent the different size of clusters. $J_I$ is the ECI energy, $m_I$ is the corresponding multiplicity. The details should refer to any review work for the introduction of CEM. This can also be extended onto the larger basic cluster approximation easily by just further expanding it and should keep the consistency. As the total energy should be conserved and would not depend on how we expand it:
\begin{equation}
\begin{split}
 E_{FYL-CVM} =E_{CEM}
    \end{split}
    \label{eq:confwf2}
\end{equation}

As they all represent the total energy, the way to transform the energy information from ECI into cluster energy is to dig the connection between the basic cluster probability and the correlation function. As we have assumed, both of them are only expanded within the same range to keep consistency. a general linear transformation connects them with the help of V-matrix \cite{sanchez1978fee,mohri1985short,ceder1991alloy,mohri2013cluster}. The detailed general connection between basic cluster probability and the correlation function could usually be expressed as \cite{ceder1991alloy}:
\begin{equation}
\begin{split}
p_{\sigma} = \frac{1}{2^n}\left(1+\sum_{l'} V_{\sigma} (l,l') \cdot \xi_{l'}\right)
    \end{split}
    \label{eq:confwf3}
\end{equation}

Where $\sigma$ represents different cluster configurations, $l$ indicates the cluster formed by n-lattice points while $l'$ is a subcluster contained in a cluster $l$, $V$ is the V-matrix. This relation connects cluster probability and correlation functions linearly and generally. 

Then we could replace $p_{ijkl}$ with $\xi_I$ in the expression of $E_{(FYL-)CVM}$, and since $E_{(FYL-)CVM}=E_{CEM}$, we would have the equalities coming from comparing the coefficients of $\xi_I$ within $E_{(FYL-)CVM}$ and $E_{CEM}$. These equations build up the relation between ECI and cluster energy. After performing the cluster expansion with several DFT computations, we could determine the ECI from first-principles calculations. Then the only unknown variables within these linear equations are cluster energy, and solving these equations would lead to the determination of the cluster energy.

We also want to mention one more interesting observation, if we take the most straightforward cluster expansion based on all the possible ground state energy to reduce the first-principles calculation cost following some historical work \cite{mohri2013cluster}. Let’s take binary FCC system and tetrahedron approximation as an example, then we only need to perform several DFT calculations for single elements A and B, and then $L1_2$ $AAAB$, $ABBB$ and $L1_0$ $AABB$. After that, we could take their mixing energy per atom and assume all these $AAAB$, $ABBB$, and $AABB$ are ground state energy and constitute the convex hull. Then we could take a linear transformation by transforming the ECI into the ground state energies:
\begin{equation}
\begin{split}
\left(
\begin{matrix}
1&1&1&1&1\\1&0.5&0&-0.5&-1\\1&0&-1/3&0&1\\1&-0.5&0&0.5&-1\\1&-1&1&-1&1
\end{matrix}
\right)
\left(
\begin{matrix}
m_0 J_0\\m_1 J_1\\m_2 J_2\\m_3 J_3\\m_4 J_4
\end{matrix}
\right)
=\left(
\begin{matrix}
0\\E_{L1_2 AAAB}\\E_{L1_2 AABB}\\E_{L1_2 ABBB}\\0
\end{matrix}
\right)
    \end{split}
    \label{eq:confwf4}
\end{equation}
The matrix is nothing but the detailed value of the correlation function for these ground states.
For this FCC binary case at tetrahedron approximation, with the help of V-matrix to clarify the relation between correlation function and cluster probability, we could build up the linear equations like this: 

\begin{equation}
\begin{split}
\left(
\begin{matrix}
0.5&0.75&0.5\\1&0&-1\\0&-1.5&0\\-1&0&1\\-0.5&0.75&-0.5)
\end{matrix}
\right)
\left(
\begin{matrix}
E_{AAAB} \\E_{AABB}\\E_{ABBB}
\end{matrix}
\right)
=\left(
\begin{matrix}
m_0 J_0\\m_1 J_1\\m_2 J_2\\m_3 J_3\\m_4 J_4
\end{matrix}
\right)
    \end{split}
    \label{eq:confwf5}
\end{equation}

As a result, we can transform these ground state energies to cluster energies with these two linear mappings. Multiply these two matrices would result in an almost identical matrix:

\begin{equation}
\begin{split}
\left(
\begin{matrix}
0&0&0\\2&0&0\\0&2&0\\0&0&2\\0&0&0
\end{matrix}
\right)
    \end{split}
    \label{eq:confwf6}
\end{equation}
Then the corresponding energy equalities would lead to:

\begin{equation}
\begin{split}
\left(
\begin{matrix}
0&0&0\\2&0&0\\0&2&0\\0&0&2\\0&0&0
\end{matrix}
\right)
\left(
\begin{matrix}
E_{AAAB} \\E_{AABB}\\E_{ABBB}
\end{matrix}
\right)
 = 
\left(
\begin{matrix}
0\\E_{L1_2 AAAB}\\E_{L1_2 AABB}\\E_{L1_2 ABBB}\\0
\end{matrix}
\right)
\end{split}
\label{eq:confwf7}
\end{equation}

Here the 2 comes from the ratio between the number of tetrahedron basic clusters and the number of atoms in the same crystal. It indicates that cluster energies are exactly half of the provided ground state energies. We could hypothesize that the cluster energies are directly connected to the energies of the ground state at the specific composition. The reasoning is that the cluster energy in our formalism must follow the cluster probability description until 0K. At 0K, there is no configurational entropy based on the thermodynamic 3rd law, and all the cluster probability distribution is concentrated onto some specific configuration with the cluster probability equal to 1. At that stage, the $E_{(FYL-CVM)}$ is only related to this specific cluster and should equal to this corresponding ground state energy to keep the consistency. Then the cluster energy of this configuration is nothing but half of the ground state at the composition of this cluster configuration with the particle number conservation.

This observation reveals the physical meaning of the cluster energy. It would be directly connected to the system's calculated phase diagram(the energy convex hull) at 0K and equal to the ground state energy value up to a constant(2 in tetrahedron approximation) at the corresponding composition. As a result, if we could determine the system's energy at 0K with the CEM to produce the whole cluster energy description, we could automatically determine the cluster energy based on this workflow.

 \subsection{Electronic Contributions}

As we mentioned, the electronic contribution is considered a subtle part of influencing phase stability. We propose a potential workflow to integrate electronic free energy into our framework. The basic idea is to properly combine the cluster expansion of electronic excitation into our cluster formalism (ref.).
First, we briefly review the cluster expansion of electronic excitation. To introduce the electronic contribution, we perform the coarse graining again over the electronic level:
\begin{equation}
\begin{split}
   z_{\alpha} = \sum_{\sigma} exp\left(\frac{\mu_{\sigma}-(\epsilon_{\sigma} +F_{vib,\sigma}+F_{ele,\sigma})}{k_B T}\right) 
    \end{split}
    \label{eq:ele1}
\end{equation}
\begin{equation}
\begin{split}
   F_{tot,ele} = \sum_{\sigma} p_{\sigma} F_{\sigma,ele}
    \end{split}
    \label{eq:ele2}
\end{equation}
Where $F_{\sigma,ele}$ is the electronic free energy for the specific configuration c. This is the formalism with the cluster probability, and we could also equivalently express the free energy with the correlation function (ref.). 
\begin{equation}
\begin{split}
   F_{tot,ele} = \sum_{i} v_i m_i \xi_i
    \end{split}
    \label{eq:ele3}
\end{equation}
here $v_i$ is the ECI of electronic excitation, $m_i$ is the multiplicity. If we only limit the cluster expansion under the same level as the cluster probability, we could build up the linear relation between $F_{\sigma,ele}$ and ECI based on the definition of the correlation function and the relation between the correlation function and the cluster probability. 
To calculate the ECI for electron contributions, we need $F_e^{\alpha}$ for multiple different structures, where $\alpha$ represents the different structures. This is computable with the standard DFT method. To achieve that, we can approximately take the calculated DOS at $T=0K$ for this thermodynamic quantity. This would break down when the temperature is too high to make such kind of approximation invalid.
Firstly, we have the fundamental thermodynamic relation: 
\begin{equation}
\begin{split}
   F_e (V,T)&=E_e (V,T)-TS_e (V,T)\\
   S_e (V,T)&=-\int^{+\infty}_{-\infty} n(\epsilon,V)\left[f(\epsilon,V)lnf(\epsilon,V)+(1-f(\epsilon,V)ln(1-f(\epsilon,V)))\right]\\
   E_e (V,T)&=\int^{+\infty}_{-\infty} \epsilon n(\epsilon,V)f(\epsilon,V)d\epsilon - \int^{\epsilon_F}_{-\infty} \epsilon n(\epsilon,V)d\epsilon
    \end{split}
    \label{eq:ele4}
\end{equation}

where $f(\epsilon,V)$ is the Fermi–Dirac distribution function. Note that $E_e$ is defined to vanish at $T=0K$ to ensure only electron excitation is involved. We could arbitrarily define $\epsilon_F=0$, then the free energy here depends on $n(\epsilon,V)$ solely. As a result, we should have

\begin{equation}
\begin{split}
    F_e^{\alpha} = F[n_e^{\alpha}(\epsilon,V)]
    \end{split}
    \label{eq:ele5}
\end{equation}

Where $n_e^{\alpha} (\epsilon,V)$ could be obtained from the DFT calculation. After such kind of several DFT calculations, we could construct the ECI for electronic excitation and then $F_(e,\sigma)$ to include the electronic contribution. Note that the convergence of ECI may need to be further discussed, while this kind of method is limited at relatively low temperatures. 

After that, we would get a temperature-dependent electronic contribution for the cluster energy. We could perform second-order polynomial fitting to reach the exact analytical formula and add this into the cluster energy to get the electronic contribution. The whole workflow could be integrated together with the configuration contribution. All these determination processes only need several DFT calculations and could be automatically performed through some workflow tool, such as \cite{pizzi2016aiida, dflow2023github}. This would be helpful to accelerate the database development, and the whole illustration diagram of the current workflow is presented as figure \ref{fig:workflow}.

 \section{Step1: Equilibrium Calculation with the Optimization Algorithm}
The equilibrium calculation is usually challenging for variational thermodynamic models such as CVM, CSA, and FYL-CVM. We would first briefly review the algorithms for CVM and then try to generalize it into FYL-CVM. There are mainly three different minimization methods developed for CVM since 1950s: two of them are from the physics/materials science side, and the other is used within the computer science/AI community.

Since CVM is first created as a statistical mechanics model for the alloy/magnetic system, physicists and materials scientists have developed two ways to perform minimization. The first systematical method is from Kikuchi \autocite{kikuchi1974superposition}, the Natural Iteration Method(NIM). The key idea is to rewrite the equation based on the first derivative into a fixed-point scheme. After that, people gradually noticed that if we make use of the basic cluster probability as the minimization variables, it will cause some redundancy. To reduce this redundancy, people developed the correlation function formalism to replace the cluster probability as the variables \cite{mohri2013cluster}. With the help of this, people could prefer to use the Newton-Raphason method to calculate the solution for the non-linear equations, which come from forcing the first derivative of the thermodynamic potential into 0. Another minimization technique is the message-passing algorithm mainly used in the computer science/ML community. It is developed while people build up the connection between the cluster variation energy and the probabilistic graphical model \cite{yedidia2000generalized}. We wouldn't clearly review this in the following subsections because of its limited usage in physics. But we mention that this message-passing algorithm, or actually the belief propagation algorithm, might be the fastest available algorithm for minimizing the CVM free energy functional \cite{pelizzola2005cluster}. However, this interdisciplinary research might need further communication between the materials science community and the computer science community.

\subsection{Natural Iteration Method and Newton-Raphason Method}
We would mention two algorithms used in CVM minimization briefly. Readers could refer to some other literature to learn more about this further. The first one is the NIM originated back to Kikuchi \cite{kikuchi1974superposition}. First, we take the first derivative of the free energy function. Usually, we would transform the CVM Helmholtz free energy functional into Landau potential through the Legendre transform to work with the chemical potential space. Then we would try to deal with this first derivative by forcing these values to 0 and considering the solution of these non-linear equations as the minima. Then all we need is how to deal with these non-linear equations. Kikuchi proposed a fixed-point scheme to converge to the solution iteratively. Then we could get the solution and the minima. Before calculating the first derivative of the thermodynamic potential, Kikuchi's original idea would transform the Helmholtz free energy into Landau potential first \cite{kikuchi1974superposition}. The reason is that for such kind of variational free energy thermodynamic model, the minimization variable is basic cluster level, but the state variable is $x$ and $T$. This means the composition $x$ is challenging to reach as a variable, it would be considered as a constraint for the basic cluster probability distribution. So when performing the calculation in the $x-T$ space, the minimization would mainly become a constraint optimization. this constraint is challenging to deal with as the constraint is implicitly related to the basic cluster level variables, the basic cluster probability. If we assume the tetrahedron approximation for the basic cluster and consider the binary fcc case, we have this following form:
\begin{equation}
\begin{split}
    &min F(p_{ijkl}, x, T)\\
    s.t. &\sum_{jkl} p_{i_0jkl} +\sum_{ijk} p_{ijkl_0}+\sum_{ikl} p_{ij_0kl}+\sum_{ijl} p_{ijk_0l}  = 4x
    \end{split}
    \label{eq:NIMcm}
\end{equation}
However, if we transform the Helmholtz free energy into Landau free energy $\Phi = F -\mu \cdot x$, we shift the variable space from $x-T$ to the $\mu-T$ space. We could minimize the cluster-level variables without treating the chemical potential and temperature in this case. As a result, we have a minimization like this:
\begin{equation}
\begin{split}
    min\left(\Phi(p_{ijkl}, \mu, T) + \lambda(1-\sum_{ijkl}p_{ijkl})\right)
    \end{split}
    \label{eq:NIMncm}
\end{equation}
Where we have implemented the Lagrangian multiplier method, and the $\lambda$ is the Lagrangian multiplier. The only constraint is the probability conservation, which would be automatically satisfied in FYL-CVM due to the FYL-transform.

For the next step, we have the first derivative of the Landau potential:
\begin{equation}
\begin{split}
    \frac{\partial \Phi}{\partial p_{ijkl}} &= 2 E_{ijkl} + 2 k_B T lnp_{ijkl} + 5/4 k_B T ln(p_{i}p_{j}p_{k}p_{l})   
    \\&-k_B Tln(p_{ij}p_{ik}p_{il}p_{kl}p_{jk}p_{jl}) -1/4(\mu_i+\mu_j+\mu_k+\mu_l)  -\lambda
    \end{split}
    \label{eq:NIM1}
\end{equation}

Then we force these first derivatives to 0 to reach the equilibrium:
\begin{equation}
\begin{split}
    &2 E_{ijkl} + 2 k_B T lnp_{ijkl} + 5/4 k_B T ln(p_{i}p_{j}p_{k}p_{l})   
    \\&-k_B Tln(p_{ij}p_{ik}p_{il}p_{kl}p_{jk}p_{jl}) -1/4(\mu_i+\mu_j+\mu_k+\mu_l) -\lambda = 0
    \end{split}
    \label{eq:NIM2}
\end{equation}
and we could actually get a fixed-point scheme after transforming this equation into a cluster probability expression:
\begin{equation}
\begin{split}
    &ln(p_{ijkl}) = -E_{ijkl}-5/8ln(p_{i}p_{j}p_{k}p_{l})  +1/2ln(p_{ij}p_{ik}p_{il}p_{kl}p_{jk}p_{jl})\\&+1/8\beta(\mu_i+\mu_j+\mu_k+\mu_l)+1/2\beta\lambda
    \end{split}
    \label{eq:NIM3}
\end{equation}
where $\beta = \frac{1}{k_B T}$. Based on the superposition condition, the right-hand side could also be expressed as the function of $p_{ijkl}$. So this expression constructed a fixed-point scheme, which means the mapping would map the variables back to itself: $x_{i+1} = g(x_i)$. The only issue is the determination of $\lambda$, which could be reached through probability conservation. It means we could sum all the cluster probability, which equals 1 to derive the expression for $\lambda$:
\begin{equation}
\begin{split}
    &exp(1/2\beta\lambda) = \\
    &\frac{1}{\sum_{ijkl}exp(-E_{ijkl}-5/8ln(p_{i}p_{j}p_{k}p_{l})  +1/2ln(p_{ij}p_{ik}p_{il}p_{kl}p_{jk}p_{jl})+1/8\beta(\mu_i+\mu_j+\mu_k+\mu_l))}
    \end{split}
    \label{eq:NIM4}
\end{equation}

Kikuchi somehow has proved that this scheme would at least keep minimizing the free energy\cite{kikuchi1974superposition}. However, we have to mention that since the CVM's free energy function is non-convex, this method only guarantees to converge to the local minima but not global minima as well. 

Another standard minimization scheme shares a similar basic idea as NIM, as they are still trying to solve the non-linear equations constructed by forcing the first derivative to zero. But the formalism is all based on the correlation function and uses the Newton-Raphason method with the correlation function as the variables. The connection between the correlation function and the cluster probability is through the V-matrix \cite{ceder1993derivation}, the linear transform between cluster probability and the correlation function, and what we need to do is to replace the cluster probability with the correlation function to reduce the redundancy:
\begin{equation}
\begin{split}
    F = E({\xi}) - TS({\xi})
    \end{split}
    \label{eq:NR1}
\end{equation}
\begin{equation}
\begin{split}
    \frac{\partial F}{\partial \xi_m} = 0
    \end{split}
    \label{eq:NR2}
\end{equation}
where $\xi_m$ is the $m$th correlation function.

Besides, we could notice that the correlation function mechanism could be directly related to the composition. For example, if we limit the discussion in the binary case, we could fix $\xi_1$ to get the composition dependence:
\begin{equation}
\begin{split}
    x_i = \frac{1}{2} (1+i\cdot \xi_1)
    \end{split}
    \label{eq:NR3}
\end{equation}
Here $x_i$ is the composition for $i$-species. As a result, the minimization could be performed directly inside the $x-T$ space but not the $\mu-T$ space, which is usually used for NIM. However, one key issue reported for this Newton-Raphason scheme is its sensitivity to the initial value. It should be directly related to the non-convexity of the CVM free energy function. As a result, this scheme may not be a good choice for phase diagram calculation. 

If we look backward to the current existing algorithms, we notice that the starting point is to force the first derivative of the free energy functional to 0, and then try to solve these non-linear equations. This doesn't guarantee that the final converged point is the global minimum as the free energy functional is non-convex but would possibly be local minima or even saddle points within the high dimensional space.

\subsection{Gradient Descent Method}
We could notice that all previously mentioned methods are based on the ``solving non-linear equations" perspective but not an ``optimization" perspective. Although these are mathematically similar, we have other ideas and methods mostly from an ``optimization" perspective. First, we need to point out that the current variational functional is non-convex \cite{pelizzola2005cluster}, which would cause the main difficulty when we want to perform the minimization. This leads to considering using the (stochastic) gradient descent method as the minimizer.

Gradient descent is a first-order iterative optimization algorithm for finding a local minimum of a differentiable function. It also doesn't guarantee to converge to the global minimum since non-convex optimization is actually an NP-hard problem and requires much more computational time to reach the global minima\cite{sun2016nonconvex}. However, in most cases, the local minima would be enough and that's the main reason for the wide application of the gradient descent method in the ML community \cite{kawaguchi2016deep}. The key idea of the gradient descent method could be summarized as:
keep stepping in the opposite direction of the function's gradient (or approximate gradient) at the current point because this is the direction of the steepest descent.
In our cases, we only need to calculate the free energy functional gradient and then use the gradient descent scheme to converge to the local minima iteratively. As a result, we could have the scheme:
\begin{equation}
\begin{split}
    \mu_i^{t+1} = \mu_i^{t} - 
    \alpha \frac{\partial \Phi}{\partial \mu_i^{t}}
    \end{split}
    \label{eq:SGD2}
\end{equation}
where $\alpha$ is the step length or called learning rate in ML community. If we add some reasonable noise to the right-hand side of \eqref{eq:SGD2}, such as the Gaussian noise, it would be possible to transform into a kind of stochastic gradient descent method, which was actually widely used in current large-scale ML optimization. This is due to its high minimization efficiency. One of the exciting points of view for the stochastic gradient descent method(and actually the gradient descent method) is that itself and its related perturbed version would be somehow helpful to avoid saddle points efficiently \cite{jin2021nonconvex}. At the same time, it is believed that there would be many saddle points but not local minima for the high dimensional non-convex optimization problem \cite{dauphin2014identifying}. This may support that a gradient descent-based minimizer would be a better choice for a stable candidate algorithm for the phase equilibrium calculation of FYL-CVM.

The implementation of this method is straightforward. It only requires the gradient of the free energy functional, which could even be calculated in many current state-of-the-art numerical computational packages such as JAX \cite{jax2018github} and could be accelerated on GPU nowadays. This fits the current development of auto-differential calculation trends in the scientific computing community. Besides, our current test reveals that the initial value requires no special treatment. We could set the initial value as Gaussian noise. It would also lead to the phase diagram while some fluctuation along the boundary must exist due to the non-convex optimization, which the boundary detection algorithm might resolve.

The selection of the step length $\alpha$ is usually fixed for the original gradient descent method. However, we can develop some adaptive step lengths to accelerate the convergence further. One of the methods is the Barzilar-Borwein method. The details could be checked in any advanced optimization textbook. 

Finally, we want to mention the potential connection between the free energy gradient descent method and the kinetics evolution. However, what we have discussed up to now is all thermodynamics. We could mathematically transform the original scheme of the gradient descent method for the free energy functional into a time-dependent equation:

\begin{equation}
\begin{split}
   \frac{ \mu_i^{t+1} - \mu_i^{t}}{\alpha} =- 
 \frac{\partial F^{t}}{\partial \mu_i^{t}}
    \end{split}
    \label{eq:SGD3}
\end{equation}
If $\alpha$ is small enough, we can transform the finite difference into a time-dependent derivative:
\begin{equation}
\begin{split}
   \frac{\partial \mu_i}{\partial t} =- 
 \frac{\partial F^{t}}{\partial \mu_i^{t}}
    \end{split}
    \label{eq:SGD4}
\end{equation}

This actually shares some similarities with the time-dependent Ginzburg-Landau equation and the Allen-Cahn equation used in the phase-field study \cite{chen2002phase}. However, we have to point out that it actually doesn't involve any actual time variable but the virtual time variable related to the step length. We should not apply this to simulate the actual kinetic phenomena as the whole mechanism is fully thermodynamic controlled. But we could consider this minimization path might indicate the actual phase transition pathway of the thermodynamic process.

Let's fix the initial state by setting the corresponding equilibrium value of the site chemical potential at the specific $\mu_0$ and temperature $T_0$. We perform the gradient descent method to reach the final state with specific $\mu_1$ and temperature $T_1$, and we record the whole process during the iterations. Based on the physical meaning of the gradient descent to always go through the fastest way to reduce the free energy, it might simulate how the physical system evolves.
\begin{figure}
    \centering
\includegraphics[width=1\textwidth]{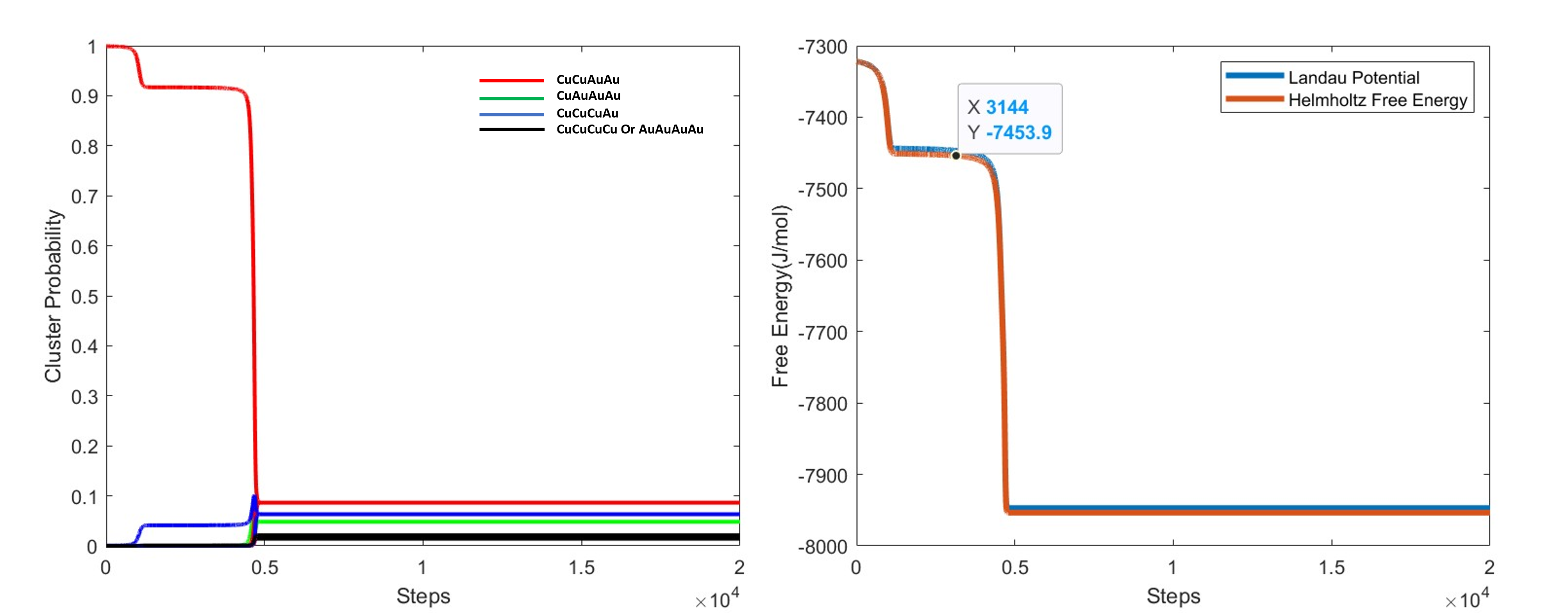}
    \caption{``Super-heating" Process simulated by the gradient descent path.}
    \label{fig:GDa}
\end{figure}

We take the Cu-Au system as an example to observe this kind of minimization path. The detailed thermodynamic assessment is performed in Chapter \ref{ch:demo}. Here we take it directly. Firstly, we set $\mu = 20 N_A k_B$, and $T=300K$ to set up its equilibrium state, the corresponding equilibrium reduced site chemical potential, as the initial condition. then perform the super-heating to $T=800K$, the result is presented in figure \ref{fig:GDa}. 
\begin{figure}
    \centering
\includegraphics[width=1\textwidth]{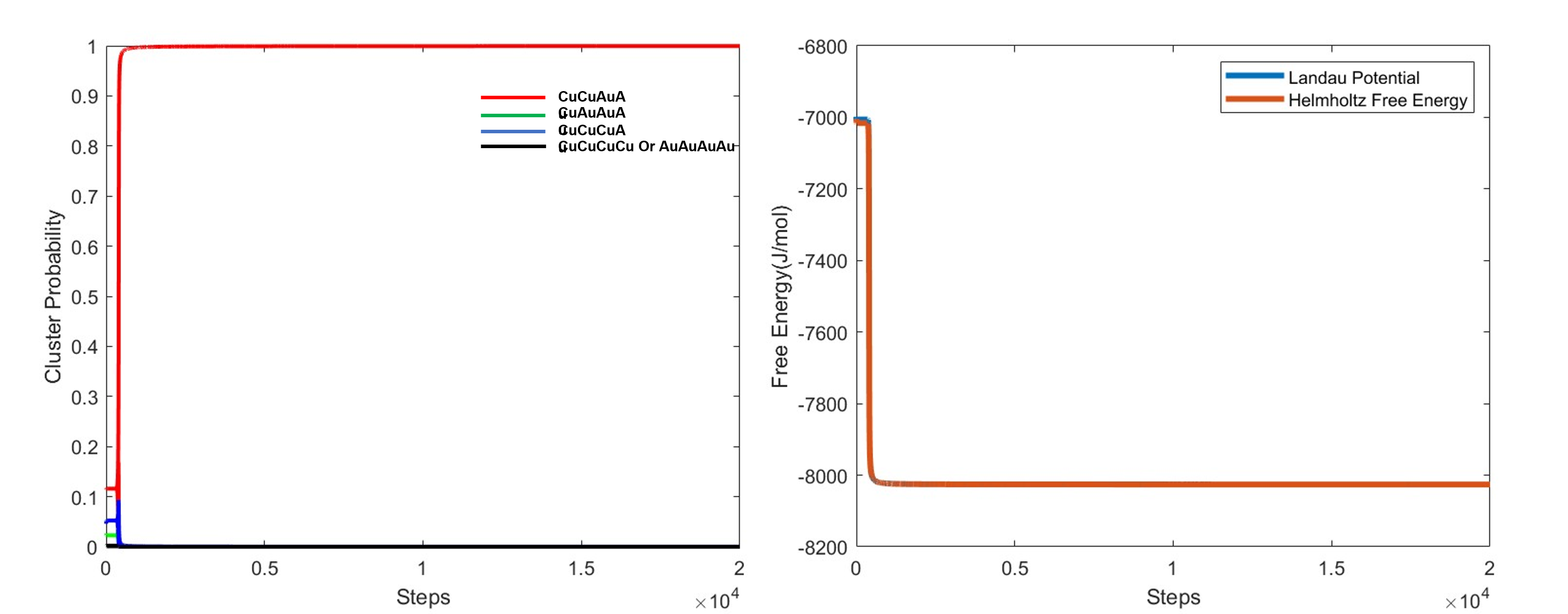}
    \caption{``Super-cooling" Process simulated by the gradient descent path.}
    \label{fig:GDb}
\end{figure}

Then we set up the backward process, still set $\mu = 20 N_A k_B$, but $T=800K$ to set up its equilibrium state, which is the corresponding equilibrium reduced site chemical potential, as the initial condition. then perform the super-cooling to $T=300K$, the result is presented in figure \ref{fig:GDb}. 

Based on the results presented, we make some observations: 

1) There is a free energy plateau in the ``super-heating" minimization process. With this plateau, the system prefers to formulate the $Cu_3 Au$ tetrahedron cluster and finally enter the final ordered state. This indicates that the $Cu_3 Au$ SRO might assist the phase transformation since the gradient descent path should go through the fastest direction to reduce the total free energy.

2) This phenomenon about the plateau doesn't appear in the ``super-cooling" process, which perhaps reveals the difference between forming the disordered and ordered phases. However, we still observe that the $Cu_3 Au$ related cluster probability would suddenly increase and decrease near the transition point. This might indicate the widely existing $L1_2$ SRO assistance for the phase transformation.

Based on both observations, we conclude that this result perhaps indicates agreement with the previous literature, which performed the molecular dynamics simulation to confirm the participation of the $L1_2$ chemical SRO during the formation of the Cu-Au solid solution \cite{desgranges2019can}. Overall it implies the importance of the chemical SRO in the phase transformation pathway. The importance of the chemical SRO in the kinetics(nucleation) of phase transformation is also discussed in many previous literature such as \cite{clouet2004nucleation,clouet2005precipitation,clouet2007classical,rappaz2020solidification}. A precise study of the nucleation process with the chemical SRO would be left for future study.

 \section{Step2: Phase Boundary Detection and its Algorithm}
After having the single point algorithms, we still need a systematic way to design how to plot the phase diagram smartly. The reason is that we only care about capturing the phase boundary when we want to understand the phase diagram of the specific system. As a result, performing the equilibrium calculation for the whole space is unnecessary. Still, only a tiny piece of them should be enough: we don't really care about the cases inside the phase region but just the situations near where the phase transition happens. this problem could be considered a ``rare event" problem: compared to the phase region, the phase boundary is rare and should be identified with the algorithms. The phase boundary is ``formed" less frequently than the whole $\mu-T$ space.

To deal with rare event problems generally, people usually use some unique sampling technique. Actually, this rare event problem happens in molecular dynamics simulation. Some meta-stable states are essential for the system but cannot be detected as they don't occur frequently enough. People then developed the enhanced sampling method \cite{yang2019enhanced} to deal with this problem. The fundamental idea of it is to improve the sampling probability of the low-frequency or low degeneracy areas. Motivated by this method, we can notice that the key is to increase the sampling probability around the possible phase boundary area and decrease the sampling probability inside the phase region. However, we cannot do the uniformly sampling among the whole $\mu-T$ space. 

Another point of motivation for us to explore the corresponding phase boundary detection algorithm is that the phase boundary would follow some fundamental physics rule. So we can bias our sampling direction based on this information from fundamental physics. For example, we could infer the ordered phase must collapse around its specific composition at a very low temperature since the configurational entropy is very small at a low temperature. We can even derive the temperature relation along the phase boundary direction \cite{van2002self}. The reasoning is that for the phase boundary, we should have $\phi^{\alpha}/kT = \phi^{\gamma}/kT$ along the phase boundary, then take the total differential calculation of both sides. Here$\phi$ is the landau free energy, $\alpha$ and $\gamma$ represent different phases. Finally, we could reach a derivative relation along the phase boundary path in $\mu-T$ space:
\begin{equation}
\begin{split}
    \frac{d\mu}{d\beta} = \frac{E^{\gamma}-E^{\alpha}}{\beta(x^{\gamma}-x^{\alpha})} - \frac{\mu}{\beta}
    \end{split}
    \label{eq:PBD1}
\end{equation}

Here $\beta = 1/kT$, $E$ is the energy, and $x$ is the corresponding composition. The detailed derivation can be found in \cite{van2002self}. This relation should also be able to be generalized to the multi-component case. With such kind of relation, we could start from the phase boundary at low temperature to roughly infer the phase boundary at high temperature in a relatively small region. With the information extracted from the basic thermodynamic relation in the $\mu-T$ space, we could save lots of computational resources to get the phase boundary by only searching the region indicated by the equation \eqref{eq:PBD1}.

We would have two goals for the algorithm: First, reduce the computational cost to detect the phase boundary efficiently; Second, make the calculation more robust to get the stable boundary to denoise the fluctuation caused by non-convexity.

As a result, we briefly summarize the proposed phase boundary detection algorithm here:

1) entirely searching the ground state around the corresponding composition at the lowest temperature.

2) performing the loop to keep increasing the temperature and calculate the corresponding derivative based on equation \eqref{eq:PBD1}. Then, the finite difference method calculates the potential phase boundary at this temperature. Set up the proper error bar to determine the range of the searching domain.

3) Using the binary searching algorithm at each searching domain to make sure we could detect the exact phase boundary position.

4) keep performing the loop to reach the required temperature and get the complete phase boundary.

Based on this proposed algorithm, we could save much computing time with this biased searching to avoid entirely searching the whole space to get the phase boundary.

 \section{Step3: Parameter Optimization and the Challenge of its Algorithm}
As we mentioned, CALPHAD should be considered as a specialized data mining task for thermodynamics. We need an algorithm to help determine these thermodynamic parameters more automatically. As a result, people have developed some modules widely used, such as PARROT in the Thermo-Calc. It's essentially the least-squares fitting coupling with Powell’s method \cite{jansson1984evaluation}. Some recent methods, ESPEI, use Markov Chain Monte Carlo to help solve the equations \cite{bocklund2019espei} and would lead to the uncertainty quantification to benefit the thermodynamic assessment.

However, we firmly believe most parameter optimization algorithms are designed for the relatively simple case. At the same time, Markov Chain Monte Carlo is a complete sampling method free of the gradient. The parameter optimization would be even more challenging for the variational model in CALPHAD, and we mainly want to demonstrate the challenge here and provide the potential method to calculate it. The main reason for this challenge is that the variational model would cause the parameter optimization to become a bi-level optimization problem \cite{colson2007overview}, which is challenging to deal with. Historically, CALPHAD people may use the terminology ``two-step optimization" but hasn't identified this bi-level optimization, which is widely appeared in many other area\cite{otis2023}. This means we have to deal with the equilibrium calculation every time first and then use the data to determine the energy parameters. As a result, it looks like most of the current prevail parameter optimization for solid solution thermodynamic models is done manually by hand. The automatic parameter optimization of the cluster energy based on the experimental data is still very tricky and challenging in practice.

Let's compare this case to the ML task. We can notice the similarity between parameter optimization and hyper-parameter optimization for the NN model, which is considered as a bi-level optimization. this hyper-parameter optimization is trying to optimize some hyperparameters within the NN model, in which the NN needs to be optimized with respect to the inside parameters to reach the local minima of the constructed loss function. The hyperparameters are fixed and not optimized through the backpropagation algorithm in its minimization process. So there is a bi-level optimization structure appeared. To achieve the goal of hyperparameter optimization, ML community is developing the corresponding algorithm to address the challenge of bi-level optimization. We consider this algorithm would benefit our case and would briefly mention this as well.

This section will first illustrate the correspondence between parameter optimization for the variational model in CALPHAD and the bi-level optimization. This would indicate the challenge we have.

 Here we would check how the parameter optimization for the variational thermodynamics model should be considered as the bi-level optimization \cite{franceschi2018bilevel}. At the same time, it is named as two-step parameter optimization in the CALPHAD community but is not widely known.
In CALPHAD modeling, the thermodynamics of any variational model, including CVM, CSA, and FYL-CVM, could be generally expressed as follow:
\begin{equation}
\begin{split}
   F(x,T,\epsilon_{cluster}) = min_{\mathbf{\lambda}}\mathcal{F}(\mathbf{\lambda},T,\epsilon_{cluster}), s.t. x
    \end{split}
    \label{eq:BL1}
\end{equation}
Here $\lambda$ is the variational variable, which could be considered as some cluster-level chemical potential-related variable. $\epsilon_{cluster}$ is the associated cluster energy and is the parameter in CALPHAD that needs to be optimized with the phase boundary and thermochemical data. $s.t. x$ means the calculated result should satisfy the composition constraint. As a result, we have the first layer of optimization for the variational variables.

After that, we want to fit the cluster energy for the data, particularly the phase boundary data. The parameter optimization with the $\epsilon_{cluster}$. However, we must remind the readers that when optimizing the $\epsilon_{cluster}$,  usually it will require to perform an iterative algorithm to keep changing the $\epsilon_{cluster}$. This would definitely change the equilibrium status, which means the previously optimized variational variables for equilibrium minimization are not optimized anymore due to the shift of the cluster energy. this interrelation between the inside optimization and outside optimization is difficult to deal with, as we have to optimize the cluster energy while maintaining the minimization for equilibrium calculation. Then we can formulate the optimization mathematically as follow:
\begin{equation}
\begin{split}
    min_{\epsilon_{cluster}}\{F_{CALPHAD}(x,T,\epsilon_{cluster})-F_{data}(x,T)\}\\
   F_{CALPHAD}(x,T,\epsilon_{cluster}) = min_{\mathbf{\lambda}}\mathcal{F}(\mathbf{\lambda},T,\epsilon_{cluster}), s.t. x
    \end{split}
    \label{eq:BL2}
\end{equation}
This formalism could exactly attribute to the bi-level optimization, which is usually expressed as \cite{franceschi2018bilevel}: 
\begin{equation}
\begin{split}
min\{f(\lambda):\lambda \in \Lambda \}
    \end{split}
    \label{eq:BL3}
\end{equation}
where function $f: \Lambda \to \mathbf{R}$ is defined at $\lambda \in \Lambda$ as 
\begin{equation}
\begin{split}
f(\lambda) = inf\{E(\omega_{\lambda},\lambda):\omega_{\lambda} \in arg min_{u \in R^{d}} L_{\lambda}(u)\}
    \end{split}
    \label{eq:BL4}
\end{equation}

This indicates that the parameter optimization for the variational thermodynamic model is equivalent to the bi-level optimization, and it usually requires one special algorithm to achieve the bi-level optimization.

Since we identify the parameter optimization for the variational thermodynamic model as the bi-level optimization, we definitely need to design a specific algorithm for this case. The previous parameter optimization algorithm in CALPHAD never identify this issue, and the basic idea for them is usually to use the least square fitting for the existing dataset \cite{jansson1984evaluation}. However, the case is difficult to be directly tackled with this simple method. Recently, people have proposed to take the Markovian Chain Monte Carlo method to deal with parameter optimization \cite{bocklund2019espei}, a sampling technique free of gradient which could be helpful to deal with such bi-level optimization. There are also some attempts from ML community to express the inner optimization of such bi-level optimization as a trajectory. Then transfer the bi-level optimization into one-level optimization with the inner optimization dynamics \cite{franceschi2018bilevel}:

\begin{equation}
\begin{split}
min_{\lambda}f_{T}(\lambda) = E(\omega_{T,\lambda},\lambda)
    \end{split}
    \label{eq:BL5}
\end{equation}
where $E$ is a smooth scalar function, and
\begin{equation}
\begin{split}
\omega_{0,\lambda} =\Phi_{0}(\lambda), \omega_{t,\lambda}=\Phi_{t}(\omega_{t-1,\lambda},\lambda)
    \end{split}
    \label{eq:BL6}
\end{equation}
where $\Phi$ could be a minimization mapping for the inner optimization.

Then we could express the gradient for the energy parameters to perform the optimization with a finite time loop for the inner optimization variables. This is a gradient-based method for such a bi-level optimization. More details can be found in \cite{franceschi2017forward,franceschi2018bilevel}. Some other methods by applying the implicit function theorem for the parameter gradient calculation can be referred to \cite{pedregosa2016hyperparameter,beirami2017optimal}.

 \section{Conclusions}
This Chapter explored the algorithm associated with our current thermodynamic framework. First, we construct the workflow for determining the initial value of the parameters for the configurational configuration and electronic contribution. We developed the corresponding equilibrium minimization algorithm based on the gradient descent method and discussed the algorithms for phase boundary detection and parameter optimization. With the developed algorithms and workflow in hand, we are ready to explore our novel computational thermodynamic framework for the actual system efficiently.

\chapter{Benchmark with Real Alloy Systems} \label{ch:demo}
    \section{Introduction}
Many different factors determine the solid solution thermodynamics, including but not limited to chemical configurational effect, vibrational effect, elastic effect, electronic effect, magnetic effect, etc. The currently developed FYL-CVM thermodynamic framework would mainly consider configurational, vibrational, electronic, and elastic contributions, which makes the framework ready to depict most of the alloy system.
We have developed the fundamental theoretical model and tested it on the prototype system, and we have developed the corresponding algorithms and the workflow for the practical application. We are ready to apply the model and the corresponding algorithms to the practical scenario and the thermodynamic data to perform the thermodynamic assessment and combine the cluster-based thermodynamic model and the CALPHAD methodology.

Due to the medical application of the Cu-Au-Ag ternary system, it has been studied many times during the past decades. Kusoffsky \cite{kusoffsky2002thermodynamic} reviewed the whole experiments previously made in detail and pointed out the possible reliability between the potentially inconsistent experimental information from different authors. In addition to the experimental studies, some thermodynamic descriptions for this ternary system, based on the CALPHAD method, have been published. Kusoffsky \cite{kusoffsky2002thermodynamic} performed the thermodynamic modeling based on compound energy formalism (CEF), which is essentially the sublattice model with ideal mixing approximation for the configuration entropy. This would not take the chemical SRO properly in physics. Later people take cluster/site approximation(CSA) to study this ternary ordering system and get a phase diagram\cite{cao2007thermodynamic} in good agreement with the thermodynamic data. However, CSA has its own issue with its non-interference assumption. Here we would focus on the chemical SRO structure inside the disordered phase for this ternary system. As we are interested in how chemical SRO in the disordered solid solution phase affects the mechanical properties \cite{ma2020unusual,ding2019tuning}, corrosion properties \cite{liu2018effect} but also the thermal and electronic transport \cite{jiang2021high}. As a result, it would be meaningful to quantify the chemical SRO among the whole composition space to take it as a new materials design dimension \cite{nohring2020design}. With the current developed FYL-CVM framework and the first-principles workflow for cluster energy in hand, we could efficiently construct the cluster distribution in the whole space. Then we would introduce the pairwise chemical SRO which is exactly the Warren-Cowley SRO \cite{cowley1950approximate}, and 4-point chemical SRO \cite{goff2021quantifying} to quantify the chemical SRO among the whole composition space.

To obtain the ternary thermodynamic description for the Cu-Au-Ag system, we have to explore the Cu-Au, Cu-Ag, and Au-Ag systems further one-by-one first. With the help of the constructed workflow introduced in Chapter 5, the cluster energy for these two systems could be determined automatically based on the DFT calculation. With the help of the previously determined elastic contribution in the $\epsilon-G$ setting \cite{wei1987first}, we could determine the chemical contribution and elastic contribution separately and get the close phase diagram compared to the previous calculated thermodynamic modeling. In particular, we would perform the thermodynamic assessment in detail for the Cu-Au system. After adding the configuration, elastic, vibrational, and electronic contributions in a sequence, we could observe how different physics affect the thermodynamic stability within the Cu-Au system individually. 

After the determination of the Cu-Au, Cu-Ag, and Au-Ag systems, we can directly go one step further to integrate their cluster energy parameters for the Cu-Au-Ag system. For the ternary system, it has 81 cluster energy to be determined initially but this number can be reduced to 15 based on the symmetry. 12 of them are determined based on the three binary sub-systems, and we could calculate the cluster energy $CuAuAg_2$, $CuAu_2Ag$, and $Cu_2AuAg$ to complete the parameter initial value determination. After that, we could get the thermodynamics for Cu-Au-Ag disordered solid solution phase and quantify the chemical SRO within this phase.
With all these computational efforts coupling the first-principles calculations, the constructed FYL-CVM framework, and the corresponding workflow, we could efficiently model the binary and ternary systems with the demonstrated Cu-Au and Cu-Au-Ag systems. This demonstration reveals the potential of this hybrid CVM-CALPHAD framework to become a new methodology for thermodynamic modeling that enables atomic-scale order to be exploited as a dimension for materials design, especially of the novel complex concentration alloys.


\section{Thermodynamic Modeling for Cu-Au ordering system}
    In order to reach the ternary Cu-Au-Ag system, we could start from the typical binary fcc ordering system, Cu-Au. We implement and benchmark the FYL-CVM thermodynamic model for the Cu-Au binary system. First, we perform the ground state energy calculations with the help of DFT in the last section to get the cluster energy with the developed workflow. Then we take the elastic energy constant calculated in previous literature to determine the actual chemical energy and the elastic contribution. 

  \begin{figure}
    \centering
\includegraphics[width=1\textwidth]{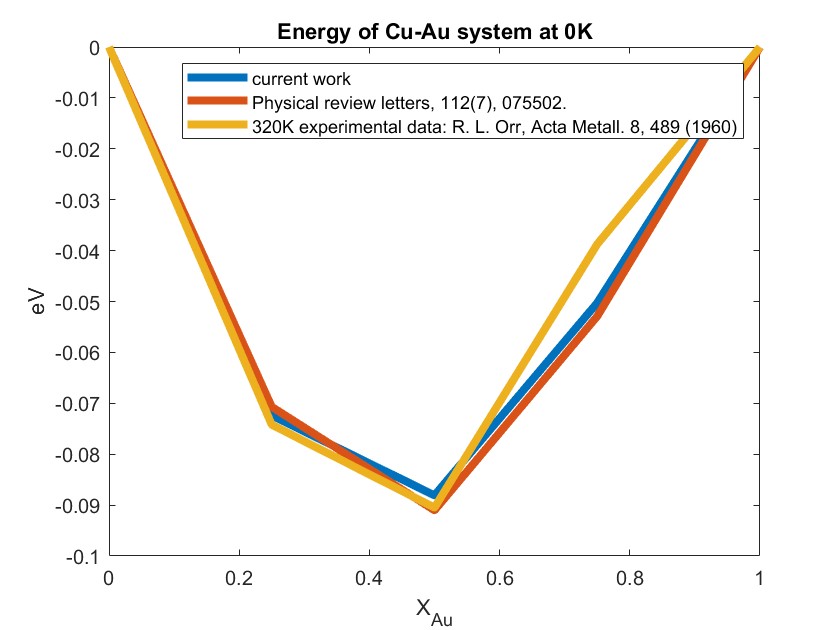}
    \caption{Energy comparison. The compared data is from \cite{zhang2014nonlocal,orr1960heats}.}
    \label{fig:cuauenergy}
\end{figure}

DFT is employed to predict the finite-temperature thermodynamic properties of phases of interest. The Vienna \textit{ab initio} simulation package (VASP) 5.4 \cite{kresse1996efficient} was used to perform the first-principles calculations. Based on the previous literature, there exists a large discrepancy between DFT (LDA or GGA)
formation energies and experimental values \cite{zhang2014nonlocal}. As a result, we could take the hybrid exchange-correlation(XC) functionals developed by Heyd, Scuseria, and Ernzerhof (HSE) \cite{heyd2003hybrid} with 500eV as the energy cutoff. As a result, we could reach the chemical compound energy and the corresponding electronic density of state to get the electronic free energy. The comparison of our calculation at 0K for the system is presented in figure \ref{fig:cuauenergy}. After comparing with the previous calculation and experimental data, it reveals our calculation is in great agreement with the existing results.

Then we handle the workflow and extract the elastic information from the previous literature \cite{ferreira1988chemical} to determine the chemical cluster energy term, the electronic contribution, and the elastic contribution together. In the binary case, we have illustrated that the cluster energy could be considered as a one-to-one mapping up to a constant ratio with the chemical energy at the corresponding composition.

  \begin{figure}
    \centering
\includegraphics[width=1\textwidth]{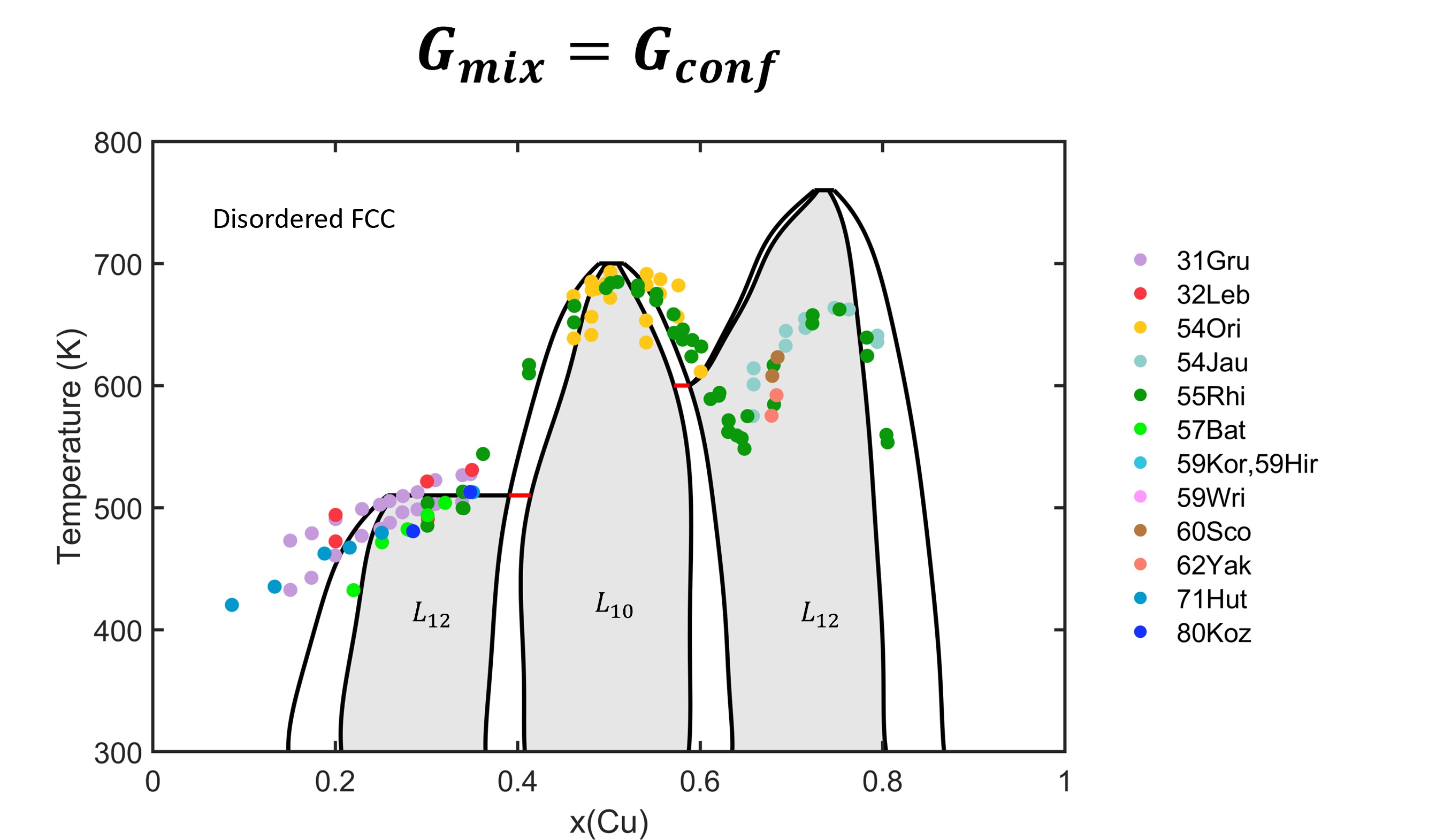}
    \caption{The Cu-Au phase diagram with only configurational contribution. The experimental data are summarized in \cite{okamoto1987cu}.}
    \label{fig:PDCUAU1}
\end{figure}

  \begin{figure}
    \centering
\includegraphics[width=1\textwidth]{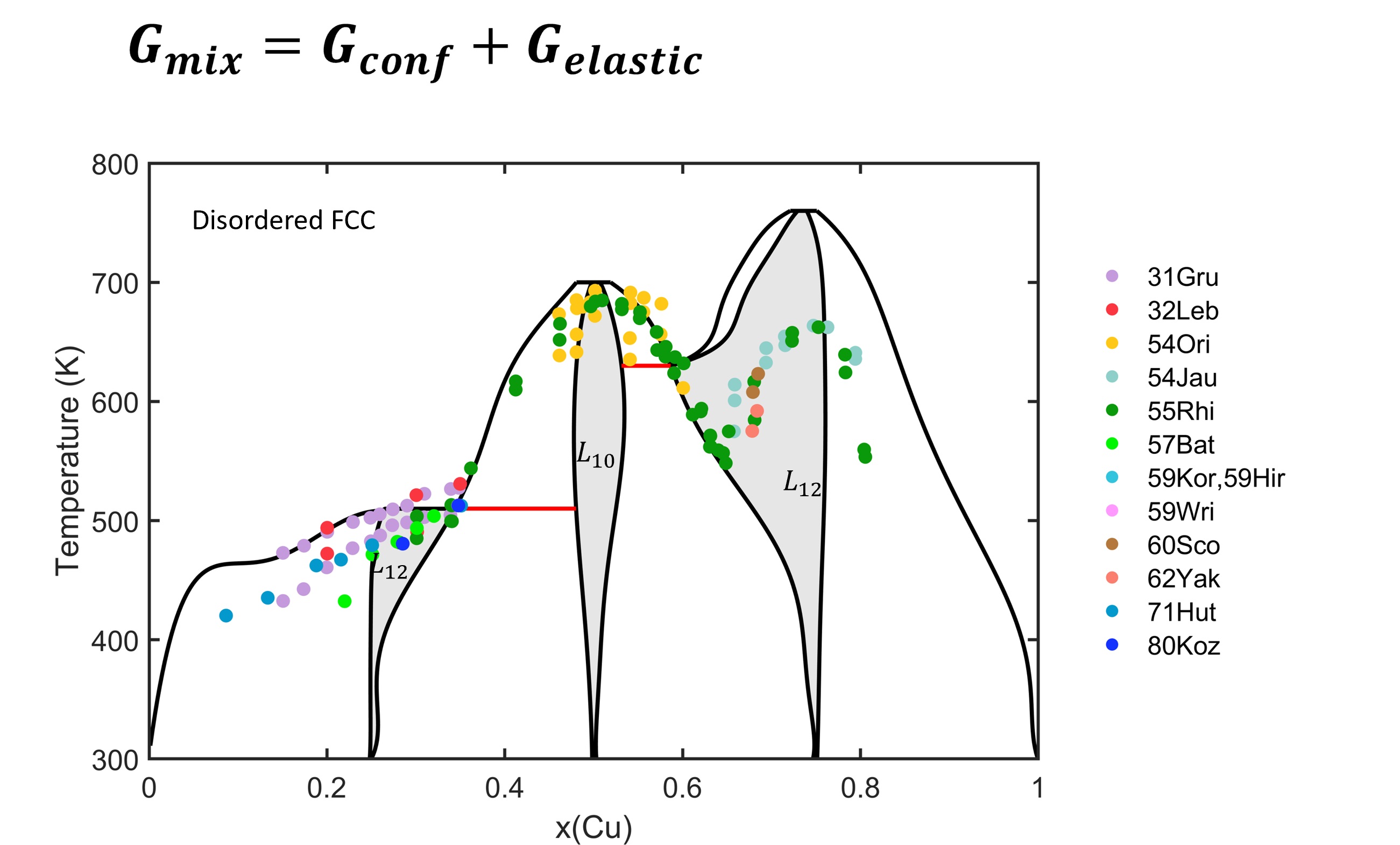}
    \caption{The Cu-Au phase diagram with elastic contribution and configurational contribution. The experimental data are summarized in \cite{okamoto1987cu}.}
    \label{fig:PDCUAU2}
\end{figure}

\begin{figure}
    \centering
\includegraphics[width=1\textwidth]{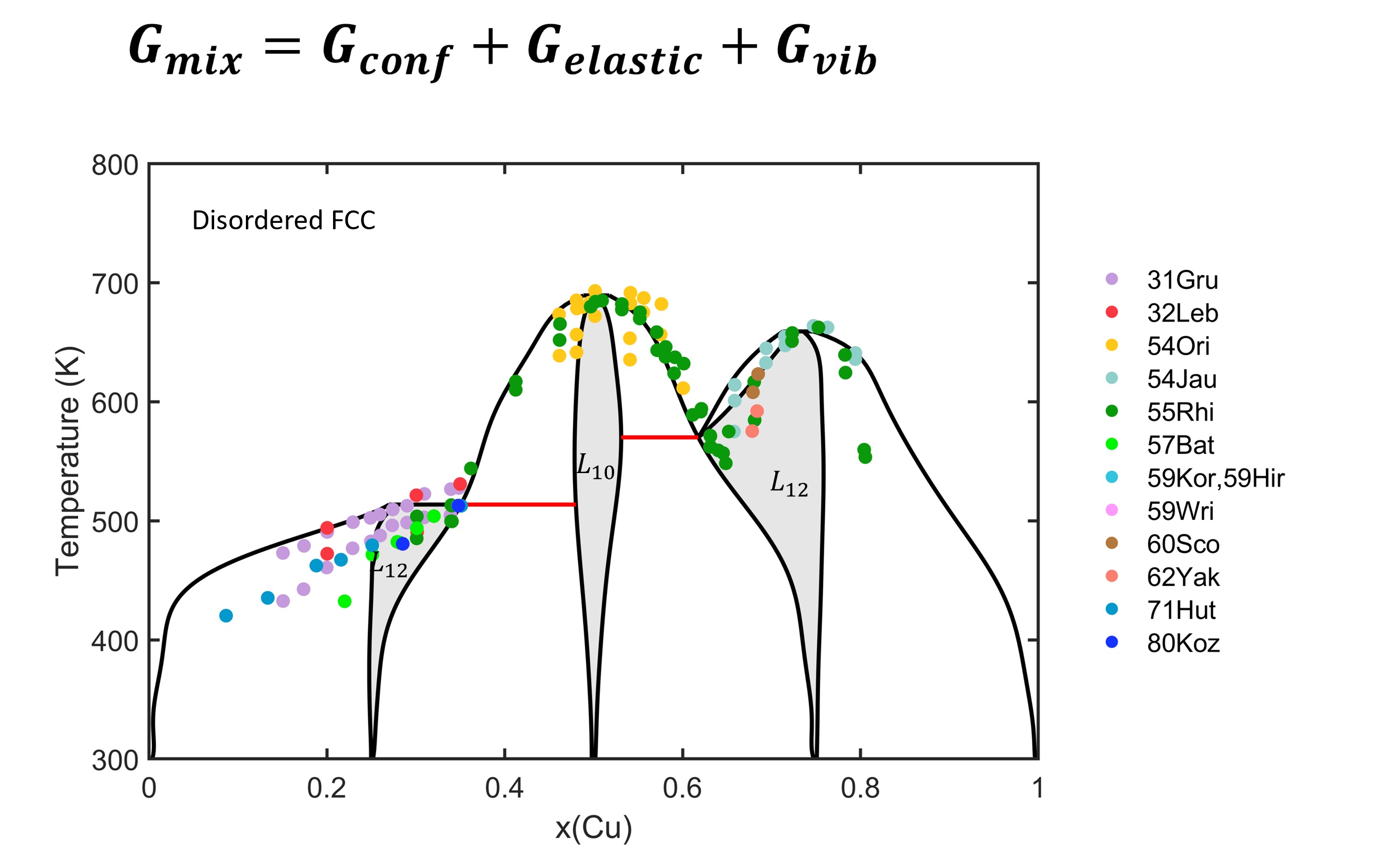}
    \caption{The Cu-Au phase diagram with vibrational contribution, elastic contribution, and configurational contribution. The experimental data are summarized in \cite{okamoto1987cu}.}
    \label{fig:PDCUAU3}
\end{figure}

\begin{figure}
    \centering
\includegraphics[width=1\textwidth]{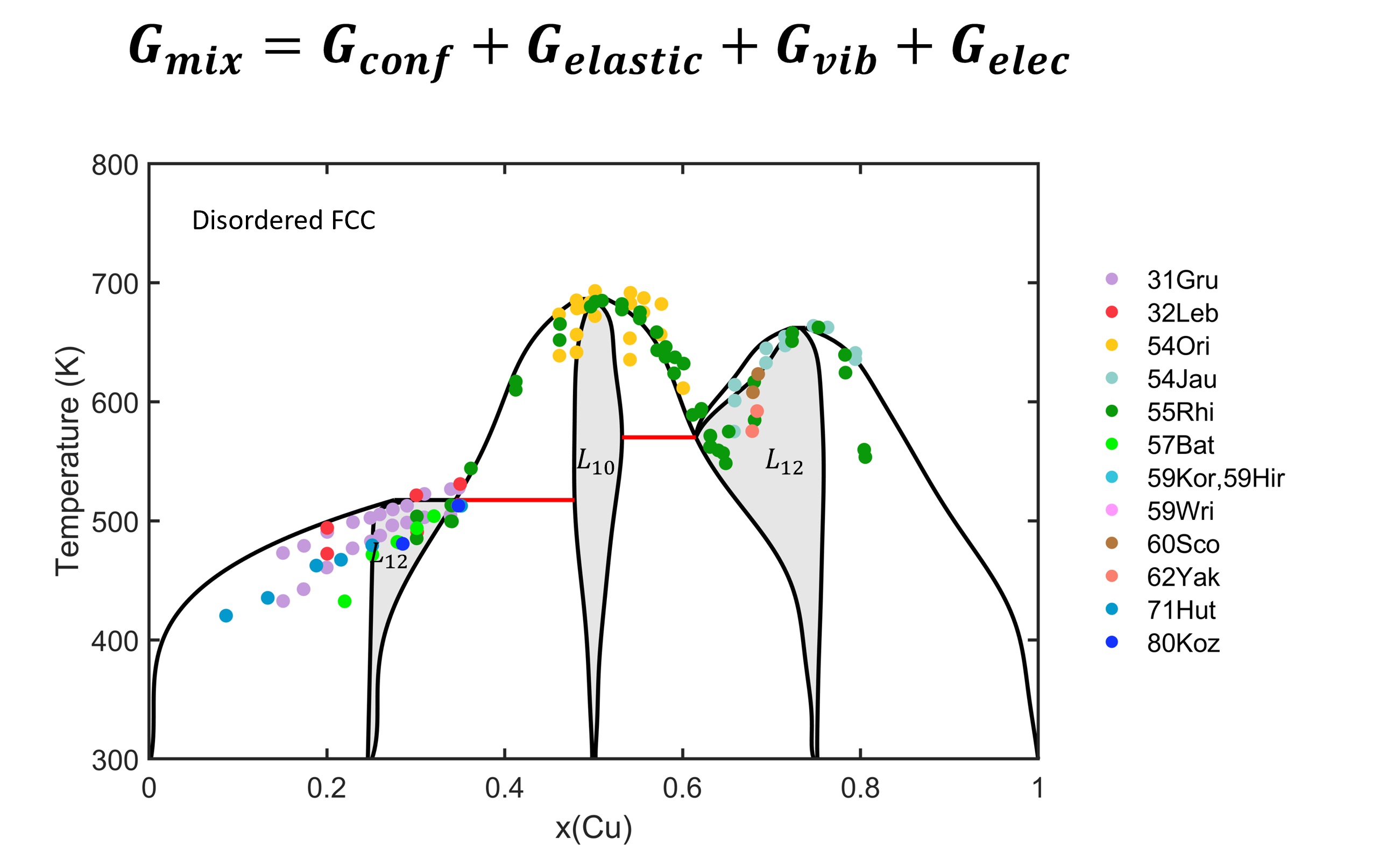}
    \caption{The Cu-Au phase diagram with electron contribution, vibrational contribution, elastic contribution, and configurational contribution. The experimental data are summarized in \cite{okamoto1987cu}.}
    \label{fig:PDCUAU4}
\end{figure}

The phase diagram of the Cu-Au system with only chemical contribution is plotted in figure \ref{fig:PDCUAU1}. In contrast, the chemical and elastic contribution phase diagram is plotted in figure \ref{fig:PDCUAU2}. We conclude that the elastic contribution would enlarge the mixing region after determining the chemical contribution under the fixed volume. On the other hand, we notice that the chemical contribution has already captured the topology of the phase diagram and is close to the experimental phase boundaries. But only the configurational contribution still cannot precisely reflect the experimental phase diagram, especially within the mixing region, and the order-disorder transition for the $Cu_3Au$ is significantly different. As a result, the next step is to implement the vibrational contribution with the extended bond proportion model and manipulate the frequencies ratio and local chemical parameters for vibration contribution to make the FYL-CVM phase diagram approach to the phase boundary data. We reached the results in agreement with the experimental data presented in figure \ref{fig:PDCUAU3} and add the electronic contribution in figure \ref{fig:PDCUAU4}. Table \ref{table:parameterCuAu} presents all the parameters used. 

Here the chemical energy, electronic energy, and elastic term are directly obtained from the first-principles calculation or the literature based on the fitting from the mechanical properties of the materials. Only three vibrational parameters are used for the phase boundary fitting of this benchmark system to finish the assessment. We conclude that: first, the electronic contribution is relatively small but still changes a little bit of the phase diagram. Second, the vibrational contribution would mostly reduce the order-disorder transitions' transition temperature. This is in good agreement with the previous first principle calculated vibrational result \cite{ozolicnvs1998first}, which indicates that the vibrational contribution will probably lower the transition temperature in the Cu-Au system.

\begin{table}[h!]
\centering
\begin{tabular}{||c || c||} 
 \hline
 Parameters & Values  \\ 
 \hline\hline
$ E_{Cu_3Au} $& -0.09070614018eV\\ 
 $E_{Cu_2Au_2}$ & -0.1167207373325eV\\
 $E_{CuAu_3}$ & -0.07964138709eV\\
 $F_{ele, Cu_3Au}$ & $ (-3.99\times 10^{-10}T^2 +-2.85375\times 10^{-7} T + 0.0001506375)$eV\\
 $F_{ele, Cu_2Au_2}$ & $(-2.7 \times 10^{-10} T^2 -4.03\times 10^{-7} T + 0.000173825)$eV \\ 
 $F_{ele, CuAu_3}$ &  $(-0.000000000857 T^2 +-0.0000004965 T + 0.0002072925)$eV\\ 
 $\Omega_{Cu,Au}$ & 0.5814eV\\
$ r_{vib} $ &  1.048\\
 $p_{lc}(Cu_3Au)$ & 1.038\\
$ p_{lc}(CuAu_3)$ & 0.985\\
 \hline
\end{tabular}
\caption{The value for parameters used in the Cu-Au system modeled with the FYL-CVM. Here $E$ refers to chemical cluster energy, and $F_{ele}$ refers to the electronic free energy at the specific cluster. $\Omega$ is the elastic constant mentioned in the elastic contribution, $r_{vib}$ is the vibrational ratio in the extended bond proportional model, and $p_{lc}$ is the local chemical parameter for the specific clusters.}
\label{table:parameterCuAu}
\end{table}

Based on the current progress, we can conclude that the proposed FYL-CVM-CALPHAD framework balances accuracy and efficiency. Determining most parameters could be automatically performed since we have constructed a clear workflow starting from the first-principles calculation. Then only three parameters of the vibrational contribution need to be adjusted during the process. Besides, the trend of each vibrational parameter is apparent in both the prototype AB system and the Cu-Au system here. When we manipulate the vibrational parameters, we would know how the phase diagram shift before the calculation. This would be helpful for people to perform the thermodynamic assessment and reach the satisfied phase diagram in a relatively short time. All these advantages would accelerate the modeling procedure for searching for the aprameters.

However, We also need to deal with thermochemical properties. As we mentioned, the CALPHAD could be considered a kind of multimodal data mining task specialized for thermodynamics, we should also check the learning results of many other thermodynamic properties such as mixing enthalpy, entropy, activity, etc. This would check the accuracy of the current framework performance with the limited training data in one modality.

\begin{figure}
    \centering
\includegraphics[width=1\textwidth]{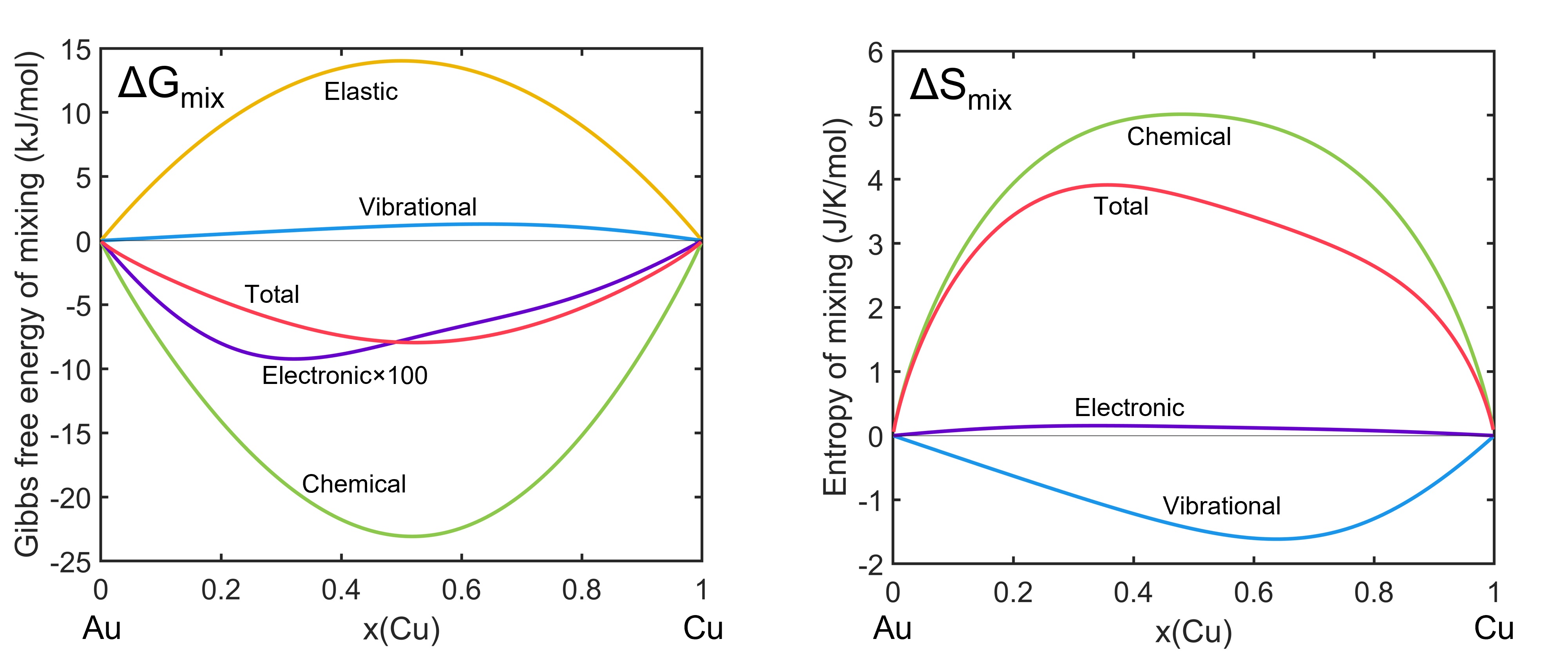}
    \caption{The thermodynamic properties(free energy and entropy) of Cu-Au at around the 800K. }
    \label{fig:800kproperty}
\end{figure}

First, we provide the calculated thermodynamics property at 800K, the results are presented in figure \ref{fig:800kproperty}. Based on the phase diagram, all disordered phases should be covered among the whole composition line. We notice that in the Cu-Au system, the large lattice mismatch would lead to large elastic energy, while electronic contribution is usually very small. However, we noticed that there is an important deviation coming from the vibrational part. The vibrational free energy should be negative while the vibrational entropy is expected to be positive \cite{ozolicnvs1998first}. We will discuss this issue later after we check other properties.

\begin{figure}
    \centering
\includegraphics[width=1\textwidth]{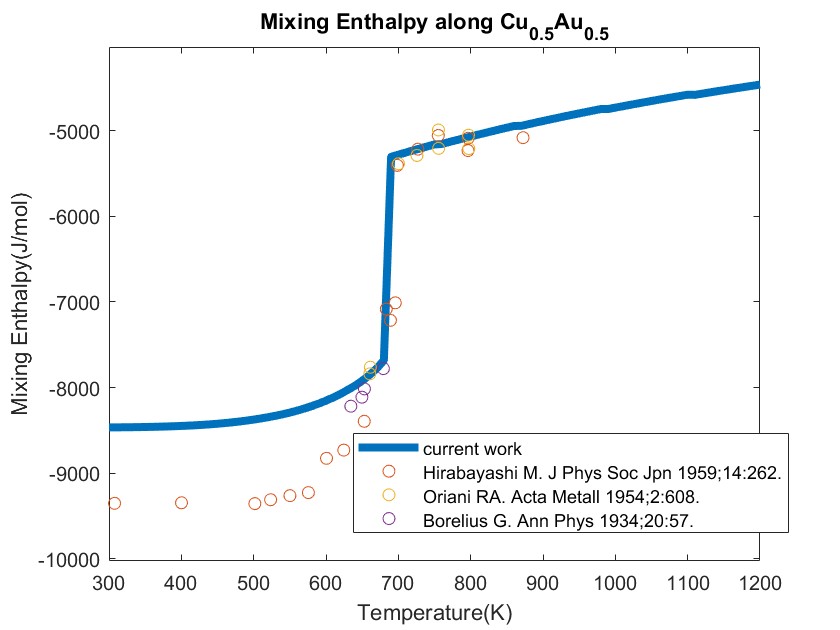}
    \caption{The change of mixing enthalpy with the temperature along the composition=0.5, the experimental data comes from \cite{oriani1954thermodynamics,hirabayashi1959electrical,borelius1934theory},The experimental data is collected by \cite{cao2007thermodynamic}.}
    \label{fig:Ecuau}
\end{figure}

\begin{figure}
    \centering
\includegraphics[width=1\textwidth]{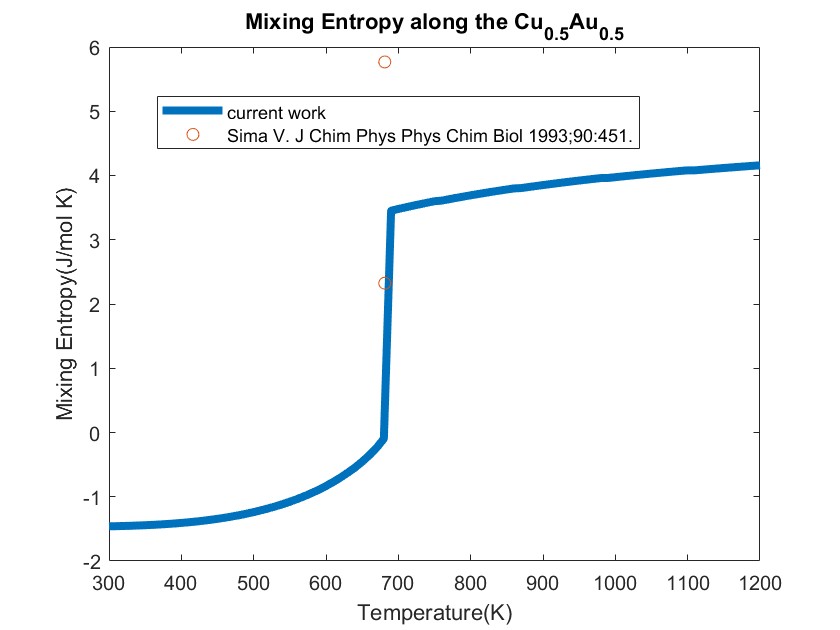}
    \caption{The change of mixing entropy with the temperature along the composition=0.5. The experimental data is collected by \cite{cao2007thermodynamic}.}
    \label{fig:Scuau}
\end{figure}

\begin{figure}
    \centering
\includegraphics[width=1\textwidth]{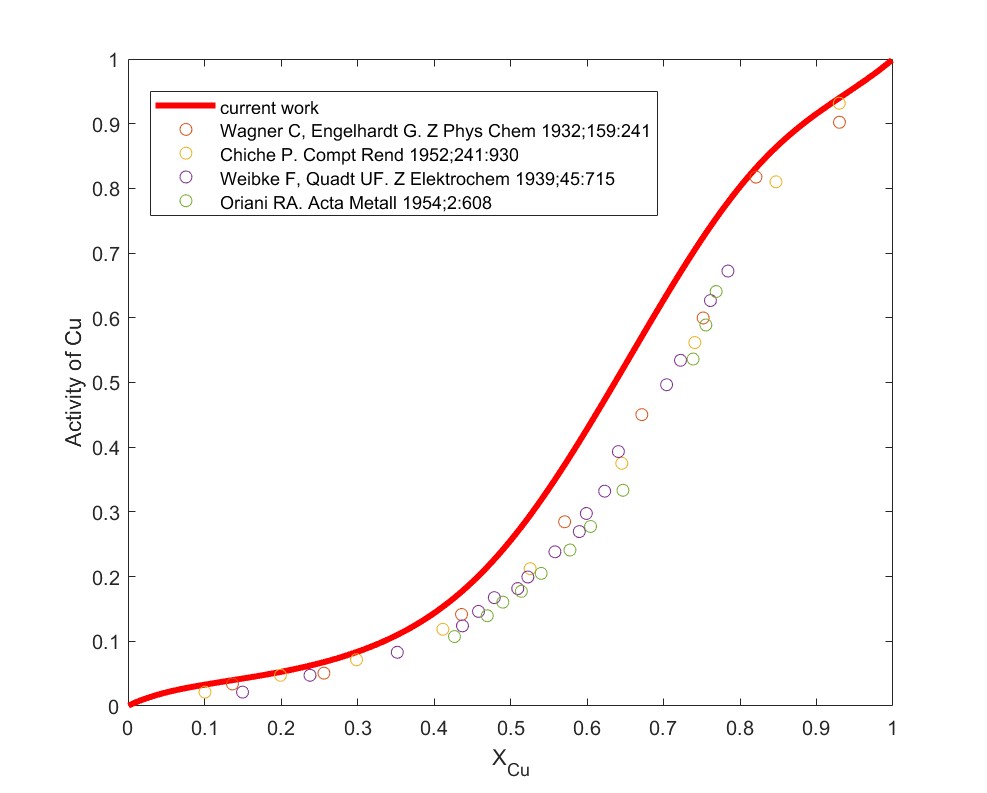}
    \caption{Activity of Cu v.s. $x_{Cu}$ at around 770K. The experimental data is collected by \cite{cao2007thermodynamic}.}
    \label{fig:activity}
\end{figure}

Second, we want to compare our calculated results with the experimental data. The change of the mixing enthalpy with temperature is presented in figure \ref{fig:Ecuau}, while the mixing entropy v.s. temperature and the activity of Cu v.s. $x_{Cu}$ are presented in figure \ref{fig:Scuau} and figure \ref{fig:activity}.

The current calculated mixing enthalpy and the experimental data matched with each other. However, the mixing entropy and the activity of Cu have a relatively large deviation compared with the experimental data. For the mixing entropy, we observe that the deviation is a uniform shift: the difference between ordered mixing entropy and disordered mixing entropy is close to the gap between the experimental mixing entropy for the ordered phase and the disordered phase at the transition temperature. This indicates a potential missing of the entropy uniformly happened. For the activity of the Cu diagram, we can notice that the current calculated activity of Cu is qualitatively in great agreement with the experimental activity data. However, there is still some deviation, especially at the middle composition around $x_{Cu} = 0.5$. This deviation indicates that the mixing free energy should be smaller at around $x_{Cu} = 0.5$.

All these calculations indicate a potential missing part of the contribution. We would consider the missing part comes from the vibrational contribution. The dilemma is that on the one side, we have already found a set of parameters with an excellent agreement based on the phase boundary data. This means the vibrational contribution should have been fully taken into account. On the other hand, several thermodynamic properties still indicate the deviation and should come from the vibrational contribution.

We must date back to the original ensemble used to derive the (FYL-)CVM to deal with this dilemma. All the CVM-related analytical model implies the volume fixed to reach the equilibrium. As a result, all the quantities in the partition function should reflect the same equilibrium volume. However, the volume change would significantly affect the vibrational contribution. When we perform the CVM calculation, it has to fix the volume based on the requirement from the ensemble: all the energy in the partition function is equilibrium volume independent. It means the mixing vibrational entropy calculated from CVM-like analytical model has not considered the volume change. We must add the vibrational contribution for volume change back into the mixing entropy to ensure all the calculation is consistent. We hypothesize that if we could quantify the vibrational entropy with the volume change for pure elements Cu and Au, we could produce the correct mixing of entropy and free energy after considering them. The related calculation is in progress to pursue.

The volume deviation of vibrational contribution comes from the ensemble we selected. The ensemble is $\mu VT$, so we fix the volume during the CVM or FYL-CVM calculation. What if we change it into some ensemble to allow volume change? Such as $\mu PT$ ensemble? Then the partition function may require adding the $PV$ term to quantify the change in volume. This is another valuable direction to avoid the issue in the CVM-like calculation caused by the volume.

With the help of the current thermodynamic framework, we can access the chemical SRO parameters among the whole composition space after we successfully assessed the thermodynamic model we constructed for the Cu-Au system. It only requires the cluster probability distribution, which the equilibrium minimization could obtain during the phase diagram calculation. As a result, the SRO diagram could be considered the byproduct after the thermodynamic assessment. 

  \begin{figure}
    \centering
\includegraphics[width=1\textwidth]{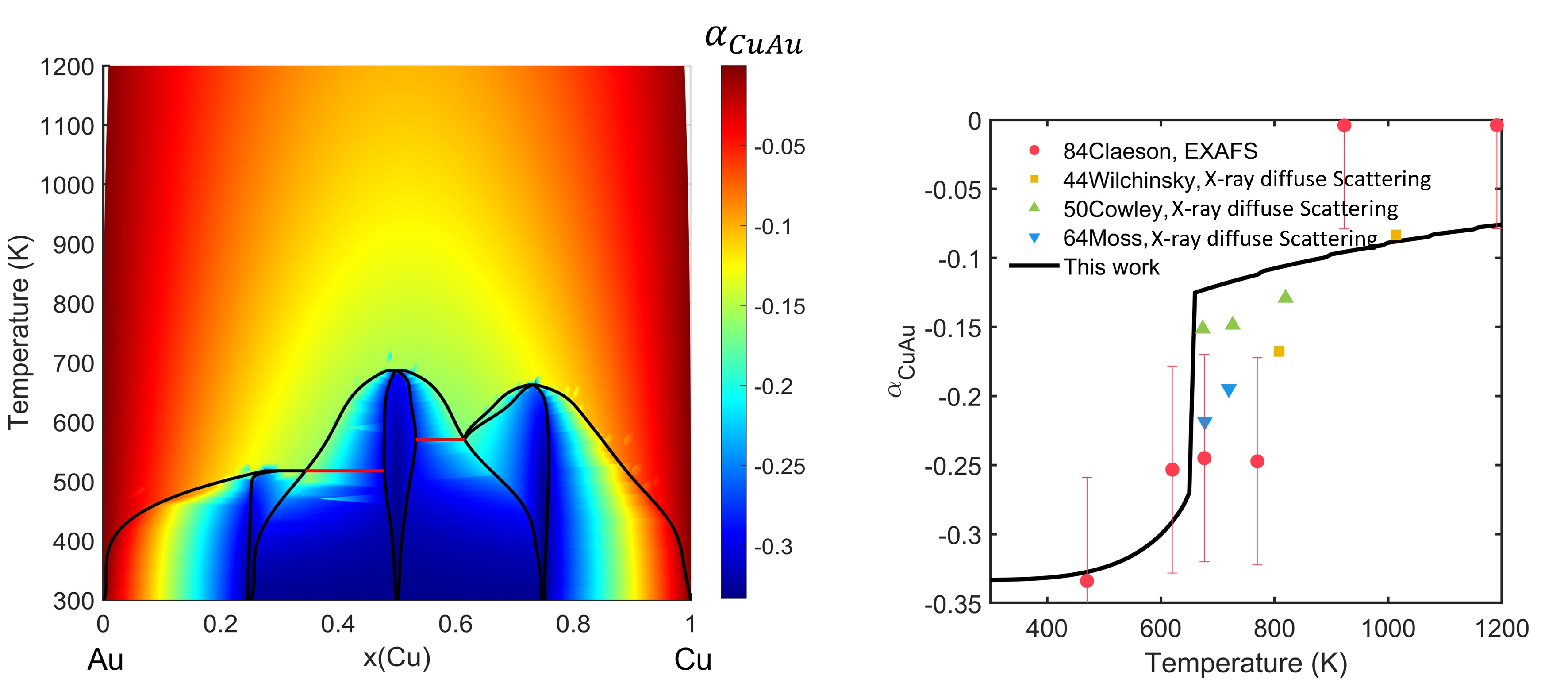}
    \caption{Left: the diagram of Warren-Cowley SRO parameter among the x-T space. Right: the Warren-Cowley SRO parameters v.s. temperature along the $Cu_3Au$ composition. The experimental data is collected from \cite{claeson1984order}, based on EXAFS and x-ray diffuse scattering.}
    \label{fig:cuausro}
\end{figure}
\begin{figure}
    \centering
\includegraphics[width=1\textwidth]{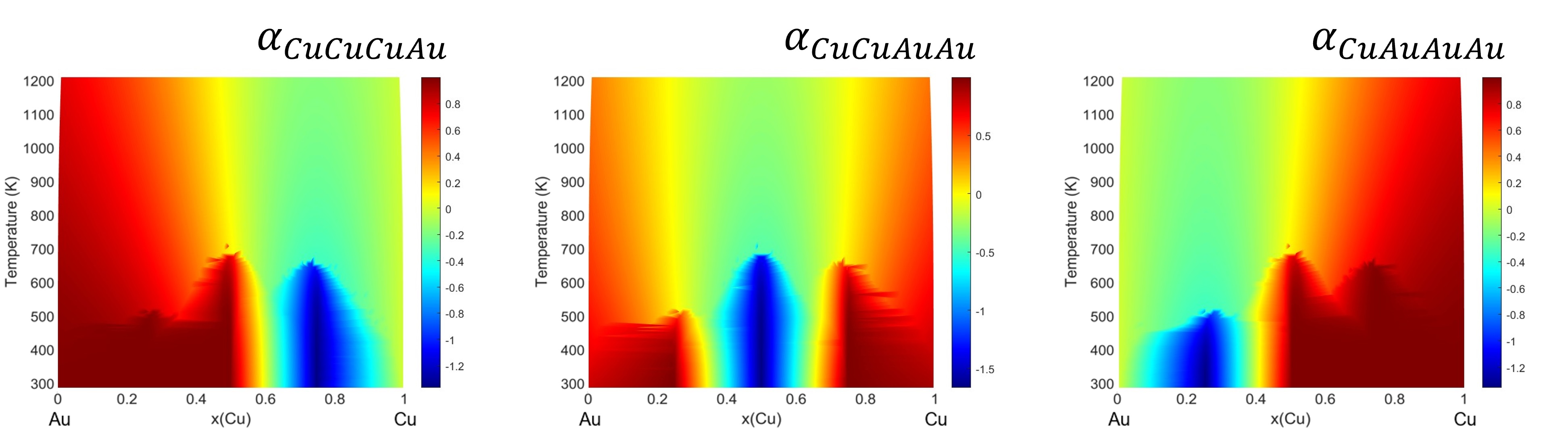}
    \caption{The four-point SRO parameters for $Cu_3Au$, $Cu_2Au_2$, and $CuAu_3$ in Cu-Au system.}
    \label{fig:SROCUAU4POINT}
\end{figure}

There are different ways to quantify the chemical SRO. One of the most famous ways might be the Warren-Cowley SRO \cite{cowley1950approximate}. The relation between this parameter and the pair probability could be:
\begin{equation}
\begin{split}
 \alpha_{ij}^{m} = 1-\frac{p_{ij}^{m}}{c_{i}c_{j}}
    \end{split}
    \label{eq:para1}
\end{equation}
where $m$ is the label for the specific shell, and $ij$ represents different occupations of these specific pairs. $c_i$ or $c_j$ is the composition of these atoms. The result of this pairwise SRO parameter is presented in figure \ref{fig:cuausro}. We also compare how the SRO parameters change with the temperature along the $Cu_3Au$ composition to the experimental data. The experimental data is collected from the previous literature \cite{claeson1984order}.
As far as we know, it is the first attempt to provide the chemical SRO parameters among the whole x-T space, which is due to the benefit of the simplified configurational model FYL-CVM. this chemical SRO parameter diagram would be taken as a computational tool to design the disordered materials with the specific chemical SRO. Based on our current assessment, the SRO parameters matched the previous EXAFS and X-ray data for quantifying the chemical SRO. This might reveal the satisfied ability of current CVM-CALPHAD for incorporating the chemical SRO into the computational thermodynamic framework as a novel and essential dimension for materials design \cite{nohring2020design}.
Our configurational model is based on the four-point tetrahedron as the basic cluster. Actually, it's even possible to extend the chemical SRO parameters from the pairwise into multi-point case \cite{goff2021quantifying}:

\begin{equation}
\begin{split}
 \gamma_{\alpha}(\sigma_{\alpha}^{k}) = 1- \frac{P(\sigma_{\alpha}^{k})}{P_{random}}
    \end{split}
    \label{eq:para2}
\end{equation}
where the order parameter $\gamma_{\alpha}$ is given in terms of the average probability of finding a cluster with
a desired occupation, $\sigma_{\alpha}^{k}$ in the crystal that is normalized by the probability of the cluster forming
with the desired occupation in a random alloy $P_{random}$.  In the random alloy, the probability of a site being occupied by a specific species, or say, is given by the atomic concentration of that species $C_i$. For example, if we select the three-point cluster $AAA$, the $P_{random}=C_A C_A C_A$. In our framework, it could be directly calculated through the cluster probability distribution after we reach the equilibrium. The four-point SRO parameter results based on our current assessment are presented in figure \ref{fig:SROCUAU4POINT} and satisfy the intuition of understanding the SRO among the whole x-T space.

\section{FYL-CVM calculated results for Cu-Ag and Au-Ag system}

Since we have built up the thermodynamic modeling for the Cu-Au system, we would forward our modeling for Cu-Ag and Au-Ag systems to prepare the SRO calculation for Cu-Au-Ag ternary system. However, people don't observe the clear order-disorder transitions in these two systems due to possible slow kinetics for Cu-Ag and phase separation for Au-Ag \cite{wei1987first}. As we have constructed the workflow to reach the cluster energy, we could perform the first-principle calculation to observe the fcc ordering phenomena in these two systems. All the first-principles calculation based on the DFT is performed through the Vienna \textit{ab initio} simulation package (VASP) 5.4 \cite{kresse1996efficient} within the generalized gradient
approximation (GGA) of Perdew, Burke, and Ernzerhof (PBE) \cite{perdew1996generalized}. The ion–electron interactions were described by the projector augmented wave method (PAW) \cite{kresse1999ultrasoft} with an energy cutoff of 500 eV for all the structures. Combined with the elastic data from the previous literature, we could get all the energetic parameters we want to construct the thermodynamic system for Cu-Ag and Au-Ag. The related parameters are presented in table \ref{table:parameterCuAuAg}.

  \begin{figure}
    \centering
\includegraphics[width=0.8\textwidth]{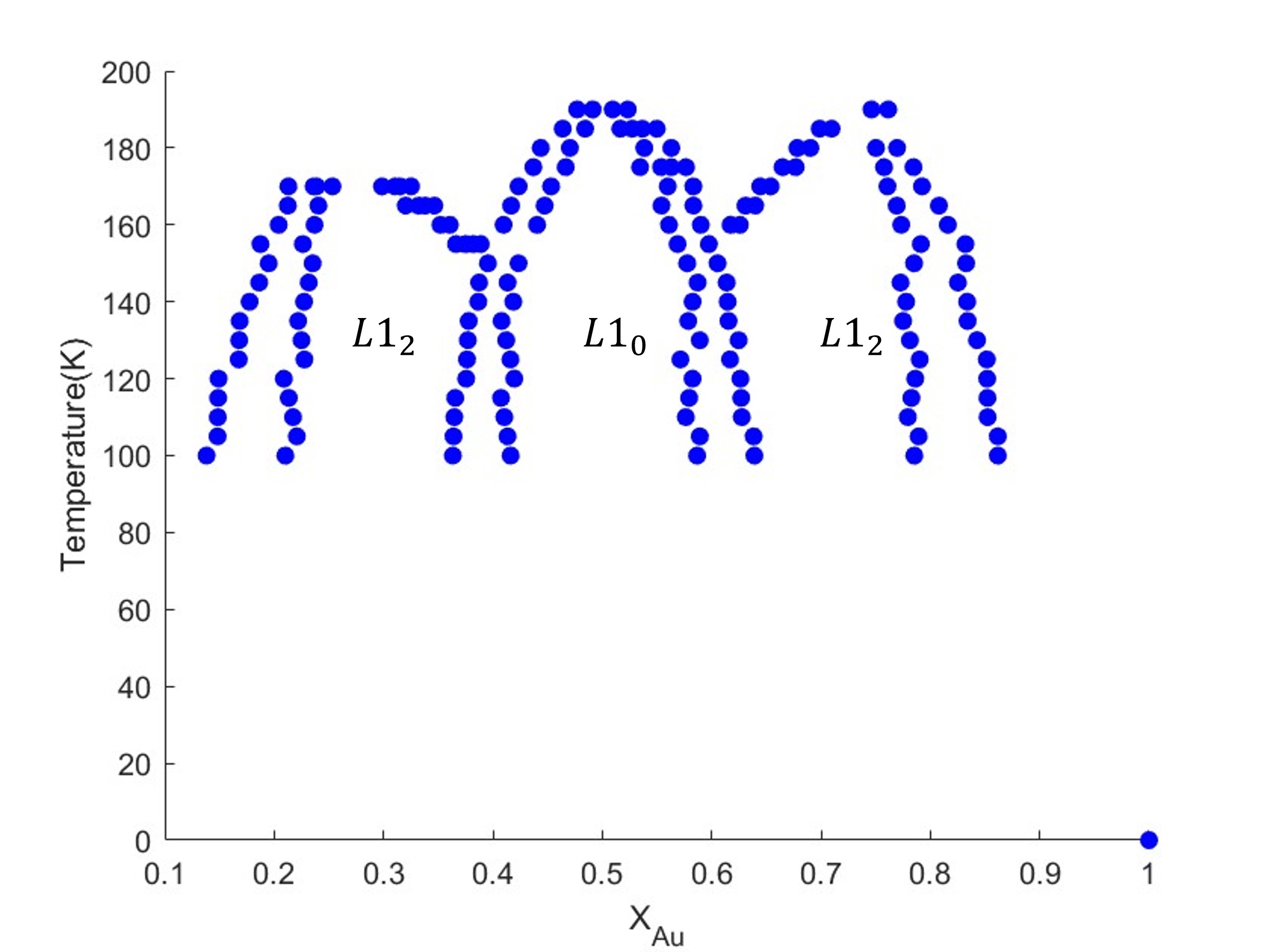}
    \caption{The raw data for Au-Ag phase diagram}
    \label{fig:PDAUAG}
\end{figure}
Based on the calculated cluster energy parameters, we could briefly analyze the Cu-Ag and Au-Ag systems' thermodynamics and then step into the thermodynamic modeling for the ternary system.

  \begin{figure}
    \centering
\includegraphics[width=0.4\textwidth]{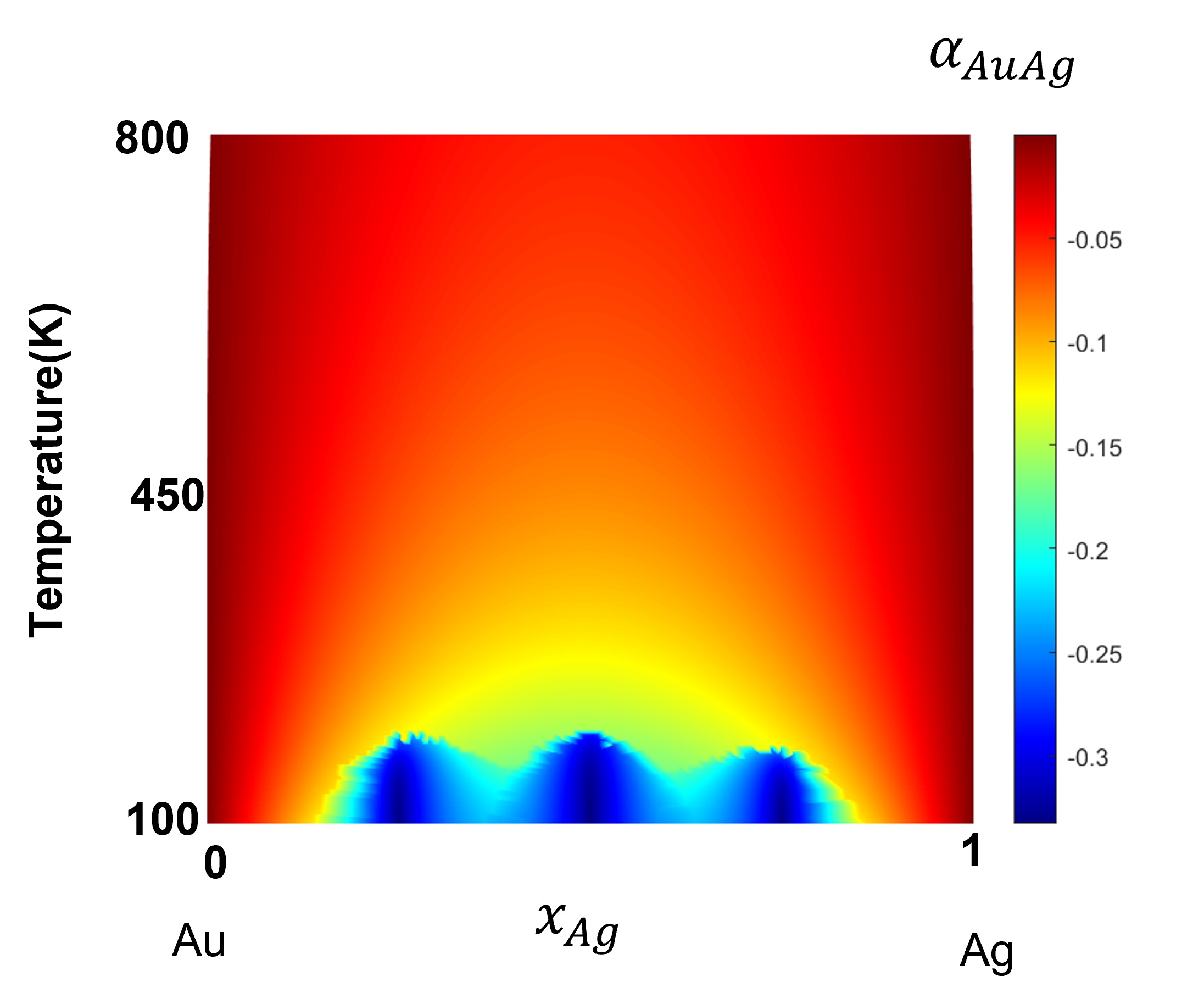}
\hspace{1in}
\includegraphics[width=0.4\textwidth]{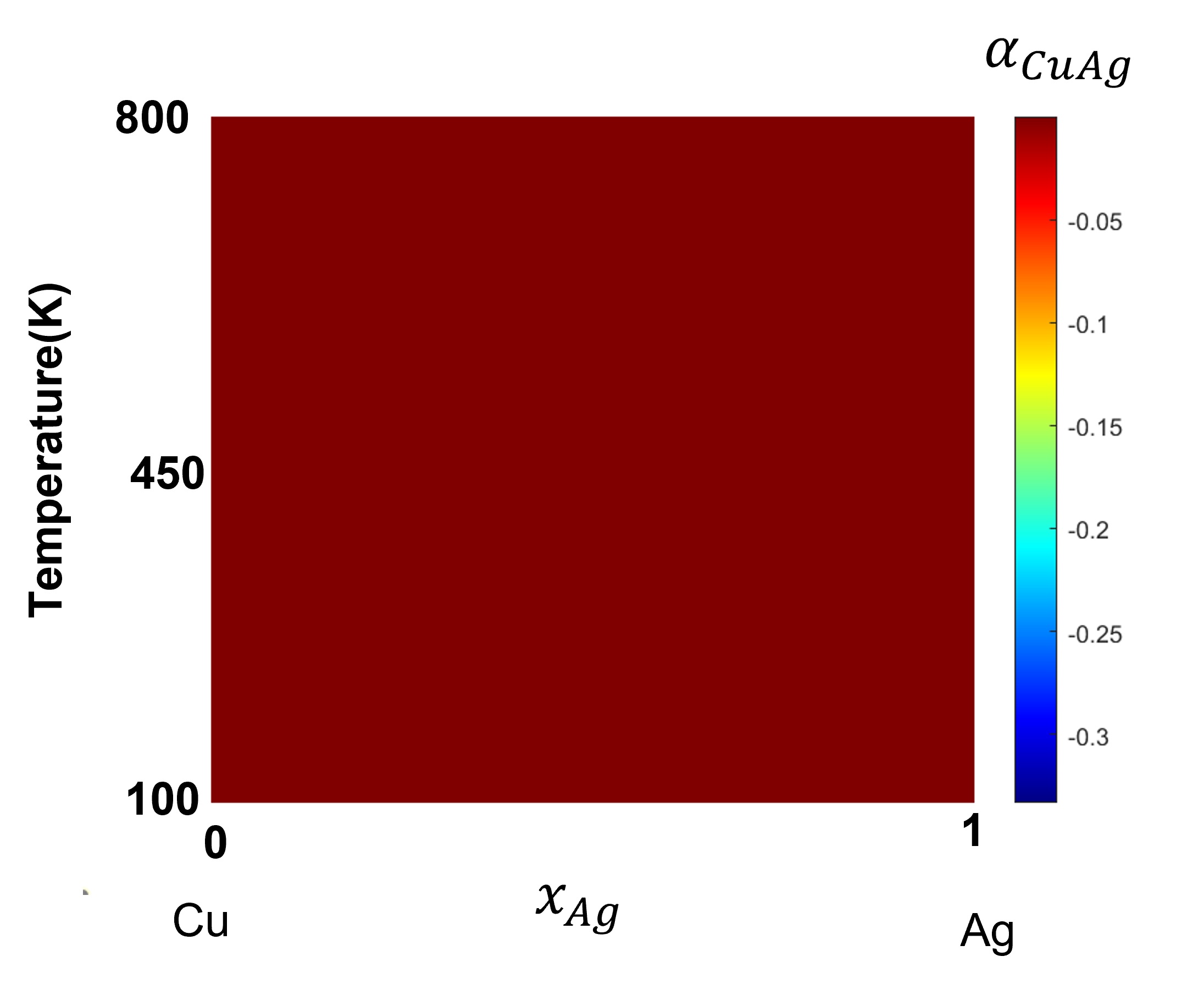}
    \caption{Pairwise SRO parameters for Au-Ag(left) and Cu-Ag(right).}
    \label{fig:SRO2PAIR}
\end{figure}

Our current phase boundary detection algorithm cannot find the phase boundary between the ordered and disordered phases of the Cu-Ag system. The equilibrium calculated results indicate all regions are disordered, and phase separation exists. The reason for the Cu-Ag system is a phase separation system should attribute to the elastic contribution. As the elastic contribution would provide a positive parabola-like free energy contribution, increasing such kind of contribution would dominate the total free energy curve. It would cause a free energy curve with a mountain peak at the middle composition and two valleys at the edge of the composition. This shape of the free energy curve would cause the phase separation. Based on the current data, the elastic constant within the Cu-Ag system based on the parameter table \ref{table:parameterCuAuAg} precisely reflects an overwhelmed elastic effect which would lead to this phase separation case.


  \begin{figure}
    \centering
\includegraphics[width=0.4\textwidth]{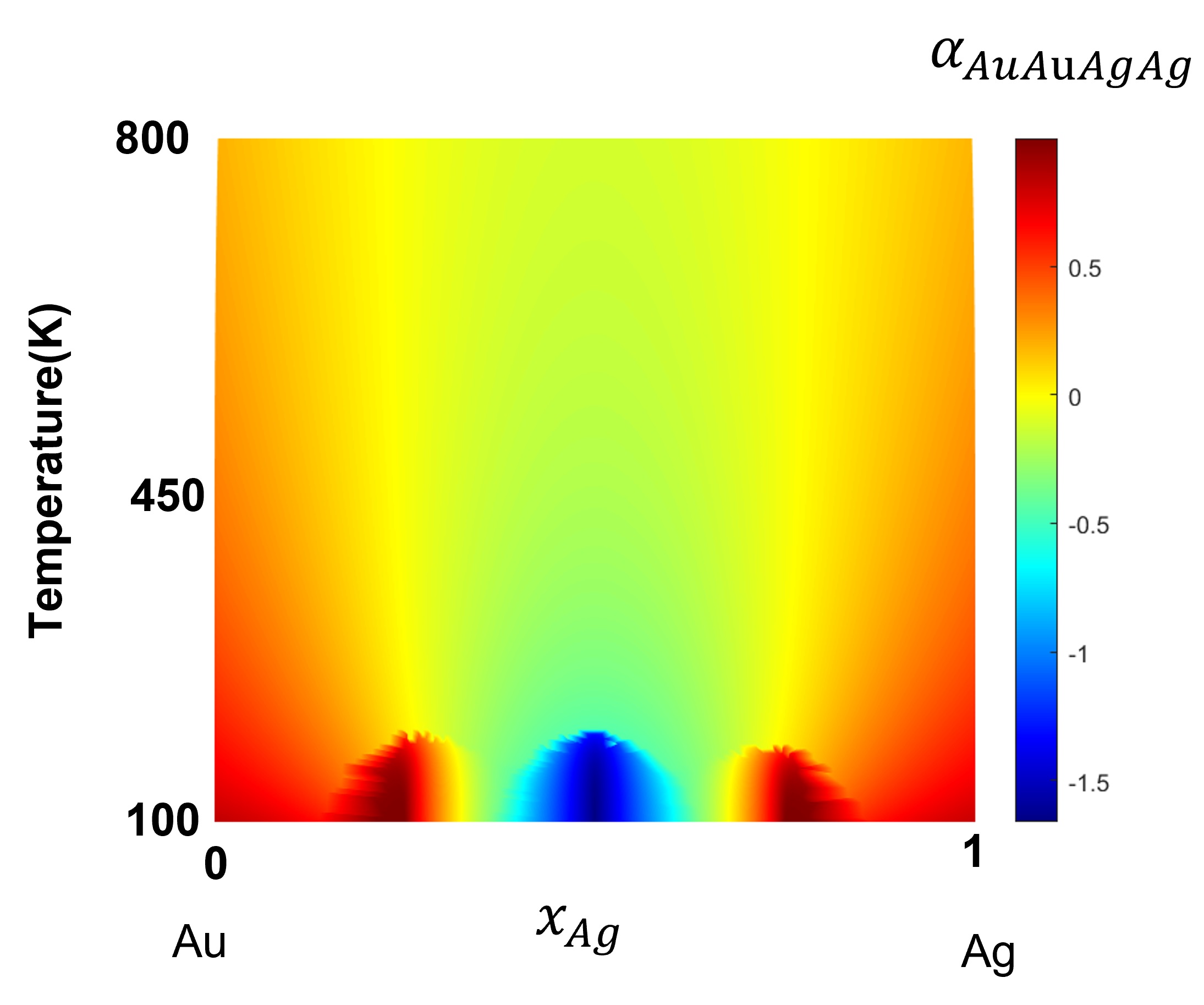}
\hspace{1in}
\includegraphics[width=0.4\textwidth]{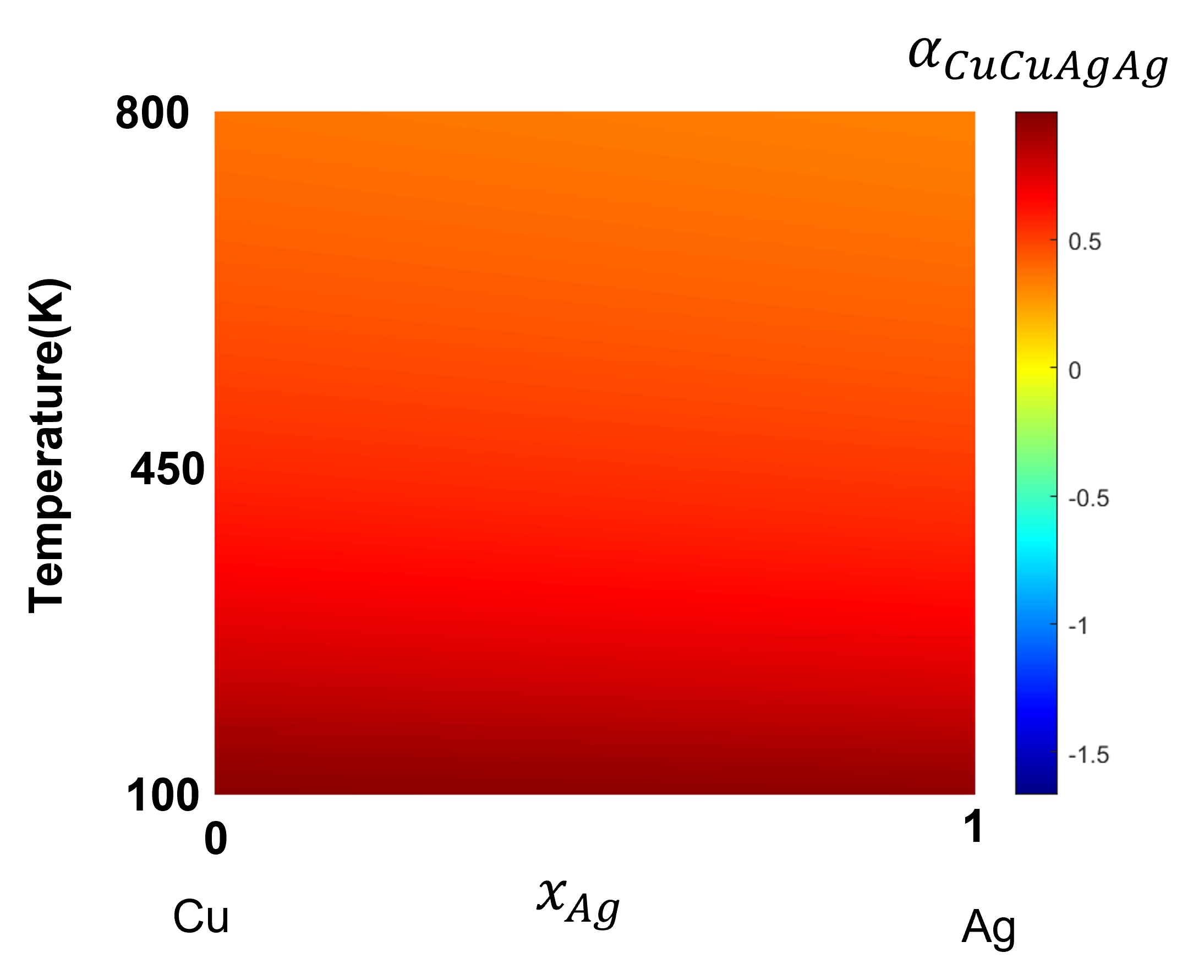}
    \caption{4-point SRO parameters for Au-Ag(left) and Cu-Ag(right).}
    \label{fig:SRO2}
\end{figure}

We could access the phase diagram through the phase boundary calculation for the Au-Ag system, presented in figure \ref{fig:PDAUAG}. The mixing region is relatively small and sharp, compatible with the small elastic contribution determined due to the good lattice mismatch. The low order-disorder transition points reflect the small chemical energy, which matched the calculation in the previous literature \cite{wei1987first}. This might be the reason why the order-disorder transition doesn't appear in the experimental phase diagram \cite{baker1992alloy}, the transition point may be too low to be detected, and the temperature is small to drive the kinetics of the phase transformation for this system.

We also perform similar SRO parameters calculations for both the Au-Ag and the Cu-Ag systems. The Warren-Cowley SRO parameters $\alpha_{AuAg}$ and $\alpha_{CuAg}$ are presented in \ref{fig:SRO2PAIR}, and the 4-point SRO parameters $\alpha_{AuAuAgAg}$ and $\alpha_{CuCuAgAg}$ are presented in \ref{fig:SRO2}. Based on the SRO figures among the whole $x-T$ space, we can observe their thermodynamic features as well: the Au-Ag system has low order-disorder transition points and Cu-Ag has the phase separation for the disordered phase.

\begin{table}[h!]
\centering
\begin{tabular}{||c || c||} 
 \hline
 Parameters & Values  \\ 
 \hline\hline
$ E_{Cu_3Au} $& -0.09070614018eV\\ 
 $E_{Cu_2Au_2}$ & -0.1167207373325eV\\
 $E_{CuAu_3}$ & -0.07964138709eV\\
 $\Omega_{Cu,Au}$ & 0.5814eV\\

$ E_{Cu_3Ag} $& -0.0407248133212249eV\\ 
 $E_{Cu_2Ag_2}$ & -0.0505371784805403eV\\
 $E_{CuAg_3}$ & -0.0386328871779974eV\\
 $\Omega_{Cu,Ag}$ & 0.792696eV\\

 $ E_{Au_3Ag} $& -0.0238639153827193eV\\ 
 $E_{Au_2Ag_2}$ & -0.0321734984592632eV\\
 $E_{AuAg_3}$ & -0.0233339448061753eV\\
 $\Omega_{Au,Ag}$ & 0.01041eV\\

$ E_{Cu_2AuAg} $& -0.0855780562499998eV\\ 
$ E_{CuAu_2Ag} $& -0.0746512812500001eV\\ 
$ E_{CuAuAg_2} $& -0.0650539531249997eV\\ 
 \hline
\end{tabular}
\caption{The value for parameters used in the Cu-Au-Ag system. Here $E$ refers to chemical cluster energy, $\Omega_{ij}$ is the elastic constant mentioned in the elastic contribution for $ij$component pairs.The elasticity-related data is extracted from \cite{wei1987first}.}
\label{table:parameterCuAuAg}
\end{table}

\section{SRO Parameters Diagram for the Disordered stable/metastable Phase of Cu-Au-Ag system}
The SRO in the ordered phase is relatively trivial since the ordered phase has already been dominated by the specific long-range order, which would overwhelm the SRO. On the other hand, our current code has already been available to quickly access the thermodynamics for the disordered phase for any component materials after determining the cluster energy. The only challenge left for the complete thermodynamic description is to deal with the ordered phase, as the ordered phase may need further symmetry consideration during the minimization and requires more effort to implement the algorithms for the multi-component system. Since we are primarily interested in the chemical SRO of the disordered stable or metastable phase to understand how to make use of this feature to manipulate the properties of the material within the disordered phase, we would mainly present the chemical SRO for the disordered phase of the Cu-Au-Ag system to display the performance of FYL-CVM to quantify the chemical SRO in disordered solid solution. 

To keep consistency, all the first-principles calculation based on the DFT is performed within the GGA of PBE \cite{perdew1996generalized}. The ion–electron interactions were described by PAW \cite{kresse1999ultrasoft} with an energy cutoff of 500 eV for all the structures. The determined cluster energy parameters are all presented in the table \ref{table:parameterCuAuAg}.

\begin{figure}
    \centering
\includegraphics[width=0.9\textwidth]{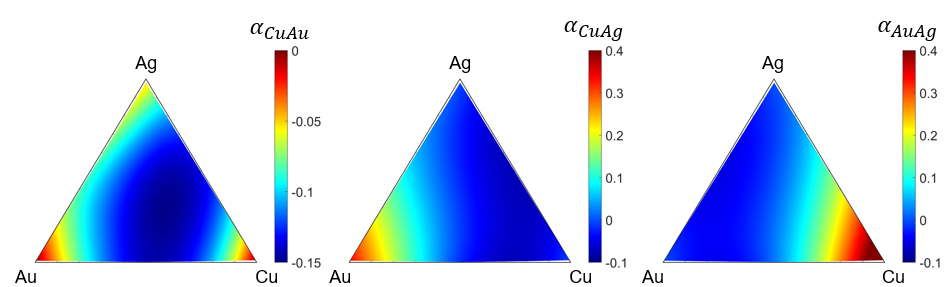}
    \caption{SRO parameter for Cu-Au, Cu-Ag, Au-Ag pairs at 800K within the disordered phase of Cu-Au-Ag system}
    \label{fig:6-4}
\end{figure}

After determining the cluster energy parameters in the previous sections, we could easily reach the chemical SRO of Cu-Au, Cu-Ag, and Au-Ag pairwise SRO parameters in 800K presented in figure \ref{fig:6-4}. This represents whether specified pairs would like or dislike such kind of composition and temperature.

\begin{figure}
    \centering
\includegraphics[width=1\textwidth]{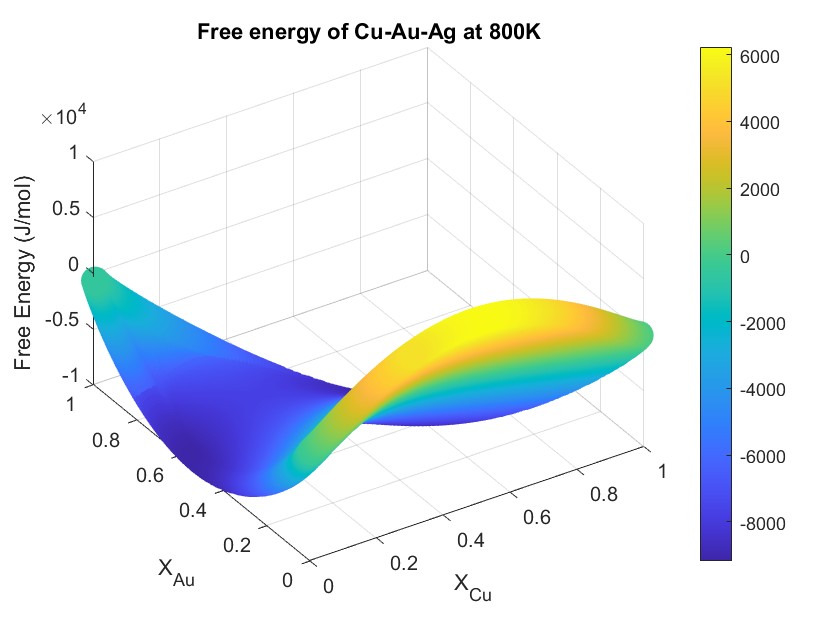}
    \caption{Free energy surface for the disordered phase of Cu-Au-Ag system }
    \label{fig:6-42}
\end{figure}

We notice that for the Cu-Ag pairs at the Cu-Ag edge, the SRO parameters indicate they should be ordered. However, based on our calculation for Cu-Ag, the cluster energy parameters determined should make the system become phase separation but not the ordering. 
Notice that what we calculated is for the disordered phase. The reason for this inconsistency comes from the miscibility gap within the disordered phase. Using the electron probe microanalysis, Ntukogu and Cadoff investigated the miscibility gap in the disordered phase \cite{ntukogu1986effect} at temperatures 500, 600, 700, and 750 degrees. We can also confirm this miscibility gap by directly depicting the free energy surface in three dimensions plotted in figure \ref{fig:6-42}. In the Cu-Ag edge, there is a clear parabola that indicates the existence of the miscibility gap. After the convex hull projection, we could directly determine the miscibility gap region in the phase diagram at this temperature.

    \section{Conclusions}
In this Chapter, we integrate our theoretical framework, first-principles calculation, the developed minimization algorithm and workflow, and the CALPHAD methodology to produce the thermodynamic model for Cu-Au, and Cu-Au-Ag systems as the benchmark. We compare different sources of contributions to construct the complete phase diagram for the Cu-Au system and calculate the system's thermodynamic properties and chemical SRO parameters. We compare our calculated short-range order parameters with the experiments as well. We also directly generalized our calculation into the ternary disordered phase and calculated the corresponding chemical SRO diagram for the Cu-Au-Ag system.

On the one hand, we observed that the performance of the current FYL-CVM thermodynamic framework could incorporate the chemical SRO successfully into our consideration. This cluster-based thermodynamic framework plays a vital role in presenting the SRO parameters among the whole $x-T$ space, which reveals its potential to be a convenient materials design tool for future usage. On the other hand, the non-configurational contributions, such as vibrational, elastic, and electronic, are integrated very well to produce a high-quality phase diagram. All these physical contributions would increase the reliability of our framework and could help identify the different sources of the contributions for thermodynamics. One more thing is that our current code has the ability to support any SRO parameter calculations efficiently for the disordered phase of any multicomponent($\geq 3$) system.

Since our framework puts more physics including configurational contribution, vibrational contribution, elastic contribution, and electronic contribution into CALPHAD and presents excellent agreement with experimental data in the benchmark system, we would expect this framework could be more reliable for extrapolation to multi-component systems and open the path of the physics-based thermodynamic model to the next generation CALPHAD.

	\chapter{Final Conclusions and Future Work} \label{ch:future&con}
We have figured out the basic computational thermodynamics framework with the help of the cluster model for the ordered and disordered phases. We have clarified different contributions, including configurational, elastic, vibrational, and electronic contributions. We also designed the algorithms and workflow to fit the practical calculation while implementing it for demonstration within real systems. In this Chapter, we will conclude the dissertation and discuss the potential future directions to develop further the current framework for the computational thermodynamics of the solid solution.

\section{Final Conclusions}

  \begin{figure}
    \centering
\includegraphics[width=1\textwidth]{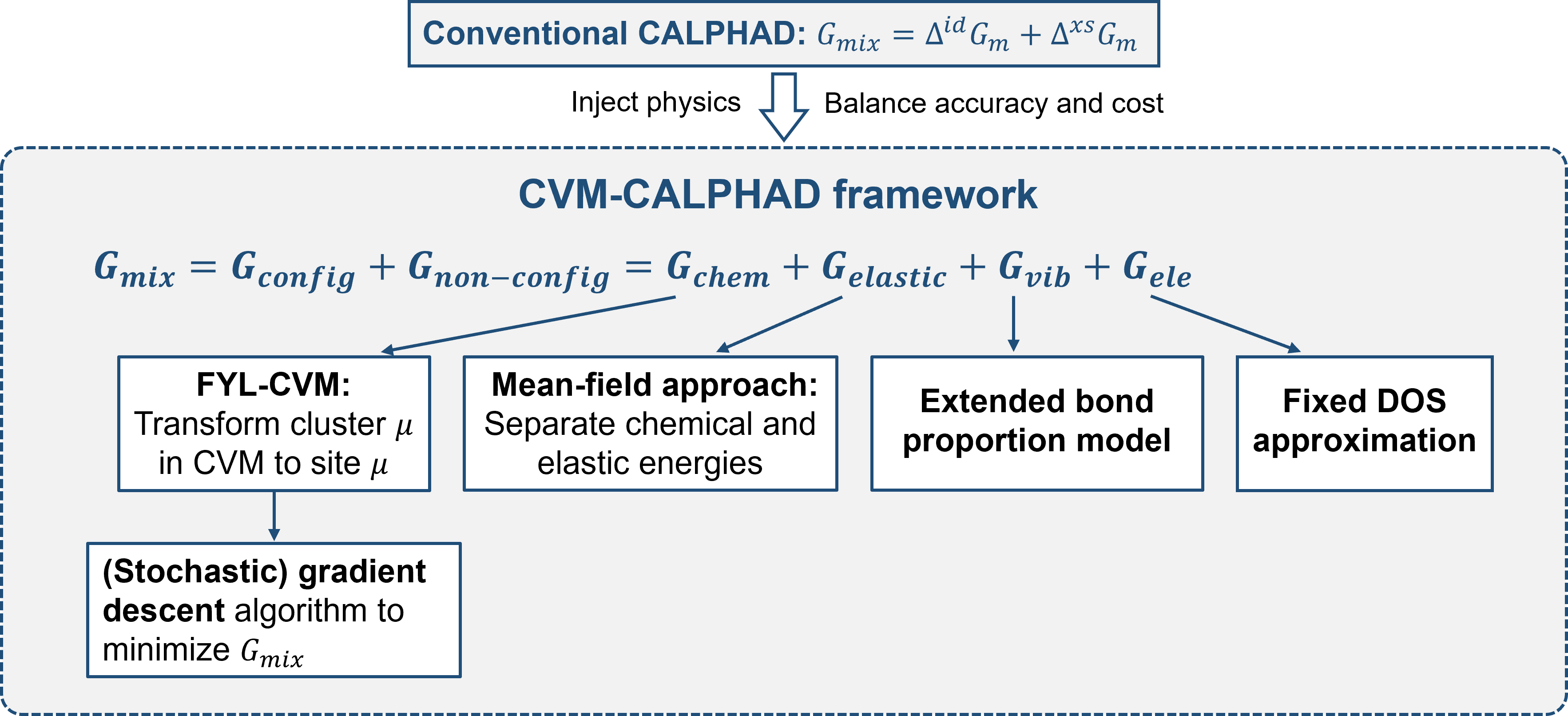}
    \caption{Summary of the work}
    \label{fig:sum}
\end{figure}

After all the efforts, we have constructed a comprehensive thermodynamic framework with intrinsic chemical SRO. This opens up the pathway to put more physics into CALPHAD, a data-driven computational thermodynamics methodology. Based on this framework, we have built up the related algorithms and workflow for the use of this near-future computational tool. We have performed benchmark tests using the Cu-Au and Cu-Au-Ag systems with our framework and obtained good agreement with the experimental data. Several considerations can be drawn:

1) With the help of FYL-transform, we construct a simpler cluster-based thermodynamic model to unify the current cluster variational understanding of solid solutions. With the CALPHAD approach, this FYL-transform simplifies and essentially reduces the dimensionality of the variable space but still captures the paramount accuracy of the cluster variation approach.

2) After we constructed the variational thermodynamic model with the statistical mechanics technique systematically, we could quickly implement many other contributions to the entropy to study how they affect the chemical configuration, such as the vibrational, electronic, and elastic contributions. It is suggested that magnetic contribution should be modelled in the future.

3) We propose the gradient descent method based on the optimization perspective to address the minimization in our current model. Besides, we emphasize the point of view that CALPHAD is a kind of specialized supervised learning task for thermodynamics. This point of view leads us to identify the challenge of parameter optimization in CALPHAD, especially for variational thermodynamics.

4) We successfully test our framework using the AB prototype system and some typical FCC ordering systems such as the Cu-Au system and the Cu-Au-Ag system as the benchmark. We present how the phase diagram of the Cu-Au system is influenced by different contributions(configurational, vibrational, elastic, and electronic). The SRO parameters among the whole x-T space are presented for the Cu-Au system as well and compare the results with the experimental data. Similarly, we generalize the SRO parameters calculation to the ternary Cu-Au-Ag system in an efficient way with the help of this FYL-CVM computational thermodynamic framework. This demonstrates our ability to deal with SRO parameter calculations of disordered phases in multicomponent cases.

\section{Future Work Directions}
        \subsection{Better Efficiency}
        To really make the framework practical for industry, computational efficiency is one of the critical components. Currently, computational efficiency is still limited by many different factors. There might be several different directions to improve the efficiency:

        1) First, implement the workflow automation tool. Recently, there are many workflow automation tools taken used in computational materials science such as AiiDA \cite{pizzi2016aiida}, aflow++ \cite{oses2023aflow++}, DFLOW \cite{dflow2023github}, etc. With the help of these workflow automation tools, we expect to generate the cluster energy automatically and directly connect the first-principles data into the cluster model we developed for computational thermodynamics and speed up the CALPHAD database development.

        2) Second, develop a code-efficient computational package. Integrating all the models and algorithms into a code-efficient Python package to deliver to other researchers would benefit the whole community. This should involve the current thermodynamics model coupling with all the algorithms and workflow we mentioned. It could also involve the data mining module to analyze the CALPHAD database better. The post-process module could be used to analyze the higher-order chemical SRO beyond the pairwise level. Including the phase diagram visualization and thermodynamics quantities analysis in the post-process module is also necessary. The comparison between calculated analytical results and experimental results could also be involved in the post-process module by generating the experimental observables from our calculations. In conclusion, this package with optimized coding would largely accelerate the calculation and help the community to study the thermodynamics of the multi-component materials with the chemical SRO effect. 

        3) Third, consider parallel computing and other new computing technique. The phase diagram calculation is a proper task exactly for parallel computing. It requires many equilibrium calculations at different conditions but a similar scheme. The single-point calculation is fast enough, but the whole phase diagram needs to be figured out at so many conditions. With the help of parallel computing, we could ultimately speed up the phase diagram calculation. Besides, it is also possible to use high-performance GPU or other hardware to accelerate the computation.

Besides, a further simplified model is also developed, calculated even without the variational calculation. This is illustrated in the appendix \ref{app:appendixA}. This might help determine the initial value to accelerate the equilibrium calculation.
        
\subsection{Wider Applications}
        The success of coupling the cluster model and the CALPHAD would not only benefit the computational thermodynamic modeling but also could influence many other research areas, including interface thermodynamics,  kinetics such as nucleation, and the electronic transport with the chemical SRO.

1) The interface thermodynamics.
Many phenomena, such as nucleation, happen on the interface or relate to the interface thermodynamics. The estimation of the interfacial free energy is non-trivial here. However, this challenge could be tackled through the cluster-based thermodynamic model. Previously, Kikuchi and Cahn have done some early work \cite{kikuchi1979theory}, we consider FYL-CVM coupling CALPHAD would have the chance to make the interface thermodynamics calculation more practical to combine it with the CALPHAD database.

2) The kinetics. Precipitation and nucleation are essential in the diffusional phase transformation between solid phases. Some early work indicated the significance of the chemical SRO within the precipitation and nucleation \cite{clouet2004nucleation,clouet2005precipitation,clouet2007classical}. It has been mentioned in previous work that only the volume-free energy included in the SRO would provide coherent results with the data produced by the kinetic Monte Carlo method \cite{clouet2004nucleation}. This study implies the potential chance for our proposed CVM-CALPHAD in nucleation theory study to consider the impact of SRO during the nucleation process.

3) The electronic structure or electronic transport within high entropy alloys we mentioned recently attracts more studies due to the introduction of chemical disordering \cite{mu2018electronic}. Due to the break of the translational symmetry, it is challenging to study the electronic properties of these alloys. Besides, the chemical SRO in these disordered alloys indicates that imposing the entirely random to study the quantum transport is not fully correct. The proposed CVM-CALPHAD may provide the chance for help, as we could determine the detailed statistical information of the microstructure. 

4) Drive the development of better measurement techniques. Since our current theoretical model could provide the chemical short-range order beyond the pair-wise bonding, it would provide the need to characterize the short-range order beyond the pair probability but larger basic clusters \cite{goff2021quantifying}. However, these multi-point correlation functions' measurements remain a challenge \cite{lemieux1999investigating,pedrini2013two,lehmkuhler2014detecting}. It would be possible that this research with the intrinsic SRO beyond pairs could motivate the development of the experiments for the multi-point correlations within the materials.


   
	\printbibliography[heading=bibintoc]

@article{baltruvsaitis2018multimodal,
  title={Multimodal machine learning: A survey and taxonomy},
  author={Baltru{\v{s}}aitis, Tadas and Ahuja, Chaitanya and Morency, Louis-Philippe},
  journal={IEEE transactions on pattern analysis and machine intelligence},
  volume={41},
  number={2},
  pages={423--443},
  year={2018},
  publisher={IEEE}
}

@article{kaufman1970computer,
  title={Computer calculation of phase diagrams. With special reference to refractory metals},
  author={Kaufman, Larry and Bernstein, Harold},
  year={1970}
}

@book{sundman2007computational,
  title={Computational thermodynamics: the Calphad method},
  author={Sundman, Bo and Lukas, HL and Fries, SG},
  year={2007},
  publisher={Cambridge university press New York}
}

@book{liu2016computational,
  title={Computational thermodynamics of materials},
  author={Liu, Zi-Kui and Wang, Yi},
  year={2016},
  publisher={Cambridge University Press}
}

@article{hubbard1996thermodynamic,
  title={Thermodynamic stability of binary oxides in contact with silicon},
  author={Hubbard, KJ and Schlom, DG},
  journal={Journal of Materials Research},
  volume={11},
  number={11},
  pages={2757--2776},
  year={1996},
  publisher={Cambridge University Press}
}

@article{liu2001thermodynamics,
  title={Thermodynamics of the Mg--B system: Implications for the deposition of MgB 2 thin films},
  author={Liu, Zi-Kui and Schlom, DG and Li, Qi and Xi, XX},
  journal={Applied Physics Letters},
  volume={78},
  number={23},
  pages={3678--3680},
  year={2001},
  publisher={American Institute of Physics}
}

@article{zhu2002linking,
  title={Linking phase-field model to CALPHAD: application to precipitate shape evolution in Ni-base alloys},
  author={Zhu, JZ and Liu, ZK and Vaithyanathan, Venu and Chen, LQ},
  journal={Scripta Materialia},
  volume={46},
  number={5},
  pages={401--406},
  year={2002},
  publisher={Elsevier}
}

@article{wu2004simulating,
  title={Simulating interdiffusion microstructures in Ni--Al--Cr diffusion couples: a phase field approach coupled with CALPHAD database},
  author={Wu, Kaisheng and Chang, YA and Wang, Yunzhi},
  journal={Scripta materialia},
  volume={50},
  number={8},
  pages={1145--1150},
  year={2004},
  publisher={Elsevier}
}

@article{xia2016precipitation,
  title={Precipitation evolution and hardening in MgSmZnZr alloys},
  author={Xia, Xiangyu and Sun, Weihua and Luo, Alan A and Stone, Donald S},
  journal={Acta Materialia},
  volume={111},
  pages={335--347},
  year={2016},
  publisher={Elsevier}
}

@article{weinan2020integrating,
  title={Integrating machine learning with physics-based modeling},
  author={Weinan, E and Han, Jiequn and Zhang, Linfeng},
  journal={arXiv preprint arXiv:2006.02619},
  year={2020}
}

@article{jumper2021highly,
  title={Highly accurate protein structure prediction with AlphaFold},
  author={Jumper, John and Evans, Richard and Pritzel, Alexander and Green, Tim and Figurnov, Michael and Ronneberger, Olaf and Tunyasuvunakool, Kathryn and Bates, Russ and {\v{Z}}{\'\i}dek, Augustin and Potapenko, Anna and others},
  journal={Nature},
  volume={596},
  number={7873},
  pages={583--589},
  year={2021},
  publisher={Nature Publishing Group}
}

@article{brunton2020machine,
  title={Machine learning for fluid mechanics},
  author={Brunton, Steven L and Noack, Bernd R and Koumoutsakos, Petros},
  journal={Annual Review of Fluid Mechanics},
  volume={52},
  pages={477--508},
  year={2020},
  publisher={Annual Reviews}
}

@article{carleo2017solving,
  title={Solving the quantum many-body problem with artificial neural networks},
  author={Carleo, Giuseppe and Troyer, Matthias},
  journal={Science},
  volume={355},
  number={6325},
  pages={602--606},
  year={2017},
  publisher={American Association for the Advancement of Science}
}

@article{KIKUCHI200233,
  title={Cluster variation method in the computational materials science},
  author={Kikuchi, R and Masuda-Jindo, K},
  journal={Calphad},
  volume={26},
  number={1},
  pages={33--54},
  year={2002},
  publisher={Elsevier}
}

@article{PERRON201416,
title={Understanding sigma-phase precipitation in a stabilized austenitic stainless steel (316Nb) through complementary CALPHAD-based and experimental investigations},
  author={Perron, A and Toffolon-Masclet, C and Ledoux, X and Buy, F and Guilbert, T and Urvoy, S and Bosonnet, S and Marini, B and Cortial, F and Texier, G and others},
  journal={Acta materialia},
  volume={79},
  pages={16--29},
  year={2014},
  publisher={Elsevier}
}

@article{wang2004ab,
  title={Ab initio lattice stability in comparison with CALPHAD lattice stability},
  author={Wang, Y and Curtarolo, S and Jiang, Chao and Arroyave, R and Wang, T and Ceder, G and Chen, L-Q and Liu, Z-K},
  journal={Calphad},
  volume={28},
  number={1},
  pages={79--90},
  year={2004},
  publisher={Elsevier}
}

@article{liu2009first,
  title={First-principles calculations and CALPHAD modeling of thermodynamics},
  author={Liu, Zi-Kui},
  journal={Journal of phase equilibria and diffusion},
  volume={30},
  number={5},
  pages={517--534},
  year={2009},
  publisher={Springer}
}

@article{zhu2017efficient,
  title={Efficient approach to compute melting properties fully from ab initio with application to Cu},
  author={Zhu, Li-Fang and Grabowski, Blazej and Neugebauer, J{\"o}rg},
  journal={Physical Review B},
  volume={96},
  number={22},
  pages={224202},
  year={2017},
  publisher={APS}
}

@article{rogal2014perspectives,
  title={Perspectives on point defect thermodynamics},
  author={Rogal, Jutta and Divinski, Sergiy V and Finnis, Mike W and Glensk, Albert and Neugebauer, Joerg and Perepezko, John H and Schuwalow, Sergej and Sluiter, Marcel HF and Sundman, Bo},
  journal={physica status solidi (b)},
  volume={251},
  number={1},
  pages={97--129},
  year={2014},
  publisher={Wiley Online Library}
}

@article{fultz2010vibrational,
  title={Vibrational thermodynamics of materials},
  author={Fultz, Brent},
  journal={Progress in Materials Science},
  volume={55},
  number={4},
  pages={247--352},
  year={2010},
  publisher={Elsevier}
}

@article{van2002automating,
  title={Automating first-principles phase diagram calculations},
  author={van de Walle, Axel and Ceder, Gerbrand},
  journal={Journal of Phase Equilibria},
  volume={23},
  number={4},
  pages={348--359},
  year={2002},
  publisher={Springer}
}

@article{becker2014thermodynamic,
  title={Thermodynamic modelling of liquids: CALPHAD approaches and contributions from statistical physics},
  author={Becker, Chandler A and {\AA}gren, John and Baricco, Marcello and Chen, Qing and Decterov, Sergei A and Kattner, Ursula R and Perepezko, John H and Pottlacher, Gernot R and Selleby, Malin},
  journal={physica status solidi (b)},
  volume={251},
  number={1},
  pages={33--52},
  year={2014},
  publisher={Wiley Online Library}
}

@article{aldegunde2016quantifying,
  title={Quantifying uncertainties in first-principles alloy thermodynamics using cluster expansions},
  author={Aldegunde, Manuel and Zabaras, Nicholas and Kristensen, Jesper},
  journal={Journal of Computational Physics},
  volume={323},
  pages={17--44},
  year={2016},
  publisher={Elsevier}
}

@article{bogojeski2020quantum,
  title={Quantum chemical accuracy from density functional approximations via machine learning},
  author={Bogojeski, Mihail and Vogt-Maranto, Leslie and Tuckerman, Mark E and M{\"u}ller, Klaus-Robert and Burke, Kieron},
  journal={Nature communications},
  volume={11},
  number={1},
  pages={1--11},
  year={2020},
  publisher={Nature Publishing Group}
}

@article{chen2020deepks,
  title={DeePKS: A Comprehensive Data-Driven Approach toward Chemically Accurate Density Functional Theory},
  author={Chen, Yixiao and Zhang, Linfeng and Wang, Han and E, Weinan},
  journal={Journal of Chemical Theory and Computation},
  volume={17},
  number={1},
  pages={170--181},
  year={2020},
  publisher={ACS Publications}
}

@article{zhang2018deep,
  title={Deep potential molecular dynamics: a scalable model with the accuracy of quantum mechanics},
  author={Zhang, Linfeng and Han, Jiequn and Wang, Han and Car, Roberto and Weinan, E},
  journal={Physical review letters},
  volume={120},
  number={14},
  pages={143001},
  year={2018},
  publisher={APS}
}

@article{oates1996putting,
  title={On putting more physics into calphad solution models},
  author={Oates, WA and Wenzl, H and Mohri, T},
  journal={Calphad},
  volume={20},
  number={1},
  pages={37--45},
  year={1996},
  publisher={Elsevier}
}

@article{otis2017high,
  title={High-throughput thermodynamic modeling and uncertainty quantification for ICME},
  author={Otis, Richard A and Liu, Zi-Kui},
  journal={JOM},
  volume={69},
  number={5},
  pages={886--892},
  year={2017},
  publisher={Springer}
}

@article{paulson2019quantified,
  title={Quantified uncertainty in thermodynamic modeling for materials design},
  author={Paulson, Noah H and Bocklund, Brandon J and Otis, Richard A and Liu, Zi-Kui and Stan, Marius},
  journal={Acta Materialia},
  volume={174},
  pages={9--15},
  year={2019},
  publisher={Elsevier}
}

@inproceedings{yedidia2000generalized,
  title={Generalized belief propagation},
  author={Yedidia, Jonathan S and Freeman, William T and Weiss, Yair and others},
  booktitle={NIPS},
  volume={13},
  pages={689--695},
  year={2000}
}

@article{yedidia2005constructing,
  title={Constructing free-energy approximations and generalized belief propagation algorithms},
  author={Yedidia, Jonathan S and Freeman, William T and Weiss, Yair},
  journal={IEEE Transactions on information theory},
  volume={51},
  number={7},
  pages={2282--2312},
  year={2005},
  publisher={IEEE}
}

@article{pelizzola2005cluster,
  title={Cluster variation method in statistical physics and probabilistic graphical models},
  author={Pelizzola, Alessandro},
  journal={Journal of Physics A: Mathematical and General},
  volume={38},
  number={33},
  pages={R309},
  year={2005},
  publisher={IOP Publishing}
}

@article{niu2020ab,
  title={Ab initio phase diagram and nucleation of gallium},
  author={Niu, Haiyang and Bonati, Luigi and Piaggi, Pablo M and Parrinello, Michele},
  journal={Nature communications},
  volume={11},
  number={1},
  pages={1--9},
  year={2020},
  publisher={Nature Publishing Group}
}

@article{zhu2020performance,
  title={Performance of the standard exchange-correlation functionals in predicting melting properties fully from first principles: Application to Al and magnetic Ni},
  author={Zhu, Li-Fang and K{\"o}rmann, Fritz and Ruban, Andrei V and Neugebauer, J{\"o}rg and Grabowski, Blazej},
  journal={Physical Review B},
  volume={101},
  number={14},
  pages={144108},
  year={2020},
  publisher={APS}
}

@article{sola2009melting,
  title={Melting of iron under Earth’s core conditions from diffusion Monte Carlo free energy calculations},
  author={Sola, Ester and Alfe, Dario},
  journal={Physical Review Letters},
  volume={103},
  number={7},
  pages={078501},
  year={2009},
  publisher={APS}
}

@article{dorner2018melting,
  title={Melting Si: beyond density functional theory},
  author={Dorner, Florian and Sukurma, Zoran and Dellago, Christoph and Kresse, Georg},
  journal={Physical review letters},
  volume={121},
  number={19},
  pages={195701},
  year={2018},
  publisher={APS}
}

@article{rang2019first,
  title={First-principles study of the melting temperature of MgO},
  author={Rang, Max and Kresse, Georg},
  journal={Physical Review B},
  volume={99},
  number={18},
  pages={184103},
  year={2019},
  publisher={APS}
}

@article{10.1145/3386252,
  title={Generalizing from a few examples: A survey on few-shot learning},
  author={Wang, Yaqing and Yao, Quanming and Kwok, James T and Ni, Lionel M},
  journal={ACM computing surveys (csur)},
  volume={53},
  number={3},
  pages={1--34},
  year={2020},
  publisher={ACM New York, NY, USA}
}

@article{kikuchi1951theory,
  title={A theory of cooperative phenomena},
  author={Kikuchi, Ryoichi},
  journal={Physical review},
  volume={81},
  number={6},
  pages={988},
  year={1951},
  publisher={APS}
}

@article{kikuchi1951theory2,
  title={A theory of cooperative phenomena. II. Equation of states for classical statistics},
  author={Kikuchi, Ryoichi},
  journal={The Journal of Chemical Physics},
  volume={19},
  number={10},
  pages={1230--1241},
  year={1951},
  publisher={American Institute of Physics}
}

@article{kurata1953theory3,
  title={A theory of cooperative phenomena. III. Detailed discussions of the cluster variation method},
  author={Kurata, Michio and Kikuchi, Ryoichi and Watari, Tatsuro},
  journal={The Journal of Chemical Physics},
  volume={21},
  number={3},
  pages={434--448},
  year={1953},
  publisher={American Institute of Physics}
}

@article{hillert2001compound,
  title={The compound energy formalism},
  author={Hillert, Mats},
  journal={Journal of Alloys and Compounds},
  volume={320},
  number={2},
  pages={161--176},
  year={2001},
  publisher={Elsevier}
}

@article{miracle2017critical,
  title={A critical review of high entropy alloys and related concepts},
  author={Miracle, Daniel B and Senkov, Oleg N},
  journal={Acta Materialia},
  volume={122},
  pages={448--511},
  year={2017},
  publisher={Elsevier}
}

@article{george2019high,
  title={High-entropy alloys},
  author={George, Easo P and Raabe, Dierk and Ritchie, Robert O},
  journal={Nature reviews materials},
  volume={4},
  number={8},
  pages={515--534},
  year={2019},
  publisher={Nature Publishing Group UK London}
}

@article{liang2018high,
  title={High-content ductile coherent nanoprecipitates achieve ultrastrong high-entropy alloys},
  author={Liang, Yao-Jian and Wang, Linjing and Wen, Yuren and Cheng, Baoyuan and Wu, Qinli and Cao, Tangqing and Xiao, Qian and Xue, Yunfei and Sha, Gang and Wang, Yandong and others},
  journal={Nature communications},
  volume={9},
  number={1},
  pages={4063},
  year={2018},
  publisher={Nature Publishing Group UK London}
}

@article{miracle2020refractory,
  title={Refractory high entropy superalloys (RSAs)},
  author={Miracle, Daniel B and Tsai, Ming-Hung and Senkov, Oleg N and Soni, Vishal and Banerjee, Rajarshi},
  journal={Scripta Materialia},
  volume={187},
  pages={445--452},
  year={2020},
  publisher={Elsevier}
}

@article{ma2018chemical,
  title={Chemical short-range orders and the induced structural transition in high-entropy alloys},
  author={Ma, Yue and Wang, Qing and Li, Chunling and Santodonato, Louis J and Feygenson, Mikhail and Dong, Chuang and Liaw, Peter K},
  journal={Scripta Materialia},
  volume={144},
  pages={64--68},
  year={2018},
  publisher={Elsevier}
}

@article{singh2015atomic,
  title={Atomic short-range order and incipient long-range order in high-entropy alloys},
  author={Singh, Prashant and Smirnov, Andrei V and Johnson, Duane D},
  journal={Physical Review B},
  volume={91},
  number={22},
  pages={224204},
  year={2015},
  publisher={APS}
}

@article{zhang2020short,
  title={Short-range order and its impact on the CrCoNi medium-entropy alloy},
  author={Zhang, Ruopeng and Zhao, Shiteng and Ding, Jun and Chong, Yan and Jia, Tao and Ophus, Colin and Asta, Mark and Ritchie, Robert O and Minor, Andrew M},
  journal={Nature},
  volume={581},
  number={7808},
  pages={283--287},
  year={2020},
  publisher={Nature Publishing Group UK London}
}

@article{ding2019tuning,
  title={Tuning element distribution, structure and properties by composition in high-entropy alloys},
  author={Ding, Qingqing and Zhang, Yin and Chen, Xiao and Fu, Xiaoqian and Chen, Dengke and Chen, Sijing and Gu, Lin and Wei, Fei and Bei, Hongbin and Gao, Yanfei and others},
  journal={Nature},
  volume={574},
  number={7777},
  pages={223--227},
  year={2019},
  publisher={Nature Publishing Group UK London}
}

@article{nohring2020design,
  title={Design using randomness: a new dimension for metallurgy},
  author={N{\"o}hring, Wolfram Georg and Curtin, WA},
  journal={Scripta Materialia},
  volume={187},
  pages={210--215},
  year={2020},
  publisher={Elsevier}
}

@article{van2013methods,
  title={Methods for first-principles alloy thermodynamics},
  author={van de Walle, Axel},
  journal={Jom},
  volume={65},
  pages={1523--1532},
  year={2013},
  publisher={Springer}
}

@article{de1979configurational,
  title={Configurational thermodynamics of solid solutions},
  author={De Fontaine, Didier},
  journal={Solid state physics},
  volume={34},
  pages={73--274},
  year={1979},
  publisher={Elsevier}
}

@article{mohri2017cluster,
  title={Cluster variation method as a theoretical tool for the study of phase transformation},
  author={Mohri, Tetsuo},
  journal={Metallurgical and materials transactions A},
  volume={48},
  pages={2753--2770},
  year={2017},
  publisher={Springer}
}

@incollection{de1994cluster,
  title={Cluster approach to order-disorder transformations in alloys},
  author={De Fontaine, Didier},
  booktitle={Solid state physics},
  volume={47},
  pages={33--176},
  year={1994},
  publisher={Elsevier}
}

@article{blander1987thermodynamic,
  title={Thermodynamic analysis of binary liquid silicates and prediction of ternary solution properties by modified quasichemical equations},
  author={Blander, Milton and Pelton, Arthur D},
  journal={Geochimica et Cosmochimica Acta},
  volume={51},
  number={1},
  pages={85--95},
  year={1987},
  publisher={Elsevier}
}

@article{yang1945generalization,
  title={A Generalization of the Quasi-Chemical Method in the Statistical Theory of Superlattices},
  author={Yang, CN},
  journal={The Journal of Chemical Physics},
  volume={13},
  number={2},
  pages={66--76},
  year={1945},
  publisher={American Institute of Physics}
}

@article{yang1947general,
  title={General theory of the quasi-chemical method in the statistical theory of superlattices},
  author={Yang, CN and Li, Y},
  journal={Chinese Journal of Physics},
  volume={11},
  number={2},
  pages={59--71},
  year={1947}
}

@article{li1949quasi,
  title={Quasi-chemical method in the statistical theory of regular mixtures},
  author={Li, Yin-Yuan},
  journal={Physical Review},
  volume={76},
  number={7},
  pages={972},
  year={1949},
  publisher={APS}
}

@article{li1949quasi2,
  title={Quasi-Chemical Theory of Order for the Copper Gold Alloy System},
  author={Li, Yin-Yuan},
  journal={The Journal of Chemical Physics},
  volume={17},
  number={5},
  pages={447--454},
  year={1949},
  publisher={American Institute of Physics}
}

@article{balabin2000thermodynamics,
  title={Thermodynamics of (Zn, Fe) S sphalerite. A CVM approach with large basis clusters},
  author={Balabin, AI and Sack, RO},
  journal={Mineralogical Magazine},
  volume={64},
  number={5},
  pages={923--943},
  year={2000},
  publisher={Mineralogical Society of Great Britain and Ireland}
}

@article{oates1996cluster,
  title={The cluster/site approximation for multicomponent solutions--a practical alternative to the cluster variation method},
  author={Oates, WA and Wenzl, H},
  journal={Scripta materialia},
  volume={35},
  number={5},
  year={1996}
}

@article{oates1999improved,
  title={Improved cluster-site approximation for the entropy of mixing in multicomponent solid solutions},
  author={Oates, WA and Zhang, F and Chen, SL and Chang, YA},
  journal={Physical Review B},
  volume={59},
  number={17},
  pages={11221},
  year={1999},
  publisher={APS}
}

@article{zunger1990special,
  title={Special quasirandom structures},
  author={Zunger, Alex and Wei, S-H and Ferreira, LG and Bernard, James E},
  journal={Physical review letters},
  volume={65},
  number={3},
  pages={353},
  year={1990},
  publisher={APS}
}

@article{van2017software,
  title={Software tools for high-throughput CALPHAD from first-principles data},
  author={van de Walle, Axel and Sun, Ruoshi and Hong, Qi-Jun and Kadkhodaei, Sara},
  journal={Calphad},
  volume={58},
  pages={70--81},
  year={2017},
  publisher={Elsevier}
}

@article{sundman1998thermodynamic,
  title={A thermodynamic assessment of the Au-Cu system},
  author={Sundman, Bo and Fries, Suzana G and Oates, W Alan},
  journal={Calphad},
  volume={22},
  number={3},
  pages={335--354},
  year={1998},
  publisher={Elsevier}
}

@article{sigli1985theoretical,
  title={Theoretical description of phase equilibrium in binary alloys},
  author={Sigli, C and Sanchez, JM},
  journal={Acta Metallurgica},
  volume={33},
  number={6},
  pages={1097--1104},
  year={1985},
  publisher={Elsevier}
}

@article{kikuchi1974superposition,
  title={Superposition approximation and natural iteration calculation in cluster-variation method},
  author={Kikuchi, Ryoichi},
  journal={The Journal of Chemical Physics},
  volume={60},
  number={3},
  pages={1071--1080},
  year={1974},
  publisher={American Institute of Physics}
}

@software{jax2018github,
  author = {James Bradbury and Roy Frostig and Peter Hawkins and Matthew James Johnson and Chris Leary and Dougal Maclaurin and George Necula and Adam Paszke and Jake Vander{P}las and Skye Wanderman-{M}ilne and Qiao Zhang},
  title = {{JAX}: composable transformations of {P}ython+{N}um{P}y programs},
  url = {http://github.com/google/jax},
  version = {0.3.13},
  year = {2018},
}

@article{colson2007overview,
  title={An overview of bilevel optimization},
  author={Colson, Beno{\^\i}t and Marcotte, Patrice and Savard, Gilles},
  journal={Annals of operations research},
  volume={153},
  pages={235--256},
  year={2007},
  publisher={Springer}
}

@inproceedings{franceschi2018bilevel,
  title={Bilevel programming for hyperparameter optimization and meta-learning},
  author={Franceschi, Luca and Frasconi, Paolo and Salzo, Saverio and Grazzi, Riccardo and Pontil, Massimiliano},
  booktitle={International Conference on Machine Learning},
  pages={1568--1577},
  year={2018},
  organization={PMLR}
}

@article{bragg1934effect,
  title={The effect of thermal agitation on atomic arrangement in alloys},
  author={Bragg, William Lawrence and Williams, Evan James},
  journal={Proceedings of the Royal Society of London. Series A, Containing Papers of a Mathematical and Physical Character},
  volume={145},
  number={855},
  pages={699--730},
  year={1934},
  publisher={The Royal Society London}
}

@article{bethe1935statistical,
  title={Statistical theory of superlattices},
  author={Bethe, Hans A},
  journal={Proceedings of the Royal Society of London. Series A-Mathematical and Physical Sciences},
  volume={150},
  number={871},
  pages={552--575},
  year={1935},
  publisher={The Royal Society London}
}

@article{takagi1941statistical,
  title={Statistical Theory of Binary Alloys. I},
  author={TAKAGI, Yutaka},
  journal={Proceedings of the Physico-Mathematical Society of Japan. 3rd Series},
  volume={23},
  pages={44--65},
  year={1941},
  publisher={THE PHYSICAL SOCIETY OF JAPAN, The Mathematical Society of Japan}
}

@article{morita1957cluster,
  title={Cluster variation method of cooperative phenomena and its generalization I},
  author={Morita, Tohru},
  journal={Journal of the Physical Society of Japan},
  volume={12},
  number={7},
  pages={753--755},
  year={1957},
  publisher={The Physical Society of Japan}
}

@article{morita1972general,
  title={General structure of the distribution functions for the Heisenberg model and the Ising model},
  author={Morita, Tohru},
  journal={Journal of Mathematical Physics},
  volume={13},
  number={1},
  pages={115--123},
  year={1972},
  publisher={American Institute of Physics}
}

@article{morita1984consistent,
  title={Consistent relations in the method of reducibility in the cluster variation method},
  author={Morita, T},
  journal={Journal of statistical physics},
  volume={34},
  pages={319--328},
  year={1984},
  publisher={Springer}
}

@article{schlijper1983convergence,
  title={Convergence of the cluster-variation method in the thermodynamic limit},
  author={Schlijper, AG},
  journal={Physical Review B},
  volume={27},
  number={11},
  pages={6841},
  year={1983},
  publisher={APS}
}

@article{an1988note,
  title={A note on the cluster variation method},
  author={An, Guozhong},
  journal={Journal of Statistical Physics},
  volume={52},
  pages={727--734},
  year={1988},
  publisher={Springer}
}

@article{rota1964foundations,
  title={On the foundations of combinatorial theory I. Theory of M{\"o}bius functions},
  author={Rota, Gian-Carlo},
  year={1964}
}

@article{morita1990cluster,
  title={Cluster variation method and M{\"o}bius inversion formula},
  author={Morita, T},
  journal={Journal of Statistical Physics},
  volume={59},
  pages={819--825},
  year={1990},
  publisher={Springer}
}

@article{sanchez1984generalized,
  title={Generalized cluster description of multicomponent systems},
  author={Sanchez, Juan M and Ducastelle, Francois and Gratias, Denis},
  journal={Physica A: Statistical Mechanics and its Applications},
  volume={128},
  number={1-2},
  pages={334--350},
  year={1984},
  publisher={Elsevier}
}

@article{mohri2013cluster,
  title={Cluster variation method},
  author={Mohri, Tetsuo},
  journal={Jom},
  volume={65},
  pages={1510--1522},
  year={2013},
  publisher={Springer}
}

@article{kikuchi1966path,
  title={The path probability method},
  author={Kikuchi, Ryoichi},
  journal={Progress of Theoretical Physics Supplement},
  volume={35},
  pages={1--64},
  year={1966},
  publisher={Oxford Academic}
}

@article{kikuchi1980theoretical,
  title={Theoretical calculation of the CU AG AU coherent phase diagram},
  author={Kikuchi, R and Sanchez, JM and De Fontaine, D and Yamauchi, Hisao},
  journal={Acta Metallurgica},
  volume={28},
  number={5},
  pages={651--662},
  year={1980},
  publisher={Elsevier}
}

@article{morita1966application,
  title={Application of the cluster variation method to the Heisenberg model with arbitrary spin and range of exchange},
  author={Morita, Tohru and Tanaka, Tomoyasu},
  journal={Physical Review},
  volume={145},
  number={1},
  pages={288},
  year={1966},
  publisher={APS}
}

@article{lim1994cvm,
  title={CVM model calculation of the Al Mg Ag phase diagram},
  author={Lim, SS and Rossiter, PL},
  journal={Calphad},
  volume={18},
  number={2},
  pages={113--123},
  year={1994},
  publisher={Elsevier}
}

@article{finel1986phase,
  title={On the phase diagram of the FCC Ising model with antiferromagnetic first-neighbour interactions},
  author={Finel, A and Ducastelle, F},
  journal={Europhysics Letters},
  volume={1},
  number={3},
  pages={135},
  year={1986},
  publisher={IOP Publishing}
}

@article{lawrence1986chemical,
  title={Chemical and magnetic interactions in FCC Fe-Ni alloys using the cluster variation method},
  author={Lawrence, PJ and Rossiter, PL},
  journal={Journal of Physics F: Metal Physics},
  volume={16},
  number={5},
  pages={543},
  year={1986},
  publisher={IOP Publishing}
}

@article{wei1987first,
  title={First-principles calculations of the phase diagrams of noble metals: Cu-Au, Cu-Ag, and Ag-Au},
  author={Wei, S-H and Mbaye, AA and Ferreira, LG and Zunger, Alex},
  journal={Physical Review B},
  volume={36},
  number={8},
  pages={4163},
  year={1987},
  publisher={APS}
}

@article{asta1993theoretical,
  title={Theoretical study of alloy phase stability in the Cd-Mg system},
  author={Asta, Mark and McCormack, Ryan and de Fontaine, Didier},
  journal={Physical Review B},
  volume={48},
  number={2},
  pages={748},
  year={1993},
  publisher={APS}
}

@article{kikuchi1988calculation,
  title={Calculation of phase diagrams of some oxide systems using the cluster variation method},
  author={Kikuchi, Ryoichi and Burton, Benjamin P},
  journal={Physica B+ C},
  volume={150},
  number={1-2},
  pages={132--141},
  year={1988},
  publisher={Elsevier}
}

@article{kikuchi1997continuous,
  title={The continuous displacement CVM treatment of alloy systems},
  author={Kikuchi, Ryoichi and Masuda-Jindo, Kinichi},
  journal={Computational materials science},
  volume={8},
  number={1-2},
  pages={1--7},
  year={1997},
  publisher={Elsevier}
}

@article{tetot1994evaluation,
  title={Evaluation of Defect-Defect Pair Interactions in Nonstoichiometric Oxides by Cvm and Monte Carlo Calculations},
  author={T{\'e}tot, R and Nacer, B and Giaconia, C and Boureau, G},
  journal={Statics and Dynamics of Alloy Phase Transformations},
  pages={577--580},
  year={1994},
  publisher={Springer}
}

@article{finel1994cluster,
  title={The cluster variation method and some applications},
  author={Finel, A},
  journal={Statics and dynamics of alloy phase transformations},
  pages={495--540},
  year={1994},
  publisher={Springer}
}

@article{shannon1948mathematical,
  title={A mathematical theory of communication},
  author={Shannon, Claude E},
  journal={The Bell system technical journal},
  volume={27},
  number={3},
  pages={379--423},
  year={1948},
  publisher={Nokia Bell Labs}
}

@book{koller2009probabilistic,
  title={Probabilistic graphical models: principles and techniques},
  author={Koller, Daphne and Friedman, Nir},
  year={2009},
  publisher={MIT press}
}

@article{fowler1940statistical,
  title={Statistical thermodynamics of super-lattices},
  author={Fowler, Ralph Howard and Guggenheim, Edward Armand},
  journal={Proceedings of the Royal Society of London. Series A. Mathematical and Physical Sciences},
  volume={174},
  number={957},
  pages={189--206},
  year={1940},
  publisher={The Royal Society London}
}

@article{zhang2001cluster,
  title={Cluster/site approximation calculation of the ordering phase diagram for Cd--Mg alloys},
  author={Zhang, J and Oates, WA and Zhang, F and Chen, SL and Chou, KC and Chang, YA},
  journal={Intermetallics},
  volume={9},
  number={1},
  pages={5--8},
  year={2001},
  publisher={Elsevier}
}

@article{zhang2003application,
  title={Application of the cluster-site approximation (CSA) model to the fcc phase in the Ni--Al system},
  author={Zhang, F and Chang, YA and Du, Y and Chen, S-L and Oates, WA},
  journal={Acta materialia},
  volume={51},
  number={1},
  pages={207--216},
  year={2003},
  publisher={Elsevier}
}

@book{cao2006application,
  title={Application of the cluster/site approximation to calculation of multicomponent alloy phase diagrams and coherent interphase energies},
  author={Cao, Weisheng},
  year={2006},
  publisher={The University of Wisconsin-Madison}
}

@article{cao2007thermodynamic,
  title={Thermodynamic modeling of the Cu--Ag--Au system using the cluster/site approximation},
  author={Cao, W and Chang, YA and Zhu, J and Chen, S and Oates, WA},
  journal={Intermetallics},
  volume={15},
  number={11},
  pages={1438--1446},
  year={2007},
  publisher={Elsevier}
}

@article{kikuchi1987grain,
  title={Grain boundaries with impurities in a two-dimensional lattice-gas model},
  author={Kikuchi, Ryoichi and Cahn, John W},
  journal={Physical Review B},
  volume={36},
  number={1},
  pages={418},
  year={1987},
  publisher={APS}
}

@article{kikuchi1980grain,
  title={Grain-boundary melting transition in a two-dimensional lattice-gas model},
  author={Kikuchi, Ryoichi and Cahn, John W},
  journal={Physical Review B},
  volume={21},
  number={5},
  pages={1893},
  year={1980},
  publisher={APS}
}

@article{kikuchi1979theory,
  title={Theory of interphase and antiphase boundaries in FCC alloys},
  author={Kikuchi, Ryoichi and Cahn, John W},
  journal={Acta Metallurgica},
  volume={27},
  number={8},
  pages={1337--1353},
  year={1979},
  publisher={Elsevier}
}

@book{pearl1988probabilistic,
  title={Probabilistic reasoning in intelligent systems: networks of plausible inference},
  author={Pearl, Judea},
  year={1988},
  publisher={Morgan kaufmann}
}

@article{freeman2000learning,
  title={Learning low-level vision},
  author={Freeman, William T and Pasztor, Egon C and Carmichael, Owen T},
  journal={International journal of computer vision},
  volume={40},
  pages={25--47},
  year={2000},
  publisher={Springer}
}

@article{kabashima2004statistical,
  title={Statistical mechanics of low-density parity-check codes},
  author={Kabashima, Yoshiyuki and Saad, David},
  journal={Journal of Physics A: Mathematical and General},
  volume={37},
  number={6},
  pages={R1},
  year={2004},
  publisher={IOP Publishing}
}

@article{krogh2001predicting,
  title={Predicting transmembrane protein topology with a hidden Markov model: application to complete genomes},
  author={Krogh, Anders and Larsson, Bj{\"o}rn and Von Heijne, Gunnar and Sonnhammer, Erik LL},
  journal={Journal of molecular biology},
  volume={305},
  number={3},
  pages={567--580},
  year={2001},
  publisher={Elsevier}
}

@article{zhou2015spin,
  title={Spin Glass and Message Passing},
  author={Zhou, HJ},
  journal={Science, Beijing},
  year={2015}
}

@article{ma2020towards,
  title={Towards a Mathematical Understanding of Neural Network-Based Machine Learning: what we know and what we don't},
  author={Ma, Chao and Wojtowytsch, Stephan and Wu, Lei and others},
  journal={arXiv preprint arXiv:2009.10713},
  year={2020}
}

@article{chen2018database,
  title={Database development and Calphad calculations for high entropy alloys: Challenges, strategies, and tips},
  author={Chen, Hai-Lin and Mao, Huahai and Chen, Qing},
  journal={Materials Chemistry and Physics},
  volume={210},
  pages={279--290},
  year={2018},
  publisher={Elsevier}
}

@article{batzner20223,
  title={E (3)-equivariant graph neural networks for data-efficient and accurate interatomic potentials},
  author={Batzner, Simon and Musaelian, Albert and Sun, Lixin and Geiger, Mario and Mailoa, Jonathan P and Kornbluth, Mordechai and Molinari, Nicola and Smidt, Tess E and Kozinsky, Boris},
  journal={Nature communications},
  volume={13},
  number={1},
  pages={2453},
  year={2022},
  publisher={Nature Publishing Group UK London}
}

@article{ury2022generalized,
  title={Generalized method of sensitivity analysis for uncertainty quantification in Calphad calculations},
  author={Ury, Nicholas and Otis, Richard and Ravi, Vilupanur},
  journal={Calphad},
  volume={79},
  pages={102504},
  year={2022},
  publisher={Elsevier}
}

@article{bocklund2019espei,
  title={ESPEI for efficient thermodynamic database development, modification, and uncertainty quantification: application to Cu--Mg},
  author={Bocklund, Brandon and Otis, Richard and Egorov, Aleksei and Obaied, Abdulmonem and Roslyakova, Irina and Liu, Zi-Kui},
  journal={MRS Communications},
  volume={9},
  number={2},
  pages={618--627},
  year={2019},
  publisher={Cambridge University Press}
}

@article{kohn1965self,
  title={Self-consistent equations including exchange and correlation effects},
  author={Kohn, Walter and Sham, Lu Jeu},
  journal={Physical review},
  volume={140},
  number={4A},
  pages={A1133},
  year={1965},
  publisher={APS}
}

@article{wang2004thermodynamic,
  title={Thermodynamic properties of Al, Ni, NiAl, and Ni3Al from first-principles calculations},
  author={Wang, Y and Liu, Z-K and Chen, L-Q},
  journal={Acta Materialia},
  volume={52},
  number={9},
  pages={2665--2671},
  year={2004},
  publisher={Elsevier}
}

@article{shang2010first,
  title={First-principles thermodynamics from phonon and Debye model: Application to Ni and Ni3Al},
  author={Shang, Shun-Li and Wang, Yi and Kim, DongEung and Liu, Zi-Kui},
  journal={Computational Materials Science},
  volume={47},
  number={4},
  pages={1040--1048},
  year={2010},
  publisher={Elsevier}
}

@article{jiang2020ab,
  title={Ab initio study and thermodynamic modeling of the Pd-Si-C system},
  author={Jiang, Chao and van Rooyen, Isabella J and Meher, Subhashish},
  journal={Computational Materials Science},
  volume={171},
  pages={109238},
  year={2020},
  publisher={Elsevier}
}

@article{car1985unified,
  title={Unified approach for molecular dynamics and density-functional theory},
  author={Car, Richard and Parrinello, Mark},
  journal={Physical review letters},
  volume={55},
  number={22},
  pages={2471},
  year={1985},
  publisher={APS}
}

@article{dinsdale1991sgte,
  title={SGTE data for pure elements},
  author={Dinsdale, Alan T},
  journal={Calphad},
  volume={15},
  number={4},
  pages={317--425},
  year={1991},
  publisher={Elsevier}
}

@article{pelton2000modified,
  title={The modified quasichemical model I—Binary solutions},
  author={Pelton, Arthur D and Degterov, Sergey A and Eriksson, Gunnar and Robelin, Christian and Dessureault, Yves},
  journal={Metallurgical and Materials Transactions B},
  volume={31},
  pages={651--659},
  year={2000},
  publisher={Springer}
}

@article{pelton2001modified,
  title={The modified quasi-chemical model: Part II. Multicomponent solutions},
  author={Pelton, Arthur D and Chartrand, Patrice},
  journal={Metallurgical and Materials Transactions A},
  volume={32},
  pages={1355--1360},
  year={2001},
  publisher={Springer}
}

@article{chartrand2001modified,
  title={The modified quasi-chemical model: Part III. Two sublattices},
  author={Chartrand, Patrice and Pelton, Arthur D},
  journal={Metallurgical and Materials Transactions A},
  volume={32},
  pages={1397--1407},
  year={2001},
  publisher={Springer}
}

@article{redlich1948algebraic,
  title={Algebraic representation of thermodynamic properties and the classification of solutions},
  author={Redlich, Otto and Kister, AT},
  journal={Industrial \& Engineering Chemistry},
  volume={40},
  number={2},
  pages={345--348},
  year={1948},
  publisher={ACS Publications}
}

@article{sundman2018review,
  title={A review of Calphad modeling of ordered phases},
  author={Sundman, Bo and Chen, Qing and Du, Yong},
  journal={Journal of Phase Equilibria and Diffusion},
  volume={39},
  pages={678--693},
  year={2018},
  publisher={Springer}
}

@article{ma2020unusual,
  title={Unusual dislocation behavior in high-entropy alloys},
  author={Ma, Evan},
  journal={Scripta Materialia},
  volume={181},
  pages={127--133},
  year={2020},
  publisher={Elsevier}
}

@article{jiang2021high,
  title={High-entropy-stabilized chalcogenides with high thermoelectric performance},
  author={Jiang, Binbin and Yu, Yong and Cui, Juan and Liu, Xixi and Xie, Lin and Liao, Jincheng and Zhang, Qihao and Huang, Yi and Ning, Shoucong and Jia, Baohai and others},
  journal={Science},
  volume={371},
  number={6531},
  pages={830--834},
  year={2021},
  publisher={American Association for the Advancement of Science}
}

@article{pizzi2016aiida,
  title={AiiDA: automated interactive infrastructure and database for computational science},
  author={Pizzi, Giovanni and Cepellotti, Andrea and Sabatini, Riccardo and Marzari, Nicola and Kozinsky, Boris},
  journal={Computational Materials Science},
  volume={111},
  pages={218--230},
  year={2016},
  publisher={Elsevier}
}

@software{dflow2023github,
  title = {DFLOW},
  url = {https://github.com/deepmodeling/dflow},
  version = {1.6.121},
  year = {2023},
}

@article{oses2023aflow++,
  title={aflow++: A C++ framework for autonomous materials design},
  author={Oses, Corey and Esters, Marco and Hicks, David and Divilov, Simon and Eckert, Hagen and Friedrich, Rico and Mehl, Michael J and Smolyanyuk, Andriy and Campilongo, Xiomara and van de Walle, Axel and others},
  journal={Computational Materials Science},
  volume={217},
  pages={111889},
  year={2023},
  publisher={Elsevier}
}

@article{rowlands2009short,
  title={Short-range correlations in disordered systems: nonlocal coherent-potential approximation},
  author={Rowlands, Derwyn A},
  journal={Reports on Progress in Physics},
  volume={72},
  number={8},
  pages={086501},
  year={2009},
  publisher={IOP Publishing}
}

@article{chen2002phase,
  title={Phase-field models for microstructure evolution},
  author={Chen, Long-Qing},
  journal={Annual review of materials research},
  volume={32},
  number={1},
  pages={113--140},
  year={2002},
  publisher={Annual Reviews 4139 El Camino Way, PO Box 10139, Palo Alto, CA 94303-0139, USA}
}

@book{jansson1984evaluation,
  title={Evaluation of parameters in thermochemical models using different types of experimental data simultaneously},
  author={Jansson, Bo},
  year={1984}
}

@article{yang2019enhanced,
  title={Enhanced sampling in molecular dynamics},
  author={Yang, Yi Isaac and Shao, Qiang and Zhang, Jun and Yang, Lijiang and Gao, Yi Qin},
  journal={The Journal of chemical physics},
  volume={151},
  number={7},
  pages={070902},
  year={2019},
  publisher={AIP Publishing LLC}
}

@article{kusoffsky2002thermodynamic,
  title={Thermodynamic evaluation of the ternary Ag--Au--Cu system—including a short range order description},
  author={Kusoffsky, A},
  journal={Acta materialia},
  volume={50},
  number={20},
  pages={5139--5145},
  year={2002},
  publisher={Elsevier}
}

@article{liu2018effect,
  title={The effect of short-range order on passivation of Fe-Cr alloys},
  author={Liu, Minglu and Aiello, Ashlee and Xie, Yusi and Sieradzki, Karl},
  journal={Journal of The Electrochemical Society},
  volume={165},
  number={11},
  pages={C830},
  year={2018},
  publisher={IOP Publishing}
}

@article{cowley1950approximate,
  title={An approximate theory of order in alloys},
  author={Cowley, JM},
  journal={Physical Review},
  volume={77},
  number={5},
  pages={669},
  year={1950},
  publisher={APS}
}

@article{goff2021quantifying,
  title={Quantifying multipoint ordering in alloys},
  author={Goff, James M and Li, Bryant Y and Sinnott, Susan B and Dabo, Ismaila},
  journal={Physical Review B},
  volume={104},
  number={5},
  pages={054109},
  year={2021},
  publisher={APS}
}

@article{van2002self,
  title={Self-driven lattice-model Monte Carlo simulations of alloy thermodynamic properties and phase diagrams},
  author={Van De Walle, Axel and Asta, Mark},
  journal={Modelling and Simulation in Materials Science and Engineering},
  volume={10},
  number={5},
  pages={521},
  year={2002},
  publisher={IOP Publishing}
}

@article{jin2021nonconvex,
  title={On nonconvex optimization for machine learning: Gradients, stochasticity, and saddle points},
  author={Jin, Chi and Netrapalli, Praneeth and Ge, Rong and Kakade, Sham M and Jordan, Michael I},
  journal={Journal of the ACM (JACM)},
  volume={68},
  number={2},
  pages={1--29},
  year={2021},
  publisher={ACM New York, NY, USA}
}

@article{dauphin2014identifying,
  title={Identifying and attacking the saddle point problem in high-dimensional non-convex optimization},
  author={Dauphin, Yann N and Pascanu, Razvan and Gulcehre, Caglar and Cho, Kyunghyun and Ganguli, Surya and Bengio, Yoshua},
  journal={Advances in neural information processing systems},
  volume={27},
  year={2014}
}

@article{wang2020generalizing,
  title={Generalizing from a few examples: A survey on few-shot learning},
  author={Wang, Yaqing and Yao, Quanming and Kwok, James T and Ni, Lionel M},
  journal={ACM computing surveys (csur)},
  volume={53},
  number={3},
  pages={1--34},
  year={2020},
  publisher={ACM New York, NY, USA}
}

@article{ntukogu1986effect,
  title={Effect of palladium on the tarnishing of Cu-Ag-Au alloys},
  author={Ntukogu, TO and Cadoff, IB},
  journal={Journal of the Less Common Metals},
  volume={125},
  pages={197--205},
  year={1986},
  publisher={Elsevier}
}

@article{kresse1996efficient,
  title={Efficient iterative schemes for ab initio total-energy calculations using a plane-wave basis set},
  author={Kresse, Georg and Furthm{\"u}ller, J{\"u}rgen},
  journal={Physical review B},
  volume={54},
  number={16},
  pages={11169},
  year={1996},
  publisher={APS}
}

@article{perdew1996generalized,
  title={Generalized gradient approximation made simple},
  author={Perdew, John P and Burke, Kieron and Ernzerhof, Matthias},
  journal={Physical review letters},
  volume={77},
  number={18},
  pages={3865},
  year={1996},
  publisher={APS}
}

@article{kresse1999ultrasoft,
  title={From ultrasoft pseudopotentials to the projector augmented-wave method},
  author={Kresse, Georg and Joubert, Daniel},
  journal={Physical review b},
  volume={59},
  number={3},
  pages={1758},
  year={1999},
  publisher={APS}
}

@article{ferreira1988chemical,
  title={Chemical and elastic effects on isostructural phase diagrams: The \ensuremath{\varepsilon}-G approach},
  author={Ferreira, LG and Mbaye, AA and Zunger, Alex},
  journal={Physical Review B},
  volume={37},
  number={18},
  pages={10547},
  year={1988},
  publisher={APS}
}

@article{okamoto1987cu,
  title={The Au- Cu (gold-copper) system},
  author={Okamoto, H and Chakrabarti, DJ and Laughlin, DE and Massalski, TB},
  journal={Journal of Phase Equilibria},
  volume={8},
  pages={454--474},
  year={1987},
  publisher={Springer}
}

@article{zhang2014nonlocal,
  title={Nonlocal first-principles calculations in Cu-Au and other intermetallic alloys},
  author={Zhang, Yongsheng and Kresse, Georg and Wolverton, C},
  journal={Physical review letters},
  volume={112},
  number={7},
  pages={075502},
  year={2014},
  publisher={APS}
}

@article{heyd2003hybrid,
  title={Hybrid functionals based on a screened Coulomb potential},
  author={Heyd, Jochen and Scuseria, Gustavo E and Ernzerhof, Matthias},
  journal={The Journal of chemical physics},
  volume={118},
  number={18},
  pages={8207--8215},
  year={2003},
  publisher={American Institute of Physics}
}

@article{orr1960heats,
  title={Heats of formation of solid Au Cu alloys},
  author={Orr, RL},
  journal={Acta Metallurgica},
  volume={8},
  number={7},
  pages={489--493},
  year={1960},
  publisher={Elsevier}
}

@article{ozolicnvs1998first,
  title={First-principles theory of vibrational effects on the phase stability of Cu-Au compounds and alloys},
  author={Ozoli{\c{n}}{\v{s}}, V and Wolverton, C and Zunger, Alex},
  journal={Physical Review B},
  volume={58},
  number={10},
  pages={R5897},
  year={1998},
  publisher={APS}
}

@article{oriani1954thermodynamics,
  title={Thermodynamics of ordering alloys, II. The gold-copper system},
  author={Oriani, RA},
  journal={Acta Metallurgica},
  volume={2},
  number={4},
  pages={608--615},
  year={1954},
  publisher={Elsevier}
}

@article{hirabayashi1959electrical,
  title={Electrical resistivity and superstructure of CuAu3},
  author={Hirabayashi, Makoto},
  journal={Journal of the Physical Society of Japan},
  volume={14},
  number={3},
  pages={262--273},
  year={1959},
  publisher={The Physical Society of Japan}
}

@article{borelius1934theory,
  title={Theory of transitions of metallic mixed phases},
  author={Borelius, G},
  journal={Ann. Phys. Berlin},
  volume={20},
  pages={57--74},
  year={1934}
}

@article{claeson1984order,
  title={Order-disorder transformation in Au-Cu alloys studied by extended x-ray-absorption fine structure},
  author={Claeson, T and Boyce, JB},
  journal={Physical Review B},
  volume={29},
  number={4},
  pages={1551},
  year={1984},
  publisher={APS}
}

@article{george2020high,
  title={High entropy alloys: A focused review of mechanical properties and deformation mechanisms},
  author={George, Easo P and Curtin, WA and Tasan, Cemal Cem},
  journal={Acta Materialia},
  volume={188},
  pages={435--474},
  year={2020},
  publisher={Elsevier}
}

@article{scully2020controlling,
  title={Controlling the corrosion resistance of multi-principal element alloys},
  author={Scully, John R and Inman, Samuel B and Gerard, Angela Y and Taylor, Christopher D and Windl, Wolfgang and Schreiber, Daniel K and Lu, Pin and Saal, James E and Frankel, Gerald S},
  journal={Scripta Materialia},
  volume={188},
  pages={96--101},
  year={2020},
  publisher={Elsevier}
}

@article{baker1992alloy,
  title={Alloy phase diagrams, ASM handbook},
  author={Baker, Hugh and Okamoto, Hiroaki},
  journal={ASM International (The Materials Information Society), USA},
  year={1992}
}

@article{el2019outstanding,
  title={Outstanding radiation resistance of tungsten-based high-entropy alloys},
  author={El-Atwani, Osman and Li, Nan and Li, Meimei and Devaraj, Arun and Baldwin, JKS and Schneider, Matthew M and Sobieraj, Damian and Wr{\'o}bel, Jan S and Nguyen-Manh, Duc and Maloy, Stuart A and others},
  journal={Science advances},
  volume={5},
  number={3},
  pages={eaav2002},
  year={2019},
  publisher={American Association for the Advancement of Science}
}

@article{oates2007configurational,
  title={Configurational entropies of mixing in solid alloys},
  author={Oates, WA},
  journal={Journal of phase equilibria and diffusion},
  volume={28},
  pages={79--89},
  year={2007},
  publisher={Springer}
}

@book{baxter2016exactly,
  title={Exactly solved models in statistical mechanics},
  author={Baxter, Rodney J},
  year={2016},
  publisher={Elsevier}
}

@article{colinet2001applications,
  title={Applications of the cluster variation method to empirical phase diagram calculations},
  author={Colinet, C},
  journal={Calphad},
  volume={25},
  number={4},
  pages={607--623},
  year={2001},
  publisher={Elsevier}
}

@article{mohri1985binary,
  title={Binary ordering prototype phase diagrams in the cluster variation approximation},
  author={Mohri, T and Sanchez, JM and De Fontaine, D},
  journal={Acta Metallurgica},
  volume={33},
  pages={1171--1185},
  year={1985}
}

@book{landau2013statistical,
  title={Statistical Physics: Volume 5},
  author={Landau, Lev Davidovich and Lifshitz, Evgenii Mikhailovich},
  volume={5},
  year={2013},
  publisher={Elsevier}
}

@book{sethna2021statistical,
  title={Statistical mechanics: entropy, order parameters, and complexity},
  author={Sethna, James P},
  volume={14},
  year={2021},
  publisher={Oxford University Press, USA}
}

@article{ackermann1986ordering,
  title={On the ordering of face-centered-cubic alloys with nearest neighbour interactions},
  author={Ackermann, H and Crusius, S and Inden, G},
  journal={Acta Metallurgica},
  volume={34},
  number={12},
  pages={2311--2321},
  year={1986},
  publisher={Elsevier}
}

@article{inden2001atomic,
  title={Atomic ordering},
  author={Inden, Gerhard},
  journal={Phase transformations in materials},
  pages={519--581},
  year={2001},
  publisher={Wiley Online Library}
}

@article{sanchez1982comparison,
  title={Comparison of approximate methods for the study of antiferromagnetism in the fcc lattice},
  author={Sanchez, JM and De Fontaine, D and Teitler, W},
  journal={Physical Review B},
  volume={26},
  number={3},
  pages={1465},
  year={1982},
  publisher={APS}
}

@article{shockley1938theory,
  title={Theory of order for the copper gold alloy system},
  author={Shockley, W},
  journal={The Journal of Chemical Physics},
  volume={6},
  number={3},
  pages={130--144},
  year={1938},
  publisher={American Institute of Physics}
}

@article{ferreira1998evaluating,
  title={Evaluating and improving the cluster variation method entropy functional for Ising alloys},
  author={Ferreira, Luiz G and Wolverton, C and Zunger, Alex},
  journal={The Journal of chemical physics},
  volume={108},
  number={7},
  pages={2912--2918},
  year={1998},
  publisher={American Institute of Physics}
}

@article{andersson2002helander,
  title={Helander. T, Hsglund. L, et a1. TheFmo. Calc\&DICTRA, computational tools for materials science},
  author={Andersson, JO},
  journal={Calphad},
  volume={26},
  number={2},
  pages={273--312},
  year={2002}
}

@article{polgreen1984monte,
  title={Monte Carlo simulation of the fcc antiferromagnetic Ising model},
  author={Polgreen, Thomas L},
  journal={Physical Review B},
  volume={29},
  number={3},
  pages={1468},
  year={1984},
  publisher={APS}
}

@incollection{soffa2014diffusional,
  title={Diffusional phase transformations in the solid state},
  author={Soffa, WA and Laughlin, David E},
  booktitle={Physical metallurgy},
  pages={851--1020},
  year={2014},
  publisher={Elsevier}
}

@article{wu2016cluster,
  title={Cluster expansion method and its application in computational materials science},
  author={Wu, Qu and He, Bing and Song, Tao and Gao, Jian and Shi, Siqi},
  journal={Computational Materials Science},
  volume={125},
  pages={243--254},
  year={2016},
  publisher={Elsevier}
}

@article{asta1991effective,
  title={Effective cluster interactions from cluster-variation formalism. I},
  author={Asta, M and Wolverton, C and De Fontaine, D and Dreyss{\'e}, H},
  journal={Physical review B},
  volume={44},
  number={10},
  pages={4907},
  year={1991},
  publisher={APS}
}

@article{wolverton1991effective,
  title={Effective cluster interactions from cluster-variation formalism. II},
  author={Wolverton, C and Asta, M and Dreyss{\'e}, H and De Fontaine, D},
  journal={Physical Review B},
  volume={44},
  number={10},
  pages={4914},
  year={1991},
  publisher={APS}
}

@article{puchala2013thermodynamics,
  title={Thermodynamics of the Zr-O system from first-principles calculations},
  author={Puchala, B and Van der Ven, A},
  journal={Physical review B},
  volume={88},
  number={9},
  pages={094108},
  year={2013},
  publisher={APS}
}

@article{natarajan2016early,
  title={On the early stages of precipitation in dilute Mg--Nd alloys},
  author={Natarajan, Anirudh Raju and Solomon, Ellen LS and Puchala, Brian and Marquis, Emmanuelle A and Van der Ven, Anton},
  journal={Acta Materialia},
  volume={108},
  pages={367--379},
  year={2016},
  publisher={Elsevier}
}

@article{wu2016lithium,
  title={Lithium--Boron (Li--B) monolayers: first-principles cluster expansion and possible two-dimensional superconductivity},
  author={Wu, Chao and Wang, Hua and Zhang, Jiajia and Gou, Gaoyang and Pan, Bicai and Li, Ju},
  journal={ACS Applied Materials \& Interfaces},
  volume={8},
  number={4},
  pages={2526--2532},
  year={2016},
  publisher={ACS Publications}
}

@article{geng2017first,
  title={First-principles study of the Cu-Pd phase diagram},
  author={Geng, Feiyang and Boes, Jacob R and Kitchin, John R},
  journal={Calphad},
  volume={56},
  pages={224--229},
  year={2017},
  publisher={Elsevier}
}

@article{wang2020first,
  title={First-principles investigation of the phase stability and early stages of precipitation in Mg-Sn alloys},
  author={Wang, Kang and Cheng, Du and Fu, Chu-Liang and Zhou, Bi-Cheng},
  journal={Physical Review Materials},
  volume={4},
  number={1},
  pages={013606},
  year={2020},
  publisher={APS}
}

@article{oates2007compound,
  title={Is it a compound or cluster energy formalism?},
  author={Oates, W Alan},
  journal={International journal of materials research},
  volume={98},
  number={9},
  pages={780--785},
  year={2007},
  publisher={De Gruyter}
}

@article{lass2006correlation,
  title={Correlation between CALPHAD data and the Cahn--Hilliard gradient energy coefficient $\kappa$ and exploration into its composition dependence},
  author={Lass, Eric A and Johnson, William C and Shiflet, Gary J},
  journal={Calphad},
  volume={30},
  number={1},
  pages={42--52},
  year={2006},
  publisher={Elsevier}
}

@article{natarajan2017symmetry,
  title={Symmetry-adapted order parameters and free energies for solids undergoing order-disorder phase transitions},
  author={Natarajan, Anirudh Raju and Thomas, John C and Puchala, Brian and Van der Ven, Anton},
  journal={Physical Review B},
  volume={96},
  number={13},
  pages={134204},
  year={2017},
  publisher={APS}
}

@article{tsao2017high,
  title={The high temperature tensile and creep behaviors of high entropy superalloy},
  author={Tsao, Te-Kang and Yeh, An-Chou and Kuo, Chen-Ming and Kakehi, Koji and Murakami, Hideyuki and Yeh, Jien-Wei and Jian, Sheng-Rui},
  journal={Scientific reports},
  volume={7},
  number={1},
  pages={12658},
  year={2017},
  publisher={Nature Publishing Group UK London}
}

@article{soffa1989decomposition,
  title={Decomposition and ordering processes involving thermodynamically first-order order→ disorder transformations},
  author={Soffa, WA and Laughlin, DE},
  journal={Acta Metallurgica},
  volume={37},
  number={11},
  pages={3019--3028},
  year={1989},
  publisher={Elsevier}
}

@article{wallace2021modeling,
  title={Modeling the high-temperature phase coexistence region of mixed transition metal oxides from ab initio calculations},
  author={Wallace, Suzanne K and van Roekeghem, Ambroise and Bochkarev, Anton S and Carrasco, Javier and Shapeev, Alexander and Mingo, Natalio},
  journal={Physical Review Research},
  volume={3},
  number={1},
  pages={013139},
  year={2021},
  publisher={APS}
}

@article{ma2015ab,
  title={Ab initio thermodynamics of the CoCrFeMnNi high entropy alloy: Importance of entropy contributions beyond the configurational one},
  author={Ma, Duancheng and Grabowski, Blazej and K{\"o}rmann, Fritz and Neugebauer, J{\"o}rg and Raabe, Dierk},
  journal={Acta Materialia},
  volume={100},
  pages={90--97},
  year={2015},
  publisher={Elsevier}
}

@article{van2002effect,
  title={The effect of lattice vibrations on substitutional alloy thermodynamics},
  author={Van De Walle, Axel and Ceder, Gerbrand},
  journal={Reviews of Modern Physics},
  volume={74},
  number={1},
  pages={11},
  year={2002},
  publisher={APS}
}

@article{ozolicnvs2001large,
  title={Large vibrational effects upon calculated phase boundaries in Al-Sc},
  author={Ozoli{\c{n}}{\v{s}}, V and Asta, M},
  journal={Physical review letters},
  volume={86},
  number={3},
  pages={448},
  year={2001},
  publisher={APS}
}

@article{burton2006first,
  title={First-principles phase diagram calculations for the system NaCl--KCl: the role of excess vibrational entropy},
  author={Burton, Benjamin P and Van de Walle, A},
  journal={Chemical Geology},
  volume={225},
  number={3-4},
  pages={222--229},
  year={2006},
  publisher={Elsevier}
}

@article{hua2018first,
  title={First-principles study of vibrational entropy effects on the PbTe-SrTe phase diagram},
  author={Hua, Xia and Hao, Shiqiang and Wolverton, Chris},
  journal={Physical Review Materials},
  volume={2},
  number={9},
  pages={095402},
  year={2018},
  publisher={APS}
}

@article{shulumba2016lattice,
  title={Lattice vibrations change the solid solubility of an alloy at high temperatures},
  author={Shulumba, Nina and Hellman, Olle and Raza, Zamaan and Alling, Bj{\"o}rn and Barrirero, Jenifer and M{\"u}cklich, Frank and Abrikosov, Igor A and Od{\'e}n, Magnus},
  journal={Physical review letters},
  volume={117},
  number={20},
  pages={205502},
  year={2016},
  publisher={APS}
}

@article{wolverton2001entropically,
  title={Entropically favored ordering: The metallurgy of Al 2 Cu revisited},
  author={Wolverton, Christopher and Ozoli{\c{n}}{\v{s}}, V},
  journal={Physical review letters},
  volume={86},
  number={24},
  pages={5518},
  year={2001},
  publisher={APS}
}

@article{ferreira1987effect,
  title={Effect of chemical and elastic interactions on the phase diagrams of isostructural solids},
  author={Ferreira, LG and Mbaye, AA and Zunger, Alex},
  journal={Physical Review B},
  volume={35},
  number={12},
  pages={6475},
  year={1987},
  publisher={APS}
}

@article{garbulsky1994effect,
  title={Effect of lattice vibrations on the ordering tendencies in substitutional binary alloys},
  author={Garbulsky, GD and Ceder, G},
  journal={Physical Review B},
  volume={49},
  number={9},
  pages={6327},
  year={1994},
  publisher={APS}
}

@book{ceder1991alloy,
  title={Alloy theory and its applications to long period superstructure ordering in metallic alloys and high-temperature superconductors},
  author={Ceder, Gerbrand},
  year={1991},
  publisher={University of California, Berkeley}
}

@article{ceder1993derivation,
  title={A derivation of the Ising model for the computation of phase diagrams},
  author={Ceder, G},
  journal={Computational Materials Science},
  volume={1},
  number={2},
  pages={144--150},
  year={1993},
  publisher={Elsevier}
}

@article{anthony1994magnitude,
  title={Magnitude and origin of the difference in vibrational entropy between ordered and disordered Fe 3 Al},
  author={Anthony, L and Nagel, LJ and Okamoto, JK and Fultz, B},
  journal={Physical review letters},
  volume={73},
  number={22},
  pages={3034},
  year={1994},
  publisher={APS}
}

@article{fultz1995phonon,
  title={Phonon densities of states and vibrational entropies of ordered and disordered Ni 3 Al},
  author={Fultz, B and Anthony, L and Nagel, LJ and Nicklow, RM and Spooner, S},
  journal={Physical Review B},
  volume={52},
  number={5},
  pages={3315},
  year={1995},
  publisher={APS}
}

@article{althoff1997vibrational,
  title={Vibrational spectra in ordered and disordered Ni 3 Al},
  author={Althoff, Jeffrey D and Morgan, Dane and de Fontaine, Didier and Asta, Mark and Foiles, SM and Johnson, DD},
  journal={Physical Review B},
  volume={56},
  number={10},
  pages={R5705},
  year={1997},
  publisher={APS}
}

@article{garbulsky1996contribution,
  title={Contribution of the vibrational free energy to phase stability in substitutional alloys: Methods and trends},
  author={Garbulsky, GD and Ceder, G},
  journal={Physical Review B},
  volume={53},
  number={14},
  pages={8993},
  year={1996},
  publisher={APS}
}

@article{van1998first,
  title={First-principles computation of the vibrational entropy of ordered and disordered Ni 3 Al},
  author={Van de Walle, A and Ceder, G and Waghmare, UV},
  journal={Physical review letters},
  volume={80},
  number={22},
  pages={4911},
  year={1998},
  publisher={APS}
}

@article{van2000first,
  title={First-principles computation of the vibrational entropy of ordered and disordered Pd 3 V},
  author={Van de Walle, A and Ceder, G},
  journal={Physical Review B},
  volume={61},
  number={9},
  pages={5972},
  year={2000},
  publisher={APS}
}

@article{morgan2000vibrational,
  title={Vibrational thermodynamics: coupling of chemical order and size effects},
  author={Morgan, Dane and van de Walle, Axel and Ceder, Gerbrand and Althoff, Jeffrey D and de Fontaine, Didier},
  journal={Modelling and Simulation in Materials Science and Engineering},
  volume={8},
  number={3},
  pages={295},
  year={2000},
  publisher={IOP Publishing}
}

@article{mu2018electronic,
  title={Electronic transport and phonon properties of maximally disordered alloys: From binaries to high-entropy alloys},
  author={Mu, Sai and Pei, Zongrui and Liu, Xianglin and Stocks, George M},
  journal={Journal of Materials Research},
  volume={33},
  number={19},
  pages={2857--2880},
  year={2018},
  publisher={Cambridge University Press}
}

@article{wang2023generalization,
  title={Generalization of the mixed-space cluster expansion method for arbitrary lattices},
  author={Wang, Kang and Cheng, Du and Zhou, Bi-Cheng},
  journal={npj Computational Materials},
  volume={9},
  number={1},
  pages={75},
  year={2023},
  publisher={Nature Publishing Group UK London}
}

@article{morgan1998local,
  title={Local environment effects in the vibrational properties of disordered alloys: an embedded-atom method study of Ni 3 Al and Cu 3 Au},
  author={Morgan, D and Althoff, JD and De Fontaine, D},
  journal={Journal of phase equilibria},
  volume={19},
  pages={559--567},
  year={1998},
  publisher={Springer}
}

@article{delaire2004negative,
  title={Negative entropy of mixing for vanadium-platinum solutions},
  author={Delaire, O and Swan--Wood, T and Fultz, B},
  journal={Physical review letters},
  volume={93},
  number={18},
  pages={185704},
  year={2004},
  publisher={APS}
}

@article{kawaguchi2016deep,
  title={Deep learning without poor local minima},
  author={Kawaguchi, Kenji},
  journal={Advances in neural information processing systems},
  volume={29},
  year={2016}
}

@book{sun2016nonconvex,
  title={When are nonconvex optimization problems not scary?},
  author={Sun, Ju},
  year={2016},
  publisher={Columbia University}
}

@article{desgranges2019can,
  title={Can ordered precursors promote the nucleation of solid solutions?},
  author={Desgranges, Caroline and Delhommelle, Jerome},
  journal={Physical Review Letters},
  volume={123},
  number={19},
  pages={195701},
  year={2019},
  publisher={APS}
}

@article{clouet2004nucleation,
  title={Nucleation of Al 3 Zr and Al 3 Sc in aluminum alloys: From kinetic Monte Carlo simulations to classical theory},
  author={Clouet, Emmanuel and Nastar, Maylise and Sigli, Christophe},
  journal={Physical Review B},
  volume={69},
  number={6},
  pages={064109},
  year={2004},
  publisher={APS}
}

@article{clouet2005precipitation,
  title={Precipitation kinetics of Al3Zr and Al3Sc in aluminum alloys modeled with cluster dynamics},
  author={Clouet, Emmanuel and Barbu, Alain and La{\'e}, Ludovic and Martin, Georges},
  journal={Acta Materialia},
  volume={53},
  number={8},
  pages={2313--2325},
  year={2005},
  publisher={Elsevier}
}

@article{clouet2007classical,
  title={Classical nucleation theory in ordering alloys precipitating with L 1 2 structure},
  author={Clouet, Emmanuel and Nastar, Maylise},
  journal={Physical Review B},
  volume={75},
  number={13},
  pages={132102},
  year={2007},
  publisher={APS}
}

@article{rappaz2020solidification,
  title={Solidification of metallic alloys: Does the structure of the liquid matter?},
  author={Rappaz, Michel and Jarry, Ph and Kurtuldu, G{\"u}ven and Zollinger, Julien},
  journal={Metallurgical and Materials Transactions A},
  volume={51},
  pages={2651--2664},
  year={2020},
  publisher={Springer}
}

@article{sanchez1978fee,
  title={The fee Ising model in the cluster variation approximation},
  author={Sanchez, JM and De Fontaine, D},
  journal={Physical Review B},
  volume={17},
  number={7},
  pages={2926},
  year={1978},
  publisher={APS}
}

@article{mohri1985short,
  title={Short range order diffuse intensity calculations in the cluster variation method},
  author={Mohri, T and Sanchez, JM and De Fontaine, D},
  journal={Acta Metallurgica},
  volume={33},
  number={8},
  pages={1463--1474},
  year={1985},
  publisher={Elsevier}
}

@misc{otis2023,title={Privative communication at CALPHAD 2023},AUTHOR={Otis,Richard}
}

@inproceedings{franceschi2017forward,
  title={Forward and reverse gradient-based hyperparameter optimization},
  author={Franceschi, Luca and Donini, Michele and Frasconi, Paolo and Pontil, Massimiliano},
  booktitle={International Conference on Machine Learning},
  pages={1165--1173},
  year={2017},
  organization={PMLR}
}

@article{beirami2017optimal,
  title={On optimal generalizability in parametric learning},
  author={Beirami, Ahmad and Razaviyayn, Meisam and Shahrampour, Shahin and Tarokh, Vahid},
  journal={Advances in Neural Information Processing Systems},
  volume={30},
  year={2017}
}

@inproceedings{pedregosa2016hyperparameter,
  title={Hyperparameter optimization with approximate gradient},
  author={Pedregosa, Fabian},
  booktitle={International conference on machine learning},
  pages={737--746},
  year={2016},
  organization={PMLR}
}

@article{lemieux1999investigating,
  title={Investigating non-Gaussian scattering processes by using nth-order intensity correlation functions},
  author={Lemieux, P-A and Durian, DJ},
  journal={JOSA A},
  volume={16},
  number={7},
  pages={1651--1664},
  year={1999},
  publisher={Optica Publishing Group}
}

@article{pedrini2013two,
  title={Two-dimensional structure from random multiparticle X-ray scattering images using cross-correlations},
  author={Pedrini, Bill and Menzel, Andreas and Guizar-Sicairos, Manuel and Guzenko, Vitaliy A and Gorelick, Sergey and David, Christian and Patterson, Bruce D and Abela, Rafael},
  journal={Nature communications},
  volume={4},
  number={1},
  pages={1647},
  year={2013},
  publisher={Nature Publishing Group UK London}
}

@article{lehmkuhler2014detecting,
  title={Detecting orientational order in model systems by X-ray cross-correlation methods},
  author={Lehmk{\"u}hler, Felix and Gr{\"u}bel, Gerhard and Gutt, Christian},
  journal={Journal of Applied Crystallography},
  volume={47},
  number={4},
  pages={1315--1323},
  year={2014},
  publisher={International Union of Crystallography}
}

	\appendix

\begin{appendices}

\chapter{The details of Cluster Ansatz Method}
		\label{app:appendixA}

			\section{Motivation} \label{app:appendixA_1}
In the main content, we have mentioned that the FYL-transform is a special ansatz. With this ansatz, the disordered phase under the FYL-CVM formalism actually doesn't involve any real variational space to be minimized. It is a one-to-one mapping between basic cluster distribution and the composition(point cluster probability) due to the disordered phase's higher symmetry. As a result, we could use a direct non-linear solver to deal with the disordered phase to calculate the free energy at $x-T$ space, but not necessary to perform the optimization. This makes the computation simpler. It also motivates me: how about the ordered phase? Could we also make some further ansatz to simplify the ordered phase to make it fully ansatz determined without the variational calculation?

Actually, we could do this by enforcing some site chemical potential values onto the non-related sites. Still, only one site type is left and directly connected to the composition. With this idea, there would be no necessity to perform any minimization but to calculate the relation between the site's chemical potential and composition, which requires a non-linear equation solver. We name it as cluster ansatz method.
				
			\section{Detailed Explanation} \label{app:appendixA_2}
Here we clarify more details. To implement this method, we should use the traditional CALPHAD method to deal with the phases. We need to assign the different phases with different thermodynamic functions. In FYL-CVM, it would be easy to achieve: we assign different symmetry of site chemical potential to determine the different phases' function.

Let's mainly concentrate on the ordered phase. As we know, every element except the reference element would have one site chemical potential variable on one site in FYL-CVM formalism. However, as we here are only considering the ordered phase, the occupation of the specific site should be directly assigned ahead. As a result, we could take all the related parameters of these elements into two sets for one kind of element. One set is still the variational variables determined by the composition, and another set of variables would be limited to several specific values depending on the design.
 It means we should already have A on sites 1, 2, 3, and B on site 4 if we consider the binary $L1_2$ phase as an example. Considering this, the primary variable is the site chemical potential on sites 1, 2, and 3 for A. The site chemical potential on 4 is not related to this ordered phase, as this site should be occupied by element B. Then we need to enforce the site chemical potential of A on site 4 to some value to reflect this point. It could be -10, for example. In real cases, we would assign -10, 0, and 10 as the candidate values to reflect this site's state: entirely unoccupied, intermediate state, or fully occupied. The specific value is adjustable. The different states would be related to the material composition and could be assigned by hand smartly. But this three-energy level model simplified the case for the ordered phase. The three site chemical potentials on sites 1,2,3 of element A would be the same as each other due to the symmetry and would be entirely determined by the composition of element A. Then we need to solve a non-linear equation to determine the free energy without any non-convex optimization consideration.
 
For this newly proposed method, first of all, it would be much fast. It only required to address a very small number of nonlinear equations; the number is about the number of components to deal with all the phases, not just the disordered phase. Second, there would be no variational calculation, so the non-convexity and the bi-level optimization are unnecessary. Third, all the analysis can be directly performed in the $x-T$ space, which would be compatible with the previous CALPHAD thermodynamic model. Fourth, we will see in the next section that the calculated phase transition point results for the prototype system can be better than CVM due to error cancellation between energy and entropy. However, the disadvantage side also has several concerns: first, it doesn't involve any clear physics for the further imposed Ansatz. Second, the ordered phase in our test would quickly become unstable in the low-temperature case. As a result, we decided not to present this proposed cluster Ansatz method but leave this in the appendix to be considered for future potential usage. However, we have to notice that even though this new method represents the not fully minimized case, it still presents a clear enough phase boundary result indicating the intrinsic physics within the FYL-CVM free energy functional.

\section{Some Results} 
Here we present some results calculated with this cluster Ansatz method. The different colors represent the different phase boundary. Red means the phase boundary with a disordered phase, green indicates the phase boundary with the $L1_2$ phase, and blue means the phase boundary with the $L1_0$ phase.

The first one is the benchmark test on the prototype AB system. The second one is the Cu-Au system with only configurational contribution. The third one is the Cu-Au system with total contributions, including configurational, elastic, and vibrational contributions. The scattered point is the experimental point. The vibrational contribution is briefly adjusted based on the experimental data. All these results reveal the great success of this method, as the calculated phase diagram has great topology and the transition temperature on the phase boundary. This method represents the non-fully minimized results in the original FYL-CVM. However, the calculated phase diagram is already good enough for use. Besides, the computational efficiency is further improved with this cluster ansatz method, as we reduced the minimization calculation to solving only one equation for a binary case. However, the enhanced calculated accuracy should be caused by the error cancellation, while the ordered phase would become unstable very quickly in the low-temperature case. These calculated results reveal both this method's advantages and disadvantages.
  \begin{figure}
    \centering
\includegraphics[width=0.8\textwidth]{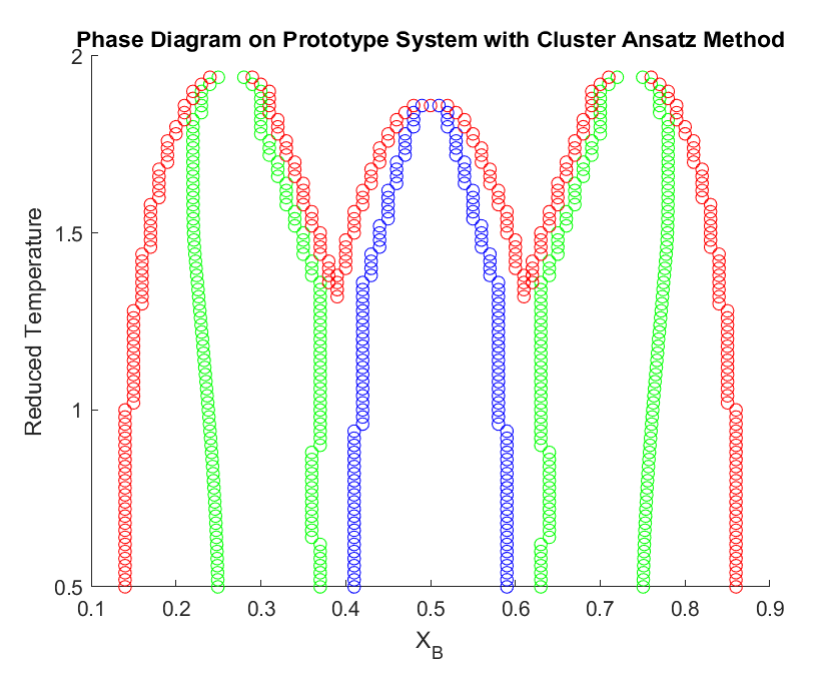}
    \caption{Cluster Ansatz Method on prototype AB system}
    \label{fig:CAM1}
\end{figure}

  \begin{figure}
    \centering
\includegraphics[width=0.8\textwidth]{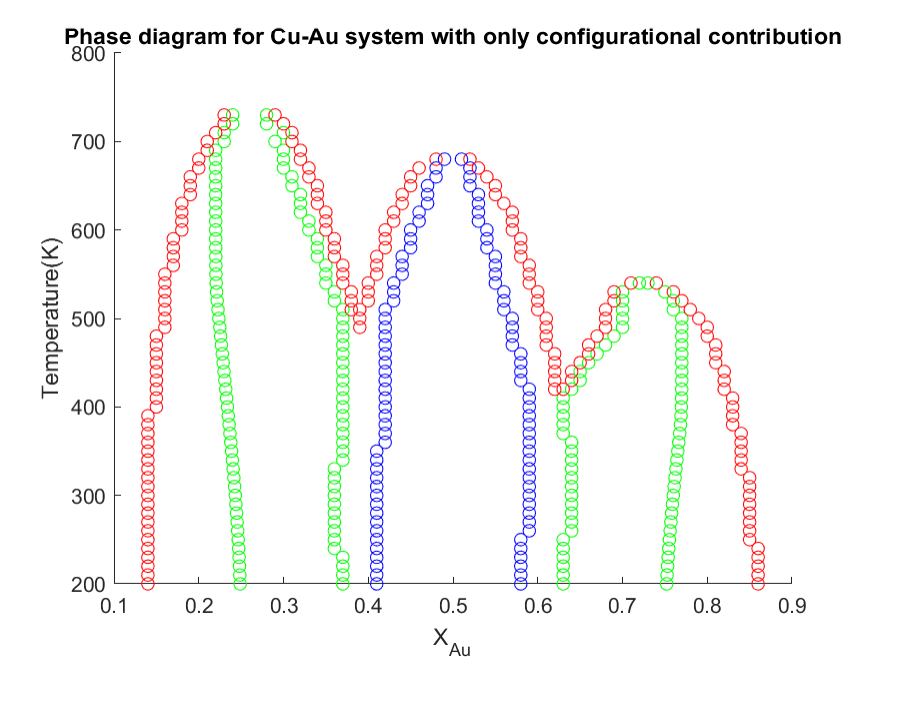}
    \caption{Cluster Ansatz Method on Cu-Au system with only configurational contribution}
    \label{fig:CAM2}
\end{figure}

  \begin{figure}
    \centering
\includegraphics[width=0.8\textwidth]{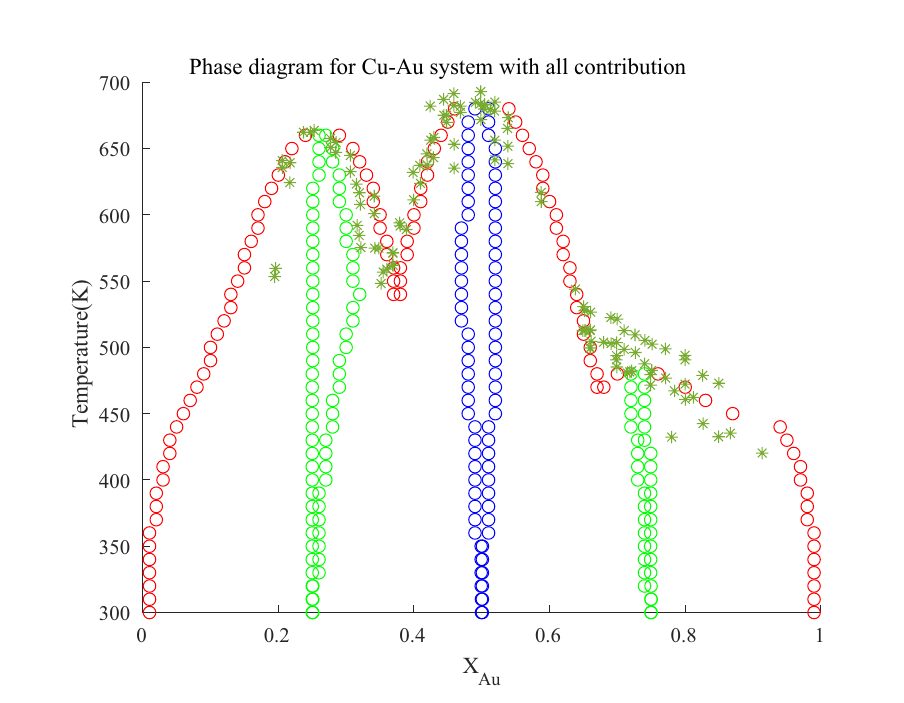}
    \caption{Cluster Ansatz Method on Cu-Au system with full contribution}
    \label{fig:CAM3}
\end{figure}

\chapter{FYL-transform and correlation functions}
Here we attempt to relate the site variables from FYL transform with correlation functions. Let’s limit our discussion to the tetrahedron approximation for FCC binary system. For FYL-CVM, we have
\begin{equation}
\begin{split}
z_{\alpha}  =\sum_{ijkl} \lambda^{(1)}\lambda^{(2)}\lambda^{(3)}\lambda^{(4)}exp\left[\frac{-\epsilon_{ijkl}}{k_B T}\right]
\end{split}
    \label{eq:AB-1}
\end{equation}
\begin{equation}
\begin{split}
\rho_{ijkl} =\frac{1}{z_{\alpha}} \lambda^{(1)}\lambda^{(2)}\lambda^{(3)}\lambda^{(4)}exp\left[\frac{-\epsilon_{ijkl}}{k_B T}\right]
\end{split}
    \label{eq:AB-2}
\end{equation}

For the correlation function formulated in the cluster expansion method \cite{wu2016cluster}, they can be connected to the cluster probability as well. What we want is to see the relation between $\lambda$ and $\xi$.
\begin{equation}
\begin{split}
\rho_{ijkl} = \frac{1}{2^4}[1+(i+j+k+l)\xi_1 +(ij+ik+il+kj+kl+jl)\xi_2 \\ + (ijk+ijl+ikl+jkl)\xi_3 +ijkl \xi_4 ]
\end{split}
    \label{eq:AB-3}
\end{equation}

However, this is for the disordered phase. If it’s the ordered phase, we have

\begin{equation}
\begin{split}
\rho_{ijkl} =\frac{1}{2^4}[1+(i\xi_1^i+j\xi_1^j+k\xi_1^k+l\xi_1^l) +(ij\xi_2^{ij}+ik\xi_1^{ik}+il\xi_1^{il}+kj\xi_1^{kj}+kl\xi_1^{kl}+jl\xi_1^{jl}) \\ +(ijk\xi_3^{ijk}+ijl\xi_3^{ijl}+ikl\xi_3^{ikl}+jkl\xi_3^{jkl}) +ijkl \xi_4 ]
\end{split}
    \label{eq:AB-4}
\end{equation}
Then there would be 16 equations to determine $\lambda$ from $\xi$. Assume we have coefficients matrix $A$ for $\xi$ to $\lambda$, then we have:
\begin{equation}
\begin{split}
A\Vec{\xi} + \frac{1}{16} = \vec{b}
\end{split}
    \label{eq:AB-5}
\end{equation}

where $\vec{b} = (\rho_{ijkl})_{ijkl} $ and $\vec{\xi}$ is organized as the vector as well. Solve these linear equations with 15 variables, then we could get the $\xi(\lambda)$.

\end{appendices}

\end{document}